\documentclass{article}
\usepackage[utf8]{inputenc}
\usepackage{url,xcolor}
\usepackage{graphicx}
\usepackage{geometry}
\usepackage{svg}
\usepackage{pifont}
\usepackage{mathtools}
\usepackage{thmtools}
\usepackage{thm-restate}
\usepackage{hyperref} 
\usepackage{cleveref}
\usepackage{tikz}
\usepackage{multirow}
\usepackage{tabularx}
\usepackage{amssymb}
\usepackage{amsmath}
\usepackage{bm}  
\usepackage{authblk}
\usepackage{algorithm}
\usepackage{algpseudocode}
\usepackage{tabularray}
\usepackage{xstring}
\usepackage{enumerate}
\usetikzlibrary {shapes.symbols}
\usepackage{wasysym}

\usepackage{enumitem}
\newlist{todolist}{itemize}{2}
\setlist[todolist]{label=$\square$}
\usepackage{pifont}

\algnewcommand{\algorithmicand}{\textbf{ and }}
\algnewcommand{\algorithmicor}{\textbf{ or }}
\algnewcommand{\OR}{\algorithmicor}
\algnewcommand{\AND}{\algorithmicand}
\usepackage{subcaption} 
\usepackage[normalem]{ulem}
\usepackage{soul}
\usetikzlibrary{decorations.pathreplacing}

\usepackage[bb=dsserif]{mathalpha}
\usepackage{amssymb}

\usetikzlibrary{shapes.geometric}
\usetikzlibrary{dsp}
\usetikzlibrary{chains}

\declaretheorem[name=Corollary]{cor}
\declaretheorem[name=Definition]{definition}
\declaretheorem[name=Example]{example}
\declaretheorem[name=Remark]{remark}

\newcommand{\mt}{\mathtt}

\usetikzlibrary{arrows,chains,matrix,positioning,scopes,patterns,decorations.pathreplacing, calc, positioning,fit,decorations.pathreplacing,calligraphy}

\usepackage[T1]{fontenc}
\usepackage{xparse}
\usepackage{enumitem}
\setlist[description]{
  font={\sffamily\bfseries},
  labelsep=0pt,
  leftmargin=\transcriptlen
}

\newlength{\transcriptlen}

\NewDocumentCommand {\setspeaker} { mo } {%
  \IfNoValueTF{#2}
  {\expandafter\newcommand\csname#1\endcsname{\item[#1:]}}%
  {\expandafter\newcommand\csname#1\endcsname{\item[#2:]}}%
  \IfNoValueTF{#2}
  {\settowidth{\transcriptlen}{#1}}%
  {\settowidth{\transcriptlen}{#2}}%
}

\setspeaker{Sender}[Sender]
\setspeaker{Receiver}[Receiver]
\setspeaker{Common}[Common]
\setspeaker{sinit}[Sender Init]
\setspeaker{rinit}[Receiver Init]

\newcommand{\dontcare}{2}
\renewcommand{\implies}{\Rightarrow}
\newcommand{\entails}{\vdash}

\newcommand{\sls}[1]{\mathcal{L}_{#1}}

\newcommand{\receiver}{ \mathtt{r} }
\newcommand{\receiverrnd}{ \mathtt{R} }
\newcommand{\sender}{ \mathtt{s} }
\newcommand{\senderhat}{ \mathtt{\hat{s}} }
\newcommand{\senderhatm}{ \mathtt{\hat{s}_m} }
\newcommand{\senderrnd}{ \mathtt{S}}
\newcommand{\senderrndhat}{ \mathtt{\hat{S}}}

\newcommand{\query}{ \mathtt{q} }
\newcommand{\queryrnd}{\mathtt{Q}}
\newcommand{\queryapo}{\mathtt{q^{\prime}}}
\newcommand{\queryapoapo}{\mathtt{q^{\prime \prime}}}
\newcommand{\queryrndapo}{\mathtt{Q^{\prime}}}

\newcommand{\Hbin}[1]{ H_{\text{bin}}\left(#1\right) }
\newcommand{\Hreg}[1]{H(#1)}

\newcommand{\kernelspace}{\mathcal{M}}
\newcommand{\surfacespace}{\mathcal{L}}

\newcommand{\qed}{{$\hfill\square$}}

\newcommand{\md}{\mu} 
\newcommand{\Mds}{M}  

\newtheorem{convention}{Convention}

\newcommand{\len}[1]{{\mathbf{len}(#1)}}
\newcommand{\algebraicset}[1]{{\kappa(#1)}}
\newcommand{\kv}[1]{\vec{\kappa}(#1)}
\newcommand{\kvv}{\vec{\kappa}}
\newcommand{\ellv}[1]{\vec{\ell}(#1)}
\newcommand{\Kpolyring}{K[x_1,\ldots,x_m]}
\newcommand{\GFpolyring}{GF(2)[x_1,\ldots,x_m]}

\newcommand{\surfaceand}[2]{#1 \mathbf{  \land  } #2}
\newcommand{\surfacenot}[1]{\mathbf{\lnot  } #1}

\newcommand{\forceindent}{\leavevmode{\parindent=1em\indent}}

\title{Towards a Unification of Logic and Information Theory}

\author[1]{Luis A.\ Lastras}
\author[1]{Barry Trager}
\author[1]{Jonathan Lenchner}
\author[2]{Wojciech Szpankowski}
\author[1]{Chai Wah Wu}
\author[1]{Mark S.\ Squillante}
\author[3]{Ron Fagin}
\author[4]{Alexander Gray}

\affil[1]{IBM T.J. Watson Research Center}
\affil[2]{Purdue University}
\affil[2]{IBM Almaden Research Center}
\affil[4,2]{Centaur AI Institute}

\definecolor{coolblue}{HTML}{3B75AF}

\newcommand{\thmattributes}[4]{
\begin{tabular}{cc} \IfEqCase{#1}{{y}{\CIRCLE} {n}{\Circle}} & $\receiver$ is available to Bob\\
 \IfEqCase{#2}{{y}{\CIRCLE} {n}{\Circle}} & $\receiver$ is available to Alice \\
 \IfEqCase{#3}{{y}{\CIRCLE} {n}{\Circle}} & $\query$  is available to Alice\\ 
 \IfEqCase{#4}{{y}{\CIRCLE} {n}{\Circle}} & $\sender$ entails $\receiver$ 
\end{tabular}
}

\newcommand{\tree}[4]{%
\scalebox{0.48}{
\begin{tikzpicture}[baseline=(current bounding box.center)]
\begin{scope}
\clip (0,0) rectangle (8,5);
\draw (0,0)[color=black,fill=white] rectangle (8,5);

\IfEqCase{#2}{
  {q}{ \draw(3.6,-0.1)[rotate=45,fill=white] ellipse(2.5 and 1.5);
       \draw(3.6,3.8) node {\scalebox{2.0}{$\algebraicset{\query}$}};
     }
  {no_q}{}
}
\IfEqCase{#3}{
   {shat_arrow}{ 
    \draw(2.5,2.2)[rotate around={45:(2.5,2.2)},draw=none,fill=coolblue!50!white] ellipse(1.9 and 1.35);
           \draw(2.8,3.1) node {\scalebox{2.0}{$\algebraicset{\senderhat}$}};
           \draw(3.8,-0.25)[dashed,rotate=45,color=orange,line width=0.5mm] ellipse(2.5 and 1.5);
           \draw(2,1.8)[fill=green!35!white,draw=none] circle(1.0);
           \draw [->,line width=0.75mm](4.6,4.2) -- (3.9,3.7);
         }
         [rotate around={90:(6,2.5)}]
    {shat}{ \draw(2.5,2.2)[rotate around={45:(2.5,2.2)},draw=none,fill=coolblue!50] ellipse(1.9 and 1.35);
           \draw(2.8,3.1) node {\scalebox{2.0}{$\algebraicset{\senderhat}$}};
           \draw(3.8,-0.25)[dashed,rotate=45] ellipse(2.5 and 1.5);
           \draw(2,1.8)[dashed,color=black] circle(1.0);
           \draw(2,1.8)[fill=green!35!white] circle(1.0);
         }
    {shat_normal}{ \draw(1.8,1.8)[fill=green!35!white] circle(1.0);  
       \draw(1.8,2.1) node {\scalebox{2.0}{$\algebraicset{\senderhat}$}};
     }
   {no_shat}{}
}
\IfEqCase{#4}{
  {s}{ \draw(1.8,1.8)[fill=green!35!white] circle(1.0);  
       \draw(1.8,2.1) node {\scalebox{2.0}{$\algebraicset{\sender}$}};
     }
  {no_s}{}
}
\IfEqCase{#1}{
  {r}{\draw(0.3,0.25) rectangle(5., 4.75);
      \draw(4.0,4.2) node {\scalebox{2.0}{$\algebraicset{\receiver}$}};
     }
  {r_partial}{\draw(2.5,0.25) rectangle(7.2, 4.75);
      \draw(6.5,4.0) node {\scalebox{2.0}{$\algebraicset{\receiver}$}};
     }
  {r_misinf}{\draw(3.,0.25) rectangle(7.7, 4.75);
      \draw(6.9,4.1) node {\scalebox{2.0}{$\algebraicset{\receiver}$}};
     }
  {no_r}{}
}
\end{scope}
\end{tikzpicture}}
}

\begin{document}

\maketitle

\begin{abstract}

Today, the vast majority of the world's digital information is represented using the fundamental assumption, introduced by Claude Shannon in 1948, that  ``...the semantic aspects of communication are irrelevant to the engineering problem (of the design of communication systems)...''. It is hard to overestimate the extraordinary positive impact of this assumption, which has allowed the design of flexible efficient and reliable communication systems that operate regardless of the intended meaning of our messages.

Consider, nonetheless, the observation that we, individuals, as well as our computing devices, often combine a received message with other information in order to deduce new facts (and hopefully make better decisions), thereby expanding the value of the originally received message. It is noteworthy that to-date, no rigorous theory of communication has been put forth which postulates the existence of deductive capabilities on the receiver's side.

The purpose of this paper is to present a proposal that combines information theory and logic at a fundamental level. We formally model such deductive capabilities using logic reasoning, and present a rigorous theory which covers the following generic scenario: Alice and Bob each have knowledge of some logic sentence, and they wish to communicate as efficiently as possible with the shared goal that, following their communication, Bob should be able to deduce a particular logic sentence that Alice knows to be true, but that Bob currently cannot prove. Many variants of this general setup are considered in this article; in all cases  we are able to provide  sharp upper and lower bounds phrased in terms of an entropy-like function that we call $\Lambda$, in reference to its apparent connection to problems of communication involving logic. Our contribution includes the identification of the most fundamental requirements that we place on a logic and associated logical language for all of our results to apply; an example is Propositional Logic over a finite number of propositions. Practical algorithms that are in some cases asymptotically optimal are provided, and we illustrate the potential practical value of the design of communication systems that incorporate the assumption of deductive capabilities at the receiver using experimental results that suggest significant possible gains compared to classical systems.
\end{abstract}

\section{Introduction and summary of contributions} \label{sec:intro}

It is well known that a significant contribution of Shannon \cite{shannon:BST48} was to provide a definition for information that is independent of the semantics of the message being conveyed. This abstraction is one of the most successful concepts in the computing and communication revolution, as it has allowed us to build flexible machines that process and communicate information in a standardized way, while the messages and intentions behind those messages remain removed from the operation of those machines. 
This aspect sometimes feels counter to one's intuition of what we think of as information, as articulated in 1949 by Warren Weaver, then Research Director of the Rockefeller Foundation, who breached the subject of semantics in information theory in an oft-cited commentary \cite{weaver:commmentary_math_comm}. To capture some intuitive feel for the difference between Shannon's classical  information and some desired notion of ``semantic'' information, we borrow from \cite{CALVANESESTRINATI2021107930}:  ``As a very simple example, pressing the keys of a computer keyboard at random generates a message that has a high syntactic information, because the generated symbols are approximately independent and uniformly distributed, so that their entropy (average information in Shannon’s sense) is maximum. However, most likely, the generated message carries zero semantic information, as it does not carry any meaningful content.''

Although Shannon's original theory is semantics-free, it was not because he had not been thinking about semantics. Six years prior to his 1948 paper, Shannon and fellow Bell Labs mathematician John Riordan considered the question of how compactly one could express a given Boolean function on $n$-bit inputs in Propositional Logic \cite{riordanlower}. We therefore know that Shannon was thinking about the transmission of semantic information even before he developed his theory of communication.
 
Shortly after Shannon's seminal 1948 paper, debate began about what might constitute a satisfactory theory that addressed the semantic content of communication. An elegant proposal for incorporating semantics in a theory of information was made by Carnap and Bar-Hillel \cite{carnap53}, which carries particular gravitas due to Carnap's status as one of the pioneers of the modern formal treatment of semantics via mathematical logic \cite{Carnap1942}, along with the likes of Tarski \cite{0ea4add0-94e6-3877-bd4e-0c67171ac249} and Kripke \cite{Kripke1963-KRISCO}.  This work systematically identified desirable properties for what semantic information should be, including how to measure it, and is the most seminal reference in all subsequent treatments of this problem. 

Another early example of interest on the subject of semantic information can be found in a little-known paper due to Shannon himself \cite{shannon:lattice}. To motivate his viewpoint, Shannon observed that a sentence may be encoded in multiple different ways, each recoverable from the other; for example, imagine a sentence and its translation to Morse code. Shannon then observed that ``For most purposes of communication, any of these forms is equally good and may be considered to contain the same information''. Shannon then concluded that  ``Thus we are led to define the actual information of a stochastic process as that which is common to all stochastic processes which may be obtained from the original by reversible encoding operations''. In both \cite{carnap53, shannon:lattice} we find the idea that if a sentence can be {\em deduced} from the  sentences one already knows, it conveys zero information, but the more general implications of this observation to communication systems were left unexplored by these authors \cite{ayg-spa}. Later on, Shannon introduced Rate-Distortion theory \cite{shannon:rd}, a general extension of information theory that allows for lossy compression. Although rate-distortion theory has had most of its practical influence on the compression of media such as images, video and audio, it is an extraordinarily general theory and correspondingly, it has found itself at the center at most of the efforts to extend classic information theory to semantics \cite{liu:rd_semantic,liu_shao_zhang_poor:indirect_rd_semantic, Shao2022ATO, guo2022semantic, stavrou:goal_semantic_rd, gunduz:beyond_bits,niu2024mathematicaltheorysemanticcommunication}. Additionally, the fundamental observation in \cite{shannon:lattice} that information may admit partial ordering has received renewed interest \cite{yu:information_lattice_learning,Yu2024SemanticCW}.

To illustrate  the main driving point of view in our article, we will rely on an unrelated, but insightful quote. At the beginning of his Lectures on Physics \cite{feynman_1965_flp}, Feynman posed a hypothetical situation where all scientific knowledge is destroyed, and a sentence that has the most information in the fewest words needs to be chosen. His choice was ``all things are made of atoms'', which assumes that scientists would be able, through experimentation, induction and deduction, to reconstruct vast amounts of our scientific knowledge from the one sentence. In the absence of a deductive process, effectively conveying all scientific knowledge would seem to require a large quantity of bits to be transmitted. Yet Feynman's sentence can be transmitted with just a few dozen bits. In this paper we aim to provide the theoretical foundations for understanding this phenomenon -- how something that can be so succinctly described can have such a profound consequence when paired with deductive reasoning.

Despite the success of classical information theory, pragmatic concerns have revived interest in the possibility that focusing on transmitting meaning accurately might offer savings over transmitting bits accurately \cite{ayg-spa}, \cite{ks08}.
 Motivated by the continued massive increase in the world's data and the need for next-generation network systems to somehow keep up,  influential vision papers such as  \cite{CALVANESESTRINATI2021107930} have 
set off a recent explosion of interest in the promise of ``semantic communication''.  However, surprisingly few have leveraged the deep original insights of Carnap/Bar-Hillel and arguably Shannon himself by invoking the power of {\em deduction}.  This is likely due to the deep cross-disciplinarity needed to do so, requiring both sufficient depth in information-theoretic tools and in the formalisms of logic.  The fact that mathematical logic serves as a foundation for much of computer science \cite{wiki:logicincs} and mathematics \cite{wiki:mathlogic} can serve as some testament to its depth of development and difficulty to penetrate for the casual non-expert.  Similarly, information theory is heavily developed mathematically and reliant on very different ideas, rooted in probability theory.  A fundamental challenge in writing this paper has been to somehow make it accessible to its multiple possible audiences.

\begin{figure}
\begin{center}
\begin{tabular}{c}
\begin{tikzpicture}

	\matrix[row sep=2.5mm, column sep=5mm](A)
	{	tm

		\node[dspnodeopen,dsp/label=left] (mm00) {message};  &
		\node[dspsquare]         (mm01) {f};     &
            \node[coordinate,label=bits] (mm02) {}; &
		\node[dspsquare]          (mm03) {g};     &
		\node[dspnodeopen,dsp/label=right]          (mm03p5) {recovered message};     &
  \\
  };

	\begin{scope}[start chain]
		\chainin (mm00);
		\chainin (mm01) [join=by dspconn];
		\chainin (mm03) [join=by dspconn];
		\chainin (mm03p5) [join=by dspconn];
        \end{scope}

        
\end{tikzpicture}\end{tabular}  \\ 
\begin{tabular}{c}
\begin{tikzpicture}

	\matrix[row sep=2.5mm, column sep=5mm](A)
	{	tm

		\node[dspnodeopen,dsp/label=left] (mm00) {logic sentence};  &
		\node[dspsquare]         (mm01) {f};     &
            \node[coordinate,label=bits] (mm02) {}; &
		\node[dspsquare]          (mm03) {g};     &
		\node[coordinate,label={[align=left]some recovered\\logic sentence}]          (mm03p5) {};     &
        \node[dspcloud,dsp/label=above] (mm04) {deductive\\machinery};    & 
		\node[dspnodeopen,dsp/label=right] (mm06) {deduced facts};  \\
  };

	\begin{scope}[start chain]
		\chainin (mm00);
		\chainin (mm01) [join=by dspconn];
		\chainin (mm03) [join=by dspconn];
		\chainin (mm04) [join=by dspconn];
		\chainin (mm06) [join=by dspconn];
        \end{scope}

\end{tikzpicture}\end{tabular} 
\end{center}
\caption{In the top, Shannon's original digital communication model. In the bottom, a sketch of our proposed extension of Shannon's model.}
\label{fig:extension_shannon_model}
\end{figure}
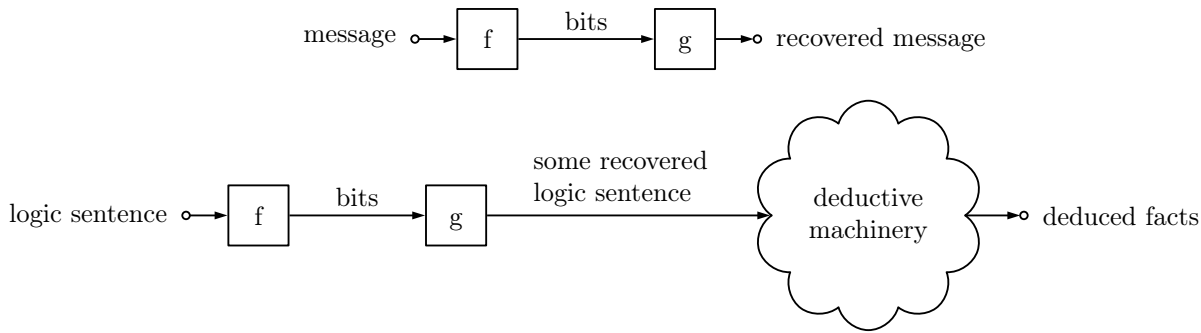

\subsection{Our contributions} 
If nothing else, we ask the reader to remember this article through Figure~\ref{fig:extension_shannon_model}, where at the top, we illustrate the famous digital communication model that Shannon introduced in \cite{shannon:BST48}, and in the bottom we sketch a version of this model where messages are replaced with logic sentences and where we assume the existence of a mechanism that allows the receiver to derive additional conclusions from whatever  has been received. The remainder of the paper can be seen as a proposal for rigorously using logic to mathematically model such deductive mathematical machinery and how to derive Shannon-style bounds for the communication cost under a variety of different goals for the communication.

We begin our investigation connecting information theory to logic by thinking about a scenario where there are two parties, a sender (Alice) and a receiver (Bob), each of whom have in general different logic sentences, but such that the Alice's sentence entails that of Bob's. Furthermore, Alice wishes to convince Bob of the truth of a given logic sentence, that Bob cannot, with the information known to him, prove by himself. Furthermore, the problem is for Alice to send the \textit{minimum} amount of information (in terms of bits sent) so that the Bob can prove what needs to be proved. A number of variants of the setting above are treated in this article -- for example, Alice and Bob may share some common logic sentences, or each may know something the other doesn't know, or the goal of the communication may be for Bob to only prove a subset of what Alice can prove. In yet another aspect we treat, there may be logical inconsistencies between Alice and Bob's sentences, leading to a rudimentary model of misinformation. 

Building on top of the foundations established by Carnap and Bar-Hillel \cite{carnap53}, Shannon's Rate-Distortion theory \cite{shannon:rd}, and the theories of source coding with side information due to Slepian-Wolf \cite{slepian_wolf:coding} and Wyner-Ziv \cite{wyner_ziv:coding} coding, we provide, for the first time, a rigorous theory that incorporates deductive reasoning directly in the communication process, providing sharp upper and lower bounds on communication cost under a wide variety of scenarios often showing significant efficiency gains compared to classic approaches. We also provide preliminary evidence of practical systems realizing a fraction of these possible gains. A pattern we found and subsequently used to guide our problem selection is the realization that the solution to all these problems had in common a simple scaled version of conditional entropy (see beginning of Section \ref{sec:fundamental}).
Our article is intended to serve as a bridge between the traditionally separate communities of information theory and logic. As a result, we devote special attention to the identification of the basic definitions that we found play a role in bridging between these fields, especially in logic and, in particular, in the sub-field of logic known as model theory. In an effort to provide as general conditions as possible for our information-theoretic results to apply, we first define the abstract notion of a \emph{logic} -- something that is rarely assayed in textbooks on the subject. 
We then introduce the notion of a \emph{Logic System}, which is a logic taken together with a set of models of the different sentences of the logic, along with two maps, one taking a logical sentence to a set of models, and the other taking a set of models to a sentence -- with the requirement that the latter map  behave like a pseudo-inverse of the former map. Our development of the notions of a logic and of a logic system has some similarities to the treatment of model theoretic logics,  introduced by Barwise and others in the 1980s \cite{barwise-model-theoretic-logics-1985}. However, our definitions are also unique and idiosyncratic since they are tailored to achieving our goal of providing the minimum set of conditions for our information-theoretic results to apply. As part of that goal, we needed for Propositional Logic and First-Order Logic to be united under a common umbrella, again something that is rarely done, and something that required us to introduce a non-standard vocabulary for Propositional Logic.
Further, the Propositional Proof System we provide is slightly different from the usual ones, and we need to make distinctions that are not always made between soundness and strong soundness, and between completeness and strong completeness. In the latter case we are led to introduce a new finitary variant of strong completeness that we have dubbed $\omega$-strong completeness.

Finally, before settling on the traditional information-theoretic concept of bits as the communication efficiency metric, we studied other paradigms, including the idea of communicating to a receiver only logic sentences that he could not derive himself already; this then posed the interesting problem of how one could construct such sentences.
To solve this problem, we explore connections beyond information theory and logic to include a third area of mathematics, which is the algebra of multivariate polynomials with variables and coefficients belonging to a finite field.
Our insight in doing so is that Propositional Logic sentences can be represented using polynomials, which in turn provides us access to powerful mathematical tools such as Gr\"obner bases. 
We exploit these mathematical tools to establish fundamental results that support our general approach for synchronizing the knowledge between a receiver and a sender in a communication optimal way by having the receiver add new non-trivial sentences in its knowledge base of logic sentences (without changing its old sentences).
More specifically, our general approach consists of first converting the original logic sentences into polynomials, then exploiting the foregoing mathematical tools to perform reduction and decomposition transformations in the polynomial domain, and finally converting the resulting polynomials back to logic expressions.
We note that the basic technique of our general mathematical framework is also very general and has applicability beyond this article. 

\subsection{Relation to other treatments of semantic information}

While reliance on mathematical logic is not universally leveraged in the field of semantic information, many authors do start with Carnap/Bar-Hillel's  \emph{logical probability}, one of many concepts introduced in their seminar work \cite{carnap53}. For the purposes of establishing contrast, here we too start with it assuming that the reader is familiar with the generalities of Propositional Logic. By means of example, assume two binary-valued propositions $\mathtt{X_1}$ and $\mathtt{X_2}$ and consider the logical sentence
\begin{eqnarray}
\mathtt{s} = \lnot \mathtt{X_1} \lor \mathtt{X_2} .
\label{eq:simple_example}
\end{eqnarray}
There are four possible choices for these propositional variables $\mathcal{M} = \{ 00, 01, 10, 11\}$ but only for three of those does $\mathtt{s}$ happen to be true, namely $\{ 00, 01, 11\}$. In general, the subset of $\mathcal{M}$ where a sentence $\mathtt{s}$ is true is defined in \cite{carnap53} to be the \emph{range} of that sentence; in our article, a generalization of this concept to general logics will be called the \emph{kernel} of $\texttt{s}$ and will be denoted by $\algebraicset{\mathtt{s}}$. Assume a distribution $P_{\mu}$ over $\mathcal{M}$ and let $\mu$ be drawn according to such a distribution. Then the \emph{logical probability} of $\texttt{s}$, relative to the distribution of $\mu$, is given by
\begin{eqnarray}
    P_{\mu}\left( \left[ \mu \in \algebraicset{s}\right] \right) ,
\end{eqnarray}
which informally is sometimes referred to as ``the probability that the logic sentence is true'', and a measure of the semantic information in $\mathtt{s}$, denoted in \cite{carnap53} by $\mathtt{inf}$, is defined by
\begin{eqnarray}
    \mathtt{inf}(\mathtt{s}) \stackrel{\Delta}{=} \log_2 \left( \frac{1}{P_{\mu}\left( \left[ \mu \in \algebraicset{s}\right] \right)} \right).
\label{eq:msi}
\end{eqnarray}
In some treatments of semantic information (see, for example, Bao, Basu, et al.~\cite{bao:semantic,bao:semantic_extended,BASU2014188}), each element of $\mathcal{M}$ is regarded as a possible ``meaning''; in our earlier example (\ref{eq:simple_example}), ``01'' is such a meaning. In this article \emph{we make no such choice}. For us, one has conveyed the semantic content of a logic sentence $\mathtt{s}$ if a receiver is able to infer from whatever is conveyed exactly the same that a sender can infer; as we will develop rigorously in this article, this can only be done if and only if $\algebraicset{\mathtt{s}}$ is reproducible by the receiver; in this case, we state that $\algebraicset{\sender} = \{00,01,11\}$. 

For now, informally, define $\mathcal{L}$ to be the set of all possible sentences. Instead of assuming a distribution $P_{\mu}$ over $\mathcal{M}$, assume a distribution $P$ over \emph{subsets} of $\mathcal{M}$. Then, an optimal code length (in bits) for sending such essentials in $\mathtt{s}$ in the above sense, relative to $P$, is given by 
\begin{eqnarray}
    \log_2 \left( \frac{1}{P( \algebraicset{\mathtt{s}})} \right).
\label{eq:alternate}
\end{eqnarray}
This quantity has a strong operational significance in the sense of measuring the optimal cost of transmission if one wishes to send to a receiver the information necessary for it to be able to  infer whatever the sender can infer from $\mathtt{s}$ (Theorem \ref{thm:logicinfo}), and is a reasonable example of an early (but not the only) idea in our paper. It is important to note that \emph{no other such definition will carry this strong operational significance.} By taking the approach of defining the semantics of a sentence via kernels, we have been able to rigorously derive a large collection of results where a receiver is able to reproduce all or a subset of the mathematical facts that a sender can infer, under a variety of assumptions of what prior logic sentences each has access to, including even settings where the sender is unaware of what the receiver knows. 

In contrast, a large segment of the community approaches the problem of semantic information without a direct linkage to logic. For us, a particularly relevant set of prior works are those who use Shannon's rate-distortion theory to set up problems involving semantics; in fact one could reasonably argue that Shannon's original foray into lossy coding \cite{shannon:rd} is an early example of exploiting semantics in compression. This theory is so general that almost any conception of semantics can be retrofitted to it. The work of Liu et al.~\cite{liu_shao_zhang_poor:indirect_rd_semantic, liu:rd_semantic} as well as the follow-up work by Stavrou and Kontouris \cite{stavrou:goal_semantic_rd} and Guo et al.~\cite{guo2022semantic}, for example, explicitly model a data source as comprising \emph{intrinsic} (unobservable) and \emph{extrinsic} (observable) components and proceed to derive information-theoretic bounds for desired approximations to either of these using general distortion measures. Another  example is Shao et al. \cite{Shao2022ATO} who model an end-to-end semantic communication process that starts with some intended meaning which is (stochastically) transformed into some expressed language, which may then experience a form of semantic noise as it is received; the authors then argue for the use of joint source-channel coding techniques for designing optimal communication systems. In contrast our work is singularly focused on exploring in rigorous depth communication assuming the existence of reasoning engines. Recently, a general theory of semantic information that draws parallels to Shannon's lossless source, channel and lossy source coding theorems has been proposed by Niu and Zhang \cite{9864327,niu2024mathematicaltheorysemanticcommunication}. Our work can be thought of as a depth-first, rather than breadth-first, work where the angle being explored is characterized by the possibility of deductive inference at the receiver's side. Our approach is rewarded with very sharp insight in this context, including the discovery of the role of the $\Lambda$ function in the characterization of a broad set of problems involving various communication paradigms.

In this last respect, the reader will notice that for some of our work involving settings where Alice is not fully aware of what logic sentence Bob possesses, we rely on the idea of multiple rounds of communication. This setup has similarities to problems involving communication complexity, famously introduced by Yao \cite{10.1145/800135.804414} in 1979; see also Papadimitriou and Sipser \cite{10.1145/800070.802192} and related work on interactive communication, for example Orlitsky \cite{orlitsky:interactive}. To establish contrast, we note that in our work, there is no pre-agreed function that Alice (or Bob) wants to compute. 

A rather different take on the problems above can be found in the work of Juba and Sudan \cite{juba:universal_semantic}, who consider the problem of communication between sender and receiver when there has been no previous agreement on protocol and where the main difference between them lies on their computational power. Compared to \cite{juba:universal_semantic}, our work follows much more closely the usual conventions in information theory where sender and receiver do agree on elementary matters such as how information will be encoded and decoded.

In this subsection, we have only covered a subset of relevant works. We refer the reader to G\"und\"uz et al.~\cite{gunduz:beyond_bits} for a broad survey on the subject.

\subsection{Outline of the remainder of the paper}

The Mathematical Preliminaries (Section \ref{sec:preliminaries}) provide the fundamental notion of a Logic System, fully discussed in the Logical Underpinnings (Section \ref{sec:logical-underpinnings}), and also include the basic definitions of how the communication system is set up. We then provide a summary of our information-theoretic contributions (Section \ref{sec:fundamental}), leaving the formal statement and proof of these results to the latter Section \ref{sec:proofs}. Theorem \ref{thm:logicinfo} contains the result supporting the discussion above surrounding \eqref{eq:alternate}; for clarity of exposition, the rest of the theorems in Section \ref{sec:fundamental} solely contain the simpler single-letter upper bounds which are tight when additional i.i.d.\ assumptions are made. We develop practical methods, including linear and nonlinear codes, in Section \ref{sec:practical}, where we also include experiments on synthetic data demonstrating significant possible gains over classical systems. We then provide a treatment of logic and a form of efficient communication from the standpoint of the algebra of polynomials on finite fields in Section \ref{sec:linkage_algebra}. Speculative future directions are included in Section \ref{sec:speculations}, followed by concluding thoughts in Section \ref{sec:concluding}.

\section{Mathematical preliminaries}

\label{sec:preliminaries}
\subsection{Logic Systems} \label{subsec:logic_systems}

Suppose we are given a logic $L$ (for example, Propositional Logic on a fixed set of $\mathtt{m}$ variables, or the First-Order Logic of graphs). We shall give a more complete definition of what we mean by a logic in Section \ref{sec:logical-underpinnings}, but for the time being it suffices to note that a logic $L$ specifies a syntax, or set of rules for constructing valid logic sentences starting from a particular logical vocabulary, $\tau$, and also provides a proof system, or rules of inference, for deducing the truth of new sentences assuming the truth of other sentences. Let us denote the set of well-formed sentences for the logic $L$ by $\surfacespace$. We call $\surfacespace$ the \emph{language} associated with the logic $L$.
We will use the $\mathtt{typewriter~font}$ to denote a logic sentence $\mathtt{s} \in \surfacespace$.  
The logic $L$ may come equipped with a set of axioms, $\sigma$, or sentences of $\surfacespace$, that are assumed to be true without proof.  
For two logic sentences $\mathtt{s_1, s_2} \in \surfacespace$, we write $\mathtt{s_1} \entails \mathtt{s_2}$ if, assuming the truth of $\mathtt{s_1}$, it is possible to prove $\mathtt{s_2}$ in the logic $L$, possibly with the assistance of some of the sentences in $\sigma$.  The empty sentence is considered to be well-formed and always true. We therefore write $\entails \mathtt{s}$ if and only if (iff) $s$ can be proven directly in $L$, starting from the axioms $\sigma$. Then $\entails$ is a relation defined among pairs of elements of $\surfacespace$. We call $\entails$ the ``entailment'' relation for the logic $L$. The definition of entailment extends naturally to \emph{sets} of sentences $\mathcal{S}_1, \mathcal{S}_2 \subseteq \surfacespace$ so that $\mathcal{S}_1 \entails \mathcal{S}_2$ iff it is possible to prove every sentence $\mathtt{s} \in \mathcal{S}_2$ assuming the truth of every sentence in $\mathcal{S}_1$. Thus, $\entails$ is both a relation among pairs of sentences in $\surfacespace$ and among pairs of \emph{sets} of sentences in $\surfacespace$.

\begin{definition}
 Let $\lambda = (L,  \kernelspace, \kappa, \ell)$, where $L$ is a logic with associated language $\surfacespace$ and entailment relation $\entails$, $\kernelspace$ is a set, $\mathcal{P}(\kernelspace)$ is its power set (i.e., set of all subsets), $\kappa:\surfacespace \rightarrow \mathcal{P}(\kernelspace)$ and $\ell:\mathcal{P}(\kernelspace) \rightarrow \surfacespace$ are both functions. We call $\lambda$ a \textbf{Logic System} if $\kappa, \ell$, and $\entails$ additionally satisfy the conditions that, for all $\Mds \subseteq \kernelspace$ and for all $\mathtt{s_1}, \mathtt{s_2} \in \surfacespace$, one has
\begin{eqnarray}
   &&\kappa(\ell(\Mds)) = \Mds,  \label{pseduo-inverse-rel} \\
   &&\mathtt{s_1} \entails \mathtt{s_2} \mbox{ if and only if } \algebraicset{\mathtt{s_1}} \subseteq \algebraicset{\mathtt{s_2}}. \label{fundamental-rel}
\end{eqnarray}
We refer to $\kappa$ as the \textbf{kernel function} associated with $\lambda$, and for a given sentence $\mathtt{s} \in \surfacespace$, we call $\kappa(\mathtt{s})$ the \textbf{kernel} of $\mathtt{s}$.
\label{def:core}
\end{definition}
This article deals exclusively with cases where $\kernelspace$ is finite, and thus $\mathcal{P}(\kernelspace)$ is a finite set\footnote{It is also possible to consider the case where $\kernelspace$ is not a set but rather a \emph{proper class}, in which case $\mathcal{P}(\kernelspace)$ is then called the \emph{power class}. See the footnote to Theorem \ref{thm:fundamental-equivalence} for a brief discussion.}.
For the vast majority of the results of this article, we shall not need to assume anything further about the Logic Systems with which we work. The underlying logical vocabularies can be arbitrary and need not include any of the usual logical operators, so long as conditions (\ref{pseduo-inverse-rel}) and (\ref{fundamental-rel}) are satisfied. For Theorem \ref{thm:master} in Subsection~\ref{ss:alice-does-not-know}, however, we will have to assume the presence of the standard logical operators $\lor, \land$ and $\lnot$, and, moreover, that they have a certain natural set-theoretic behavior with respect to the kernel function $\kappa$.

\begin{definition} \label{def:proper-logic-system}
A Logic System  $\lambda = (L, \kernelspace, \kappa, \ell)$ is said to be \textbf{proper} if the logical vocabulary of $L$ includes the operators $\lor, \land$ and $\lnot$, and, moreover, for every $\mathtt{s,t} \in \surfacespace$, where $\surfacespace$ is the language associated with $L$, the following hold:
\begin{enumerate}
\item $\kappa(\mathtt{s} \lor \mathtt{t}) = \kappa(\mathtt{s}) \cup \kappa(\mathtt{t})$, \label{kr1}
\item $\kappa(\mathtt{s} \land \mathtt{t}) = \kappa(\mathtt{s}) \cap \kappa(\mathtt{t})$, \label{kr2}
\item $\kappa(\lnot \mathtt{s}) = \kappa(\mathtt{s})^c$. \label{kr3} 
\end{enumerate}
\end{definition}

\begin{figure} [t]
\centerline{\scalebox{0.55}{\includegraphics{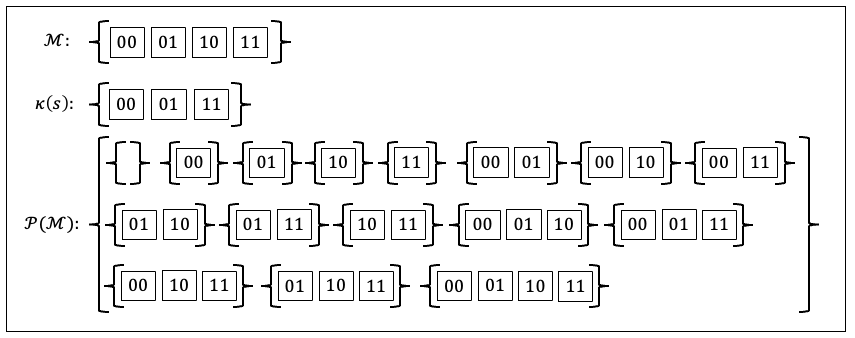}}}
\caption{Pictorial representation of the objects $\mathcal{M}, \algebraicset{\mathtt{s}}$ and $\mathcal{P}(\mathcal{M})$ for the case of Propositional Logic on 2 variables. The displayed kernel, $\algebraicset{\mathtt{s}}$, is for the sentence $\mathtt{s} = \lnot \mathtt{X_1} \lor \mathtt{X_2}$. The notation $\fbox{01}$, for example, corresponds to the truth value assignment $\mathtt{X_1}=\textrm{False}, \mathtt{X_2}=\textrm{True}$.}
\label{fig:SimplePropLogicExamples}
\end{figure}

\begin{example}[Propositional Logic]  \label{ex:prop_logic}
Let us consider the case where $L$ is (classical) Propositional Logic on a fixed number, $\mathtt{m}$, of propositional variables. The vocabulary $\tau$ consists of the logical connectives  $\lor, \land, $ and $\lnot$, the $\mathtt{m}$ propositional variables $\mathtt{X_1},...,\mathtt{X_m}$, as well as parentheses $(\cdot)$ to aid in grouping. Any single standalone propositional variable $\mathtt{X_i}$ is considered to be a well-formed sentence. We also consider the empty sentence, denoted alternatively by $\top$, to be a well-formed sentence. If $\mathtt{s}, \mathtt{t}$ are well-formed sentences, then so are $\mathtt{s} \lor \mathtt{t}, \mathtt{s} \land \mathtt{t}, \neg \mathtt{s},$ and $(\mathtt{s})$. We typically write $\bot$ in lieu of $\lnot \top$. The symbols $\lor$ and $\land$ are understood to apply in left-to-right order. In other words, the sentence $\mathtt{r} \lor \mathtt{s} \land \mathtt{t}$ is syntactically equivalent to $(\mathtt{r} \lor \mathtt{s}) \land \mathtt{t}$. 

In this case, we let the set $\kernelspace$ be the set of the $2^\mathtt{m}$ different truth-value assignments to the $\mathtt{m}$ propositional variables. 
Further, we let $\kappa: \surfacespace \rightarrow \mathcal{P}(\kernelspace)$ be the function that maps each sentence $\mathtt{s} \in \surfacespace$ to the set of truth-value assignments that make $\mathtt{s}$ true. 
In Subsection~\ref{ss:pl} we will see that one can define a function $\ell:\mathcal{P}(\kernelspace) \rightarrow \surfacespace$ such that, for all $M \subseteq \kernelspace$, we have $\kappa(\ell(M)) = M$, and thus (\ref{pseduo-inverse-rel}) holds. We will also see that for $\kappa$ as we have defined it, as long as we equip $L$ with a standard propositional proof system (look ahead to Definition \ref{def:classical-pps}), condition (\ref{fundamental-rel}) holds. Furthermore, from the definition of $\kappa$,
it is an elementary exercise to verify that conditions \ref{kr1}--\ref{kr3} of Definition~\ref{def:proper-logic-system} all hold, so that Propositional Logic on a fixed number of variables can thus be turned into a proper Logic System.
\end{example}
Figure~\ref{fig:SimplePropLogicExamples} depicts several of the objects described in the above example for the case of Propositional Logic on 2 variables and the sentence $\mathtt{s} = \lnot \mathtt{X_1} \lor \mathtt{X_2}$. The set $\kernelspace$ consists of all truth value assignments to the variables $\mathtt{X_1}$ and $\mathtt{X_2}$. The kernel of $\mathtt{s}, \kappa(\mathtt{s})$, is the set of all truth value assignments to $\mathtt{X_1}$ and $\mathtt{X_2}$ making $\mathtt{s}$ true, and $\mathcal{P}(\kernelspace)$ is the set of all subsets of $\kernelspace$, in other words, the set of all possible kernels of sentences in the two variables $\mathtt{X_1}$ and $\mathtt{X_2}$.

\medskip
The basic preliminaries on logic systems in this subsection are sufficient for the majority of what follows.
As previously noted, in Section \ref{sec:logical-underpinnings} we provide a more complete definition of logic systems. 

Subsection~\ref{ss:logic} describes more formally what we mean by a \emph{logic}. Subsection~\ref{ss:models} provides a brief introduction to the branch of logic known as Model Theory and describes what it means for a mathematical object to be a model of a given set of logic sentences. Using just a small amount of model-theoretic formalism, we will then be able to show that the condition (\ref{fundamental-rel}) for being a Logic System is satisfied by virtually all logics we care about.  Lastly, Subsections~\ref{ss:pl} and \ref{ss:fol} provide examples of proper Logic Systems, first for Propositional Logic and then for First-Order Logic.

\subsection{Elementary information-theoretic definitions and notation}

For a scalar $0 \leq p \leq 1$, we denote Shannon's binary entropy by
\begin{eqnarray*}
\Hbin{p} =-p \log_2 p -(1-p) \log_2 (1-p) .
\end{eqnarray*} 
For a random variable $X$ on any arbitrary \emph{discrete} alphabet, governed by a distribution $p_X$, we define
\begin{eqnarray*}
    \Hreg{X} = E_{X} \left[ \log_2 \frac{1}{p_X(X)} \right] ,
\end{eqnarray*}
where $p_X$ denotes the probability mass function of the discrete random variable $X$. In either case, entropy is expressed in bits, as we are using the logarithm base 2. Given two discrete random variables $X,Y$, we define conditional entropy and mutual information as
\begin{eqnarray*}
    \Hreg{X|Y} &=& E_{X,Y} \left[ \log_2 \frac{1}{p_{X|Y}(X|Y)} \right], \\
    I(X;Y) &=& E_{X,Y} \left[ \log_2 \frac{p_{X|Y}(X|Y)}{p_Y(Y)} \right].
\end{eqnarray*}    
It is also the case that
\begin{eqnarray*}
   I(X;Y) &=& \Hreg{X} + \Hreg{Y} - \Hreg{X,Y}, \\
          &=& \Hreg{X} - \Hreg{X|Y} = \Hreg{Y} - \Hreg{Y|X}
\end{eqnarray*}
when the corresponding individual entropies are finite.

For three random variables $A,B,C$ we say that they from a Markov chain if given $B$, $A$ and $C$ are statistically independent, and we write
\begin{eqnarray*}
    A \rightarrow B \rightarrow C.
\end{eqnarray*}
If $A \rightarrow B \rightarrow C$ then the data processing inequality states that 
\begin{eqnarray*}
    I(A;B ) \geq I(A;C).
\end{eqnarray*}
In addition,
\begin{eqnarray}
    H(A|BC) = H(A|B).
\label{eq:markov_chain_conditional_entropy}
\end{eqnarray}
 
\subsection{Communication setup}

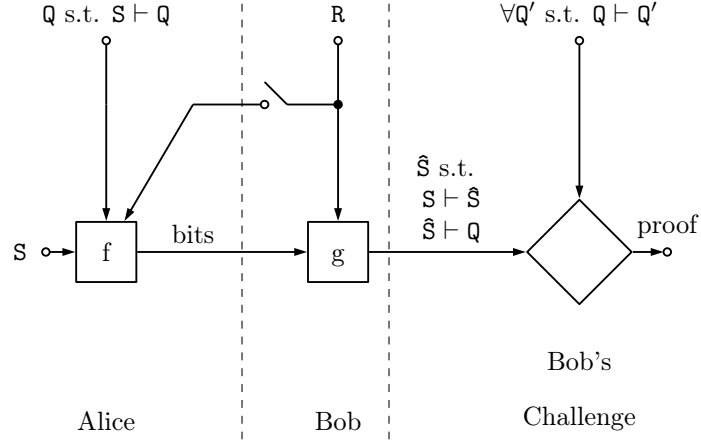
\begin{figure}
    \centering
    \begin{tikzpicture}[baseline=(current bounding box.center)]
	\matrix[row sep=2.5mm, column sep=2.5mm]
	{tm		
  

  		\node[coordinate] (h0) {};     &
     	\node[coordinate] (h1) {};     &	
		\node[coordinate] (h2) {};      &
        \node[coordinate] (h3){}; &
        \node[coordinate] (h3a){}; &
        \node[coordinate] (h4) {}; &
        \node[coordinate] (h5) {};    &
		\node[coordinate] (h7) {};    &
		\node[coordinate] (h8) {}; \\

        \node[coordinate]         (tm00) { };     &
		\node[dspnodeopen,dsp/label=above,overlay]         (tm01) { $\queryrnd  \mbox{ s.t. } \senderrnd \entails \queryrnd$ };     &
		\node[coordinate,label=] (tm02) {};      &
        \node[coordinate] (tm022) {}; &
        \node[coordinate] (tm025) {}; &
        \node[coordinate] {}; &
		\node[dspnodeopen,dsp/label=above]          (tm03) {$\receiverrnd$};    &
		\node[coordinate]          (tm03p5) {};    &
		\node[coordinate]          (tm03p7) {};    &
        \node[dspnodeopen,dsp/label=above,overlay]         (tm05) { $\forall \queryrndapo \mbox{ s.t. } \queryrnd \entails \queryrndapo$ };     &
		\node[coordinate]          (tm06) {}; \\ \\ 

  		\node[coordinate]         (mid0) {};     &
     	\node[coordinate]         (mid1) {};     &	
		\node[coordinate]         (mid2) {};      &
        \node[coordinate]         (mid3){}; &
        \node[coordinate]         (mid3a){}; &
        \node[coordinate]         (mid4) {}; &
        \node[coordinate]         (mid5) {};    &
        \node[coordinate]         (mid6) {};    &
        \node[coordinate]         (mid7) {};    &
		\node[coordinate]         (mid8) {}; \\

  		\node[coordinate]         (fm00) {};     &
     	\node[coordinate]         (fm01) {};     &	
		\node[coordinate]         (fm02) {};      &
		\node[coordinate]         (fm02z) {};      &
        \node[dspnodeopen]        (fm02a){}; &
        \node[coordinate]         (fm02b) {}; &
        \node[dspnodefull]        (fm03) {};    &
		\node[coordinate]         (fm04) {};    &
        \node[coordinate]         (fm05) {};    &
		\node[coordinate]         (fm06) {}; \\ \\

	    \node[dspnodeopen,dsp/label=left] (mm00) {$\senderrnd$};  &
	    \node[dspsquare]          (mm01) {f};     &
        \node[coordinate,label=bits] (mm02) {}; &
        \node[coordinate] {}; &
        \node[coordinate] {}; &
        \node[coordinate] {}; &
        \node[dspsquare]          (mm03) {g};     &
	    \node[coordinate] (mm03p5) {};     &
        \node[coordinate,label={[align=center]$\senderrndhat \mbox{ s.t. }$\\$\senderrnd \entails \senderrndhat$ \\ $\senderrndhat \entails \queryrnd$}] (mm03p7) {}; &
        \node[dspdiamond]         (mm05) {};  &
	    \node[dspnodeopen,dsp/label=above] (mm06) {proof};   \\ \\

  		\node[coordinate]         {}; &
     	\node[coordinate]         {}; &	
		\node[coordinate]         {}; &
		\node[coordinate]         {}; &
        \node[coordinate]         {}; &
        \node[coordinate]         {}; &
        \node[coordinate]         {}; &
        \node[coordinate]         {}; &
		\node[coordinate]         {}; &
		\node[coordinate,label=Bob's]         {}; \\ \vspace{-0.7in}

  		\node[coordinate]         {};     &
     	\node[coordinate,label=Alice]         {};     &	
		\node[coordinate]          {};      &
		\node[coordinate]  (aha)        {};      &
        \node[coordinate]         {}; &
        \node[coordinate]          {}; &
        \node[coordinate,label=Bob]         {};    &
        \node[coordinate] (bottom) {}; &
		\node[coordinate]         {};    &
		\node[coordinate,label=Challenge]         {}; \\ \\ 
  };

    \draw[dashed] (h3)--(aha);
    \draw[dashed] (h7)--(bottom);

	\begin{scope}[start chain]
		\chainin (mm00);
		\chainin (mm01) [join=by dspconn];
		\chainin (mm03) [join=by dspconn];
		\chainin (mm05) [join=by dspconn];
		\chainin (mm06) [join=by dspconn];
        \end{scope}
        \begin{scope}[start chain]
            \chainin (tm05);
            \chainin (mm05) [join=by dspconn];
        \end{scope}

        \begin{scope}[start chain]
          \chainin (tm01);
        \chainin (fm01) [join=by dspline];
          \chainin (mm01) [join=by dspconn];
        \end{scope}

        \begin{scope}[start chain]
          \chainin (tm03);
          \chainin (mm03) [join=by dspconn];
        \end{scope}

        \begin{scope}[start chain]
            \chainin (fm02a);
            \chainin (fm02) [join=by dspline];
            \chainin (mm01) [join=by dspconn];
        \end{scope}
        \begin{scope}[start chain]
        \chainin (fm03);
        \chainin (fm02b) [join=by dspline];
        \chainin (mid3a) [join=by dspline];
        \end{scope}

\end{tikzpicture}
    \caption{General communication diagram treated in this work.}
    \label{fig:general_diagram}
\end{figure}

In this subsection, we describe the fundamentals of our communication setup, also summarized in Figure~\ref{fig:general_diagram}.

\subsubsection{Notation}
We use the letter $\sender$ to denote a logic sentence known to Alice, the sender. Similarly, we use $\receiver$ to denote a logic sentence known to Bob, the receiver. In many, but not all, of our setups $\receiver$ is also known to Alice. Once Bob decodes whatever information he receives  from Alice, possibly in combination with $\receiver$, Bob deduces $\senderhat$, which he uses for the purpose of deducing $\query$, the problem that tests the success of this communication endeavor. To model various kinds of uncertainties in what Alice and Bob know about each other's knowledge,  we will introduce random versions of the logic sentences by using upper case notation $\senderrnd, \receiverrnd, \queryrnd$.

For any given kernel $k \in \kernelspace$, we let $|k|$ denote the size of the set; obviously  $0 \leq |k| \leq |\kernelspace|$. We will often refer in our article to the \emph{normalized expected kernel size} of some random sentence, defined, in this example, as follows:
\begin{eqnarray}
    \frac{1}{|\kernelspace|} E \left[ |\algebraicset{\senderrnd}|\right].
\label{eq:neks}
\end{eqnarray}

\subsubsection{Initial meeting}
Alice and Bob meet ahead of time, and settle on a Logic System (see Definition \ref{def:core}). They agree on the general conditions of a future communication: that Alice will have access to $\senderrnd$, whether Bob will have access to  $\receiverrnd$; if so, whether Alice herself will have access to it as well. They also agree that the goal is for Bob to prove the truth of a logic sentence $\queryrnd \in \surfacespace$ using information that Alice will provide employing pre-agreed upon encoding and decoding functions, and that the output of the decoding function, called $\senderrndhat$, must be entailed by $\senderrnd$. The sentences $\senderrnd, \receiverrnd, \queryrnd$ are not known at the time of this meeting, and will be revealed to the relevant parties later. We will describe these functions shortly. It is important to note that any logic sentences that can be deduced within the Logic System are assumed to be known to be true by both Alice and Bob as a result of this meeting. No communication cost whatsoever is levied against any exchange that happens during this meeting.

\subsubsection{Correlated world observations}
\label{ss:cwo}
After the initial meeting, Alice and Bob go their own ways; Alice, the sender, obtains knowledge about the world summarized in a logic sentence $\senderrnd \in \surfacespace$ whereas Bob, the receiver, obtains $\receiverrnd \in \surfacespace$. We consider both settings where Alice knows and doesn't know $\receiverrnd$. We assume that there is consistency between Alice's and Bob's observations but that Alice has a potentially sharper view of the world:
\begin{eqnarray}
    \senderrnd \entails \receiverrnd .
    \label{eq:sir}
\end{eqnarray}
There is one exception to this assumption when we treat a misinformation scenario, which will be clear during that discussion. We assume that the query $\queryrnd$ that Bob will be able to prove after the communication takes place is provable using Alice's knowledge:
\begin{eqnarray}
    \senderrnd \entails \queryrnd .
    \label{eq:siq}
\end{eqnarray}
This is universally true in all of our results, including those of misinformation. Finally, we make an assumption that is more technical in nature:
\begin{eqnarray}
    \queryrnd \entails \receiverrnd .
\label{eq:technical}
\end{eqnarray}
In the case that Alice knows $\receiverrnd$, the assumption above is justified in light of the following result.

\begin{restatable}{lem}{sufficiencyqentailsr}
Given a Logic System $ (L,  \kernelspace, \kappa, \ell)$, for $\sender, \query, \receiver \in \surfacespace$, if $\sender \entails \query$ and $\sender \entails \receiver$, then there exists a $\query^{\prime} \in \surfacespace$, given by $\query^{\prime}= \ell( \algebraicset{\query} \cap \algebraicset{\receiver})$, such that $\query^{\prime} \entails \query$, $\sender \entails \query^{\prime}$ and $\query^{\prime} \entails \receiver$.
\label{lem:sufficiencyqentailsr}
\end{restatable}
The Lemma follows from the definition of a Logic System. Thus, in the case both Alice and Bob share $\receiverrnd$, without loss of essential generality, the query that Alice is attempting to ensure Bob can prove can be assumed to satisfy (\ref{eq:technical}). 

The case that Alice does not know $\receiverrnd$ splits in two cases. In one case, $\queryrnd = \senderrnd$ and (\ref{eq:technical}) simply reduces to (\ref{eq:sir}); this is a very interesting setting in practice. If in general $\queryrnd$ is a weaker sentence than $\senderrnd$, then the assumption (\ref{eq:technical}) is too strong since it does not reduce to (\ref{eq:sir}) and thus we believe it to be of reduced practical interest. For mathematical completeness, we do provide a result (Theorem \ref{thm:master}) under such an assumption but do not rely on it to make the main points of our paper. 

\subsubsection{Communication}
To communicate, Alice and Bob rely on the functions agreed upon during the initial meeting. The nature of these functions depend on the nature of the specific situation Alice and Bob have planned for. In the simplest of settings, neither Bob nor Alice have access to $\receiverrnd$ and Alice will be communicating to Bob a message that ensures he prove all that she can prove.

The encoding function is generally denoted $f$, and in this simple case it maps $\senderrnd$ to a finite sequence of bits:
\begin{eqnarray*}
    f : \surfacespace \rightarrow \{0,1\}^{*} ,
\end{eqnarray*}
where the notation $\{0,1\}^{*}$ is meant to signify the set of finite binary strings. In turn a receiver will decode the information being send by the sender using a decoding function denoted by $g$:
\begin{eqnarray*}
    g : \{0,1\}^{*} \rightarrow \surfacespace .
\end{eqnarray*}
The output of $g$ is generally denoted by $\senderrndhat$. 
More complex situations augment the arguments that $f$ can take on to include $\queryrnd, \receiverrnd$ as relevant; similarly $g$ may also depend on $\receiverrnd$. In even more complex situations, the communication involves a conversation where sender and receiver take turns.

The function $f$ in any of the settings under consideration (Figure~\ref{fig:summary}) is capable of producing a variable number of bits, as this is a more flexible setting than assuming a fixed number of bits. However, an additional complication is that it may not be easy to determine when these bits start and finish in an otherwise arbitrary bit sequence. To resolve this matter, we will rely on the standard concept from information theory of prefix-free codes. Let $\mathcal{C} \subseteq \{0,1\}^*$ be a set of codewords. We say that $\mathcal{C}$ is prefix-free if, for all distinct $c_1, c_2 \in \mathcal{C}$, $c_1$ is not a prefix of $c_2$. We will assume that the image of the encoder is prefix free; this will be mathematically explicit when the theorems are stated and proved.

Finally, we will use the expected number of transmitted bits as a performance metric for any proposed system. In the simple example described above, such a metric is
\begin{eqnarray*}
    E_{\senderrnd}[\len{f(\senderrnd)}] ,
\end{eqnarray*}
where $\len{}$ is the function that maps a finite string to its length. 
 
\subsubsection{ Challenge and deduction}
After the communication takes place, Bob is challenged with any sentence that can be proven by $\queryrnd$ (including possibly $\queryrnd$ itself), and Bob is able to produce a proof for that query starting from the logic sentence~$\senderrndhat$, which in turn we assume is entailed by $\senderrnd$. Mathematically, for the system to have succeeded, it must be the case these conditions hold: $\senderrnd \entails \senderrndhat$, $\senderrndhat \entails \queryrnd$.

\subsubsection{Probabilistic model}
We will present theoretical results in the form of upper and lower bounds on the total number of expected bits. Our upper bounds are applicable to any possible distribution over $\senderrnd, \queryrnd, \receiverrnd$ as long as the entailment conditions described in Subsection~\ref{ss:cwo} are met, and are phrased in terms of normalized expected kernel sizes (see Equation (\ref{eq:neks})). We next make a definition that we will rely on when we formally state our theorems:

\begin{definition}[probability laws for kernels]
We say that the random logic sentence $\mathtt{A} \in \surfacespace$ has a kernel that follows a $p_a$-law if $|\kernelspace|^{-1}E\left[ | \algebraicset{\mathtt{A}}| \right]=p_a$. We say that the random logic sentences $\mathtt{A}, \mathtt{B} \in \surfacespace$ have kernels that follow a ($p_a, p_{b}$)-law, with $p_b \geq p_a$, if $\mathtt{A}$ has a kernel that follows a $p_a$-law, $\mathtt{B}$ has a kernel that follows a $p_b$-law, and $\mathtt{A} \entails \mathtt{B}$. We say that the random logic sentences $\mathtt{A}, \mathtt{B},\mathtt{C} \in \surfacespace$ have their kernels follow a ($p_a, p_{b},p_c$)-law, with $p_c \geq p_b \geq p_a$, if $\mathtt{A}$ follows a $p_a$-law, $\mathtt{B}$ follows a $p_b$-law, $\mathtt{C}$ follows a $p_c$-law, $\mathtt{A} \entails \mathtt{B}$ and  $\mathtt{B} \entails \mathtt{C}$.
\label{def:kernel_law}
\end{definition}

For our lower bounds, we use stronger assumptions that involve independent and identically distributed (i.i.d.) random variables. Throughout the paper, the reader will notice that whenever we use the i.i.d.\ assumption, which is otherwise never assumed in the upper bounds, the upper and lower bounds will match asymptotically. This pattern follows similar patterns in information theory, where ``the i.i.d.\ source is the hardest to compress'' (amongst all sources with the same marginal statistics).

\section{Summary of information-theoretic contributions}
\label{sec:fundamental}
In this section, we provide a summary of the core information-theoretic results that we have obtained, intended as a guide to understand the actual formal  result statements and proofs found in Section \ref{sec:proofs}, and also the practical results in Section \ref{sec:experiments}.

\begin{figure}
    \centering
    \scalebox{0.8}{\includegraphics{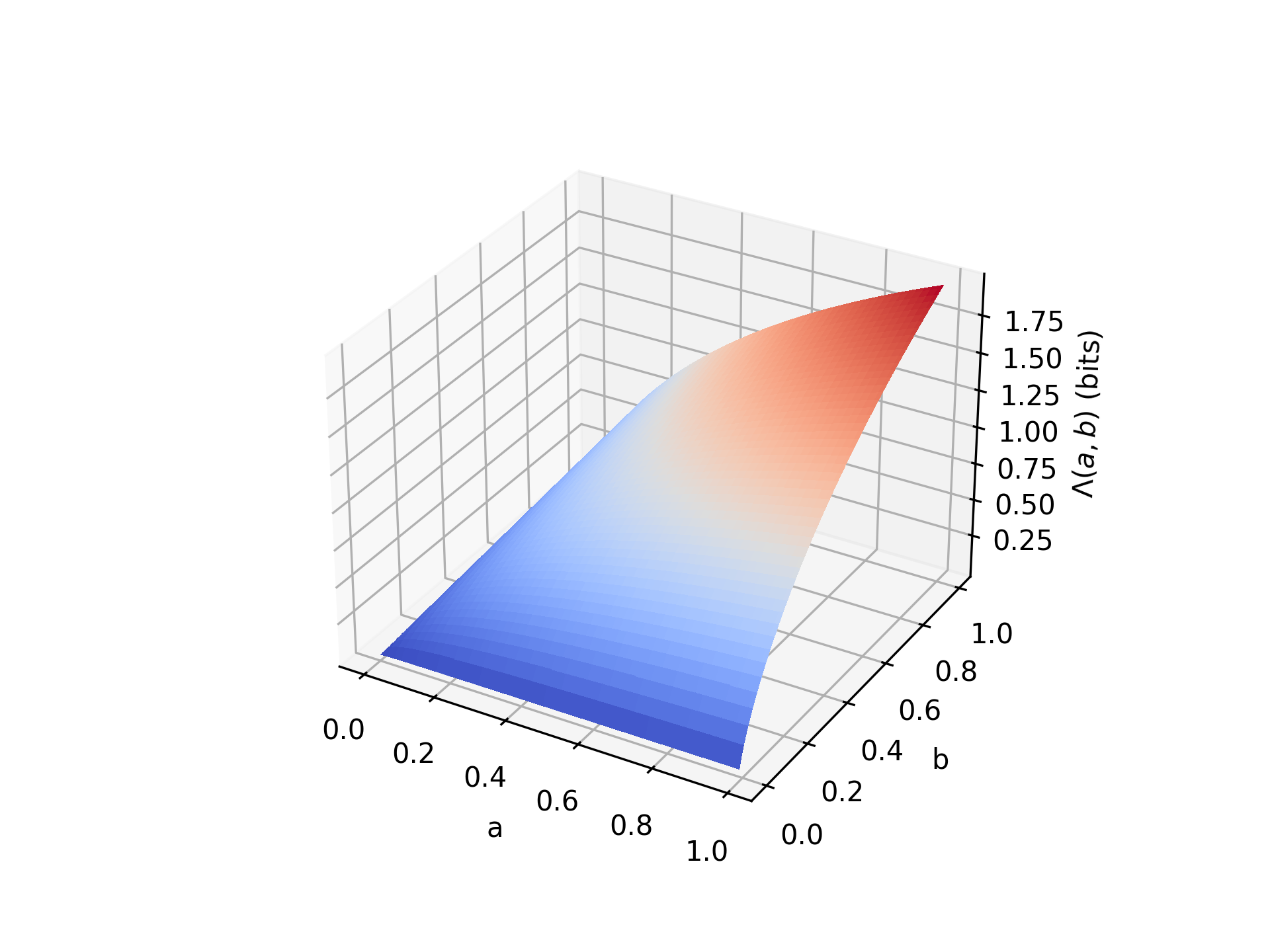}}

    \caption{Illustration of $\Lambda(a,b)$}
    \label{fig:lambda}
\end{figure}

All of our information-theoretic results are expressed using a function with two arguments $\Lambda(a,b)$, defined this way: for any $a,b \geq 0$,
\begin{eqnarray*}
\Lambda(a,b) = a \log_2 \left( \frac{a+b}{a} \right) + b \log_2 \left( \frac{a+b}{b}  \right) = (a+b)\Hbin{ \frac{a}{a+b} } = (a+b)\Hbin{ \frac{b}{a+b} },
\end{eqnarray*}
where we additionally define $\Lambda(0,b) = \Lambda(a,0) = 0$. The greek letter $\Lambda$ is chosen for this function in reference to its apparent emergence in problems involving logic. This function is illustrated in Figure~\ref{fig:lambda}.

The reader is not expected to appreciate, at the present moment, the intuition behind why the $\Lambda(a,b)$ function is relevant to our problem. The way we first encountered this function was as the solution to a variational problem that is at the core of the proof of Theorem~\ref{thm:shannonpartition}, which in turn is the basis for how Theorem~\ref{thm:less_is_more_simple} is proven. Subsequently, we noticed that all of our information-theoretic results could be rewritten in terms of this function. This expression satisfies the following basic properties, which are proved in the Appendix.
\begin{restatable}[Elementary properties of $\Lambda(a,b)$]{lem}{elementarylambda}
The function $\Lambda(a,b)$ is concave over the domain $[0,+\infty) \times [0,+\infty)$. If $\Delta_a, \Delta_b \geq 0$ with at least one of them being strictly positive, then $\Lambda(a + \Delta_a ,b + \Delta_b) > \Lambda(a,b)$.  If $a + b < 1$, then for any mixture parameter $\lambda \in [0,1]$, $\Lambda(a,b) < \Hbin{ \lambda a + (1-\lambda) b }$. For any $\xi$, $\xi \Lambda(a,b) = \Lambda(\xi a, \xi b)$.
\label{lem:elementarylambda}
\end{restatable}

We believe that this article is the first to point out the relevance of this particular form of entropy to communication problems involving logic.  
To this end, given a Logic System $(L, \surfacespace, \kernelspace, \kappa, \ell, \entails)$ and a chosen ordering of the elements of $\kernelspace = \{\mu_1,\ldots,\mu_{|\kernelspace|}\}$, we define $\kvv:\surfacespace \rightarrow \{0,1\}^{|\kernelspace|}$ via
\begin{eqnarray}
\kv{\sender}_i & \stackrel{\Delta}{=} &  \left\{ \begin{array}{cc}
                               1 & \mbox{ if } \mu_i \in \algebraicset{\sender},  \\
                               0 & \mbox{otherwise;}
                           \end{array} \right. \\
                          |\kv{\sender}| & \stackrel{\Delta}{=}  & |\algebraicset{\sender}| .
\label{eq:kvdef}
\end{eqnarray}
Note that the function $\kvv$ is just another way of thinking about the function $\kappa$, e.g., as an indicator function. Similarly, define the function 
\begin{eqnarray}
    \vec{\ell} :  \{0,1\}^{|\kernelspace|} \rightarrow \surfacespace
\end{eqnarray}
that accepts a set indicator vector, recovers the corresponding subset of $\kernelspace$, and then passes that subset to $\ell$.

Our main result, to be interpreted in the context of a given Logic System $(L, \surfacespace, \kernelspace, \kappa, \ell, \entails)$ and the communication setup in Section~\ref{sec:preliminaries} (as illustrated in Figure~\ref{fig:general_diagram}), is stated next.

\begin{restatable}{thm}{core}
Given a Logic System $(L,  \kernelspace, \kappa, \ell)$, 
for any distribution over $(\senderrnd, \queryrnd, \receiverrnd)$ meeting the entailment conditions $\senderrnd \entails \queryrnd$ and $\queryrnd \entails \receiverrnd$, if the corresponding kernels have normalized sizes $p_s,p_q,p_r$, respectively, then an algorithm exists with a normalized average cost in total bits exchanged that is upper bounded by $\Lambda(p_s,p_r-p_q) + O(|\kernelspace|^{-1} \log_2|\kernelspace| )$. If additionally the random variables $\{ ( \kv{\senderrnd}, \kv{\queryrnd}, \kv{\receiverrnd} )_j \}_{j=1}^{|\mathcal{\kernelspace}|}$ are i.i.d.\ and $\receiverrnd \rightarrow \algebraicset{\receiverrnd} \rightarrow (\algebraicset{\senderrnd},\algebraicset{\queryrnd})$, then the normalized average cost of any such algorithm is lower bounded by $\Lambda(p_s,p_r-p_q)$. The theorem statement holds true regardless of whether Alice knows $\receiverrnd$ or not.
\label{thm:aggregate}
\end{restatable}

We remark that this theorem is a consequence of Theorems \ref{thm:less_is_more_background} and \ref{thm:master}, and in fact, we will not provide a direct proof for it. Nonetheless, the mathematical machinery developed to address the scenarios addressed by those theorems is unnecessarily complex and thus for didactic purposes, we chose to gradually build the sophistication of our results over a series of theorems so as to allow the key ideas to settle more easily and firmly.

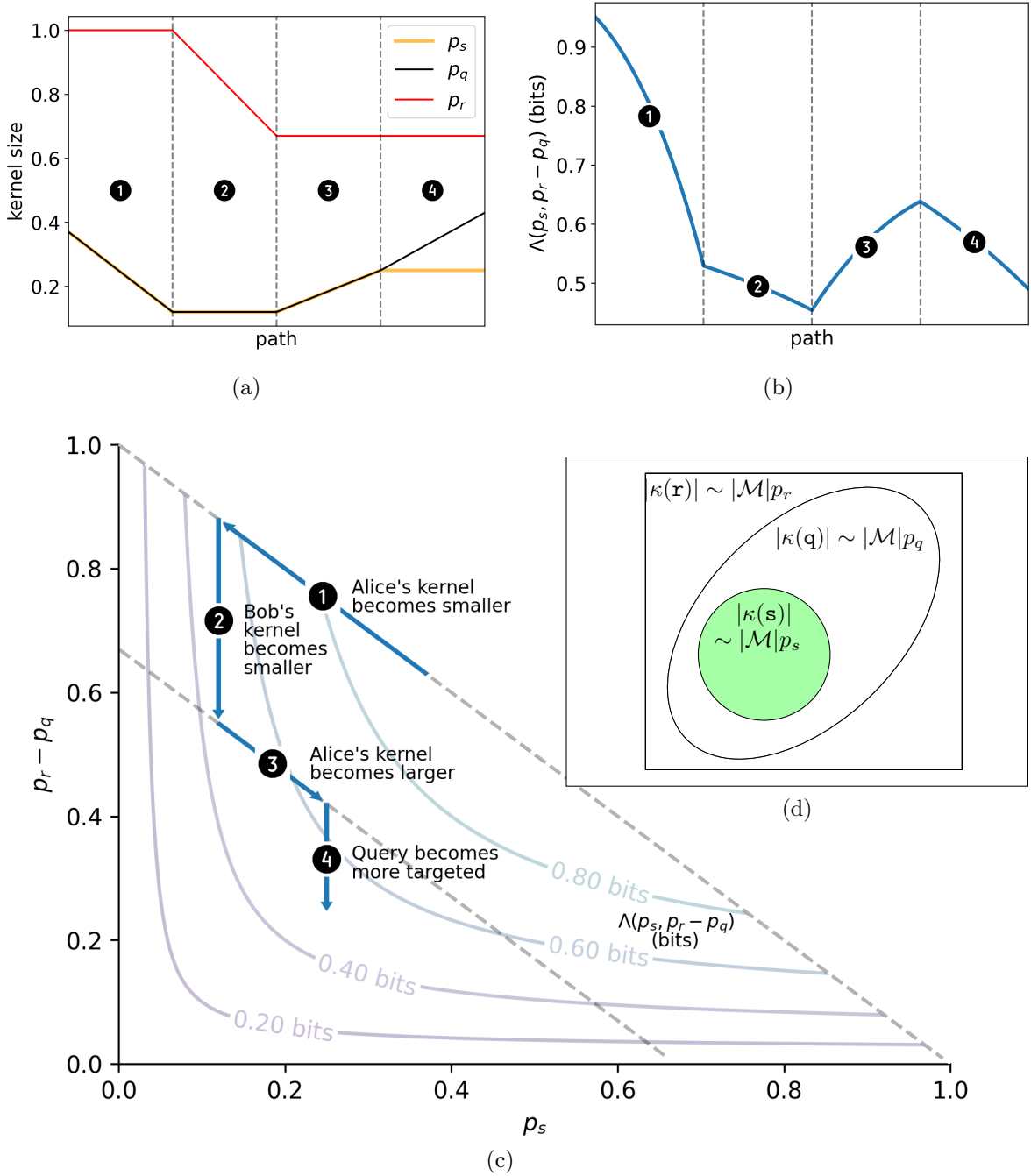
\begin{figure}
\begin{tblr}{Q[c,m]Q[c,m]}
 \scalebox{0.5}{\includegraphics{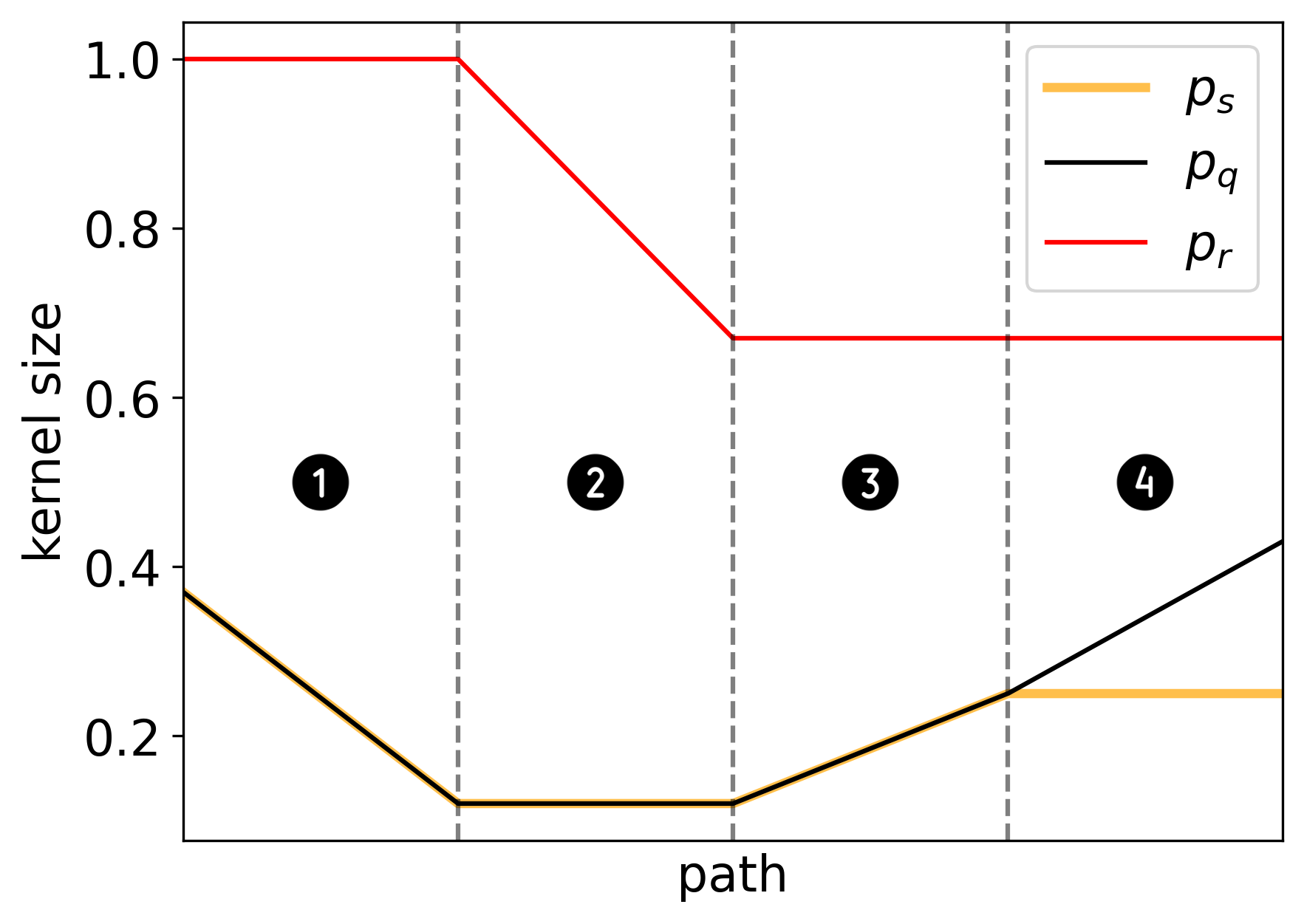}} & \scalebox{0.52}{\includegraphics{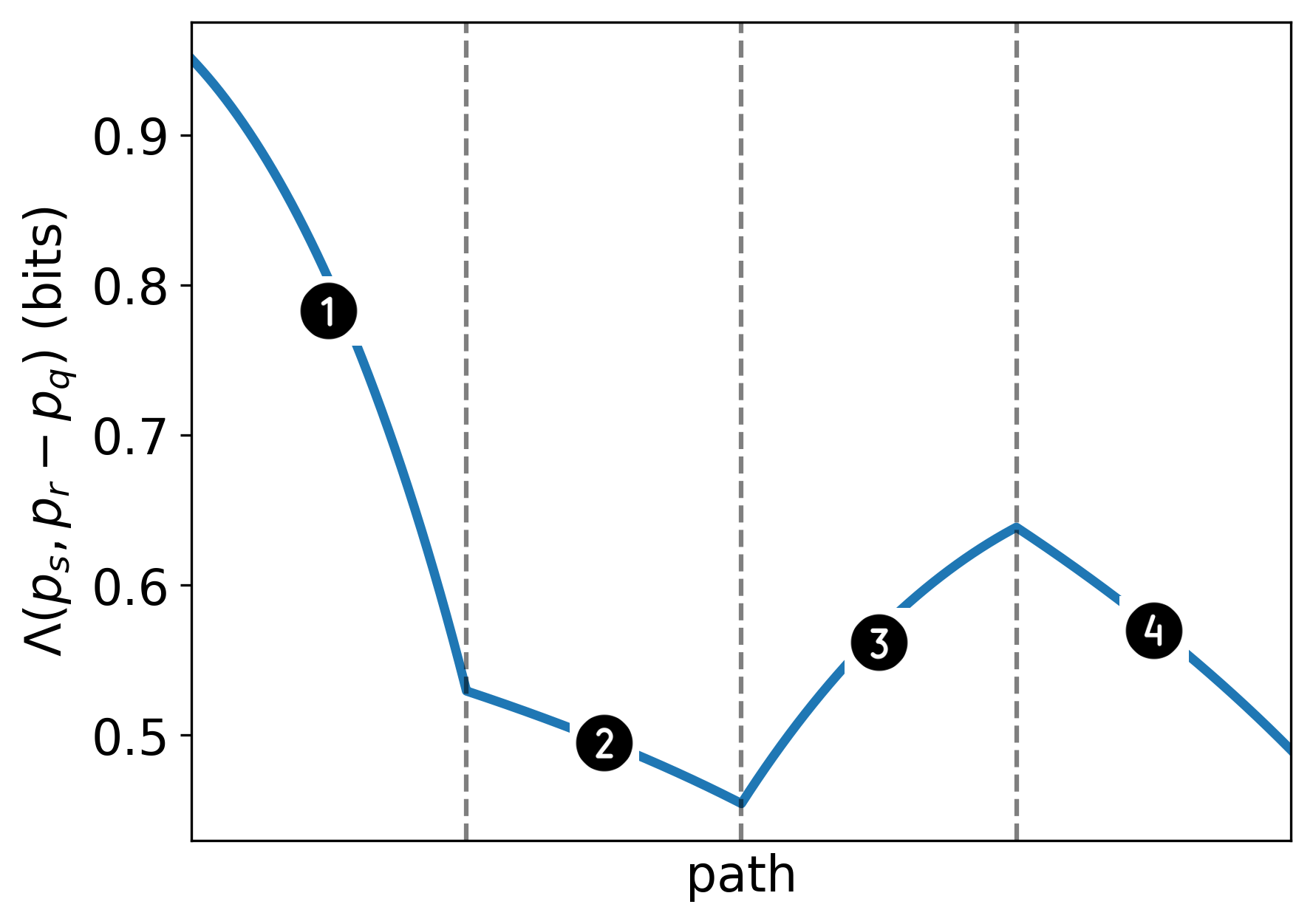}}  \\ 
 (a) & (b) \\
    \SetCell[c=2]{c} 
    
\begin{tikzpicture}
\node (myfirstpic) at (0,0) {\includegraphics{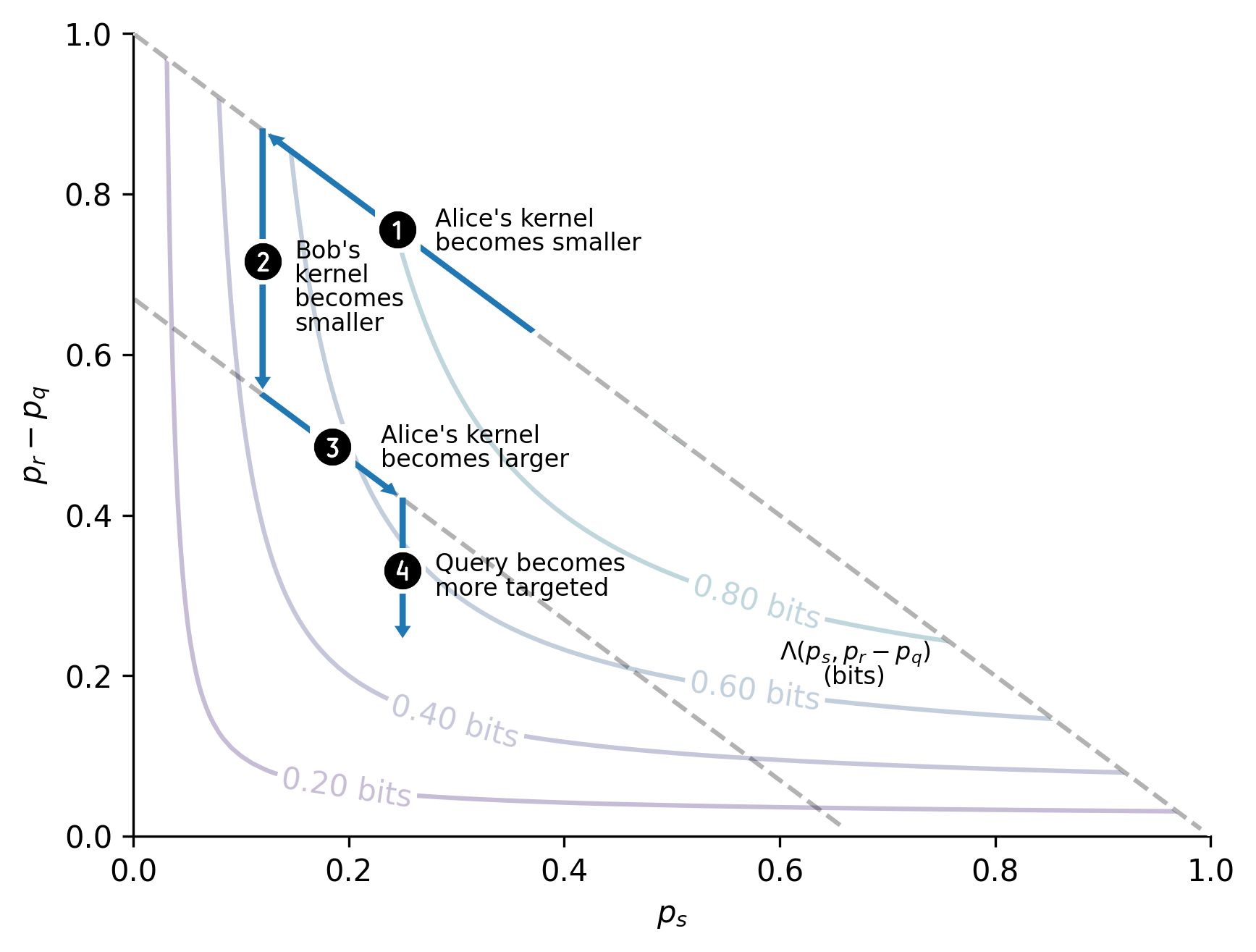}};
\begin{scope}
    \clip (0,0) rectangle (8,5);
\draw (1,0)[color=black,fill=white] rectangle (8,5);
\draw(2.2,0.25) rectangle(7, 4.7500);
\draw(5,-1.5)[rotate=45,fill=white] ellipse(2.5 and 1.5);
\begin{scope}
\clip(2.2,0.25) rectangle(7, 4.75);
\draw(2.2,0.25) rectangle(7, 4.75);
\draw(3.3,4.5) node {$|\algebraicset{\receiver}|\sim |\kernelspace| p_r$};
\draw(5,-1.5)[rotate=45,fill=white] ellipse(2.5 and 1.5);
\draw(5.3,3.75) node {$|\algebraicset{\query}| \sim |\kernelspace| p_q$};
\draw(4,2)[fill=green!35!white] circle(1.0);  
\draw(4.0,2.6) node {$|\algebraicset{\sender}|$};
\draw(3.9,2.2) node {$\sim |\kernelspace| p_s$};
\end{scope}
\draw(2.2,0.25) rectangle(7, 4.75);
\end{scope}

\draw(0,-5.7) node {(c)};
\draw(4.5,-0.36) node {(d)};

\end{tikzpicture}    
\end{tblr}

       \caption{An illustration of several scenarios where Alice's knowledge entails that of Bob's. The contours are those of $\Lambda(p_s,p_r-p_q)$. As the parameters $p_s,p_q, p_r$ are changed (a), the bound changes in value (b), also seen as as paths on a contour plot (c). The general relationship between the kernels is illustrated in (d). In all cases, Alice need not know $\receiver$ for the result to hold.}
\label{fig:cool_figure}
\end{figure}

For the following discussion, the reader is referred to Figures~\ref{fig:summary}, \ref{fig:summary_2} and~\ref{fig:cool_figure}. The first figure is in essence a case-by-case expansion of Figure~\ref{fig:general_diagram}, which explicitly links each case of interest to its assumptions and the corresponding Theorem that treats it. While in Figure~\ref{fig:summary} we emphasize the end-to-end nature of our setup, where Alice and Bob experience sentences from the language $\surfacespace$ and where Bob ends up with an updated sentence after the communication takes place, much of our work is predicated on the analysis of the underlying kernels. Figure~\ref{fig:summary_2} illustrates the assumptions of each notable result as set relations between the various kernels involved in the communication setup, including kernels that each of Alice and Bob know separately at the time of communication as well as the kernel that Bob has in his hands after the communication takes place. Figure~\ref{fig:cool_figure} elaborates upon Figure~\ref{fig:lambda} by replacing the $3$-dimensional plot with a contour plot and illustrating how changes in the $p_s, p_q, p_r$ values result in different theoretical bounds based on $\Lambda(p_s,p_r-p_q)$. The point of this last figure is to ``put it all together'' -- in spite of the different set of assumptions that we will walk the reader through next, at the end all the results can be expressed in terms of $\Lambda(p_s,p_r-p_q)$.

\subsection{Full ignorance -- Theorem \ref{thm:logicinfo}}
In this setup (Figures~\ref{fig:summary}-a, \ref{fig:summary_2}-a) the goal is for Bob to be able to prove any mathematical sentence that Alice can prove, whilst Bob has access to no logic sentence, and hence the reference to ``full ignorance'', understood here as Bob's state relative to Alice's knowledge. A single parameter determines the results that we have in this scenario, namely the normalized expected kernel size of Alice's sentence, denoted by $p_s = |\kernelspace|^{-1} E | \algebraicset{\senderrnd}|$. In the context of our more general result (Theorem \ref{thm:aggregate}), this scenario corresponds to the setting where $p_r=1, p_q = p_s$. In Figure~\ref{fig:cool_figure} we illustrate a contour plot of $\Lambda(p_s,p_r-p_q)$; full ignorance is then represented by the top-most negative 1 slope line. We note that $\Lambda(p_s,1-p_s) = \Hbin{p_s}$, and thus this result agrees with the intuition that the optimal cost in this case is the entropy of Alice's kernel.  A smaller kernel, in our setup, is associated with a more informative logic sentence, since it has ruled out more of $\kernelspace$ as impossible. However, for $p_s < 0.5$, smaller kernels are in fact \emph{cheaper} to send, contradicting the intuition that they somehow correspond to ``more knowledge''. The opposite happens nonetheless for $p_s > 0.5$, where smaller kernels are indeed more expensive to send. The lesson here is that one should not necessarily equate the notion of the amount of knowledge facts with information bits.

We stress that this result, in and by itself, is not particularly surprising given how we have defined the Logic System and the kernel of a sentence, but it is a useful baseline to understand our general result as well as our proof techniques. 

We also reinforce that  $\Lambda(p_s,1-p_s)$ \emph{is loose} when the special additional i.i.d.\ conditions in the theorem are not met. In fact Theorem \ref{thm:logicinfo} includes a generally tight bound which states that, not surprisingly,
\begin{eqnarray}
 \Hreg{\algebraicset{\senderrnd}}
\label{eq:tight-bound}
\end{eqnarray}
is the ultimate compression bound; this is to be connected to our early discussion leading to (\ref{eq:alternate}). This type of strong, ultimate bound calculation is presently not provided for the rest of the Theorems in an effort to emphasize so-called ``single letter results'', such as the bound given by $\Lambda(p_s,1-p_s)$, which are tight under special conditions and which are often held in special esteem in the information theory field as they are much simpler to state and reason about, and thus yield more early insight. 

\subsection{Partial ignorance -- Theorem \ref{thm:background_log_info}}
A straightforward way to improve upon the full-ignorance setting is to assume that at the time of communication a sentence $\receiver$ is revealed to both Alice and Bob, in addition to $\sender$ being revealed to Alice only.  This is represented by the two new arrows in Figure~\ref{fig:summary}-b as well as the the new kernel with a rectangular shape in Figure~\ref{fig:summary_2}-b. Note that this rectangular kernel shows up in both the sender and receiver diagrams, as it is available to both during the communication act. As disclosed earlier, we assume that $\sender \entails \receiver$ and thus $\receiver$ does not allow Alice to prove any more sentences than she could with $\sender$ alone, however it gives her significant context to what Bob is aware of, thus reducing the total cost of communication. In this setting, two parameters determine the scenario: $p_s \leq p_r$, and in the more general context of Theorem \ref{thm:aggregate}, the additional condition is that $p_q=p_s$, since it is still the goal for Bob to be able to prove anything that Alice can. In Figure~\ref{fig:cool_figure} this scenario is illustrated with negative 1 sloped lines that are strictly below the top-most such line.

In the spirit of highlighting Bob's state with respect to that of Alice's, we say that Bob is partially ignorant. The corresponding bound in this scenario is $\Lambda(p_s,p_r-p_s)$. We remind the reader that $\Lambda$ is monotonically increasing on either of its two arguments, and therefore as $p_r$ decreases while keeping $p_s$ fixed, the bound strictly decreases. As we discussed earlier, a smaller kernel is associated with a more informative logic sentence and thus unlike in the full ignorance case, in this case the result does agree with intuition: the more informative is the logic sentence that is shared by Alice and Bob, the lower the communication cost.

\subsection{Less is More -- Theorem \ref{thm:less_is_more_simple}}
\label{ss:less_is_more}
For this scenario, we return to the full-ignorance setting, but add a twist: the goal is not for Bob to prove all that Alice can, but rather, to prove a more targeted query $\query$ that can be derived from Alice's $\sender$, but in general is not logically equivalent to $\sender$. In the context of Theorem \ref{thm:aggregate}, this scenario corresponds to the scenario $p_q > p_s$ and $p_r=1$, with the bound being $\Lambda(p_s,1-p_q)$. We introduce it in  Figure~\ref{fig:summary}-c with a query $\query$ that is given to Alice at the time of communication, but not to Bob. Correspondingly, in Figure~\ref{fig:summary_2}-c we introduce an oval shaped kernel which includes that of Alice's logic sentence, but which is unavailable to Bob. 

\begin{figure*}[!ht]
\begin{center}
\scalebox{0.75 }{
\begin{tikzpicture}
    \node (mysecpic) at (0,0) {\includegraphics{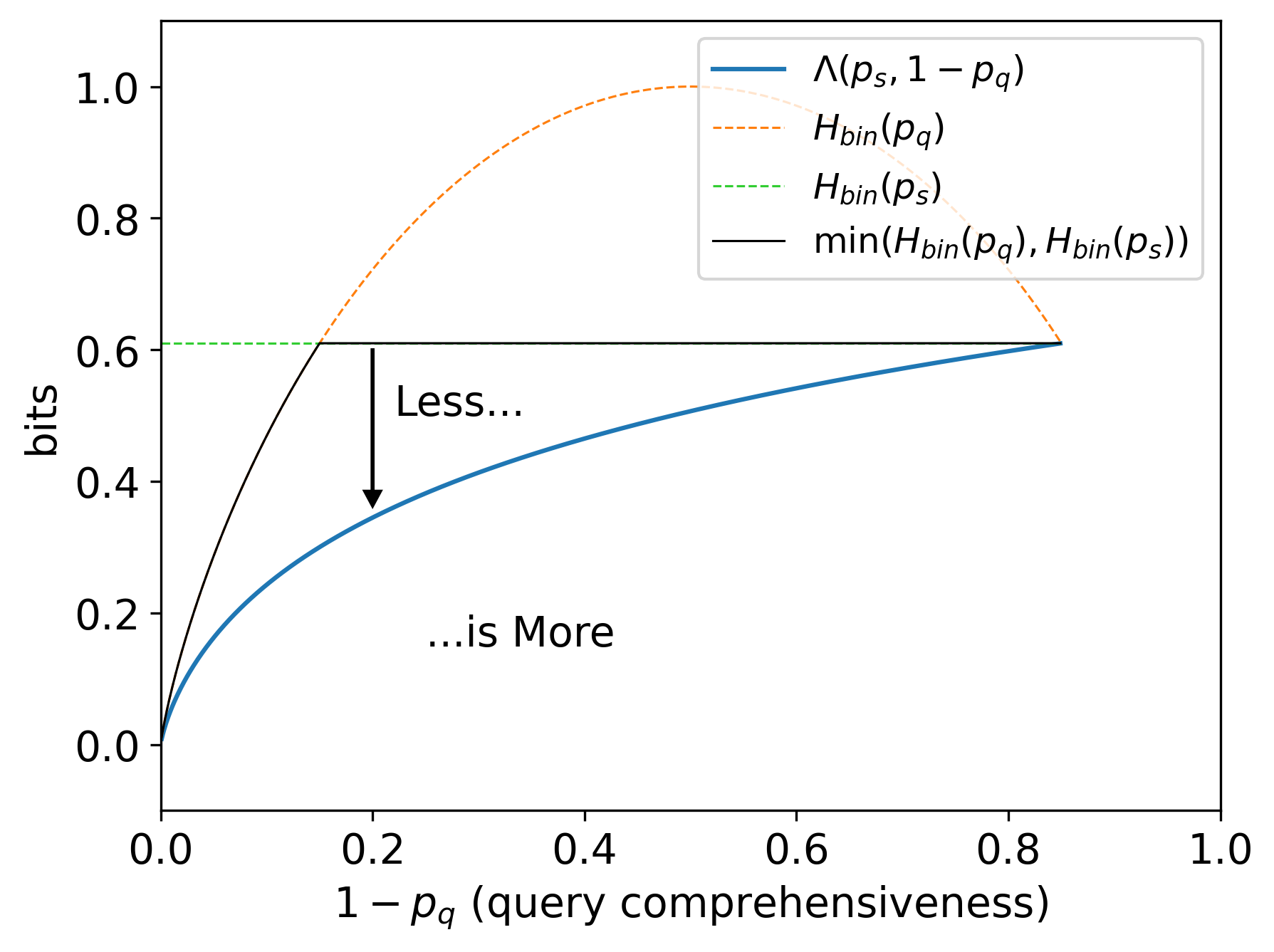}};
    \node (myfirstpic) at (3.5,-1.8) {\scalebox{1.6}{\tree{no_r}{no_q}{shat_arrow}{no_s}}};
\end{tikzpicture}}
\end{center}
\caption{In blue, the ultimate communication limit $\Lambda$  for the case $p_r=1$, as the query ranges from trivial ($p_q=1$) to coinciding with the sender's information ($p_q=p_s=0.15$). $\Lambda$ is cheaper (Less...) than the two obvious strategies, yet the kernel size received by Bob is smaller than that of the query, showing Bob can prove even more things (is More...) than required. A similar picture will hold for any $p_r$. }
\label{fig:less_is_more_paradox}
\end{figure*}

The reader's first instinct may be, why don't we either send the kernel of $\query$ or the kernel of $\sender$, whatever is least expensive? The normalized average cost for this strategy is
\begin{eqnarray}
    \min\{\Hbin{p_s},\Hbin{p_q}\} > \Lambda(p_s,1-p_q) ,
\end{eqnarray}
where this last inequality is a consequence of Lemma \ref{lem:elementarylambda}. Thus this strategy is strictly speaking suboptimal. We give the insight as to why. In reference to the third column of Figure~\ref{fig:summary_2}-c, notice that any $\senderhat$ with the property that
\begin{eqnarray}
    \algebraicset{\sender} \subseteq \algebraicset{\senderhat} \subseteq{\algebraicset{\query}}
\end{eqnarray}
will allow Bob to prove $\query$ with an $\senderhat$ with the property that $\sender \entails \senderhat$; this is a consequence of Definition \ref{def:core} of a Logic System. Thus  Alice has many more options to meet this goal than sending the kernel of $\query$ or that of $\sender$, and it is possible to create an efficient listing of those options to cover all the possibilities for $\query$ and $\sender$. We illustrate this in Figure~\ref{fig:less_is_more_paradox}, which illustrates what we call the ``less is more'' paradox. Notice that for Bob to be able to prove $\query$, the bit cost was smaller than either sending the kernels of $\query$ or $\sender$ (Less...). But notice that in general $\algebraicset{\senderhat}$ may be a strict subset of $\algebraicset{\query}$. Because it is still the case that $\algebraicset{\sender} \subseteq \algebraicset{\senderhat}$, it follows that Bob is able to prove even more facts than he needed to prove  (...is more). 

Of note, this also has potential implications for security -- being as efficient as one can to allow Bob to prove $\query$ using facts consistent with Alice's $\sender$ results in revealing more than $\query$. One may say that one needs to say more to say less.

\subsection{No need to know -- Theorem \ref{thm:slepian_wolf_sender_unaware}}
\label{ss:no_need_to_know}
We now return to the Partial-ignorance setting, but eliminate Alice's ability to directly observe $\receiver$ (see the  difference between Figures~\ref{fig:summary}-b and~\ref{fig:summary}-d). Since Alice doesn't know $\receiver$, she cannot use the strategy that we described under Partial ignorance which leverages the kernel of $\receiver$, illustrated as a rectangle in the first and third columns of Figure~\ref{fig:summary_2}-b, in order to reduce the total bit cost (observe the absence of the rectangular kernel in the first column of Figure~\ref{fig:summary_2}-d). We stress that for simplicity reasons, Figure~\ref{fig:summary}-d 
only shows a single turn of communication where Alice is the sender and Bob is the receiver. In our work, the communication pattern is more complex -- multiple turns are allowed. To keep evaluation as consistent as possible, the total sum of the average bits exchanged in any direction is the figure of merit in this setup.

The surprising result here is that exactly the same achievable limit as in the case of Partial ignorance applies -- i.e., $\Lambda(p_s,p_r-p_s)$ -- hence the ``Alice does \emph{not have a need to know}'' reminder in the title of this subsection. The main proof mechanism borrows from the theories of Slepian/Wolf \cite{slepian_wolf:coding} and Wyner/Ziv \cite{wyner_ziv:coding} the idea of hashing, which in this case is applied to the kernel of $\sender$; however, specialized arguments are introduced in this article that allow us to prove an upper bound under very general assumptions on how $\senderrnd, \receiverrnd$ are distributed, and a lower bound is also introduced that accounts for the potential of multiple turns as well. The details can be found in Section \ref{sec:proofs}.

\subsection{Misinformation -- Theorem \ref{thm:less_is_more_background}}
\label{ss:misinformation}

To discuss this subsection, we depart from the Partial-ignorance setup, and replace the assumption that $\sender \entails \receiver$ with an assumption that, instead, the sentences $\sender$ and $\receiver$ are logically inconsistent -- $\algebraicset{\sender} \cap \algebraicset{\receiver} = \emptyset$ -- see Figure~\ref{fig:summary_2}-e. 
Bob still wants to be able to prove all that Alice can, and thus we regard Bob as being in a state of misinformation, albeit in a highly cooperative situation.

\begin{figure}
\begin{center}
\scalebox{0.75}{\includegraphics{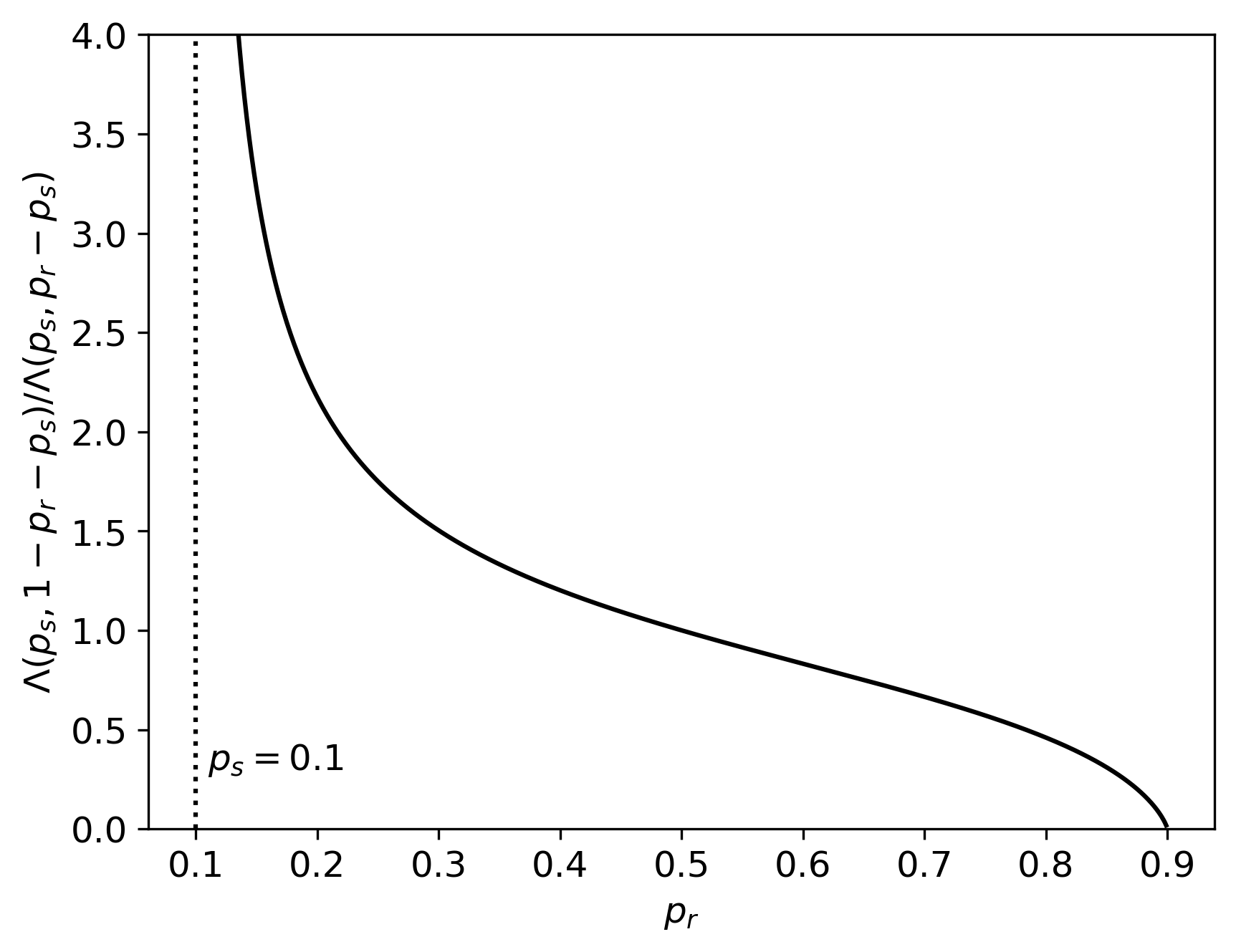}}
\end{center}
\caption{Illustration of the ratio of the cost of misinformation over ignorance in a cooperative setting, which grows as $p_r$ becomes smaller (Bob becomes more opinionated), and tends to infinity as $p_r \longrightarrow p_s=0.1$. }
\label{fig:ratio_plot}
\end{figure}

The fundamental limit in this situation is in fact quite easy to derive from the arguments for Partial ignorance, or as a direct consequence of the much more general setup of the subsequent Subsection~\ref{ss:general_alice_knows_bob}, and for this reason no separate  proof of it is provided in this article. The corresponding bound, under the assumption that $p_s \leq 1-p_r$, is given by
\begin{eqnarray*}
    \Lambda(p_s,1-p_r-p_s) ,
\end{eqnarray*}
which is quite intuitive since it simply replaces $p_r$ with $1-p_r$ in the bound for Partial ignorance. Some thought provoking ideas can be derived from the result above. Consider the ratio of the cost of misinformation vs.\ ignorance
\begin{eqnarray*}
    \frac{\Lambda(p_s,p_r-p_s)}{\Lambda(p_s,1-p_r-p_s)} ,
\end{eqnarray*}
which makes sense only under the restrictions $p_s \leq 1-p_r, p_s \leq p_r, p_s \leq 1/2$. In Figure~\ref{fig:ratio_plot} we plot this ratio for the case $p_s=0.1$. Note that the curve is monotonically decreasing with increasing $p_r$. Also note that as $p_r \longrightarrow p_s=0.1$ from the right, the curve diverges to infinity. We may then colloquially state that, in a cooperative misinformation setting, the relative cost of correcting misinformation vs.\ correcting ignorance grows unbounded as the receiver becomes more opinionated. This agrees with one's intuition of what we would expect should happen.

\begin{figure}
\begin{center}
\scalebox{0.9}{
\includegraphics{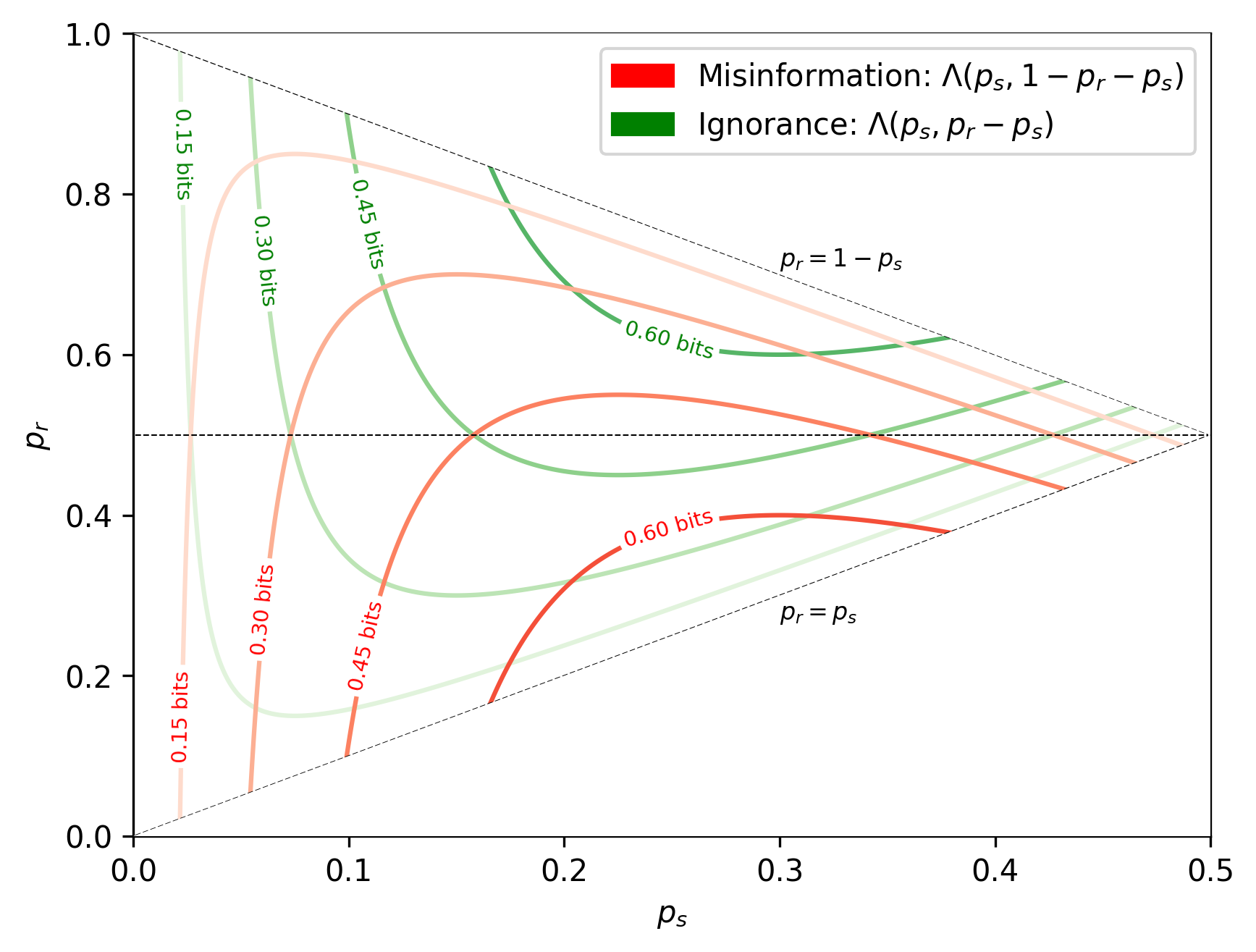}}
\end{center}
\caption{Contour plots for $\Lambda(p_s,p_r-p_s)$ (ignorance) and $\Lambda(p_s,1-p_r-p_s)$ (misinformation). The plot accentuates the symmetry of these around $p_r=1/2$; what is cheap for ignorance is expensive for misinformation and vice versa.   }
\label{fig:ignorant_vs_misinformed}
\end{figure}

Figure~\ref{fig:ignorant_vs_misinformed} illustrates contour plots of $\Lambda(p_s,p_r-p_s)$ and $\Lambda(p_s,1-p_r-p_s)$, under the restriction $p_s \leq p_r$, $p_s \leq 1-p_r$, $p_s \leq 1/2$. These functions are obviously symmetric under the transformation $p_r \leftrightarrow 1-p_r$.  Notice that, as the cost of misinformation is kept constant and one approaches  the regime where Bob is highly opinionated ($p_r$ close to $p_s$), one ``cuts'' through contours for the ignorance case that are ever decreasing in bit cost value. This 
helps explain the unbounded growth shown in Figure~\ref{fig:ratio_plot}.

Notice also that exactly the same behavior occurs with roles reversed in the upper part of Figure~\ref{fig:ignorant_vs_misinformed}. Even though near the upper part (the line $p_r=1-p_s$) one might feel tempted to regard $\receiver$ as the most uninformative for a given $p_s$, this in fact is not true: in this case, the complement of the kernel of $\receiver$ is close to the kernel of $\sender$ (said differently, Bob can simply negate $\receiver$ and thus obtain a sentence that is close, logically, to $\sender$). Thus in reality the most uninformative sentences $\receiver$ are those associated with $p_r=1/2$.

\subsection{General setup when Alice knows what Bob knows -- Theorem \ref{thm:less_is_more_background}}
\label{ss:general_alice_knows_bob}
In this subsection, we present the most general result we have been able to obtain in the case where both Alice and Bob share knowledge of $\receiver$. For this setting, we drop the assumption $\sender \entails \receiver$; correspondingly this is shown as stricken in
illustrated in Figure~\ref{fig:summary}-e. We do allow for any query $\query$ with the property that it is provable using $\sender$ ($\sender \entails \receiver$).

The additional ways in which various kernels may relate to each other when $\sender \entails \receiver$ is dropped (but we keep $\sender \entails \query$) are illustrated in Figure~\ref{fig:summary_2}-e, where the rectangular kernel may only be partially overlapping the kernel of $\sender$ and the kernel of $\query$. Correspondingly, the result is more complex, with a sum of two terms involving the $\Lambda$ function. The reader is not expected to immediately understand how to interpret these bounds, since additional notation has been introduced that is only discussed in Section \ref{sec:proofs}. Having said this, in spite of all its apparent complexity, for the upper bound all that is really happening here is that two communication paths are being established: one to address ignorance, and one to address misinformation, and by themselves, these actually do not introduce fundamentally new ideas beyond those already introduced in our other results. This result completely subsumes the result on misinformation described in Subsection~\ref{ss:misinformation}.

\subsection{General setup when Alice does not know what Bob knows -- Theorem \ref{thm:master}}

To complete our set of results, in this subsection we present a result where the conditions $\sender \entails \query$, $\query \entails \receiver$ (and hence $\sender \entails \receiver$) are assumed. Critically, we do not assume that Alice knows $\receiver$. Thus this setting can be seen as a general result that subsumes those of Subsections~\ref{ss:less_is_more} and \ref{ss:no_need_to_know}. The mathematical aspects of this result are the most complex in the paper, incorporating every technique we developed elsewhere. Yet, one should be cautious in interpreting the practical significance of this result beyond what we have already argued in Subsections~\ref{ss:less_is_more} and \ref{ss:no_need_to_know}. We have already made this observation in the discussion subsequent to the sentence of Lemma \ref{lem:sufficiencyqentailsr}; in a more practical version of this setting, the assumption $\query \entails \receiver$ is dropped altogether. We conjecture that in that case, the estimate $\Lambda(p_s,p_r-p_q)$ is too low -- the actual bit cost is in effect higher.

\begin{figure}
\scalebox{0.85}{
\input{all_comm_diagrams}}
\caption{ 
Communication diagrams for the scenarios covered in this article. The circle stands for the knowledge that the receiver has after the transmission has taken place. The diamond represents a computational device that given $\hat{\sender}$ and a query $\query$, is capable of producing a proof of $\query$ as long as $\hat{\sender} \entails \query$. Throughout all the diagrams, we assume that if $\query$ is such that $\sender \entails \query$, then $\sender \entails \senderhat$ and $\senderhat \entails \query$. For (b,d,f) we additionally assume that  $\sender \entails \receiver$ and $\query \entails \receiver$; these assumptions are crucially omitted in (e). }
\label{fig:summary}
\end{figure}

\begin{figure}
\begin{center}
\scalebox{0.85}{
\begin{tblr}{
  colspec = {cX[c]X[c]X[c]X[c,m]},
  stretch = 0,
  rowsep = 3pt
}
& Sender (before) & Receiver (before)  & Receiver (after) & Achievable limit \\ \hline
(a) & \tree{no_r}{no_q}{no_shat}{s} & \tree{no_r}{no_q}{no_shat}{no_s} & \tree{no_r}{no_q}{shat_normal}{no_s} & {$\Lambda(p_s,1-p_s)$\\ Theorem \ref{thm:logicinfo} \\ (full ignorance)} \\
(b) & \tree{r}{no_q}{no_shat}{s} & \tree{r}{no_q}{no_shat}{no_s} & \tree{no_r}{no_q}{shat_normal}{no_s} & {$\Lambda(p_s,p_r-p_s)$ \\ Theorem \ref{thm:background_log_info} \\ (partial ignorance) }  \\
(c) & \tree{no_r}{q}{no_shat}{s} & \tree{no_r}{no_q}{no_shat}{no_s} & \tree{no_r}{no_q}{shat}{no_s} & {$\Lambda(p_s,1-p_q)$ \\ Theorem \ref{thm:less_is_more_simple} \\ (less is more)} \\
(d) & \tree{no_r}{no_q}{no_shat}{s} & \tree{r}{no_q}{no_shat}{no_s} & \tree{no_r}{no_q}{shat_normal}{no_s} & {$\Lambda(p_s,p_r-p_s)$ \\ Theorem \ref{thm:slepian_wolf_sender_unaware} \\ (no need to know) }\\
(e) & \tree{r_misinf}{no_q}{no_shat}{s} & \tree{r_misinf}{no_q}{no_shat}{no_s} & \tree{no_r}{no_q}{shat_normal}{no_s} & {$\Lambda(p_s,1-p_r-p_s)$\\ Theorem \ref{thm:less_is_more_background} \\ (misinformation)} \\
(f) & \tree{r_partial}{q}{no_shat}{s} & \tree{r_partial}{no_q}{no_shat}{no_s} & \tree{no_r}{no_q}{shat}{no_s} & {$\Lambda(p_{s^*},p_r-p_{q^*}) + \Lambda(p_{s^{**}},1-p_r-p_{q^{**}})$ \\ Theorem \ref{thm:less_is_more_background} \\ (Bob's sentence may \\ not be entailed by Alice's)}\\
(g) & \tree{no_r}{q}{no_shat}{s} & \tree{r}{no_q}{no_shat}{no_s} & \tree{no_r}{no_q}{shat}{no_s} & {$\Lambda(p_s,p_r-p_q)$ \\ Theorem \ref{thm:master} \\ (general setup when\\Alice does not know\\what Bob knows)}
\end{tblr}}
\end{center}
\caption{Kernel relationships for the communication scenarios considered in this paper. }
\label{fig:summary_2}
\end{figure}

\section{Practical algorithms and experimental results}
\label{sec:practical}

In this section, we develop two practical algorithms that are  components of our proposed logical semantic communication system, aimed at the targeted query scenario of Theorem \ref{thm:less_is_more_simple} and the scenario where Alice does not know what Bob knows addressed in Theorem \ref{thm:slepian_wolf_sender_unaware}. These algorithms are based on linear coding concepts over the Galois Field with two elements $\{0,1\}$, denoted $GF(2)$, and are shown to be optimal on specific settings.  We then present experimental results comparing the performance of an ensemble of methods, including the ones developed here, and contrast them with the $\Lambda$ bound as well as optimized methods that treat the logic expressions as strings in the classic information-theoretic sense.

\subsection{Linear codes for targeted queries}
\label{ss:linear_targeted}
Let $X_1^n \in \{0,1,\dontcare\}^n$ denote a random vector that represents three sets: $\{i : X_i = 0\}$, $\{i : X_i = 1\}$ and $\{i : X_i = 2\}$.  The problem is to transmit to a receiver a vector $\hat{X}_1^n \in \{0,1\}^n$ which agrees with $X_1^n$ on all the positions $\{i : X_i \in \{0,1\}\}$ as efficiently as possible, where efficiency is quantified as expected number of bits transmitted. One way to see this is that we are trying to efficiently send a partition that splits the sets $\{i : X_i = 0\}$, $\{i : X_i = 1\}$. The connection to the problem of semantic logic communication for the targeted queries scenario (Theorem \ref{thm:less_is_more_simple}) emerges from the observation that sending a partition that splits the sets $\algebraicset{\senderrnd}$ and $\algebraicset{\queryrnd}^c$ is the key step in that problem.

We now present a general algorithm for partition compression based on linear codes over $GF(2)$. This algorithm can be applied even when we do not have any statistical model of the underlying sets for which we are creating a partition. In the special case where the entries of $X_1^n$ are drawn 
i.i.d.\
from $\{0,1,\dontcare\}$ using probability masses ($p_0,p_1,1- p_0-p_1$), we will show that this algorithm is asymptotically optimal if $p_0 = p_1$. The attractiveness of linear codes stems from the fact that they are easier to implement in practice.

For ease of analysis, we assume that the encoder and decoder share a random matrix with
i.i.d.\
entries drawn from $GF(2)$ uniformly at random, and  with $n$ columns and an infinite number of rows, which we shall refer to as $G$; nonetheless, the number of rows we will effectively be using is only slightly larger than $n(p_0 + p_1)$.  Define
\begin{eqnarray}
    \Psi = \{ i : X_i \in \{0,1\}\} .
\end{eqnarray}

Let $G_r$ denote the matrix obtained by extracting the first $r$ rows from $G$, and for any given vector $x \in \{0,1\}^n$ and set of indices $\xi \subset \{0,1,\ldots,n-1\}$, let $x_\xi$ denote the $|\xi|$-long vector obtained by extracting from $x$ only the indices given by $\xi$.  The algorithm starts by finding the smallest positive integer $J$ such that the equation 
\begin{eqnarray}
[M \cdot G_J]_{\Psi} = [X]_{\Psi}
\label{eq:want}
\end{eqnarray}
can be solved for some $M \in GF(2)^{1\times J}$.
Exploiting $\delta$ Elias coding~\cite{elias:integers}, the sender then sends the integer $J$ to the receiver using 
\begin{eqnarray*}
\len{\mbox{elias}_{\delta}(J)}
\end{eqnarray*}
bits, followed by the $J$ bits in the message $M$. The receiver then decodes $J$ and computes $M \cdot G_J$ to retrieve the partition.

We now analyze the expected performance of this algorithm:
\begin{eqnarray}
n^{-1} E_{J} \left[ \len{\mbox{elias}_{\delta}(J)}  \right] &\leq& E_{J} \left[ J + \log_2 J + 2 \log_2(\log_2 J )+ 3 \right] \nonumber \\ 
&=& n^{-1} E_{\Psi} \left[ E_{J} \left[ J + \log_2 J + 2 \log_2(\log_2 J) + 3 | \Psi  \right]    \right] \nonumber \\
& \leq & n^{-1} E_{\Psi} \left[ E_{J}\left[ J | \Psi \right]  +  \log_2 E_{J}\left[ J | \Psi\right]  + 2 \log_2( \log_2 E_{J}\left[J | \Psi \right]) + 3\right] .  \label{eq:cont}
\end{eqnarray}
We next upper bound $E_{J} \left[ J | \Psi \right]$. A sufficient condition to be able to solve Equation (\ref{eq:want}) is that $[G_J]_{\Psi}$ has full-row rank, that is, the dimension of the space spanned by the rows of $[G_J]_{\Psi}$ is exactly $|\Psi|$. 

Let $W_i$ be the smallest integer such that the row subspace spanned by $[G_{W_i}]_{\Psi}$ has dimension $i$. We can then write
\begin{eqnarray}
    W_{|\Psi|} = W_1 + \sum_{i=1}^{|\Psi|-1} W_{i+1} - W_{i} .
\end{eqnarray}
The probability that a vector drawn uniformly from $GF(2)^{|\Psi|}$ is nonzero, and therefore spans a space of dimension 1, is $1 - 2^{-|\Psi|}$; hence,
\begin{eqnarray}
    E\left[ W_1 | \Psi \right] = \frac{1}{1-2^{-|\Psi|}} .
\end{eqnarray}
We now analyze the difference $E[W_{i+1} - W_i|\Psi]$. Note that, by definition, $[G_{W_i}]_{\Psi}$ spans a subspace of dimension $i$, which must consist of exactly $2^i$ elements. If one chooses, uniformly at random, an element from $GF(2)^{|\Psi|}$, the probability that it lies within the subspace spanned by $[G_{W_i}]_{\Psi}$ is $2^{i-|\Psi|}$. As a consequence, the probability that the subspace spanned by the rows of $[G_{W_i}]_{\Psi}$ with such a random vector having dimension $i+1$ is $1-2^{i-|\Psi|}$, and thus
\begin{eqnarray}
  E [  W_{i+1} - W_{i} | \Psi ] = \frac{1}{1-2^{i-|\Psi|}}.
\end{eqnarray}
Observing that  $J \leq W_{|\Psi|}$, we obtain
\begin{eqnarray*}
E_J\left[ J | \Psi \right] \leq \sum_{i=0}^{|\Psi|-1} \frac{1}{1-2^{i-|\Psi|}} \leq |\Psi| + 2 .
\end{eqnarray*}
Continuing from inequality (\ref{eq:cont}), we obtain an upper bound on performance as
\begin{eqnarray*}
n^{-1} E_{\Psi} \left[ |\Psi| + \log_2 ( | \Psi| + 2 ) + 2 \log_2\left( \log_2 ( | \Psi | + 2 )\right) + 5 \right]  .
\end{eqnarray*}
To understand how good this bound is, we assume that the entries of $X_1^n$ are drawn 
i.i.d.\
from $\{0,1,\dontcare\}$ using probability masses ($p_0,p_1,1- p_0-p_1$) and therefore $E\left[|\Psi| \right] = n(p_0 + p_1)$, which results in the following upper bound on average performance:
\begin{eqnarray}
p_0 + p_1 + \frac{\log_2\left( n(p_0+p_1) + 2 \right) }{n} + 2 \frac{\log_2\left(\log_2\left( n(p_0+p_1) + 2 \right) \right)}{n}  +\frac{5}{n} .
\label{eq:linear_code_perf}
\end{eqnarray}
Now assume that $p_0 = p_1$. The achievable Shannon limit for this setting is given by
\begin{eqnarray*}
\Lambda(p_0,p_1) = \Lambda(p_0,p_0) = 2 p_0 ,
\end{eqnarray*}
which is the same performance as in (\ref{eq:linear_code_perf}) asymptotically as $n$ grows. If $p_0 \neq p_1$, then $\Lambda(p_0,p_1) < p_0 + p_1$ and thus our linear code construction is not optimal.

\subsection{Nonlinear codes for targeted queries}
\label{ss:nonlinear}
We next present an example of a small nonlinear code for the case $n=6$ that is optimal. In this setting, we assume that $X_1^n \in \{0,1,2\}$ with exactly one entry of $X_1^n$ equal to 0, and exactly one entry equal to 1.

One way to send a partition that separates the (single entry) sets $\{i : X_i = 0\}$ and $\{i : X_i = 1\}$ is to send an integer in the set $\{1,2,3,4,5,6\}$ identifying the one element in, say, the first set. The cost of this is $\log_2 6$ bits. Alternately, one can scan each of the 4 codewords below to find one that matches with $X_0 X_1 X_2 X_3 X_4 X_5$ on the 0s and 1s, regarding the 2s as ``don't care'':

\begin{eqnarray}
    \begin{array}{ccccccc}
    \mbox{ sender vector ($X_i \in \{0,1,2\}$) } & X_0 & X_1 & X_2 & X_3 & X_4 & X_5 \\ \hline
       \mbox{first codeword}  & 0 & 0 & 0 & 1 & 1 & 1  \\ 
       \mbox{second codeword} & 0 & 1 & 1 & 0 & 1 & 0\\
       \mbox{third codeword}  & 1 & 0 & 1 & 1 & 0 & 0 \\
       \mbox{fourth codeword} & 1 & 1 & 0 & 0 & 0 & 1
    \end{array} .
\end{eqnarray}
We observe that any two columns of  this binary matrix contain at least one row with the pattern ``0 1'' and one row with the pattern ``1 0''.  For example, if $X_0=1, X_1=2, X_2=2, X_3=2, X_4=0, X_5=2$ the third and fourth codewords are valid codewords.

Therefore, one can always find one such codeword, which can be specified using $2 < \log_2 6$ bits.

This small example can in fact be extended easily. Notice that the columns of the matrix are exactly the set of all binary patterns with 2 ones (in the parlance of coding theory, the columns have weight 2), and thus this poses the interesting question of what the properties are of matrices whose column weights are a constant. One such property is easy to deduce, as stated in the following result.

\begin{restatable}[Constant column weight codes]{lem}{ccwc}
~Let $c$ be a $t \times n$ binary matrix where every column has exactly the same weight $w$, and any two columns are different. Then the result code partitions any two sets each comprising exactly one (but different) integer in the set $\{0,\ldots,n-1\}$.
\end{restatable}
\textbf{Proof.} Let $i,j$ be the indices of any two distinct columns of the matrix $c$. The problem is to demonstrate that there is a row $k$ such that $[c_{k,i},c_{k,j}] = [0,1]$ and that there is another row $k^{\prime}$ where $[c_{k^{\prime},i},c_{k^{\prime},j}] = [1,0]$. Supposing that neither of these conditions is true, then we deduce that  $[c_{k,i},c_{k,j}] \in \{ [0,0], [1,1] \}$ for all $0 \leq k < t$ and thus necessarily the two columns are identical, which contradicts the assumption of the lemma. Suppose that, say, the first condition is true, but not the second one. Then it must be the case that the second column indexed by $j$ has a strictly larger weight than the column indexed by $i$, which is also a contradiction of the assumptions in the lemma. The case where the second condition is true but not the first one is dealt with similarly. \qed

It is possible to obtain, for any given desired length $n$, a crude bound on the minimum number of rows $t$ in a $t \times n$ binary matrix $c$ that partitions two  sets  each comprising exactly one non-overlapping integer in $\{0,\ldots,n-1\}$. Take any one row of the matrix $c$, and assume it has $n_0$ zeros and $n_1$ ones. The number of patterns with exactly one 0, one 1, and the rest don't cares, that can be handled by any one row is at most $n_0 n_1 \leq n^2/4$. Therefore the entire matrix $t$ can handle at most $t n^2/4$ patterns. There are a total of $n(n-1)$ patterns that we need to handle, and therefore the following relation must always hold:
\begin{eqnarray*}
    t  \geq 4\left(1 - \frac{1}{n}\right) .
\end{eqnarray*}
Rounding up (since $t$ is an integer), we see that at least 4 rows are needed for any value of $n>1$, showing that our $4 \times 6$ code is optimal in this sense. This bound is obviously too loose for anything other than $n=6$.
\subsection{Linear codes for the the case Alice does not know what Bob knows}\label{ss:no_need}

Let $X_1^n, Y_1^n \in GF(2)^n$ represent two random vectors with the property that $\{i : X_i=1\} \subseteq \{i : 
 Y_i=1\}$. We assume that Alice knows $X_1^n$ but not $Y_1^n$ (other than the condition above), and that Bob knows $Y^n_1$. The goal is to efficiently transmit $X_1^n$ to Bob. The connection to the problem alluded to in the title of this subsection arises by identifying $X_1^n, Y_1^n$ with the kernels of $\senderrnd$ and $\receiverrnd$, respectively.

In what follows we show a practical method based on linear codes. Let $\Delta > 0$ be an integer that is a design parameter. For convenience, we assume that Alice and Bob share a matrix $G$ with dimensions $(n + \Delta) \times n$ and entries drawn uniformly and independently at random from $GF(2)$.

Let $\Psi = \{i : Y_i = 1\}$. The method starts with Bob transmitting to Alice the integer $|\Psi|$. At this point both Alice and Bob will keep only the first $|\Psi|+\Delta$ rows of $G$, denoted $G_{|\Psi|+\Delta}$. Alice sends to Bob the bits resulting from the multiplication $G_{|\Psi|+\Delta} X_1^n$, where $X_1^n$ is interpreted as a column vector.  Let $[G_{|\Psi|+\Delta}]_{\Psi}$ denote the matrix obtained by extracting from $G_{|\Psi|+\Delta}$ the columns implied by the indices $\Psi$; note that this matrix is computable by Bob but not Alice. Bob then attempts to solve the equation
\begin{eqnarray}
    [G_{|\Psi|+\Delta}]_{\Psi} z  = G_{|\Psi|+\Delta} X_1^n
    \label{eq:hashing_condition}
\end{eqnarray}
for a unique $z$. If such a unique $z$ exists, Bob can retrieve $X_1^n$ by making use of the fact that the only entries of $X_1^n$ that could possibly be equal to 1 must have an index contained in $\Psi$ and the fact that
\begin{eqnarray*}
    [X_1^n]_{\Psi} = z ,
\end{eqnarray*}
where $[X_1^n]_{\Psi}$ denotes the entries of $X_1^n$ subset to the indices in $\Psi$. Therefore $X_1^n$ can be recovered from $z$ by lifting the latter using the indices $\Psi$.

By construction, (\ref{eq:hashing_condition}) has at least one solution. If the $|\Psi|$ columns of $[G_{|\Psi|+\Delta}]_{\Psi}$ are linearly independent, then that solution must be unique. The probability that these columns are linearly independent is given by $\Pi_{i=0}^{|\Psi|-1} ( 1 - 2^{i-|\Psi|-\Delta})$ and can be lower bounded in this manner:
\begin{eqnarray*}
    \Pi_{i=0}^{|\Psi|-1} ( 1 - 2^{i-|\Psi|-\Delta}) &=& \exp\left( \sum_{i=0}^{|\Psi|-1} \log ( 1 - 2^{i-|\Psi|-\Delta})  \right) \\
    & \geq & \exp\left( \sum_{i=0}^{|\Psi|-1} - \frac{2^{i-|\Psi|-\Delta}}{1-2^{i-|\Psi|-\Delta}}  \right) \\
        & = & \exp\left( \sum_{i=1}^{|\Psi|} - \frac{2^{-i-\Delta}}{1-2^{-i-\Delta}}  \right) \\
    & \geq & \exp\left( \sum_{i=1}^{|\Psi|} - 2^{-i-\Delta+1}  \right) \\
        & \geq & \exp\left( - 2^{-\Delta+1} \sum_{i=1}^{|\Psi|} 2^{-i}  \right) \\
    & \geq & \exp\left( - 2^{-\Delta+1}  \right) .
\end{eqnarray*}
Bob can signal to Alice success or failure in finding a unique $z$ with a single bit, and in the case of failure, Alice can simply send $X_1^n$ verbatim by sending $n$ bits.

The normalized expected number of bits transmitted in either direction is then upper bounded as follows:
\begin{eqnarray*}
    \lefteqn{\left(1 -\exp\left( - 2^{-\Delta+1}  \right) \right) + \frac{1}{n} \left( E\left[ |\Psi| \right] + \Delta + E \left[ \len{\mbox{elias}_{\delta}(|\Psi|)} \right] + 1 \right) } \\
    &\leq & 2^{-\Delta+1} + O\left( 2^{-2 \Delta} \right) + \frac{1}{n} \left( E\left[ |\Psi| \right] + \Delta +  \log_2 E\left[ | \Psi | \right] + \log_2 \log_2 E\left[| \Psi |\right] + 4 \right) .
\end{eqnarray*}
Define $q_1 \stackrel{\Delta}{=} n^{-1} E\left[ |\Psi| \right]$. With this definition, then the upper bound may be summarized as
\begin{eqnarray}
    q_1 + O\left( \frac{\log_2 n}{n} \right) + 2^{-\Delta + 1} + O\left( 2^{-2 \Delta} \right) ,
\label{eq:upper_bound_linear_code_alice_does_not_know}
\end{eqnarray}
where the $O(\cdot)$ terms are to be interpreted with respect to the $n \rightarrow \infty$ and $\Delta \rightarrow 0$ limits, respectively. 

To understand how good the above bound is, assume temporarily that $Y_1^n$ has entries drawn i.i.d.\ from $\{0,1\}$ using probability masses $\{1-q_1,q_1\}$, and that each $X_i$ is drawn using the conditional distribution
\begin{eqnarray*}
    P(X_i = 1 | Y_i = 1)  &=& p_1/q_1 , \\
    P(X_i = 0 | Y_i = 0 ) &=& 1.
\end{eqnarray*}
Under these assumptions for $X_1^n,Y_1^n$, a lower bound is given by $n^{-1} H(X_1^n | Y_1^n) = q_1 \Hbin{p_1/q_1}$. Comparing this lower bound to (\ref{eq:upper_bound_linear_code_alice_does_not_know}), we see that if $q_1 = 2 p_1$, $n$ is sufficiently large and $\Delta$ is sufficiently small, then the linear coding algorithm can arbitrarily approach the lower bound. For other choices of $q_1, p_1$, nonetheless the algorithm is not optimal.

\subsection{Experimental setup}

The goal of this section is to illustrate the possible gains that one may expect from a practical semantic communication system compared to a classical one, and to also show the existing gap between such semantic communication system with respect to the ultimate bound given by $\Lambda$.

A first problem is the fact it is possible to portray classical communication systems to be nearly arbitrarily inefficient when compared to semantic ones. The reason is the multiple different ways in which sentences can express the same underlying semantic content; classic compression systems must be faithful to the original sentence itself whereas semantic systems as regarded in this article can take advantage of the intended meaning of the symbols within the sentence. To illustrate this problem, note that  given $H(\algebraicset{\senderrnd}|\senderrnd)=0$,
\begin{eqnarray}
    H(\senderrnd) = H(\senderrnd| \algebraicset{\senderrnd}) + H(\algebraicset{\senderrnd}) \geq H(\algebraicset{\senderrnd}) .
\label{eq:gap}
\end{eqnarray}
The gap $H(\senderrnd | \algebraicset{\senderrnd})$ can be very large; consider, for example, propositional logic in two variables $X_1, X_2$ and note that 
\begin{center}
    $(X_1 \land X_2) \lor \lnot X_3$ \\
    $\lnot (\lnot (X_1 \land X_2) \land X_3)$ \\
    $\lnot ( (\lnot X_1 \lor \lnot X_2 ) \land X_3)$ \\
    $\lnot ( ( \lnot X_1 \land X_3 ) \lor ( \lnot X_2 \land X_3 ) )$ \\
    $\lnot ( ( \lnot X_1 \land X_3 ) \lor ( \lnot X_2 \land X_3 ) ) \land (X_1 \lor \lnot X_1)$
    $\lnot ( ( \lnot X_1 \land X_3 ) \lor ( \lnot X_2 \land X_3 ) ) \land (X_1 \lor \lnot X_1) \land (X_1 \lor \lnot X_1)\land (X_1 \lor \lnot X_1)\land (X_1 \lor \lnot X_1)\land (X_1 \lor \lnot X_1)$
\end{center}
are all logically equivalent sentences. To be clear, one can legitimately expect semantic communication systems to take advantage of the observation (\ref{eq:gap}), but one must exercise caution and not overstate this. Our paper's theoretical results not only leverage the phenomenon (\ref{eq:gap}), but go beyond and exploit more delicate findings on how targeted queries may admit unusually efficient representations or how, in some occasions, communication is surprisingly just as efficient when Alice doesn't know Bob's sentence $\receiverrnd$ as when she does know it.

Our experimental evaluation methodology follows the philosophy of always choosing a stronger ``classical'' (non-semantic) baseline to compare against whenever a choice is on the table, even if these baselines start to become more semantic in nature. In particular:
\begin{itemize}
    \item On purpose, we will forego the type of advantage that stems from observation (\ref{eq:gap}) even though we can legitimately claim it.
    \item Logic sentences will be optimized so that they can be represented more compactly by ``classical'' compressions systems.
\end{itemize}

\subsubsection{Scenarios}
\label{ss:scenarios}
We demonstrate two contrasting scenarios, derived from Theorem \ref{thm:less_is_more_background} and Theorem \ref{thm:slepian_wolf_sender_unaware}, respectively:

\begin{itemize}
    \item \textbf{(Targeted query with a shared statement)} After the communication, Bob will be able to prove $\queryrnd$, which satisfies $\senderrnd \entails \queryrnd$. Both Alice and Bob know $\receiverrnd$; only Alice knows $\senderrnd$; we simplify to the setting $\senderrnd \entails \receiverrnd$.
    \item \textbf{(Alice doesn't know what Bob knows)} After the communication, Bob will be able to prove all that Alice can. Only Bob knows $\receiverrnd$, and only Alice knows $\senderrnd$. We assume $\senderrnd \entails \receiverrnd$.
\end{itemize}

\subsubsection{Test case generation}

The reader shall recall that all of our theorems have an upper bound that holds in significant generality, and a corresponding lower bound that holds under additional assumptions. We have chosen distributions for $\senderrnd, \queryrnd$ and $\receiverrnd$ that meet these additional assumptions to enable comparisons with both upper and lower bounds. In particular, their distribution is chosen so that:
\begin{itemize}
\item $\senderrnd \entails \queryrnd$, $\queryrnd \entails \receiverrnd$ (and therefore $\senderrnd \entails \receiverrnd$);
    \item The random tuples $\{ ( \kv{\senderrnd}, \kv{\queryrnd}, \kv{\receiverrnd})_i \}$ are i.i.d.;
\item The Markov chain $\receiverrnd \rightarrow \algebraicset{\receiverrnd} \rightarrow (\algebraicset{\senderrnd},\algebraicset{\receiverrnd})$ holds.
\end{itemize}
A simple way to construct $\senderrnd, \queryrnd, \receiverrnd$ with these properties is to choose their underlying kernels first, and then construct sentences from those kernels. If the reader has, instead, a collection of sentences meeting the entailment constraints which are believed to come from an otherwise unknown distribution, then the reader may still obtain value from our paper by considering only the upper bounds.
We generated random kernels meeting all the three constraints above on sets of $10$ variables to test our methods in practice. For each of the kernels we produced a generalized decision tree with the given kernel as its set of satisfying truth value assignments. We then post-processed each sentence by writing it using postfix notation, compressing variable names and removing spaces. For the targeted query scenario we have chosen $p_r = 0.5$, $p_s = 0.075$ and $p_q$ by sampling the range $[p_s,p_r)$ uniformly. For the scenario where Alice does not know what Bob knows, we set $p_s = p_q = 0.075$ and let $p_r$ range in $(p_s,0.5]$.

We generated $1000$ test cases for each choice of $(p_s,p_q,p_r)$ considered. This data could be used, through appropriate subselection, to illustrate all of the theorems in this article in fullness with the exception of Theorem \ref{thm:less_is_more_background}. In this theorem the assumption $\senderrnd \entails \receiverrnd$ is simply omitted. Therefore only those subcases where it is true can be demonstrated  with this data.

\subsubsection{Method of Generating Kernels}
Since the sender's kernel is a subset of the receiver's kernel, and, moreover, the query is provable by the sender but not the receiver, it follows that the sender's kernel is contained in the kernel associated with the query, which in turn is contained in the receiver's kernel. It then follows that we can go through all the possible truth value assignments for the 10 variables and, for each assignment $\mu$, generate a random number $\eta \in [0,1]$ where, if $\eta \leq p_r$, we place $\mu$ in the receiver's kernel. If additionally $\eta \leq p_q$, then we can place $\mu$ in the query kernel, and if additionally to that $\eta \leq p_s$, then we can place $\mu$ in the sender's kernel. Pseudocode for this simple procedure is given in Algorithm \ref{alg:gen_zeroes} in Appendix \ref{app:zero-sets}.

\subsubsection{Method of Generating Generalized Decision Tree Sentences}

The gist of the decision tree approach is to recursively call a method $\mathtt{GENERATE\_GDT\_FOR\_KERNEL(\cdot)}$ to find a most balanced variable $\mt{X_b}$ among the variables $\mt{X_1,...,X_n}$ -- meaning a variable where $\mt{X_b} = 1$ and $\mt{X_b} = 0$ as close to equally as possible among the kernel elements -- and then output 
\begin{eqnarray*}
    (\mt{X_b} \land \mt{GENERATE\_GDT\_FOR\_KERNEL(pos\_kernel} )) \vee \\
    (\neg \mt{X_b} \land \mt{GENERATE\_GDT\_FOR\_KERNEL(neg\_kernel} )),
\end{eqnarray*}
where $\mt{pos\_kernel}$ is the reduced kernel on the $\mt{n-1}$ variables $\mt{X_1,...,X_n}$ but with $\mt{X_b}$ excluded, and in the original kernel $\mt{X_b} = 1$, while $\mt{neg\_kernel}$ is the reduced kernel on the $\mt{n-1}$ variables $\mt{X_1,...,X_n}$ but with $\mt{X_b}$ excluded, and in the original kernel $\mt{X_b} = 0$. The generalized version of this heuristic is a bit more nuanced so that variables that are effectively constant with respect to the kernel (in other words, there is a variable $\mt{X_i}$ such that either $\mt{X_i} = 1$ or $\mt{X_i} = 0$ for every kernel element) are efficiently split out, and when there are $\mt{k}$ variables remaining and the kernel size is either $2^\mt{k}$ or $0$ the routine immediately terminates, outputting an empty string (equivalent to outputting True) in the former case, and outputting a False indicator in the latter case. 
The pseudocode for implementing the generalized decision tree algorithm, given a kernel, along with an accompanying detailed description and worked example are provided in Appendix \ref{app:decision-trees}. There is a considerable literature on the use of decision trees to represent Boolean functions. See, for example, \cite{breitbart:size_of_binary_decision_diagrams, mehta:decsion_tree_representation_of_boolean_functions, odonnell:decision_trees_influential_variable}.

\subsection{Experimental results}
\label{sec:experiments}

\subsubsection{Semantic communication systems}
\label{ss:exp_scs}
\begin{figure}[t]
    \begin{subfigure}[h]{0.49\linewidth}
    \includegraphics[width=\linewidth]{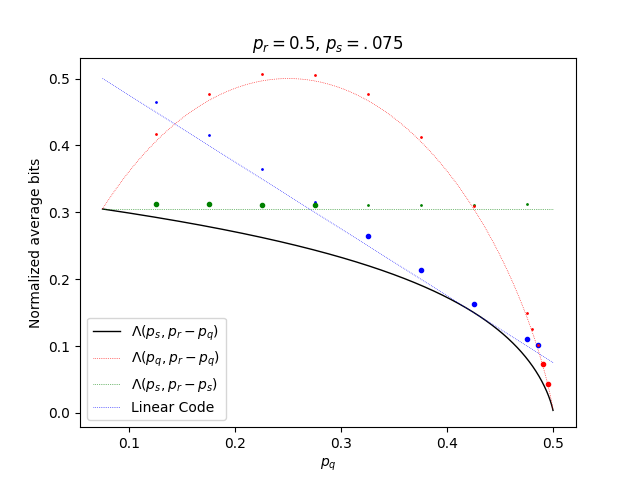}
    \caption{Semantic communication systems for targeted queries with a shared sentence with $p_r = 0.5$, $0.075 < p_q \le 0.5$}
    \label{fig:semantic_targeted_queries}
    \end{subfigure}
    \begin{subfigure}[h]{0.49\linewidth}
    \includegraphics[width=\linewidth]{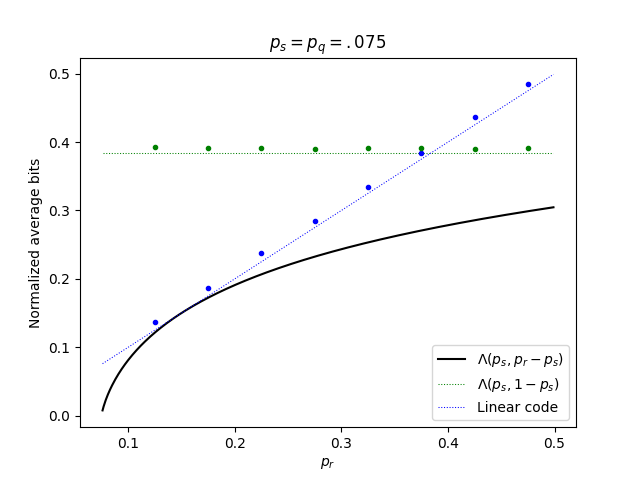}
    \caption{Semantic communication systems for when Alice doesn't know what Bob knows with $p_s = p_q = 0.075$}
    \label{fig:semantic_alice_does_not_know}
    \end{subfigure}
    \caption{Comparison of practical semantic communication methods against the Shannon bound $\Lambda(p_s,p_r-p_q)$ for two scenarios.}

    \label{fig:tests}
\end{figure}

In a first set of experiments, we aim to demonstrate the performance of our practical semantic communication techniques against the new theoretical bounds. We have implemented the linear codes described in Subsections~\ref{ss:linear_targeted} and \ref{ss:no_need} which are used to implement the two scenarios in Subsection~\ref{ss:scenarios}. Due to the fact that these linear codes are not always optimal, we augmented these systems with ``n\"aive'' semantic communication strategies based on efficient lossless transmission of kernels. In certain parameter ranges these are better than the linear codes so in those cases, we can use them instead. In the case of targeted queries, the task of allowing Bob to prove $\queryrnd$ can be alternately accomplished by sending to him $\algebraicset{\queryrnd}$ or $\algebraicset{\senderrnd}$, whichever is cheapest. The results, averaged across all 1000 test cases for each choice of parameters, can be found in Figure~\ref{fig:semantic_targeted_queries}, where in blue, we illustrate the performance of linear codes, and in green and red we illustrate the performance of sending $\algebraicset{\senderrnd}$  and $\algebraicset{\queryrnd}$ using enumerative source codes \cite{cover:enumerative}, respectively. In each of these three practical systems, we include a line of the same color that is a lower bound on performance for the specific technique. The bolder dots are those that are closest to the limiting Shannon bound, which is the bold, lowest plot shown in the figure. Similarly, in the case that Alice doesn't know what Bob knows, the task of allowing Bob to prove $\senderrnd$ can be alternately accomplished by sending to him $\algebraicset{\senderrnd}$, also using enumerative source coding \cite{cover:enumerative}. The results of this experiment can be found in Figure~\ref{fig:semantic_alice_does_not_know}. It can be appreciated that our practical codes can be quite efficient in some scenarios, but in general, more work is needed to develop practical codes that meet the Shannon bound in general.

\subsubsection{Classic communication systems}
In a second set of experiments, we want to contrast practical classic systems with semantic ones. In these experiments, we will reuse the best results from practical methods in Subsection~\ref{ss:exp_scs} and compare those with classic compression systems.

We define classic compressions systems as ones that may only use the sentence as presented to the communication system, and not perform operations on it with awareness of its underlying semantic content. These are examples of such systems:

\begin{itemize}
    \item \textbf{(Targeted query with a shared statement)} Using the Decision Tree representation of a sentence, employ a standard compression algorithm (in particular, gzip, lzma, bzip2) to compress $\senderrnd$ and $\queryrnd$; choose the best representation and send that one.
    \item \textbf{(Alice doesn't know what Bob knows)} The same as above, but considering only $\senderrnd$ as in this scenario, $\queryrnd = \senderrnd$.
\end{itemize}

\begin{figure}[t]
    \begin{subfigure}[h]{0.5\linewidth}
    \centering\captionsetup{width=.9\linewidth}
    \includegraphics[width=0.9\linewidth]{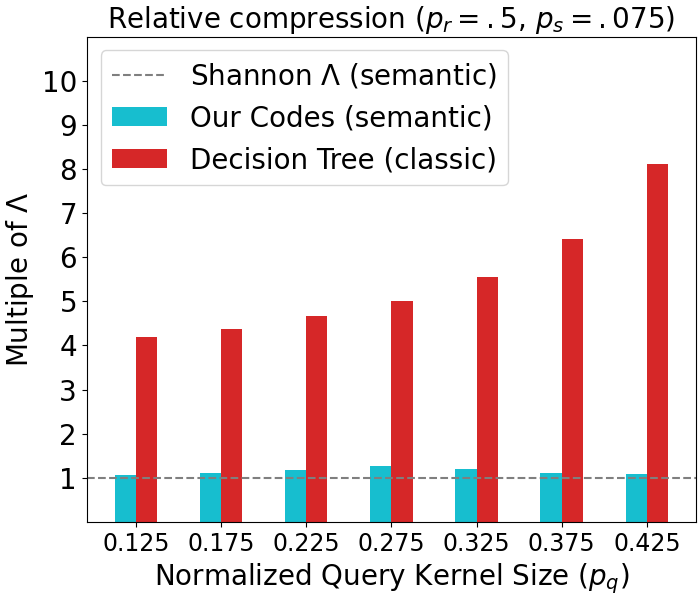}
   \caption{Targeted queries with a shared sentence ($p_r=.5$, $p_s=.075$)}
    \label{fig:compress_code}
    \end{subfigure}
    \begin{subfigure}[h]{0.5\linewidth}
    \centering\captionsetup{width=.9\linewidth}
    \includegraphics[width=0.9\linewidth]{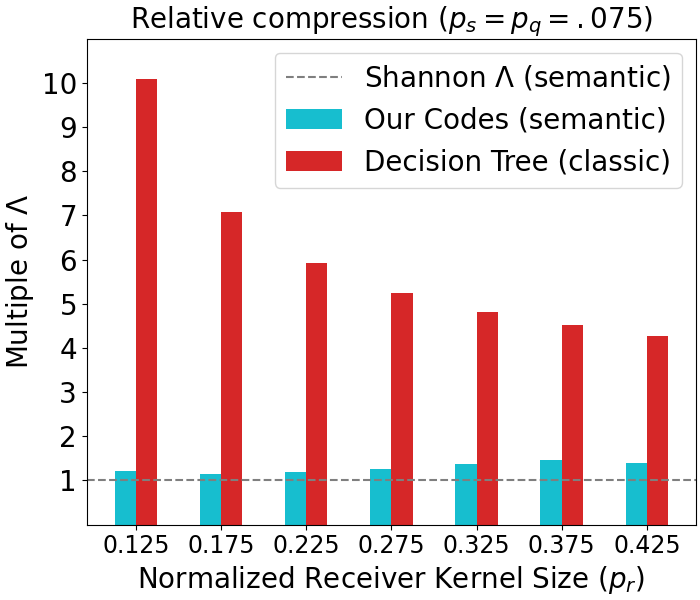}
    \caption{Alice does not know Bob's sentence($p_s=p_q=.075$)}
    \label{fig:compress_no_need}
    \end{subfigure}
    \caption{Contrasting classic and semantic communication systems relative to the semantic Shannon limit}
 \label{fig:compress_ind}
\end{figure}

We provide an additional advantage to the type of classic communication systems used above: we compress all 1000 samples simultaneously, which allow the compression algorithms described above to leverage patterns that only emerge when more data is available. In contrast, the semantic communication systems are compressing only one instance at a time, which is a much more difficult target. We reiterate that by using Decision Trees, we have already used semantic concepts to benefit the classical baseline. The results of these experiments can be found in Figures~\ref{fig:compress_code} and~\ref{fig:compress_no_need}. These plots normalize the performance of the practical semantic or classical systems against the corresponding Shannon bound $\Lambda(p_s,p_r-p_q)$ and $\Lambda(p_s,p_r-p_s)$, respectively. We acknowledge that in the realm of classical communication systems we should consider Slepian-Wolf compression algorithms \cite{slepian_wolf:coding} as a baseline. This is particularly difficult to do as we are not aware of practical instances of such algorithms that would be applicable to the complex statistical distributions present in $\senderrnd, \receiverrnd$; we leave this comparison as an open item for future research.

\section{Formal information-theoretic results}
\label{sec:proofs}
The purpose of this section is to formally state and prove each of the information-theoretic results we offer in this paper.

\subsection{Full ignorance}

In this first problem, Alice is in possession of a logic sentence and wishes to communicate to Bob with the goal that whatever logic inferences Alice can make starting from that sentence, Bob, who otherwise knows nothing, can do the same.  Crucially, Bob is not required to reproduce the particular way in which Alice's logic sentence is represented in her mind, but rather it just needs to be able to retrieve a functionally equivalent sentence. The intention of this result is mainly to introduce notation and concepts as we build up to more interesting cases. 

As with all our theorems, Theorem~\ref{thm:logicinfo} has a lower bound on performance, as well as an algorithm to achieve that performance. We sketch next that algorithm, and then formally prove both upper and lower bounds. Alice and Bob meet in preparation for a future where $\sender$ is revealed to Alice. Given that any possible query that is entailed by $\sender$ must be proved by Bob (see Figure~\ref{fig:strategy_logicinfo},  which illustrates the kernels of two possible queries $\query^{\prime}$ and $\query^{\prime}$  with the property that $\sender \entails \query^{\prime}$ and $\sender \entails \query^{\prime}$), and given that this must be done as efficiently as possible, they agree that the sensible goal is not to send her sentence $\sender$ verbatim, but to send the kernel $\algebraicset{\sender}$ instead. From this, Bob can use the function $\ell$ to recover a functionally equivalent sentence $\senderhat$.

To implement this, our suggested algorithm is for Alice to first send to Bob the \emph{size} of the kernel. Once this size is transmitted, all possible kernels can be enumerated by both Alice and Bob and then Alice can simply send the index of the kernel that she has in her possession. To implement the idea above we need two tools: a means for efficiently sending integers (to encode the size) and a way to efficiently encode the kernel indices. The same needs will recur in all of our results, so we pause here to introduce two widely known tools to accomplish this.

For integer encoding, we use 
$\delta$ Elias coding~\cite{elias:integers}. This code assigns codewords to each integer, where the length of a codeword for the integer $n \geq 1$ is given by
\begin{eqnarray*}
\len{\mbox{elias}_{\delta}(n)} &=& \lfloor \log_{2}(n)\rfloor +2\lfloor \log_{2}(1+\lfloor \log_{2}(n)\rfloor )\rfloor +1 \mbox{ bits} \\
& \leq & \log_{2}(n) +2 \log_{2}\log_{2}(n)  +3 \mbox{ bits}.
\end{eqnarray*}
Our rationale for choosing this particular code is that it is both convenient and sufficient to prove asymptotically good results. For the problem of sending a kernel index, techniques for enumerating members of a set and efficiently sending and receiving indices from such enumeration can be found in the work of Cover on enumerative source coding \cite{cover:enumerative}. This overall strategy is described at the bottom of Figure~\ref{fig:strategy_logicinfo}. 

We now formally present our main result for this case followed by its proof.

\begin{restatable}[Bob can prove all Alice is able to]{thm}{logicinfo}
~Let $(L,  \kernelspace, \kappa, \ell)$ be a Logic System. Let $\senderrnd\in \surfacespace$ have a kernel that follows a $p_s$-law. Let the encoder $f$ and decoder $g$ be functions
\begin{eqnarray*}
f : \surfacespace & \rightarrow & \{0,1\}^* ,\\
g : \{0,1\}^* & \rightarrow & \surfacespace,
\end{eqnarray*}
respectively.  Then,
\begin{eqnarray*}
\min_{f,g} |\kernelspace|^{-1} E_{\senderrnd} \left[ \len{f(\senderrnd)} \right] \leq |\kernelspace|^{-1} H(\algebraicset{\mathtt{S}}) + \frac{1}{|\kernelspace|} \leq \Hbin{p_s} + O \left( \frac{\log_2 |\kernelspace|}{|\kernelspace|}\right),
\end{eqnarray*}
where the minimization is over $f, g$ such that the image of $f$ is prefix-free and such that, if $\sender \entails \query$, then $\sender \entails g(f(\sender))$ and $g(f(\sender)) \entails \query$. Under the same assumptions for $f,g$, the following lower bound holds:
\begin{eqnarray}
|\kernelspace|^{-1}H(\algebraicset{\senderrnd})
  \leq \min_{f,g} |\kernelspace|^{-1} E_{\senderrnd} \left[ \len{f(\senderrnd)} \right] .
\label{eq:general_lower_bound}
\end{eqnarray}
Furthermore, if we additionally have that the $\{\kv{\senderrnd}_j\}_{j=1}^{|\kernelspace|}$ are i.i.d., then 
\begin{eqnarray*}
\Lambda( p_s, 1-p_s) = \Hbin{p_s}
  \leq \min_{f,g} |\kernelspace|^{-1} E_{\senderrnd} \left[ \len{f(\senderrnd)} \right] .
\end{eqnarray*}

\label{thm:logicinfo}
\end{restatable}

\begin{figure}[h]
\begin{center}
\begin{tikzpicture} 
\clip(0,0) rectangle(8, 5);
\draw (0,0) rectangle (8,5);
\draw(4,2)[fill=green!25!white] circle(1.0);
\draw(4,2)[pattern=checkerboard,pattern color=blue!15!white] circle(1.0);
\draw(4.0,2.0) node {$\algebraicset{\sender}=\algebraicset{\senderhat}$};
\draw(5,-1.5)[dotted, rotate=45] ellipse(2.5 and 1.5);
\draw(5.6,4.25) node {$\algebraicset{\queryapo}$};
\draw(4.0,3.0)[dotted, rotate=-15] ellipse(2.2 and 1.3);
\draw(5.8,0.9) node {$\algebraicset{\queryapoapo}$};
\end{tikzpicture}
\end{center}
\centerline{(a)}
\begin{center}
     \begin{tikzpicture}
	\matrix[row sep=2.5mm, column sep=5mm]
	{tm		
            \node[coordinate] (fm01) {}; &
            \node[coordinate] (f00) {}; &
            \node[coordinate] (f01) {}; &
            \node[coordinate] (f02) {}; &
            \node[coordinate] (f03) {} ; &
            \node[coordinate] (f04) {}; &
            \node[coordinate] (f05) {}; & \\
		\node[dspnodeopen,dsp/label=left] (mmm01) {$\sender$};  &
        \node[] (mm00) {};  &
		\node[dspmedsquare]         (mm01) {$\kappa$};     &
            \node[coordinate,label=$\algebraicset{\sender}$] (mm02) {}; &
		\node[dspelongated]          (mm03) { ennumerative \\ code };     &
		\node[coordinate,label=]          (mm03p5) {};     &
            \node[dspbigsquare] (mm04) {decoder};    & 
            \node[coordinate,label=$\algebraicset{\sender}$] (mm05) {}; &
            \node[dspsquare]         (mm06) {$\ell$};  &
		\node[] (mm07) {}; &
		\node[dspnodeopen,dsp/label=above] (mm08) {$\senderhat$};    \\
            
            \node[coordinate] (km01) {}; &
            \node[coordinate] (k00) {}; &
            \node[coordinate] (k01) {}; &
            \node[coordinate,label=$\len{\algebraicset{\sender}}$] (k02) {}; &
            \node[dspelongated]  (k03) { Elias-$\delta$ \\ code } ; &
            \node[coordinate] (k04) {}; &
            \node[coordinate] (k05) {}; & \\

            \node[coordinate] (bm01) {}; &
            \node[coordinate] (b00) {}; &
            \node[coordinate] (b01) {}; &
            \node[coordinate] (b02) {}; &
            \node[coordinate] (b03) {} ; &
            \node[coordinate] (b04) {}; &
            \node[coordinate] (b05) {}; & \\
  };

	\begin{scope}[start chain]
		\chainin (mmm01);
		\chainin (mm01) [join=by dspconn];
		\chainin (mm03) [join=by dspconn];
		\chainin (mm04) [join=by dspconn];
		\chainin (mm05) [join=by dspline];
		\chainin (mm06) [join=by dspconn];
        \chainin (mm08) [join=by dspconn];
        \end{scope}

        \begin{scope}[start chain]
            \chainin (mm01);
            \chainin (k01) [join=by dspline];
            \chainin (k03) [join=by dspconn];
            \chainin (k05) [join=by dspline];
            \chainin (mm04) [join=by dspconn];
        \end{scope}

        \begin{scope}[start chain,dotted]
            \chainin(f04);
            \chainin(b04) [join=by dspline];
        \end{scope}
        
\end{tikzpicture}
\end{center}
\centerline{(b)}
\caption{Proof strategy for  Theorem~\ref{thm:logicinfo}.}
\label{fig:strategy_logicinfo}
\end{figure}

\textbf{Proof.}
We begin with the proof of the lower bound. Let $f, g$ satisfy the conditions for the minimization. The starting point for this proof is the classical result from information theory proved using Kraft's inequality as follows.
\begin{restatable}{lem}{kraft}
Let $\{l_i\}$ be the codeword lengths of a binary code that is prefix-free. Assume a distribution over these codewords, and let $C$ be a random codeword drawn according to that distribution. Then,
\begin{eqnarray*}
E_{C}\left[ \len{C} \right] \geq \Hreg{C}.
\end{eqnarray*}
\label{lem:vlc_kraft}
\end{restatable}
Using this result together with the assumption that the image of $f$ is prefix-free, we can write
\begin{eqnarray}
E_{\senderrnd}\left[ \len{f(\senderrnd)} \right] = E_{f(\senderrnd)}\left[ \len{f(\senderrnd)} \right] \geq \Hreg{ f(\senderrnd) } \geq \Hreg{g(f(\senderrnd))}  \geq \Hreg{\algebraicset{g(f(\senderrnd))}} = \Hreg{\algebraicset{\senderrndhat}} ,
\label{eq:ineqs}
\end{eqnarray}
where the last two inequalities follow from the fact that deterministic functions of random variables cannot increase entropy. 

For any $\sender, \query$ such that $\sender \entails \query$, we are assuming that  $\sender \entails g(f(\sender))$, $g(f(\sender)) \entails \query$ and therefore using the definition of a Logic System, we have
\begin{eqnarray}
\algebraicset{\sender} \subseteq \algebraicset{ g(f(\sender)) } \subseteq \algebraicset{\query} .
\label{eq:sand_cond}
\end{eqnarray}
In particular, choosing $\query=\sender$, we obtain $\algebraicset{\sender} = \algebraicset{ g(f(\sender)) } = \algebraicset{\senderhat}$. This is geometrically described in Figure~\ref{fig:strategy_logicinfo}-a, which illustrates the kernels of two queries satisfying $s \entails q'$, $s \entails q''$ (depicting the choice of $q$ in the right hand side of (\ref{eq:sand_cond})), as well as the conclusion that $\algebraicset{\sender} = \algebraicset{\senderhat}$. Substituting this in the right hand side of (\ref{eq:ineqs}), we conclude
\begin{eqnarray*}
E_{\senderrnd}\left[ \len{f(\senderrnd)} \right]   &\stackrel{(a)}{\geq}& \Hreg{\algebraicset{\senderrnd}} \\
&\stackrel{(b)}{=}& \Hreg{  \kv{\senderrnd}_1, \ldots, \kv{\senderrnd}_{|\kernelspace|} } \\
&\stackrel{(c)}{=}& \sum_{i=1}^{\kernelspace} \Hreg{ \kv{\senderrnd}_i } \\
&\stackrel{(d)}{=}& |\kernelspace| \Hbin{p_s} ,
\end{eqnarray*}
where (a) already proves (\ref{eq:general_lower_bound}), (b) follows from the definition of $\kv{\cdot}$ in (\ref{eq:kvdef}), (c) follows from the independence assumption, and (d) follows from the $p_s$-law assumption combined with the ``identically distributed'' assumption. This completes the proof of the lower-bound result.

To prove the upper bound, we construct a code as follows. First, the sentence $\mathtt{s}$ is mapped to its kernel, $\algebraicset{\mathtt{s}}$; let $P_K$ denote the probability distribution governing $\algebraicset{\senderrnd}$. Then we use a Shannon code to encode this kernel, which uses a code length of
\begin{eqnarray}
    \left \lceil \log_2 \left( \frac{1}{P_K(\algebraicset{\sender})}\right) \right \rceil \leq  \log_2 \left( \frac{1}{P_K(\algebraicset{\sender} )}\right) +1 .
    \label{eq:progress_progress}
\end{eqnarray}
We point the reader to the early discussion of this quantity in (\ref{eq:alternate}). When substituting a random $\senderrnd$ in lieu of $\sender$, the expectation of the right hand side of the expectation of (\ref{eq:progress_progress}) is
\begin{eqnarray}
    \Hreg{\algebraicset{\senderrnd}} + 1 .
\end{eqnarray}
One then writes
\begin{eqnarray*}
 \Hreg{\algebraicset{\senderrnd} }&\stackrel{(a)}{=}& \Hreg{  \kv{\senderrnd}_1, \ldots, \kv{\senderrnd}_{|\kernelspace|} } \\
&\stackrel{(b)}{\leq}& \sum_{i=1}^{\kernelspace} \Hreg{ \kv{\senderrnd}_i }  \\ 
&\stackrel{(c)}{=}& \sum_{i=1}^{\kernelspace} \Hbin{ P( \kv{\senderrnd}_i=1) }  \\
&\stackrel{(d)}{\leq}& |\kernelspace| \Hbin{ |\kernelspace|^{-1}\sum_{i=1}^{\kernelspace} P( \kv{\senderrnd}_i=1) } \\
&\stackrel{(e)}{=}& |\kernelspace| \Hbin{p_s} = |\kernelspace| \Lambda(p_s,1-p_s) ,
\end{eqnarray*}
where (a) follows from the definition of $\kv{\cdot}$ in (\ref{eq:kvdef}), (b) is a standard upper bound on the entropy of joint variables, (c) follows from the fact that $\kv{\senderrnd}_i$ is a binary random variable, (d) follows from the concavity of entropy, and (e) follows from the assumption of $\senderrnd$ following a $p_s$-law.

We note that this result already proves the tighter upper bound of the Theorem and that such a tighter upper bound can be included in the results of our subsequent theorems, as explained in the discussion around~\eqref{eq:tight-bound}.
We include an additional argument nonetheless, which only proves the looser upper bound, but that agrees in style with the rest of the results in the article; see Figure~\ref{fig:strategy_logicinfo}. We first specify the encoder $f$. For a given size $\xi$ of a kernel, there are a total of
\begin{eqnarray*}
{|\kernelspace| \choose \xi}
\leq 2^{|\kernelspace| \Hbin{\frac{\xi}{|\kernelspace|}}}
\end{eqnarray*}
possible kernels of the same size. Let $\mbox{enum}_{\xi} : \{ k \subseteq \kernelspace : |k| = \xi \} \rightarrow \{0,1\}^{*}$ be a function that maps each possible kernel of size $\xi$ to a fixed-length binary encoding of the integers $\left\{1,\ldots, 
\displaystyle{|\kernelspace| \choose \xi}
\right\}
$, which is an integer that uniquely determines such a set.
Hence, in particular, 
\begin{eqnarray*}
\len{ \mbox{enum}_{|\algebraicset{\senderrnd}|}(\algebraicset{\senderrnd})  } \leq |\kernelspace| \Hbin{\frac{|\algebraicset{\senderrnd}|}{|\kernelspace|}} + 1 ,
\end{eqnarray*}
where in the above the binary entropy function $\Hbin{p}=-p \log_2(p) -(1-p)\log_2(1-p)$ is evaluated on the random variable $|\algebraicset{\senderrnd}|/|\kernelspace|$, and thus the result of the evaluation is also a random variable. 

For a sentence kernel $\sender \in \surfacespace$, we define the encoder $f$ as a concatenation of two separate encodings:
\begin{eqnarray}
f(\sender) = \mbox{elias}_{\delta}( |\algebraicset{\sender}| )  \mbox{enum}_{|\algebraicset{\sender}|}(\algebraicset{\sender}) ,
\label{eq:concatenation}
\end{eqnarray}
where the above is the result of the concatenation of two codes.
Since both of the codes implied by each encoding are prefix-free, the concatenation is also prefix-free. Finally, we let $g$ be the decoder that recovers the kernel $\algebraicset{\sender}$ from the output of $f$, and evaluates $\ell$ on that kernel.

Note that, by construction, if $\query$ is such that $\sender \entails \query$, then since $\algebraicset{\sender} = \algebraicset{ g(f(\sender))}$, we conclude $g(f(\sender)) \entails \query$ as well, and thus we have met the conditions of the Theorem.

The estimate for the overall cost of the encoding can be done by separately estimating the length of the two encodings in (\ref{eq:concatenation}), namely
\begin{eqnarray*}
 |\kernelspace| \Hbin{ \frac{|\algebraicset{\senderrnd}|}{|\kernelspace|} }  + \log_2 |\algebraicset{\senderrnd}| + 2 \log_2\left(  \log_2 |\algebraicset{\senderrnd}| \right) + 4
\end{eqnarray*}
bits. Taking the expectation with respect to $\senderrnd$, using the concavity $\cap$ of the logarithm and entropy functions, and normalizing by $|\kernelspace|$, we obtain a normalized upper estimate of
\begin{eqnarray*}
\Hbin{p_s} + \frac{\log_2 \left( p_s |\kernelspace| \right) }{|\kernelspace|}  +2  \frac{\log_2\left(\log_2 \left( p_s |\kernelspace| \right) \right)}{|\kernelspace|} + \frac{4}{|\kernelspace|} .
\end{eqnarray*}
Note that, in this upper bound, at no point did we assume the more restrictive condition involving the i.i.d.\ assumptions of the lower bound, thus completing the proof of the Theorem. This pattern of the upper bound holding under much more general conditions will repeat throughout our other proofs. \qed

\subsection{Partial Ignorance }
\label{ss:bli}

In this Theorem, both Alice and Bob have access to the same background logic sentence $\receiverrnd$. It is important to emphasize that in this setup, such background information is \emph{not} known at the initial meeting between Alice and Bob, and rather, it is the result of both independently collecting the same such logic sentences from the environment. We refer the reader to Figure~\ref{fig:background_log_info}, which extends the corresponding Figure~\ref{fig:strategy_logicinfo} by incorporating such background information. In some respects, the effect that the presence of the background $\receiverrnd$ has in the problem is rather elementary: we still want to send somehow $\algebraicset{\senderrnd}$, however we can now leverage the fact that both Alice and Bob know that such a kernel must be contained within $\algebraicset{\receiverrnd}$. In the same figure we show an updated strategy: both Alice and Bob first compute $\algebraicset{\receiverrnd}$, and then Alice uses it to enumerate all possible subsets of $\algebraicset{\receiverrnd}$ of size $\len{\algebraicset{\senderrnd}}$, and sends $|\algebraicset{\senderrnd}|$ followed by the index of the kernel she has in her possession. Bob then recovers $\algebraicset{\senderrnd}$ and uses $\ell$ to recover a logic sentence.

\begin{restatable}[Alice and Bob share a logic sentence]{thm}{backgroundloginfo}
~Let $(L,  \kernelspace, \kappa, \ell)$ be a Logic System. Let $\senderrnd, \receiverrnd \in \surfacespace$ represent the sender's logic sentence and the shared logic sentence, with the property that $\senderrnd \entails \receiverrnd$ and, in particular, $\senderrnd, \receiverrnd$ have kernels that follow a ($p_s,p_r$)-law. Let the encoder $f$ and decoder $g$ be functions
\begin{eqnarray*}
f : \surfacespace^2 & \rightarrow & \{0,1\}^* , \\
g : \{0,1\}^*\times \surfacespace  & \rightarrow & \surfacespace.
\end{eqnarray*}
Then
\begin{eqnarray*}
\min_{f,g} |\kernelspace|^{-1} E_{\senderrnd,\receiverrnd} \left[ \len{f(\senderrnd,\receiverrnd)} \right] 
\leq \Lambda(p_s,p_r-p_s) + O\left( \frac{\log_2 |\kernelspace|}{|\kernelspace|} \right),
\end{eqnarray*}
where the minimization is over $f, g$ such that the image of $f(\cdot,\receiver)$ is prefix-free for any choice of $\receiver$ and such that, if $\sender \entails \receiver$ and $\sender \entails \query$, then $\sender \entails g(f(\sender,\receiver),\receiver)$ and $g(f(\sender,\receiver),\receiver) \entails \query$. Furthermore if additionally the random variables $\{ (\kv{\senderrnd}, \kv{\receiverrnd} )_j \}_{j=1}^{|\kernelspace|}$ are i.i.d.\ and $\receiverrnd \rightarrow \algebraicset{\receiverrnd} \rightarrow \algebraicset{\senderrnd}$, then
\begin{eqnarray*}
 \Lambda(p_s,p_r-p_s) \leq \min_{f,g} |\kernelspace|^{-1} E_{\senderrnd,\receiverrnd} \left[ \len{f(\senderrnd,\receiverrnd)} \right] 
.
\end{eqnarray*}
\label{thm:background_log_info}
\end{restatable}

\begin{figure}[h]
\begin{center}
\begin{tikzpicture}  
\clip (0,0) rectangle (8,5);
\draw (0,0) rectangle (8,5);
\draw(4,2)[fill=green!25!white] circle(1.0);
\draw(4,2)[pattern=checkerboard,pattern color=blue!15!white] circle(1.0);
\draw(4.0,2.0) node {$\algebraicset{\sender}=\algebraicset{\senderhat}$};
\draw(5,-1.5)[dotted, rotate=45] ellipse(2.5 and 1.5);
\draw(5.6,4.25) node {$\algebraicset{\queryapo}$};
\draw(4.0,3.0)[dotted, rotate=-15] ellipse(2.2 and 1.3);
\draw(5.8,0.9) node {$\algebraicset{\queryapoapo}$};
\draw(2.2,0.25) rectangle(7, 4.65);
\draw(2.6,3.7) node {$\algebraicset{\receiver}$};
\end{tikzpicture}
\end{center}
\centerline{(a)}
\begin{center}
     \begin{tikzpicture}
	\matrix[row sep=2.5mm, column sep=5mm]
	{tm		

             \node[coordinate] (f00) {}; &
            \node[coordinate] (f01) {}; &
            \node[coordinate] (f02) {}; &
            \node[coordinate] (f03) {} ; &
            \node[coordinate] (f04) {}; &
            \node[coordinate] (f05) {}; & \\
            
            \node[dspnodeopen,dsp/label=left] (y00) {$\receiver$}; &
            \node[dspsquare] (y01) {$\kappa$}; &
            \node[coordinate,label=$\algebraicset{\receiver}$] (y02) {}; &
            \node[coordinate] (y03) {} ; &
            \node[coordinate] (y04) {}; &
            \node[coordinate] (y05) {}; & 
            \node[coordinate,label=$\algebraicset{\receiver}$] (y06) {}; & 
            \node[dspsquare] (y07) {$\kappa$}; & 
            \node[dspnodeopen,dsp/label=above] (y08) {$\receiver$}; 
            \\

		\node[dspnodeopen,dsp/label=left] (mm00) {$\sender$};  &
		\node[dspsquare]         (mm01) {$\kappa$};     &
            \node[coordinate,label=$\algebraicset{\sender}$] (mm02) {}; &
		\node[dspelongated]          (mm03) { ennumerative \\ code };     &
		\node[coordinate,label=]          (mm03p5) {};     &
            \node[dspbigsquare] (mm04) {decoder};    & 
            \node[coordinate,label=$\algebraicset{\sender}$] (mm05) {}; &
            \node[dspsquare]         (mm06) {$\ell$};  &
		\node[dspnodeopen,dsp/label=above] (mm07) {$\senderhat$};  \\
            
            \node[coordinate] (k00) {}; &
            \node[coordinate] (k01) {}; &
            \node[coordinate,label=$\len{\algebraicset{\sender}}$] (k02) {}; &
            \node[dspelongated]  (k03) { Elias-$\delta$ \\ code } ; &
            \node[coordinate] (k04) {}; &
            \node[coordinate] (k05) {}; & \\

            \node[coordinate] (b00) {}; &
            \node[coordinate] (b01) {}; &
            \node[coordinate] (b02) {}; &
            \node[coordinate] (b03) {} ; &
            \node[coordinate] (b04) {}; &
            \node[coordinate] (b05) {}; & \\
  };

        \begin{scope}[start chain]
            \chainin (y00);
            \chainin (y01) [join=by dspconn];
            \chainin (y03) [join=by dspline];
            \chainin (mm03) [join=by dspconn];
        \end{scope}
        \begin{scope}[start chain]
            \chainin (y08);
            \chainin (y07) [join=by dspconn];
            \chainin (y05) [join=by dspline];
            \chainin (mm04) [join=by dspconn];
        \end{scope}
        \begin{scope}[start chain]
        \chainin (y08);
        \end{scope}
	\begin{scope}[start chain]
		\chainin (mm00);
		\chainin (mm01) [join=by dspconn];
		\chainin (mm03) [join=by dspconn];
		\chainin (mm04) [join=by dspconn];
		\chainin (mm05) [join=by dspline];
		\chainin (mm06) [join=by dspconn];
            \chainin (mm07) [join=by dspconn];
        \end{scope}

        \begin{scope}[start chain]
            \chainin (mm01);
            \chainin (k01) [join=by dspline];
            \chainin (k03) [join=by dspconn];
            \chainin (k05) [join=by dspline];
            \chainin (mm04) [join=by dspconn];
        \end{scope}

        \begin{scope}[start chain,dotted]
            \chainin(f04);
            \chainin(b04) [join=by dspline];
        \end{scope}
\end{tikzpicture}
\end{center}
\centerline{(b)}
\caption{Proof strategy for Theorem~\ref{thm:background_log_info}.}
\label{fig:background_log_info}
\end{figure}

\textbf{Proof.} As with Theorem~\ref{thm:logicinfo}, we begin with the proof of the lower bound. Let $f, g$ satisfy the conditions for the minimization. Paralleling the definition (\ref{eq:kvdef}), for any given $k \subseteq \mathcal{M}$, let 
\begin{eqnarray*}
    k_j &=& \left\{ \begin{array}{cc} 
            1 & \mbox{if } \mu_j \in k \\
            0 & \mbox{otherwise}
    \end{array} \right. .
\end{eqnarray*}
For the proof of the lower bound, we will rely on the following Lemma, which is provided without proof since it is elementary.
\begin{restatable}{lem}{simpleab}
    For given $0 < p_a \leq p_b \leq 1$, let the random variables $A,B \in \{0,1\}$ be such that $E A = p_a$ and $E B = p_b$, and furthermore, $A \leq B$. Then conditioned on $B=1$, the random variable $A$ is distributed according to $\{1-p_a/p_b, p_a/p_b\}$.
\label{lem:simpleab}
\end{restatable}
\allowdisplaybreaks
We now derive
\begin{eqnarray}
E_{\senderrnd,\receiverrnd}\left[ \len{f(\senderrnd,\receiverrnd)} \right] &=& E_{f(\senderrnd,\receiverrnd)}\left[ \len{f(\senderrnd,\receiverrnd)} \right] \nonumber \\ 
&\stackrel{(a)}{=}& E_{\receiverrnd} \left[ E_{f(\senderrnd,\receiverrnd)|\receiverrnd}\left[ \len{f(\senderrnd,\receiverrnd)} | \receiverrnd \right] \right] \nonumber \\ 
&\stackrel{(b)}{\geq}& \Hreg{ f(\senderrnd, \receiverrnd) | \receiverrnd } \nonumber \\
& \stackrel{(c)}{\geq}& \Hreg{\algebraicset{g(f(\senderrnd,\receiverrnd),\receiverrnd)} | \receiverrnd}  \nonumber \\
& = & \Hreg{\algebraicset{\senderrndhat} | \receiverrnd} \nonumber \\
& \stackrel{(d)}{=} & \Hreg{\algebraicset{\senderrnd} | \receiverrnd} \nonumber \\
& \stackrel{(e)}{=} & \Hreg{\algebraicset{\senderrnd} | \algebraicset{\receiverrnd}} \nonumber \\
& = & \sum_{k \subseteq \kernelspace} P(\algebraicset{\receiverrnd} = k) \Hreg{\algebraicset{\senderrnd} | \algebraicset{\receiverrnd}=k} \nonumber \\
& \stackrel{(f)}{=} & \sum_{k \subseteq \kernelspace}  P(\algebraicset{\receiverrnd} = k) \sum_j \Hreg{\kv{\senderrnd}_j | \kv{\receiverrnd}_j=k_j} \nonumber \\
& \stackrel{(g)}{=} & \sum_{k \subseteq \kernelspace}  P(\algebraicset{\receiverrnd} = k) \sum_{j : k_j = 0 } \Hreg{\kv{\senderrnd}_j | \kv{\receiverrnd}_j=0}\nonumber \\
& & + \sum_{k \subseteq \kernelspace}  P(\algebraicset{\receiverrnd} = k) \sum_{j : k_j = 1 } \Hreg{\kv{\senderrnd}_j | \kv{\receiverrnd}_j=1} \nonumber \\
& \stackrel{(h)}{=} & \sum_{k \subseteq \kernelspace}  P(\algebraicset{\receiverrnd} = k) \sum_{j : k_j = 1 } \Hreg{\kv{\senderrnd}_j | \kv{\receiverrnd}_j=1} \nonumber \\
& \stackrel{(i)}{=} & \sum_{k \subseteq \kernelspace} P(\algebraicset{\receiverrnd} = k) |k| \Hbin{ p_s / p_r}  \nonumber \\
& =& |\kernelspace| p_r \Hbin{ p_s/ p_r } \nonumber \\
& \stackrel{(j)}{=}& |\kernelspace| \Lambda(p_s,p_r-p_s) ,\label{eq:conditional_lower_bound_derivation}
\end{eqnarray}
where (a) follows from the law of total expectation, (b) follows from the assumption that the image of  $f(\cdot,\receiver)$ is prefix-free for any choice of $\receiver$ as well as the definition of conditional entropy, and (c) follows from the fact that deterministic functions of random quantities cannot increase entropy. To see (d), 
note that $\senderhat=g(f(\sender,\receiver),\receiver)$ and by assumption $\sender \entails \senderhat$, and for any $\query$ such that $\sender \entails \query$, we have $\senderhat \entails \query$. As a consequence, using the property (\ref{fundamental-rel}) of a Logic System, we conclude
\begin{eqnarray*}
\algebraicset{\sender} \subseteq \algebraicset{\senderhat} \subseteq \algebraicset{\query}.
\end{eqnarray*}
By choosing $\query = \sender$, we obtain the statement that $\algebraicset{\sender} =\algebraicset{\senderhat}$. Since $\algebraicset{\senderrnd}=\algebraicset{\senderrndhat}$, the conditional distribution of either given $\receiverrnd$ is the same, establishing (d). Step (e) follows from the assumption that  $\receiverrnd \rightarrow \algebraicset{\receiverrnd} \rightarrow \algebraicset{\senderrnd}$,  and then utilizing the result (\ref{eq:markov_chain_conditional_entropy}). Step (f) follows from the assumption of independence. Step (g) is a simple splitting of the summation in (f). To justify (h), we use the assumption that $\senderrnd \entails \receiverrnd$ which in particular means that 
\begin{eqnarray*}
    \algebraicset{\senderrnd} \subseteq \algebraicset{\receiverrnd} .
\end{eqnarray*}
Thus necessarily if $\kv{\receiverrnd}_j=0$ then $\kv{\senderrnd}_j=0$ and the corresponding conditional entropy is zero.  To justify (i), we invoke the ``identically distributed'' property, together with the assumption that $|\mathcal{M}|^{-1} E |\algebraicset{\senderrnd}| = p_s$, $ |\mathcal{M}|^{-1} E |\algebraicset{\receiverrnd}| = p_r$, as well as Lemma \ref{lem:simpleab}. The final step (j) follows from the definition of $\Lambda$. This establishes the lower bound. 

The upper bound is proved with a simple variant of the proof of Theorem~\ref{thm:logicinfo}, where $\kernelspace$ is substituted with $\algebraicset{\receiverrnd}$.  We write the proof here for completeness. For a given size $\xi$ of a kernel, there are a total of
\begin{eqnarray*}
{|\algebraicset{\receiverrnd}| \choose \xi}
\leq 2^{|\algebraicset{\receiverrnd}| \Hbin{\frac{\xi}{|\algebraicset{\receiverrnd}|}}}
\end{eqnarray*}
possible kernels of the same size, since both sender and receiver share knowledge of $\receiverrnd$. Let $\mbox{enum}_{\xi,\psi} : \{ k_s \subseteq \kernelspace : |k_s| = \xi \}  \times \{ k_r \subseteq \kernelspace : |k_r| = \psi \} \rightarrow \{0,1\}^{*}$ be a function that maps each possible pair $(k_s,k_r)$ of kernels respectively of size $\xi$ and $\psi$ such that $k_s \subseteq k_r$, to a fixed-length binary encoding of the integers
$\left\{1,\ldots, 
\displaystyle{\psi \choose \xi}
\right\}$,
which is an integer that uniquely determines $k_s$ as a subset of $k_r$. Then, in particular, 
\begin{eqnarray*}
\len{ \mbox{enum}_{|\algebraicset{\senderrnd}|,|\algebraicset{\receiverrnd}|}(\algebraicset{\senderrnd},\algebraicset{\receiverrnd})  } &\leq& |\algebraicset{\receiverrnd} | \Hbin{\frac{|\algebraicset{\senderrnd}|}{|\algebraicset{\receiverrnd}|}} + 1  \\
 &=& \Lambda(|\algebraicset{\senderrnd}|,|\algebraicset{\receiverrnd}|-|\algebraicset{\senderrnd}|) + 1.
\end{eqnarray*}
The last step follows from the definition of $\Lambda$; note that the resulting expression is a random variable. For given sentences $\sender, \receiver \in \surfacespace$, we define the encoder $f$ as a concatenation of two separate encodings:
\begin{eqnarray}
f(\sender,\receiver) = \mbox{elias}_{\delta}( |\algebraicset{\sender}| )  \mbox{enum}_{|\algebraicset{\sender}|,|\algebraicset{\receiverrnd}|}(\algebraicset{\sender},\algebraicset{\receiver})
\label{eq:concatenation_conditional} .
\end{eqnarray}
Since both of the codes implied by each encoding are prefix-free, the concatenation is also prefix-free. Finally, we let $g$ be the decoder that recovers the kernel $\algebraicset{\sender}$ from the output of $f$, and evaluates $\ell$ on that kernel.

We note that, by construction, if $\query$ is such that $\sender \entails \query$, then $g(f(\sender,\receiver),\receiver) \entails \query$ since $\algebraicset{\sender} = \algebraicset{ g(f(\sender,\receiver),\receiver)}$, and thus we have met the conditions of the Theorem. The estimate for the overall cost of the encoding can be done by separately estimating the length of the two encodings in (\ref{eq:concatenation_conditional}), namely
\begin{eqnarray*}
\Lambda( |\algebraicset{\senderrnd}|, |\algebraicset{\receiverrnd}| - |\algebraicset{\senderrnd}|)  + \log_2 |\algebraicset{\senderrnd}| + 2 \log_2\left(  \log_2 |\algebraicset{\senderrnd}| \right) + 4 .
\end{eqnarray*}
Taking the expectation with respect to $\senderrnd, \receiverrnd$, using the concavity $\cap$ of $\Lambda$ (see Lemma~\ref{lem:elementarylambda}) as well as of the  logarithm, and normalizing by $|\kernelspace|$, we obtain an upper estimate of
\begin{eqnarray*}
\Lambda(p_s,p_r-p_s) + \frac{\log_2 \left( p_s |\kernelspace| \right) }{|\kernelspace|}  +2  \frac{\log_2\left(\log_2 \left( p_s |\kernelspace| \right) \right)}{|\kernelspace|} + \frac{4}{|\kernelspace|} .
\end{eqnarray*}
As before, the upper bounds holds under the more general assumption that $\senderrnd, \receiverrnd$ have kernels that follow a ($p_s,p_r$)-law. This completes the proof of the Theorem.

\subsection{The partition compression problem}
\label{ss:partition_compression}
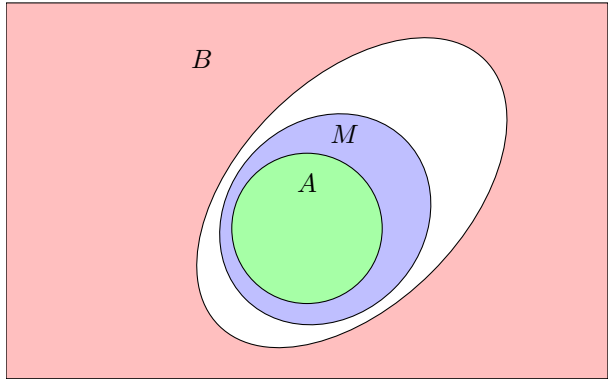
\begin{figure}
\begin{center}
\begin{tikzpicture}  
\clip(0,0) rectangle (8,5);
\draw (0,0)[fill=pink] rectangle (8,5);
\draw(5,-1.5)[rotate=45,fill=white] ellipse(2.5 and 1.5);
\draw(2.6,4.25) node {$B$};
\draw(4.5,-1.5)[rotate=45,fill=blue!25!white] ellipse(1.5 and 1.3);
\draw(4.5,3.25) node {$M$};
\draw(4,2)[fill=green!35!white] circle(1.0);  
\draw(4.0,2.6) node {$A$};
\end{tikzpicture}
  \caption{An example of a partition $\{M,M^c\}$ that separates $A$ from $B$.}
\label{fig:set_partitioning}
\end{center}
\end{figure}

In this subsection, we treat a lossy data  compression problem of central relevance to the ``targeted query'' settings of Theorems \ref{thm:less_is_more_simple}, \ref{thm:less_is_more_background} and \ref{thm:master}.  We refer the reader to Figure~\ref{fig:set_partitioning}. Imagine one has $n$ items as well as two non-intersecting subsets of those $n$ items, which we call $A$ and $B$, and one is interested in efficiently sending to a receiver a subset $M$ which contains $A$ but excludes $B$; or, alternatively stated, a partition that separates $A$ from $B$. One solution is to send $M=A$ or $M=B$, whichever is cheapest to send, but it turns out that there is generally a better solution. Let $X^n \in \{0,1,2\}^n$ be a vector with $X_i = 1$ if the $i$th element is inside of $A$, $X_i=0$ if the $i$th element is inside of $B$, and $X_i = 2$ if the $i$th element is neither in $A$ nor in $B$. Assume that the entries of $X^n$ meet the following conditions:
\begin{eqnarray}
n^{-1} E | \{i : X_i = 0 \} | &=& p_b , \nonumber \\
n^{-1} E | \{i : X_i = 1 \} | &=& p_a , \nonumber \\
n^{-1} E | \{i : X_i = 2 \} | &=& 1-p_a-p_b .
 \label{eq:vector_conditions}
\end{eqnarray}

Define a distortion metric to be a function
\begin{equation}
    \rho : \{0,1,2\} \times \{0,1\} \rightarrow \{0,1\}
\label{eq:distortion_metric}
\end{equation}
using this matrix
\begin{equation}
    \begin{array}{c|cc}
       & 0 & 1 \\ \hline
      0  & 0 & 1 \\
      1  & 1 & 0 \\
      2  & 0 & 0 
    \end{array} \;\; .
\label{eq:distortion_metric_matrix}
\end{equation}
An encoder and decoder are functions $f_n : \{0,1,2\}^n \rightarrow \{0,1\}^*$ and  $g_n : \{0,1\}^* \rightarrow \{0,1\}^n$, respectively. The problem is to find a good upper bound for
\begin{eqnarray*}
\min_{f_n,g_n} n^{-1} E_{X^n}\left[ \len{f_n(X^n) }\right]
\end{eqnarray*}
subject to the condition that, for any $x_1^n \in \{0,1,2\}^n$, 
\begin{eqnarray}
\sum_{i=0}^{n-1} \rho(x_i,g_n(f_n(x_1^n))_i) = 0 .
\label{eq:corecondition}
\end{eqnarray}
The solution to this problem can be obtained by an application of Shannon's Rate-Distortion theory.

\begin{restatable}[Shannon bounds for partition compression]{thm}{shannonpartition}
~For any $n \geq 1$, let $X_1^n \in \{0,1,2\}^n$ be a random vector with the property (\ref{eq:vector_conditions}). Let $f_n, g_n$ be encoder and decoder functions as defined earlier. Then
\begin{eqnarray*}
\min_{f_n,g_n} n^{-1} E_X\left[ \len{f_n(X_1^n) }\right]\leq \Lambda(p_a,p_b) + 2 \frac{\log_2\left( n \Lambda(p_a,p_b)\right)}{n} + \frac{3}{n} ,
\end{eqnarray*}
where the minimization is over $f_n, g_n$, such that the image of $f_n$ is prefix-free and that condition (\ref{eq:corecondition}) is met. Furthermore, if the random variables $\{X_i\}$ are additionally 
i.i.d., we have
\begin{eqnarray*}
\Lambda(p_a,p_b)  \leq \min_{f_n,g_n} n^{-1} E_X\left[ \len{f_n(X_1^n) }\right].
\end{eqnarray*}
\label{thm:shannonpartition}
\end{restatable}
\textbf{Remark:} From Lemma~\ref{lem:elementarylambda}, if $p_a + p_b < 1$, then we can deduce that $\Lambda(p_a,p_b) < \min\{ \Hbin{p_a}, \Hbin{p_b} \}$, and thus this result predicts the existence of partition compression techniques which are more efficient than the ``naive'' solution of sending the cheapest of the sets ($A$ or $B$). \\

\textbf{Proof.}
\label{ss:shannonpartition_proof}
We provide a proof of the upper bound that is simple and includes a useful second order upper bound which will play a role in the proofs of Theorems \ref{thm:less_is_more_simple}, \ref{thm:less_is_more_background}. As we will show, the expected performance of a random matrix with the density of ones appropriately tuned will asymptotically approach the Shannon limit $\Lambda(p_a,p_b)$. This implies, as per the classic random coding argument of Shannon, the existence of a deterministic code with performance at least as good as the expected performance of the random code. 

Let $X_1,\ldots,X_n \in \{0,1,\dontcare\}^n$ be the input random vector, and 
let $\Psi \subseteq \{1,\ldots,n\}$ denote the random positions where $X_i$ is taking on $0$ or $1$,
thus where we want to enforce a bit pattern. For any arbitrary $z \in \{0,1\}^n$, let $[z]_{\Psi}$ denote the $1 \times |\Psi|$ vector obtained by extracting from $z$ the columns indexed by $\Psi$. 

Let $C$ be a random binary matrix with an infinite number of rows $\{C_1,C_2,\ldots\}$ and each with $n$ columns. Assume its entries are chosen i.i.d.\ according to the distribution
\begin{eqnarray*}
P(C_{i,j} = 0) = \frac{p_a}{p_a+p_b} .
\end{eqnarray*}
The sender scans the matrix $C$ from top to bottom until it finds the first row $J$ that satisfies the following condition:
\begin{eqnarray}
[C_{J}]_{\Psi} = X_{\Psi} ,
\label{eq:match_condition}
\end{eqnarray}
and sends the index $J$ of that row. The receiver then recovers the row from the index. Let $N_0 = \{ i : X_i = 0\}$ and $N_1 = \{i : X_i = 1\}$. The  probability of a row of $C$ satisfying (\ref{eq:match_condition}), conditional on $\Psi$, is given by
\begin{eqnarray*}
P([C_{i}]_{\Psi} = X_{\Psi} | \Psi ) =  \left(\frac{p_a}{p_a+p_b}\right)^{N_0}\left(\frac{p_b}{p_a+p_b}\right)^{N_1}.
\end{eqnarray*}
Note that independence of the entries in the vector $X$ was not necessary to assert this, and instead, the way $C$ is constructed is sufficient. 
It is now easy to calculate
$$
P(J=j)=(1-P([C_{i}]_{\Psi} = X_{\Psi} | \Psi ))^{j-1} P([C_{i}]_{\Psi} = X_{\Psi} | \Psi )
$$
since $J$ is the {\it first} occurrence of the pattern $\Psi$. 
As a consequence, 
\begin{eqnarray*}
E\left[J | \Psi \right] = 1/P([C_{i}]_{\Psi} = X_{\Psi} | \Psi )=
2^{\left( - N_0 \log_2 \frac{p_a}{p_a+p_b} - N_1 \log_2 \frac{p_b}{p_a+p_b} \right)} .
\end{eqnarray*}
To send the index $J$ we will be using $\delta$ Elias coding. Note that $E_{\Psi} \left[ \log_2 \left( E\left[J | \Psi \right] \right) \right] = n \Lambda(p_a,p_b)$. We can then upper bound the performance of the code as
\begin{eqnarray*}
E_J \left[ \len{ \mbox{elias}_{\delta}(J) } \right]
&=& E_{\Psi} \left[ E_J \left[ \len{ \mbox{elias}_{\delta}(J)}
| \Psi \right] \right] \\
& \leq & E_{\Psi} \left[ E_{J} \left [\log_2 J + 2 \log_2 (\log_2 J )+ 3  \ |\Psi \right] \right] \\
& \leq & E_{\Psi} \left[ \log_2 E_{J}\left[ J | \Psi \right]  + 2 \log_2 (\log_2 E_{J} \left[ J | \Psi \right]  ) + 3 \right] \\
& \leq & E_{\Psi} \left[ \log_2 E_{J}\left[ J | \Psi \right] \right] + 2 \log_2(E_{\Psi} \left[ \log_2 E_{J} \left[ J | \Psi \right] ) + 3 \right] \\
& \leq & n\Lambda(p_a,p_b) + 2 \log_2(n \Lambda(p_a,p_b)) + 3 ,
\end{eqnarray*}
where in addition to the estimate of the performance of $\delta$ Elias coding, we used the concavity $\cap$ of the logarithm. The lower bound, which is only valid in the case $X_1^n$ is an i.i.d.\ random vector, is a straightforward consequence of existing $R(D)$ literature~\cite{berger1971rate} so it is omitted.

\subsection{Less is more}
\label{ss:lessismoreproof}

\begin{figure}
\begin{tblr}{
  colspec = {X[c,h]X[c]X[c]},
  stretch = 0,
  rowsep = 3pt
}
\scalebox{0.9}{\begin{tikzpicture}[baseline={([yshift=0ex]current bounding box.center)}]
	\matrix[row sep=2.5mm, column sep=5mm]
	{tm	
        &
        &
        &
        \node[coordinate] (bottop) {}; &
        &
        &
        & \\
        \node[dspnodeopen,dsp/label=left] (r10) {$\queryrnd$}; & 
        \node[dspsquare]                  (r11) {$\kappa^c$};       &
        \node[coordinate] (r12) {};  &
        \node[coordinate] (r14) {};  &
        \\
 	    \node[dspnodeopen,dsp/label=left] (r00) {$\senderrnd$}; & 
        \node[dspsquare]        (r01) {$\kappa$};            &
        \node[dspsettopenc]     (r02) {};                    &
        \node[coordinate]       (r04) {};                    &
        \node[dspbigsquare]          (r05) { \scriptsize partition \\ \scriptsize decode};           &
        \node[dspsquare]        (r06) {$\ell$};            &
        \node[dspnodeopen,dsp/label=right] (r07) {$\senderrndhat$};
        \\

        &
        &
        &
        \node[coordinate] (bbot) {};&
         \\
    };

   \draw[dashed] (bbot)--(bottop);
   
	\begin{scope}[start chain]
		\chainin (r00);
		\chainin (r01) [join=by dspconn];
		\chainin (r02) [join=by dspconn];
		\chainin (r04) [join=by dspline];
		\chainin (r05) [join=by dspconn];
		\chainin (r06) [join=by dspconn];
		\chainin (r07) [join=by dspconn];
    \end{scope}

    \begin{scope}[start chain]
		\chainin (r10);
		\chainin (r11) [join=by dspconn];
		\chainin (r12) [join=by dspline];
  		\chainin (r02) [join=by dspconn];
    \end{scope}


    
\end{tikzpicture}} & \scalebox{0.85}{\begin{tikzpicture}  
\clip (0,0) rectangle (8,5);
\draw (0,0)[fill=pink] rectangle (8,5);
\draw(5,-1.5)[rotate=45,fill=white] ellipse(2.5 and 1.5);
\draw(5.6,4.25) node {$\algebraicset{\query}$};
\draw(4.5,-1.5)[rotate=45,fill=blue!25!white] ellipse(1.5 and 1.3);
\draw(4.5,3.25) node {$\algebraicset{\senderhat}$};
\draw(4,2)[fill=green!35!white] circle(1.0);  
\draw(4.0,2.6) node {$\algebraicset{\sender}$};
\end{tikzpicture}} 
\\ (a) & (b)
\end{tblr}
\caption{Proof strategy for Theorem~\ref{thm:less_is_more_simple}}
\label{fig:less_is_more_simple}
\end{figure}
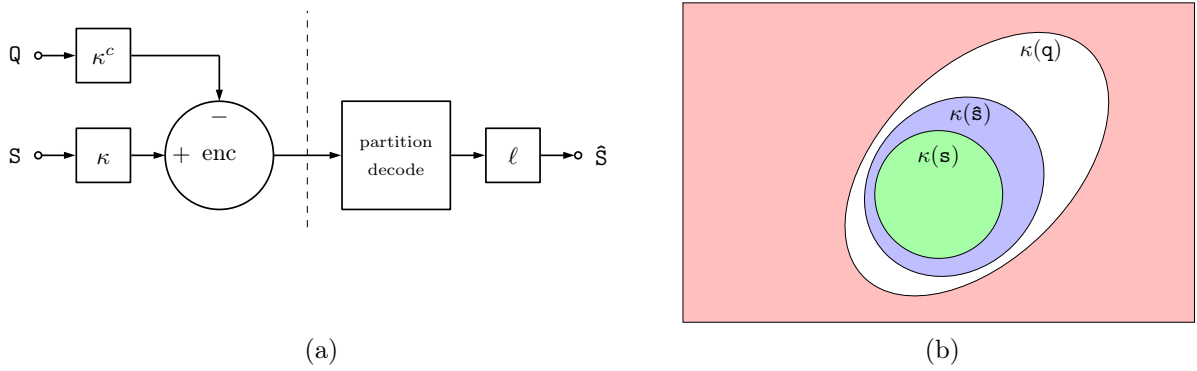

In this pattern the purpose of Alice is not to communicate to Bob enough to prove anything she can prove; rather it is to communicate to him the minimal amount of information needed to prove a \emph{specific sentence} $\query$ (and of course, having achieved that, any other sentence that is entailed by $\query$). We refer the reader to the top of Figure~\ref{fig:less_is_more_simple}. In here we illustrate $\algebraicset{\sender}$ in green, and $\algebraicset{\query}$ in white. One fairly obvious strategy to solve this problem would be to send to Bob enough information to reconstruct $\algebraicset{\query}$; then Bob will be able to prove anything that can be proved starting from $\query$. Alternately, one could send to Bob enough information to reconstruct $\algebraicset{\sender}$ itself, which would allow Bob to prove potentially even more logic sentences. For the purposes of quantifying the cost of either of these two strategies, let us assume that $\senderrnd,\queryrnd$ have kernels that follow a ($p_s,p_q$)-law (with $p_s < p_q$), then using the ideas behind Theorem \ref{thm:logicinfo}, and choosing the best from either of these two strategies, we would be spending a normalized average total of
\begin{eqnarray}
    \min \{\Hbin{p_s}, \Hbin{p_q}\}
\label{eq:to_beat}
\end{eqnarray}
bits (neglecting asymptotically vanishing terms). Yet, as we will soon show, this strategy is in general \emph{suboptimal}. This was a surprising revelation to us, so we want to equip the reader with the insight used to prove this result. From the top of Figure~\ref{fig:less_is_more_simple}, it should be apparent that not only could Alice use either of the two strategies above to communicate to Bob; in fact she has the freedom to send any possible $\algebraicset{\senderhat}$ that satisfies $\algebraicset{\sender} \subseteq \algebraicset{\senderhat} \subseteq \algebraicset{\query}$; one such example $\algebraicset{\senderhat}$ is illustrated in blue in the Figure. However, crucially, note that a set $\algebraicset{\senderhat}$ that is bigger than $\algebraicset{\sender}$, even though it may feel more complex to describe, can actually be good for many choices for $\sender$ and $\query$, \emph{and thus we may not need that big a pre-agreed collection of those}. This insight means we can beat the estimate (\ref{eq:to_beat}). In the same figure, we show an architecture with a special form of encoding which was introduced in the proof of Theorem \ref{thm:shannonpartition}.

\begin{restatable}[Goal is a targeted query]{thm}{lessismoresimple}
~Let $(L,  \kernelspace, \kappa, \ell)$ be a Logic System. Let $\senderrnd, \queryrnd \in  \surfacespace$ represent the sender's and query logic sentences, respectively, with the property that $\senderrnd \entails \queryrnd$ and in particular, $\senderrnd, \queryrnd$ have kernels that follow a ($p_s,p_q$)-law. Let the encoder $f$ and decoder $g$ be functions
\begin{eqnarray*}
f : \surfacespace^2 & \rightarrow & \{0,1\}^* , \\
g : \{0,1\}^* & \rightarrow & \surfacespace.
\end{eqnarray*}
Then
\begin{eqnarray*}
\min_{f,g} |\kernelspace|^{-1} E_{\senderrnd,\queryrnd} \left[ \len{f(\senderrnd,\queryrnd)} \right] 
\leq \Lambda(p_s,1-p_q) + O\left( \frac{\log_2|\kernelspace|}{|\kernelspace|} \right),
\end{eqnarray*}
where the minimization is over $f, g$ such that the image of $f$ is prefix-free and such that, if $\sender \entails \query$, then $\sender \entails g(f(\sender,\query))$ and $g(f(\sender,\query)) \entails \query$. If additionally, $\{(\kv{\senderrnd},\kv{\queryrnd})_j\}_{j=1}^{|\kernelspace|}$ are i.i.d., then
\begin{eqnarray*}
 \Lambda( p_s, 1-p_q) \leq \min_{f,g} |\kernelspace|^{-1} E_{\senderrnd,\queryrnd} \left[ \len{f(\senderrnd,\queryrnd)} \right] 
\end{eqnarray*}

\label{thm:less_is_more_simple}
\end{restatable}
 
\textbf{Proof.} As with Theorem~\ref{thm:logicinfo}, we start by invoking Kraft's inequality through Lemma~\ref{lem:vlc_kraft} and write
\begin{eqnarray*}
E_{\senderrnd,\queryrnd} \left[ \len{f(\senderrnd,\queryrnd)} \right] \geq \Hreg{ f(\senderrnd,\queryrnd)} \geq \Hreg{g(f(\senderrnd,\queryrnd))} .
\end{eqnarray*}
The direction of the proof now diverges with respect to that of Theorem~\ref{thm:logicinfo}, by deriving
\begin{eqnarray*}
 \Hreg{g(f(\senderrnd,\queryrnd))} &\stackrel{(a)}{=}& \Hreg{ g(f(\senderrnd,\queryrnd))} - \Hreg{ g(f(\senderrnd,\queryrnd)) | \senderrnd,\queryrnd} \\
 &\stackrel{(b)}{=}& I( \senderrnd,\queryrnd; g(f(\senderrnd,\queryrnd)) ) \\
 &\stackrel{(c)}{\geq}& I(\kv{\senderrnd},\kv{\queryrnd};  \kv{g(f(\senderrnd,\queryrnd))} ) 
\end{eqnarray*}
where (a) follows from the fact that discrete entropy is zero when conditioning on all randomness, (b) is from the definition of mutual information, and (c) follows from the data processing inequality.

Referencing Figure~\ref{fig:less_is_more_simple}, we now construct a function $\psi$ that accepts $\kv{\senderrnd}$ and $\kv{\queryrnd}$ and produces a vector of length $|\kernelspace|$ with entries in $\{0,1,\dontcare\}$ indicating whether each entry is in the red (0), green (1) or white (2) regions.  Let $\mathtt{v}, \mathtt{w} \in \{0,1\}^{|\kernelspace|}$. The construction is as follows:
\begin{eqnarray}
\psi(\mathtt{v},\mathtt{w}) = 0 \times (\underline{1} - \mathtt{w}) + 1 \times \mathtt{v} + 2 \times ( \underline{1} - \mathtt{v} )\mathtt{w} ,
\label{eq:def_psi}
\end{eqnarray}
where the $\times$ operator is multiplying a scalar times a vector element-wise, and $\underline{1}$ represents a vector of all ones, and the product of vectors in the last term is element-wise. Of course, the leftmost term is always zero, but we included it to ensure that the connection to Figure~\ref{fig:less_is_more_simple} is clear. Next define
\begin{eqnarray*}
    X_{1}^{|\kernelspace|} &=& \psi(\kv{\senderrnd},\kv{\queryrnd}) , \\
    Z_{1}^{|\kernelspace|} &=& 1 \times \kv{g(f(\senderrnd,\queryrnd))} .
\end{eqnarray*}

The relationship between $X_1^{|\kernelspace|}$ and $Z_1^{|\kernelspace|}$ is in general very complex, as we have few assumptions on $f$ and $g$. However, some key assertions can be made. First, recall that $\senderrnd,\queryrnd$ have kernels that follow a ($p_s,p_q$)-law; see Definition \ref{def:kernel_law}, and in particular recall that $\senderrnd \entails \queryrnd$. Using elementary probability,
\begin{eqnarray*}
    P(X_i = 0) &=& P([\kv{\senderrnd}_i = 0] \cap [ \kv{\queryrnd}_i = 0]) \nonumber \\
    &=&P([\kv{\senderrnd}_i = 0]) + P([ \kv{\queryrnd}_i = 0]) - P([\kv{\senderrnd}_i = 0] \cup [ \kv{\queryrnd}_i = 0]) \\
    &=&P([\kv{\queryrnd}_i = 0])
\end{eqnarray*}
where the last equality uses (\ref{fundamental-rel}) from the fundamental Definition \ref{def:core}, to conclude that $\kv{\senderrnd} \leq \kv{\queryrnd}$ where the inequality is to be interpreted element-wise. Now using the ``identically distributed'' assumption from the (and not yet independence), we obtain
\begin{eqnarray*}
    P(X_i = 0) = 1 - p_q
\end{eqnarray*}
With a far simpler argument, also without using independence and only the identical distributed assumption, we obtain
\begin{eqnarray*}
    P(X_i = 1) =  P([\kv{\senderrnd}_i = 1]) = p_s
\end{eqnarray*}
Finally we add the independence assumption, thus concluding that the $\{X_i\}$ are i.i.d.\ according to the distribution
\begin{eqnarray*}
P(X_i = 0) &=& 1-p_q, \label{eq:x01d}\\
P(X_i = 1) &=& p_s, \nonumber\\
P(X_i = 2) &=& p_q - p_s . \nonumber
\end{eqnarray*}
To continue, the assumption that $\sender \entails g(f(\sender,\query))$, $ g(f(\sender,\query)) \entails \query$  can be used to establish a useful relation between $X_i$ and $Z_i$. Recall the definition of the distortion metric $\rho$ from Equations (\ref{eq:distortion_metric}) and (\ref{eq:distortion_metric_matrix}).  Then, the assumption implies that, for all $1 \leq i \leq |\kernelspace|$,
\begin{eqnarray}
\rho( X_i, Z_i) = 0 .
\label{eq:fact} 
\end{eqnarray}
This is an important fact that will be used shortly. We now apply the data processing inequality once more, taking advantage of the definitions for $X_i$ and $Z_i$, and continue the proof with a pattern commonly found in Rate-Distortion theory:
\begin{eqnarray}
I(\kv{\senderrnd}, \kv{\queryrnd};  \kv{g(f(\senderrnd,\queryrnd))} ) &\stackrel{(d)}{\geq}& I(X_{1}, \ldots, X_{|\kernelspace|}; Z_1, \ldots, Z_{|\kernelspace|}  ) \nonumber \\
  &=& \Hreg{ X_{1}, \ldots, X_{|\kernelspace|}} - \Hreg{ X_{1}, \ldots, X_{|\kernelspace|} | Z_1, \ldots, Z_{|\kernelspace|} } \nonumber \\
 &\stackrel{(e)}{=}&  \sum_{i=1}^{|\kernelspace|} \Hreg{X_i} - \Hreg{X_i | Z_1, \ldots, Z_{|\kernelspace|}, 
  X_{1}, \ldots X_{i-1}} \nonumber \\
  & \stackrel{(f)}{\geq} & \sum_{i=1}^{|\kernelspace|} \Hreg{X_i} - \Hreg{X_i | Z_i} \nonumber  \\
    & = & \sum_{i=1}^{|\kernelspace|} I(X_i ; Z_i) \label{eq:rd_trip} 
\end{eqnarray}
where (d) follows from the data processing inequality, (e) follows from the fact that the $\{X_i\}$ are independent and from the chain rule for entropy, and (f) follows from the fact that conditioning cannot increase entropy. 

We now pause to observe that for any given random variables $V,W$, the mutual information $I(V;W)$ is an expression that can be entirely computed from $p_{V}$ and $Q_{W|V}$ as these two completely determine the distribution of $V,W$. Thus we could write $I(V,W) = \iota(p_V,Q_{W|V})$ for some function $\iota$ and in particular the step (\ref{eq:rd_trip})  can be rewritten as
\begin{eqnarray}
\sum_{i=1}^{|\kernelspace|} \iota(p_{X_i}, Q_{Z_i|X_i}) = \sum_{i=1}^{|\kernelspace|} \iota(p_{X_1}, Q_{Z_i|X_i})
\label{eq:transitioning}
\end{eqnarray}
where this last equality follows from the fact that 
the marginal for $X_i$ is identical for all $i$, but the conditionals $Q_{Z_i|X_i}$ are in general different. It is known that the function $\iota(p_{V}, Q_{W|V})$ is convex $\cup$ on $Q_{W|V}$ and therefore by Jensen's inequality, the expression (\ref{eq:transitioning}) can be lower bounded by
\begin{eqnarray*}
\iota\left(p_{X_1},\frac{1}{|\kernelspace|} \sum_{i=1}^{|\kernelspace|} Q_{Z_i|X_i}(z|x) \right).
\end{eqnarray*}
Let $X^{\prime},Z^{\prime}$ be distributed according to the marginal for $X_1$ and the averaging of conditional distribution above.  Because of (\ref{eq:fact}), it is the case that
\begin{eqnarray*}
E_{X^{\prime}, Z^{\prime}} \rho(X^{\prime},Z^{\prime}) = 0 .
\end{eqnarray*}
Therefore the following bound holds
\begin{eqnarray*}
\sum_{i=1}^{|\kernelspace|} I(X_i ; Z_i)  \geq |\kernelspace| I(X^{\prime};Z^{\prime}) \geq |\kernelspace| \min_{P_{X,Z} \in \mathcal{D}} I(X;Z),
\end{eqnarray*}
where the domain $\mathcal{D}$ for the minimization is defined by joint distributions for $X,Z$ with $X \sim (1-p_q,p_s,p_q-p_s )$ and $ E[\rho(X,Z)] = 0$. The fact that such a minimization results in $\Lambda(p_s,1-p_q)$ can be checked using standard variational methods. This concludes the proof of the lower bound.

To prove the upper bound, we construct a code as follows.  Define
\begin{eqnarray*}
X_1^{|\kernelspace|} = \psi(\kv{\senderrnd},\kv{\queryrnd}) .
\label{eq:defxvec}
\end{eqnarray*}
We note that by construction, under the weaker assumption that $\senderrnd, \queryrnd$ have kernels that follow a $(p_s, p_q)$-law.
\begin{eqnarray*}
n^{-1} \sum_{i=0}^{n-1} P(X_i = 0) &=& 1-p_q ,\nonumber \\
n^{-1} \sum_{i=0}^{n-1} P(X_i = 1) &=& p_s ,\nonumber \\
n^{-1} \sum_{i=0}^{n-1} P(X_i = 2) &=& p_q - p_s .
\end{eqnarray*}
In reference to (\ref{eq:vector_conditions}), we next invoke the upper bound of Theorem~\ref{thm:shannonpartition}, which guarantees the existence of $\hat{f}$, $\hat{g}$ such that
\begin{eqnarray*}
|\kernelspace|^{-1} E_{X_1^{|\kernelspace|}}\left[ \len{\hat{f}(X_1^{|\kernelspace|})}\right] &\leq& \Lambda(p_s,1-p_q) +  2 \frac{\log_2\left( |\kernelspace| \Lambda(p_s,1-p_q)\right)}{|\kernelspace|} + \frac{3}{|\kernelspace|} , \\
\sum_{i=1}^{|\kernelspace|} \rho(X_i,\hat{g}(\hat{f}(X_1^{|\kernelspace|}))_i) &=& 0.
\end{eqnarray*}
We define our encoder then as
\begin{eqnarray*}
f(\sender,\query) = \hat{f}( \psi(\kv{\sender},\kv{\query}) ) .
\end{eqnarray*}
Recall that $\hat{g}(\mbox{codeword}) \in \{0,1\}^{|\kernelspace|}$ and thus it can be regarded also as a kernel via the dual notation where kernels are depicted by the indicator vector of the corresponding subset of $\kernelspace$. To conclude, we define the decoder $g$ as
\begin{eqnarray*}
g( \mbox{codeword} ) = \ellv{ \hat{g}(\mbox{codeword}) } ,
\end{eqnarray*}
where $\mbox{codeword} \in \{0,1\}^*$ is meant to be precisely $f(\sender,\query)$ when the encoder and decoder are being used simultaneously. We conclude the proof by noting that, by construction, if $\sender \entails \query$, we have $\sender \entails g(f(\sender,\query))$, $ g(f(\sender,\query)) \entails \query$ and furthermore
\begin{eqnarray*}
|\kernelspace|^{-1} E_{\senderrnd,\queryrnd} \left[ \len{f(\senderrnd,\queryrnd)} \right] \leq \Lambda( p_s, 1-p_q) +  2 \frac{\log_2\left( |\kernelspace| \Lambda(p_s,1-p_q)\right)}{|\kernelspace|} + \frac{3}{|\kernelspace|}.
\end{eqnarray*} \qed

\subsection{No need to know}
\label{ss:unaware}

In this problem, Bob's logic sentence $\receiverrnd$ is unavailable to Alice, and the goal of the communication is for Bob to be able to prove anything that Alice can. To address the problem, we will need to introduce a more complex communication pattern, where Alice and Bob are able to exchange messages over several rounds, taking turns on who is sending and who is receiving. As before, we will use $f$ to denote an encoder and $g$ to denote a decoder, but the agent doing the encoding and decoding will change depending on the turn.

\begin{table}
\begin{center}
\begin{tabular}{ccccccc}
Alice's & common       &Alice's & direction & Bob's & common & Bob's  \\
private & context  & function &  & function & context & private  \\ 
context &     & &  & &  & context  \\ \hline 
$\sender$ & & $g^0$ & $\stackrel{b^0}{\longleftarrow}$ & $f^0$ & & $\receiver$ \\
$\sender$ & $b^0$ & $f^1$ & $\stackrel{a^1}{\longrightarrow}$ & $g^1$ & $b^0$ & $\receiver$ \\
$\sender$ & $b^0, a^1$ & $g^2$ & $\stackrel{b^2}{\longleftarrow}$  & $f^2$ & $b^0, a^1$ & $\receiver$\\
$\sender$ & $b^0, a^1, b^2$ & $f^3$ & $\stackrel{a^3}{\longrightarrow}$  & $g^3$ & $b^0,a^1,b^2$ & $\receiver$ \\
& & & & $\downarrow$ & \\
& & & & $\senderhat$ &  \\
& & & & ($\forall \query$ s.t. $\sender \entails \query$, &  \\
& & & &  $\senderhat \entails \query$ )& 
\end{tabular}
\end{center}
\caption{A 4-turn code for full synchronization. It is assumed that $\sender \entails \receiver$. The actual step of Bob proving queries is not shown. }
\label{tab:mt}
\end{table}

Table \ref{tab:mt} is an exemplary representation of the conversational paradigm that we are now entering. In this table, the context to which either Alice or Bob are privy is shown, split into whether the context is either common or private. The context informs what the encoder or decoder functions are allowed to depend on; in addition, every decoder is allowed to depend on the output of the corresponding encoder, which is denoted by the letter on the top of each arrow indicating the direction of the communication. For example, the $f^3$ encoder, whose output is $a^3$, may depend on $\sender, b^0, a^1$ and $b^2$. In all cases, the image for each of the $f^i$ encoders is a subset of $\{0,1\}^{*}$. The communication task finishes when $g^3$ computes its output $\senderhatm \in \surfacespace$ which can now be used to prove any $\query$ with the property that $\sender \entails \query$. To measure the efficiency of communication, we define the total normalized average cost: 
\begin{eqnarray*}
    \frac{1}{|\kernelspace|} E_{\senderrnd, \receiverrnd} \left[ \len{B^0} + \len{A^1} + \len{B^2} + \len{A^3} \right]
\end{eqnarray*}
where $B^0, A^1, B^2, A^3$ are random versions of $b^0, a^1, b^2, a^3$ when in the interaction the sender experiences $\senderrnd$ and the receiver experiences $\receiverrnd$.
The reader may recall that in previous results, we assumed that the image of an encoder was a prefix-free code. In a similar manner, we assume that the image of encoder $f^0$ is prefix-free. In turn 1, both parties have $b^0$ as common context, and therefore the assumption that we have is that the image of $f^1$, when the value of $b^0$ is kept fixed, is prefix-free. Similarly, we assume that the image of $f^2$, when the value of $b^0, a^1$ are kept fixed, is prefix-free and finally, we assume that the image of $f^3$ when $b^0,a^1,b^2$ are kept fixed is prefix-free. We call any code with the properties illustrated by Table \ref{tab:mt} and this discussion a \emph{4-turn code} for full synchronization.

Before presenting the result we introduce the main intuition behind how the upper bound of this result is proved. Alice and Bob meet ahead of time, and they agree on a family of hash functions that can be used to map sets $\algebraicset{\senderrnd}$ of various cardinalities to a bin in a many-to-one fashion.  After the receiver is presented with $\receiverrnd$ and the sender is presented with $\senderrnd$, in Turn 0, the receiver informs the sender of the cardinality of $\receiverrnd$ set, and as part of Turn 1, the sender does the same but for $\senderrnd$. As a result of this calibration, also as part of Turn 1, the sender sends the index of a hash bin where $\algebraicset{\senderrnd}$ has been mapped. Aided by $\receiverrnd$, the receiver is able to recover $\algebraicset{\senderrnd}$ from the hash bin the majority of time. Turns 2 and 3 are there to address the possible exception where the receiver fails to recover $\kappa(\senderrnd)$ in Turn 1. The communication costs in Turns 0, 2 and 3 are asymptotically negligible and thus Turn 1 dominates the total communication cost. The main task is to demonstrate that asymptotically, the normalized logarithm of the number of such bins approaches $\Lambda(p_s,p_r-p_s)$ and to demonstrate that all other communication cost is negligible.

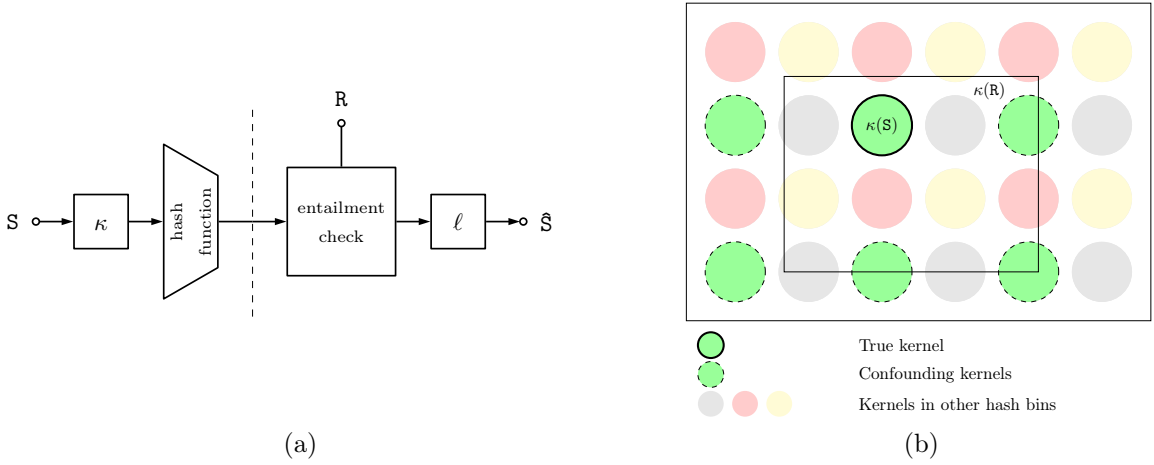
\begin{figure}
    \centering
    \begin{tblr}{
  colspec = {X[c]X[c]X[c]},
  stretch = 0,
  rowsep = 3pt
}
\scalebox{0.9}{\begin{tikzpicture}[baseline={([yshift=0ex]current bounding box.center)}]
	\matrix[row sep=2.5mm, column sep=5mm]
	{tm	

                &
        &
        &
        \node[coordinate] (bbotbot) {};&
        &
         \\
         
        \node[coordinate] (br10) {}; & 
        \node[coordinate] (br11) {}; &
        \node[coordinate] (br13) {}; &
        \node[coordinate] (br14) {}; &
        \node[dspnodeopen,dsp/label=above] (br15) {$\receiverrnd$}; &
        \\
 	    \node[dspnodeopen,dsp/label=left] (br00) {$\senderrnd$}; & 
        \node[dspsquare]        (br01) {$\kappa$}; &
        \node[dsphash]          (br03) { \scriptsize hash \\ \scriptsize function}; &
        \node[coordinate]       (br04) {}; &
        \node[dspbigsquare]          (br05) { \scriptsize entailment \\ \scriptsize check}; &
        \node[dspsquare]        (br06) {$\ell$};               &
        \node[dspnodeopen,dsp/label=right]        (br07) {$\senderrndhat$};            &
        \\
        &
        &
        &
        \node[coordinate] (bbot) {};&
        &
         \\
    };

   \draw[dashed] (bbot)--(bbotbot);
   

	\begin{scope}[start chain]
   	    \chainin (br00);
		\chainin (br01) [join=by dspconn];
  		\chainin (br03) [join=by dspconn];
		\chainin (br04) [join=by dspline];
		\chainin (br05) [join=by dspconn];
		\chainin (br06) [join=by dspconn];
		\chainin (br07) [join=by dspconn];
    \end{scope}

    \begin{scope}[start chain]
		\chainin (br15);%
		\chainin (br05) [join=by dspline];
    \end{scope}

\end{tikzpicture}} &
\scalebox{0.65}{\begin{tikzpicture}[darkstyle/.style={circle,fill=gray!20,minimum size=35},greenstyle/.style={circle,draw,dashed,fill=green!40,minimum size=35},mainstyle/.style={circle,draw,very thick,fill=green!40,minimum size=35},bluestyle/.style={circle,fill=blue!10,minimum size=35},yellowstyle/.style={circle,fill=yellow!20,minimum size=35},redstyle/.style={circle,fill=red!20,minimum size=35},smalldarkstyle/.style={circle,fill=gray!20,minimum size=15},smallgreenstyle/.style={circle,draw,dashed,fill=green!40,minimum size=15},smallmainstyle/.style={circle,draw,very thick,fill=green!40,minimum size=15},smallbluestyle/.style={circle,fill=blue!10,minimum size=35},smallyellowstyle/.style={circle,fill=yellow!20,minimum size=15},smallredstyle/.style={circle,fill=red!20,minimum size=15},baseline=(current bounding box.center)]
  \foreach \x in {0,...,5}
    \foreach \y in {0,...,3} 
      {\pgfmathtruncatemacro{\label}{(Mod(\x,2)*2)+Mod(\y,2)}
       \node [darkstyle]  (\x\y) at (1.5*\x,1.5*\y) {\label};} 

  \foreach \x in {0,...,2}
    \foreach \y in {0,...,1}
       {
       \pgfmathtruncatemacro{\label}{0}
       \pgfmathtruncatemacro{\xs}{2*\x}
       \pgfmathtruncatemacro{\ys}{2*\y}
       \pgfmathtruncatemacro{\xspo}{1+2*\x}
       \pgfmathtruncatemacro{\yspo}{1+2*\y}
       \node [greenstyle]  (\x\y) at (1.5*\xs,1.5*\ys) {};
       \node [darkstyle]    at (1.5*\xspo  ,1.5*\ys) {};
       \node [redstyle]     at (1.5*\xs  ,1.5*\yspo) {};
       \node [yellowstyle]  at (1.5*\xspo  ,1.5*\yspo) {};
       }


\node [mainstyle]  (main) at (1.5*2,1.5*2) {$\algebraicset{\senderrnd}$};

\node[smallmainstyle] at (-0.5,-1.5) {};
\node[smallgreenstyle] at (-0.5,-2.1) {};
\node[smalldarkstyle] at (-0.5,-2.7) {};
\node[smallredstyle] at (0.2,-2.7) {};
\node[smallyellowstyle] at (0.9,-2.7) {};
\node[anchor=west] at (2.4,-1.5) {True kernel};
\node[anchor=west] at (2.4,-2.1) {Confounding kernels};
\node[anchor=west] at (2.4,-2.7) {Kernels in other hash bins};

\draw (1,0)[color=black] rectangle (6.2,4);
\node at (5.2,3.7) {$\algebraicset{\receiverrnd}$}; 
\draw (-1,-1)[color=black] rectangle (8.5,5.5);
\end{tikzpicture}} 
\\ (a) & (b)
\end{tblr}
    \caption{Illustration of the general strategy for the upper bound in the proof of Theorem \ref{thm:slepian_wolf_sender_unaware}, based on sending the index of a hash bin containing the kernel of the sender's sentence, and then resolving ambiguity by testing whether  $\receiverrnd$ is entailed.}
    \label{fig:no_need_to_know}
\end{figure}

\newpage
 
\begin{restatable}[Alice doesn't know what Bob knows, a.k.a. No Need to Know]{thm}{slepianwolfish}
Let $(L,  \kernelspace, \kappa, \ell)$ be a Logic System. Let $\senderrnd, \receiverrnd \in \surfacespace$ represent the sender's logic sentence and the receiver's logic sentence, assuming $\senderrnd, \receiverrnd$ have kernels that follow a ($p_s,p_r$)-law (and thus $\senderrnd \entails \receiverrnd$). Then
\begin{eqnarray*}
\lefteqn{\min_{ \{f^i,g^i\}_{i=0}^{3}} |\kernelspace|^{-1} E_{\senderrnd, \receiverrnd} \left[ \len{B^0} + \len{A^1} + \len{B^2} + \len{A^3} \right]} \\
&\leq &\Lambda(p_s,p_r-p_s) + O\left( \frac{\log_2|\kernelspace|}{|\kernelspace|} \right).
\end{eqnarray*}
where the minimization is over all 4-turn codes for full synchronization, as defined by Table \ref{tab:mt} and the corresponding explanation of it. Furthermore, the if $\{ (\kv{ \senderrnd }, \kv{\receiverrnd})_i \}_{i=1}^{|\kernelspace|}$ are i.i.d., and furthermore, $\receiverrnd \rightarrow \algebraicset{\receiverrnd} \rightarrow \algebraicset{\senderrnd}$, then
\begin{eqnarray*}
\Lambda(p_s,p_r-p_s) &\leq& \min_{ \{f^i,g^i\}_{i=0}^{3}} |\kernelspace|^{-1} E_{\senderrnd, \receiverrnd} \left[ \len{B^0} + \len{A^1} + \len{B^2} + \len{A^3} \right].
\end{eqnarray*}
\label{thm:slepian_wolf_sender_unaware}
\end{restatable}

\textbf{Proof.} 
The lower bound is proved as follows. We start from from a general 4-turn code for full synchronization $\{f^i,g^i\}_{i=0}^3$, and let $B^0, A^1, B^2, A^3 \in \{0,1\}^*$ be the binary strings output by the encoders $f^0,f^1,f^2,f^3$ that are used by Bob and Alice, when the interaction is operating over the random quantities $\senderrnd$ and $\receiverrnd$. The expected communication cost is then lower bounded as follows:

\begin{eqnarray}
E\left[ \len{B^0} + \len{A^1} + \len{B^2} + \len{A^3} \right] & \stackrel{(a)}{=} & E\left[ \len{B^0} \right] + E\left[ E\left[\len{A^1} | B^0\right] \right] \nonumber \\
& + & E \left[ E\left[ \len{B^2} | B^0, A^1 \right] \right] + E \left[ E\left[ \len{A^3} | B^0, A^1, B^2 \right] \right] \nonumber \\
& \stackrel{(b)}{\geq} & \Hreg{B^0} + \Hreg{A^1 | B^0} + \Hreg{B^2 | B^0, A^1} + \Hreg{A^3|B^0, A^1, B^2} \nonumber \\
& = & \Hreg{B^0,A^1,B^2,A^3} \nonumber \\
& \geq & \Hreg{B^0,A^1,B^2,A^3| \receiverrnd} \nonumber \\
& \geq & \Hreg{\senderrndhat | \receiverrnd} \label{eq:first_step_lower_bound}.
\end{eqnarray}
 Step (a) is a straightforward application of the law of total expectation. Step (b) follows from the application of Lemma~\ref{lem:vlc_kraft} and the prefix-free assumption in the theorem.
 The argument continues as follows:
\begin{eqnarray*}
 \Hreg{\senderrndhat | \receiverrnd} & \geq & \Hreg{\algebraicset{\senderrndhat} | \receiverrnd} \\
& \stackrel{(c)}{=} & \Hreg{\algebraicset{\senderrnd} | \receiverrnd} \\
& \stackrel{(d)}{\geq} & |\kernelspace| \Lambda(p_s,p_r-p_s).
\end{eqnarray*}
Step (c) is justified using arguments similar to those of Subsection~\ref{ss:bli} substituting $\senderhat=g^3(\receiver, b^0,a^1,b^2,a^3)$ instead. Step (d) is justified by exactly the same derivation found in (\ref{eq:conditional_lower_bound_derivation}), starting from step (e). This establishes the lower bound. 

We now turn our attention to the upper bound. First we establish the precise domain/image of the encoder and decoders for our 4-turn mode:
\begin{eqnarray*}
f^{ 0 } : \surfacespace & \rightarrow & \{0,1\}^*, \hspace{0.1in} g^{0} : \{0,1\}^{*} \rightarrow \mathbb{Z}, \\
f^{ 1 } : \surfacespace \times \mathbb{Z} & \rightarrow & \{0,1\}^*, \hspace{0.1in} g^{ 1} : \{0,1\}^*\times \surfacespace   \rightarrow  \surfacespace \times \{ \text{success}, \text{failure} \}, \\
f^2 : \{ \text{success}, \text{failure} \} &\rightarrow& \{0,1\}, \hspace{0.1in} g^2 : \{0,1\} \rightarrow \{ \text{success}, \text{failure} \}, \\
f^{ 3} : \surfacespace & \rightarrow & \{0,1\}^{*}, \hspace{0.1in}  g^{ 3} : \{0,1\}^{*} \times \{ \text{success}, \text{failure} \} \times \surfacespace  \rightarrow  \surfacespace .
\end{eqnarray*}

Alice and Bob first interact to establish a family of hash functions. How this is done will be described shortly. The transmission protocol is as follows:

\begin{enumerate}
\item The receiver sends $|\algebraicset{ \receiverrnd}|$ to the sender using 
$\delta$ Elias coding, which is decoded at the sender side. This step defines $f^{0}$ and $g^{0}$. \label{step:rec}
\item The sender sends $|\algebraicset{\senderrnd}|$ to the receiver using $\delta$ Elias coding, which is decoded at the receiver side. This only partly defines $f^1$ and $g^1$ \label{step:send}.
    \item At this point, the receiver and sender independently can compute 
    \begin{eqnarray*}
        \text{RATE} = \Lambda\left( \frac{|\algebraicset{\senderrnd}|}{|\kernelspace|} , \frac{|\algebraicset{\receiverrnd}|}{|\kernelspace|} - \frac{|\algebraicset{\senderrnd}|}{|\kernelspace|} \right) + \frac{\log_2 |\kernelspace|}{|\kernelspace|}.
    \end{eqnarray*}
    The same hash function is then selected by both parties independently, which matches the selected rate and depends on $\kernelspace$, $|\algebraicset{\senderrnd}|$ and $|\algebraicset{\receiverrnd}|$.
     
    \label{step:rate}
    \item The sender sends the bin index $ \text{BIN}_{|\algebraicset{\senderrnd}|,|\algebraicset{\receiverrnd}|}(\algebraicset{\senderrnd})$ to the receiver using $2^{|\kernelspace| \text{RATE}}$ bits. This step completes the definition of $f^1$. \label{step:bin}
    \item The receiver attempts to decode $\algebraicset{\senderrnd}$ using the bin index, its knowledge of $\algebraicset{\receiverrnd}$ and the algorithm described below. By leveraging the function $\ell$, the outcome of this attempt results in an element of $\surfacespace$ and a ``success'' or a default, dummy element from $\surfacespace$ and a ``failure''. This completes the definition of $g^1$.
    
    \label{step:dec}
    \item The success/failure of the attempt is signaled back to the sender using a single bit. This defines both $f^2$ and $g^2$.
    \item If successful, the sender has nothing to do anymore, as the receiver expects no further communication. The receiver simply outputs the element from $\surfacespace$ computed by $g^1$, designating it as the output $\senderrndhat$, partly defining $g^3$.
    \item If unsuccessful, the sender sends $\algebraicset{\senderrnd}$ as a binary vector of length $|\kernelspace|$, which is decoded by the receiver and by passing it through $\ell$, becomes $\senderrndhat$, the designated output of the protocol, completing the definition of $f^3$ and $g^3$. \label{step:fallback}
\end{enumerate}
The reader may appreciate that the algorithm guarantees that the receiver at the end will have in possession enough information to reproduce a message $\senderrndhat$ that is functionally equivalent to $\senderrnd$. The problem that remains is demonstrating that the overall communication efficiency of this protocol is such that the normalized total bidirectional cost in bits is asymptotically $\Lambda(p_s,p_r-p_s)$.

\begin{figure}
\begin{center}
\begin{tikzpicture}
\draw (2.2,0)[color=black,fill=white] rectangle (10,7);
\draw(0.0,0.25)[fill=gray!35!white] rectangle(7, 4.00);
\draw(6.5,2.3)[fill=green!15!white] circle(0.7);
\draw(7.5,5.5)[fill=green!15!white] circle(0.7);
\draw(5.0,3)[fill=red!15!white] circle(0.7);
\draw(0.5,3.7) node {$\algebraicset{\receiverrnd}$};
\begin{scope}
\clip(0.0,0.25) rectangle(7, 4.00);
\draw(0.0,0.25) rectangle(7, 4.00);

\draw(3.2,2)[fill=yellow!35!white] circle(0.7);  
\draw(3.2,2.1) node {$\algebraicset{\senderrnd}$};
\end{scope}
\draw(0.0,0.25) rectangle(7, 4.00);
\draw (2.2,0)[color=black,fill=none] rectangle (10,7);
\draw(9.0,6.7) node {$\text{BIN}(\senderrnd)$};
\end{tikzpicture}
\end{center}
\caption{An example of a hash bin collision in Theorem~\ref{thm:slepian_wolf_sender_unaware}, where the receiver is unable to decide between the true kernel in yellow and a confounding one in red.}

\label{fig:slepian_wolf_fig}
\end{figure}
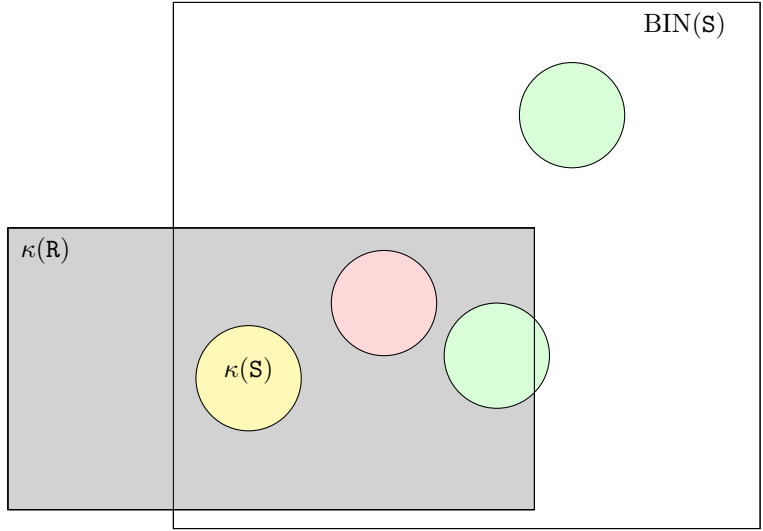

Prior to stating the protocol we disclosed that Alice and Bob met before to agree on a family of hash functions. We describe next how this agreement is arrived to. For each possible combination of potential sender and receiver kernel sizes $0 \leq s \leq r \leq |\kernelspace|$ with corresponding coding rate $\mbox{rate}(s,r) \stackrel{\Delta}{=} \Lambda\left( \frac{s}{|\kernelspace|} , \frac{r}{|\kernelspace|} - \frac{s}{|\kernelspace|}\right)+ \frac{\log_2 |\kernelspace|}{|\kernelspace|}$,  we will build a random hash function for $\{ \mathtt{s} \in \surfacespace : |\algebraicset{\mathtt{s}}|=s\}$. For each element of this set we choose a bin index independently and uniformly at random from the set $\{1,\ldots,\lceil 2^{|\kernelspace| \cdot \mbox{rate}(s,r) } \rceil \}$; call the resulting  function $\text{BIN}_{ s,r} : \{0,1\}^{|\kernelspace|} \rightarrow \{1,\ldots,\lceil 2^{|\kernelspace| \cdot \mbox{rate}(s,r)} \rceil \}$.  We use the upper case notation BIN to remind the reader that this is a random hash function. The expected, normalized cost for the transmissions associated with steps \ref{step:rec},\ref{step:send},\ref{step:bin} admit the following upper bounds, obtained using convexity arguments and the assumption that $(\senderrnd, \receiverrnd)$ have kernels that follow a ($p_s,p_r$)-law:
\begin{gather}
    |\kernelspace|^{-1} \left( \log_2 \left( |\kernelspace| p_s \right)  +  2 \log_2 \log_2 \left( |\kernelspace| p_s \right) + 3 \right) \label{eq:estimate1}, \\
    |\kernelspace|^{-1} \left( \log_2 \left( |\kernelspace| p_r \right)  +  2 \log_2 \log_2 \left( |\kernelspace| p_r \right) + 3 \right) \label{eq:estimate2}, \\
    \Lambda(p_s, p_r - p_s)+ \frac{\log_2 |\kernelspace|}{|\kernelspace|} + \frac{1}{|\kernelspace|} . \label{eq:estimate3}
\end{gather}
The last term uses Lemma~\ref{lem:elementarylambda}. We now describe the decoding algorithm behind step \ref{step:dec}. As of step \ref{step:send}, the receiver has decoded $|\algebraicset{\senderrnd}|$ and step \ref{step:bin}, the receiver has decoded the bin index $ \text{BIN}_{|\algebraicset{\senderrnd}|,|\algebraicset{\receiverrnd}|}(\algebraicset{\senderrnd})$. Next he would like to recover $\algebraicset{\senderrnd}$ itself. Recall that by assumption, $\senderrnd \entails \receiverrnd$ and therefore the receiver knows that irrespective of what $\senderrnd$ is, the following relation must hold:
\begin{eqnarray*}
    \algebraicset{\senderrnd} \subseteq \algebraicset{\receiverrnd}.
\end{eqnarray*}
Define the hypotheses set as 
\begin{eqnarray*}
    \text{hypotheses}(\senderrnd,\receiverrnd) \stackrel{\Delta}{=} \left\{ \alpha \subseteq \kernelspace :  \alpha \subseteq \algebraicset{\receiverrnd} ,  | \alpha | = |\algebraicset{\senderrnd}|, \text{BIN}_{|\algebraicset{\senderrnd}|,|\algebraicset{\receiverrnd}|}(\alpha) = \text{BIN}_{|\algebraicset{\senderrnd}|,|\algebraicset{\receiverrnd}|}(\algebraicset{\senderrnd}) \right\} .
\end{eqnarray*}
If the hypotheses set has cardinality exactly one, $g^1$ outputs the result of mapping the one element to the space of logic sentences using  $\ell$ and additionally outputs ``success''; here $\ell$ refers to the function that maps a kernel to a logic expression (c.f.\ Definition \ref{def:ell}). Otherwise, $g^1$ outputs a dummy element from $\surfacespace$ (doesn't matter which) and additionally outputs ``failure''.  We illustrate an example of a decoding failure in Figure~\ref{fig:slepian_wolf_fig}. The box in white represents $\text{BIN}_{|\algebraicset{\senderrnd}|,|\algebraicset{\receiverrnd}|}(\senderrnd)$, and the box in gray represents $\algebraicset{\receiverrnd}$. The various circles represent kernels that have been mapped to the same bin. The yellow circle represents the true kernel $\algebraicset{\senderrnd}$ that we want Bob to recover. The kernels illustrated through the green color can be eliminated by the receiver from consideration, since they are not fully included in $\algebraicset{\receiverrnd}$. The kernel illustrated in red cannot be discerned from the kernel in yellow by the receiver, and thus in this example, we have a decoding failure.

To estimate the failure probability, we upper bound the probability of this event:
\begin{eqnarray*}
 \left[ | \text{hypotheses}(\senderrnd,\receiverrnd)| \geq 2 \right].
\end{eqnarray*}
Let $I_{\alpha}$ be equal to 1 if $\alpha \in \text{hypotheses}(\senderrnd,\receiverrnd)$ and 0 otherwise. Then we can upper bound the error probability as
\begin{eqnarray}
P\left(\left[ | \text{hypotheses}(\senderrnd,\receiverrnd)| \geq 2 \right]\right) &=& P\left(  \sum_{\alpha} I_{\alpha} \geq 2 \right) \nonumber \\
&=& 
    E_{\senderrnd, \receiverrnd} \left[ P\left(  \sum_{\alpha} I_{\alpha} \geq 2 \lvert \senderrnd, \receiverrnd  \right) \right] \label{eq:totalexp}.
\end{eqnarray}
We then write
\begin{eqnarray}
    P\left(  \sum_{\alpha} I_{\alpha} \geq 2 \lvert \senderrnd, \receiverrnd  \right) &=& 1-P\left(  \sum_{\alpha :  \alpha \neq \algebraicset{\senderrnd}, \alpha \subseteq \algebraicset{\receiverrnd} ,  | \alpha | = |\algebraicset{\senderrnd}|} I_{\alpha} = 0  \lvert \senderrnd, \receiverrnd  \right) \label{eq:first} \\
    &=& 1 - \Pi_{\alpha \neq \algebraicset{\senderrnd},   \alpha \subseteq \algebraicset{\receiverrnd} ,  | \alpha | = |\algebraicset{\senderrnd}|} P(I_{\alpha} = 0 | \senderrnd, \receiverrnd) \label{eq:second} \\
       &=& 1 - \Pi_{\alpha \neq \algebraicset{\senderrnd},   \alpha \subseteq \algebraicset{\receiverrnd} ,  | \alpha | = |\algebraicset{\senderrnd}|} \left( 1 - 2^{-|\kernelspace| \text{RATE}}  \right) 
       \nonumber \\
       &\leq& 1 - \left( 1 - 2^{-|\kernelspace| \text{RATE}}  \right)^{ \left( \begin{array}{c} |\algebraicset{\receiverrnd}| \\  |\algebraicset{\senderrnd}|   \end{array} \right)  }. \label{eq:ready_for_bernoulli}
\end{eqnarray}
We describe the rationale behind this critical derivation. By construction, the hypotheses set always contains at least one element (the true kernel for the sentence in possession by Alice), and thus the first equality  (\ref{eq:first}) focuses on estimating the failure probability by estimating instead the success probability, where every kernel other than $\algebraicset{\senderrnd}$ is not mapped to the hypotheses set. The second equality (\ref{eq:second}) follows by recognizing that under the given conditioning, the terms identified in the summation are statistically independent and hence the overall probability can be reduced to a product of individual probabilities. The third equality uses the specifics on how we constructed the random hash function to compute the probability that a kernel is not mapped to the bin to which $\algebraicset{\senderrnd}$ belongs. Finally (\ref{eq:ready_for_bernoulli}) uses combinatorial counting arguments to obtain an estimate of the product. We bound the combinatorial as follows:
\begin{eqnarray}
    \left( \begin{array}{c} |\algebraicset{\receiverrnd}| \\  |\algebraicset{\senderrnd}|   \end{array} \right) &\leq& 2^{  |\kernelspace| \frac{|\algebraicset{\receiverrnd}|}{|\kernelspace|} H\left( \frac{|\algebraicset{\senderrnd}|}{ |\algebraicset{\receiverrnd}| }        \right) } \nonumber \\
    &=& 2^{|\kernelspace| \Lambda \left(\frac{|\algebraicset{\senderrnd}|}{|\kernelspace|} ,  \frac{|\algebraicset{\receiverrnd}| - |\algebraicset{\senderrnd}|}{|\kernelspace|} \right) } \nonumber \\
    &=& 2^{|\kernelspace| \text{RATE} - \log_2|\kernelspace|} \label{eq:combest}.
\end{eqnarray}
Recall now Bernoulli's inequality:
\begin{eqnarray*}
    (1-a)^n \geq 1 - an.
\end{eqnarray*}
Combining this inequality with (\ref{eq:ready_for_bernoulli}), (\ref{eq:combest}),  and (\ref{eq:totalexp}), we obtain
\begin{eqnarray*}
P\left(\left[ | \text{hypotheses}(\senderrnd,\receiverrnd)| \geq 2 \right]\right) &\leq& \frac{1}{|\kernelspace|}.
\end{eqnarray*}
With this estimate, we can account for the remainder of the bits in the protocol (going on either direction) with the expression
\begin{eqnarray*}
     \left( 1 + P\left(\left[ | \text{hypotheses}(\senderrnd,\receiverrnd)| \geq 2 \right]\right)|\kernelspace| \right)/|\kernelspace| \leq 2/|\kernelspace|.
\end{eqnarray*}
Using this estimate together with (\ref{eq:estimate1}), (\ref{eq:estimate2}), (\ref{eq:estimate3}), we obtain that the cost in bits is at most
\begin{eqnarray*}
     \Lambda(p_s, p_r - p_s)+ O\left( \frac{\log_2 |\kernelspace|}{|\kernelspace|} \right) .
\end{eqnarray*}
The proof of the upper bound is completed by noting that since the above is the expected performance of a random hash function, there must exist at least one hash function with an error probability not worse than the average. \qed

\subsection{Bob's sentence may not be entailed by Alice's
}

In Figure~\ref{fig:less_is_more_background}-a, we illustrate the same situation that we had described in Theorem~\ref{thm:less_is_more_simple}, but with the addition of background information $\receiver$ that is known to both Bob and Alice, and with the  assumption $\sender \entails \receiver$. Now suppose that it is no longer the case that $\sender \entails \receiver$, then what we obtain is the more general setup in Figure~\ref{fig:less_is_more_background}, where as it can be appreciated, $\algebraicset{\receiver}$ no longer contains $\algebraicset{\sender}$ fully and where the complement of $\algebraicset{\receiver}$ is patterned with a ``dotted'' fill. This more general setup is the subject of this subsection.

In order to derive results for this setting, we will need to make additional assumptions. For the first time in the article, we will use a \emph{proper} Logic System (see Definition \ref{def:proper-logic-system}), where we assume the availability of the standard logic operators $\lor$, $\land$, $\lnot$ as well as the set theoretic implications of the kernels resulting from such operations; see Equations (\ref{kr1}-\ref{kr3}). 
We introduce a more complex set of measurements that need to be made on the probability laws of $\senderrnd, \queryrnd, \receiverrnd$. We keep the assumption $\senderrnd \entails \queryrnd$ in \eqref{eq:siq}  but drop the assumptions  $\senderrnd \entails \receiverrnd, \queryrnd \entails \receiverrnd$ in \eqref{eq:sir},\eqref{eq:technical}. 
We measure the following expected kernel sizes: 
\begin{eqnarray*}
     p_{r} &\stackrel{\Delta}{=}& |\kernelspace|^{-1} E|\algebraicset{\receiverrnd}| , \nonumber \\ 
      p_{s^*}  &\stackrel{\Delta}{=}&|\kernelspace|^{-1} E|\algebraicset{ \senderrnd \land \receiverrnd}| , \nonumber \\ 
     p_{q^*} &\stackrel{\Delta}{=}& |\kernelspace|^{-1} E|\algebraicset{ \queryrnd \land \receiverrnd}|, \nonumber \\ 
      p_{s^{**}} &\stackrel{\Delta}{=}& |\kernelspace|^{-1} E|\algebraicset{ \senderrnd \land \lnot \receiverrnd}|, \nonumber \\ 
     p_{q^{**}} &\stackrel{\Delta}{=}& |\kernelspace|^{-1} E|\algebraicset{ \queryrnd \land \lnot \receiverrnd}|.
\label{eq:remainingassum}
\end{eqnarray*}
Our upper bounds will be phrased in terms of these expected sizes. Together with the assumption that $\senderrnd \entails \queryrnd$, we say that the kernels of $\senderrnd, \queryrnd, \receiverrnd$ follow a ($p_r, p_{s^*}, p_{q^*}, p_{s^{**}}, p_{q^{**}}$)-law. As with previous results, we briefly introduce the strategy for proving the upper bound for this result is illustrated in Figure~\ref{fig:less_is_more_background}. While the figure appears formidable, upon further examination its elements are quickly decomposed into elements that should be familiar to the reader now. At the highest level, the problem is simply split in two: because both Alice and Bob know $\receiver$, they can create a dual strategy: one to handle sending whatever piece of $\algebraicset{\sender}$ and $\algebraicset{\query}$ that will intersect with $\algebraicset{\receiver}$, and the other one to handle the same but that intersects with $\algebraicset{\receiver}^c$; this explains why there is vertical symmetry on the figure. Then, focusing on, say, only the top half of the diagram, we realize that the resulting system is in essence a combination of the strategies used in Theorems \ref{thm:background_log_info} and \ref{thm:less_is_more_simple}. As a result, the fundamental device for efficiently sending partitions, denoted by a circle with the $+$ and $-$ hooks (and fully addressed in Theorem \ref{thm:shannonpartition}) is used twice.

\begin{restatable}[Bob's sentence may not be entailed by Alice's]{thm}{lessismorebackground}
~Let $(L,  \kernelspace, \kappa, \ell)$ be a proper Logic System. Let $\senderrnd$, $\queryrnd$, $\receiverrnd$ represent the sender, query and receiver logic sentences, respectively, which we assume have kernels that follow a ($p_r, p_{s^*}, p_{q^*}, p_{s^{**}}, p_{q^{**}}$)-law. Let the  encoder $f$ and decoder $g$ be functions
\begin{eqnarray*}
f : \surfacespace^3 & \rightarrow & \{0,1\}^* \\
g : \{0,1\}^* \times  \surfacespace & \rightarrow & \surfacespace.
\end{eqnarray*}
Then 
\begin{eqnarray*}
 \min_{f,g} |\kernelspace|^{-1} E_{\senderrnd,\queryrnd,\receiverrnd} \left[ \len{f(\senderrnd,\queryrnd,\receiverrnd)}  \right] &\leq & \Lambda(p_{s^*},p_r-p_{q^*}) + \Lambda(p_{s^{**}},1-p_r-p_{q^{**}}) + O\left( \frac{\log_2 |\kernelspace|}{|\kernelspace|} \right)
\end{eqnarray*}
where the minimization is over $f, g$ such that the image of $f(\cdot,\cdot,\receiver)$ is prefix-free for any choice of  $\receiver$, and such that if $\sender \entails \query$  then $\sender \entails g(f(\sender,\query,\receiver),\receiver)$, and $g(f(\sender,\query,\receiver),\receiver) \entails \query$. Furthermore, if $\{(\kv{\senderrnd},\kv{\queryrnd}, \kv{\receiverrnd})_i\}_{i=1}^{|\kernelspace|}
$ are i.i.d.\ and $\receiverrnd \rightarrow \algebraicset{\receiverrnd} \rightarrow (\algebraicset{\senderrnd},\algebraicset{\queryrnd})$, then
\begin{eqnarray*}
\Lambda(p_{s^*},p_r-p_{q^*}) +  \Lambda(p_{s^{**}},1-p_r-p_{q^{**}}) &\leq& \min_{f,g} |\kernelspace|^{-1} E_{\senderrnd,\queryrnd,\receiverrnd} \left[ \len{f(\senderrnd,\queryrnd,\receiverrnd)}  \right]
\end{eqnarray*}

\label{thm:less_is_more_background}
\end{restatable}

\begin{figure}
\input{proof_strategy_ignorance_and_misinformation}
\caption{Proof strategy for Theorem \ref{thm:less_is_more_background}}
\label{fig:less_is_more_background}
\end{figure}

\textbf{Proof.}
The proof of this result builds upon the ideas in the proofs of Theorem \ref{thm:background_log_info} and \ref{thm:less_is_more_simple}. We start using the law of total expectation to write
\begin{eqnarray*}
E_{\senderrnd,\queryrnd,\receiverrnd} \left[ \len{f(\senderrnd,\queryrnd,\receiverrnd)}  \right]  = E_{\receiverrnd} \left[ E_{\senderrnd,\queryrnd} \left[ \len{f(\senderrnd,\queryrnd,\receiverrnd)} | \receiverrnd \right]    \right]
\end{eqnarray*}
We invoke again Kraft's inequality through Lemma \ref{lem:vlc_kraft} and write
\begin{eqnarray*}
 E_{\senderrnd,\queryrnd} \left[ \len{f(\senderrnd,\queryrnd,\receiver)} | \receiverrnd=\receiver \right]  &=&  E_{f(\senderrnd,\queryrnd,\receiver)|\receiverrnd=\receiver} \left[ \len{f(\senderrnd,\queryrnd,\receiver)} | \receiverrnd=\receiver \right]  \\
 &\geq& \Hreg{f(\senderrnd,\queryrnd,\receiver)|\receiverrnd=\receiver}
\end{eqnarray*}
where we have used the assumption that $f(\cdot,\cdot,\receiver)$ is prefix-free for any choice of $\receiver$. From here, we apply the same ideas as in Theorem \ref{thm:less_is_more_simple} to obtain
\begin{eqnarray*}
 E_{\senderrnd,\queryrnd | \receiverrnd = \receiver} \left[ \len{f(\senderrnd,\queryrnd,\receiver)} | \receiverrnd=\receiver \right]   \geq  I(\kv{\senderrnd},\kv{\queryrnd};  \kv{g(f(\senderrnd,\queryrnd,\receiver),\receiver)} | \receiverrnd=\receiver) .
\end{eqnarray*}

Recall the definition of $\psi$ in (\ref{eq:def_psi}) and write
\begin{eqnarray*}
    X_{1}^{|\kernelspace|} &=& \psi(\kv{\senderrnd},\kv{\queryrnd}) , \\
    Z_{1}^{|\kernelspace|} &=& 1 \times \kv{g(f(\senderrnd,\queryrnd,\receiver),\receiver)} .
\end{eqnarray*}
Similarly, as before  note that the assumptions that $\sender \entails g(f(\sender,\query,\receiver)$ and $\receiver)$, $g(f(\sender,\query,\receiver),\receiver) \entails \query$ imply that for all $1 \leq i \leq |\kernelspace|$,
\begin{eqnarray*}
\rho( X_i, Z_i) = 0 .
\end{eqnarray*}
We continue applying the ideas in the proof of Theorem \ref{thm:less_is_more_simple} and obtain
\begin{eqnarray*}
\lefteqn{I(\kv{\senderrnd}, \kv{\queryrnd};  \kv{g(f(\senderrnd,\queryrnd,\receiver),\receiver)} |\receiverrnd=\receiver)} \\
&\geq   & 
 I(X_{1}^{|\kernelspace|}; Z_1^{|\kernelspace|} | \receiverrnd=\receiver  ) \\
 &=  & \Hreg{X_{1}^{|\kernelspace|} | \receiverrnd = \receiver} - \Hreg{ Z_1^{|\kernelspace|} | X_{1}^{|\kernelspace|} , \receiverrnd=\receiver  } \\
& \stackrel{(a)}{=}  & \Hreg{X_{1}^{|\kernelspace|} | \kv{\receiverrnd} = \kv{\receiver}} - \Hreg{ Z_1^{|\kernelspace|} | X_{1}^{|\kernelspace|} , \receiverrnd=\receiver  } \\
&  \stackrel{(b)}{\geq}  & \Hreg{X_{1}^{|\kernelspace|} | \kv{\receiverrnd} = \kv{\receiver}} - \Hreg{ Z_1^{|\kernelspace|} | X_{1}^{|\kernelspace|} , \kv{\receiverrnd} = \kv{\receiver}  } \\
 & = & I(X_{1}^{|\kernelspace|}; Z_1^{|\kernelspace|} | \kv{\receiverrnd}=\kv{\receiver}  )
\end{eqnarray*}
where (a) follows from the assumption that the following Markov chain holds:
\begin{eqnarray*}
\receiverrnd \rightarrow \kv{\receiverrnd} \rightarrow ( \kv{\senderrnd}, \kv{\queryrnd} ),
\end{eqnarray*}
and (b) follows from the fact that conditional entropy can only increase under weaker conditioning.

We will next argue that conditioned on $\kv{\receiverrnd} = \kv{\receiver}$, the $\{X_i\}$ are independent, and distributed according to at most two distributions, which follow from the assumptions of the distribution of $\senderrnd,\receiverrnd,\queryrnd$. 

\begin{restatable}{lem}{complexabc}
Let the random variables $A,B,C \in \{0,1\}$ and assume that $A \leq B$. Then the distribution of $ \psi(A,B)$ given $C=c$
\begin{eqnarray}
\left\{P(B=0|C=c), P(A=1|C=c), 1 - P(
B=0|C=c) - P(A=1|C=c) \right\}.
\end{eqnarray}
\label{lem:complexabc}
\end{restatable}
\textbf{Proof.} Recall that
\begin{eqnarray*}
    \psi(A,B) = A + 2(1-A)B
\end{eqnarray*}
and therefore, the the following event equivalences hold:
\begin{eqnarray*}
\left[ \psi(A,B) = 1 \right] &=& \left[ A= 1\right] \\
\left[ \psi(A,B) = 0 \right] &=& \left[ A= 0, B=0\right] = \left[B=0\right]
\end{eqnarray*}
where the last equality follows from the assumption that $A \leq B$. Then given $C=c$,
\begin{eqnarray*}
    P(\psi(A,B) = 1 | C = c) &=& P(A = 1 | C = c)  \\
    P(\psi(A,B) = 0 | C = c) 
    &=& P(B = 0 | C = c) 
\end{eqnarray*}
\qed

We now write
\begin{eqnarray*}
    P(X_i = 0 | \kv{\receiverrnd}=\kv{\receiver} ) &\stackrel{(a)}{=}& P(X_i = 0 | \kv{\receiverrnd}_i=\kv{\receiver}_i) \nonumber \\
    &\stackrel{(b)}{=}& P( \kv{\queryrnd}_i = 0 | \kv{\receiverrnd}_i = \kv{\receiver}_i) \nonumber \\
    &\stackrel{(c)}{=}& P( \kv{(\queryrnd \land \receiverrnd) \lor (\queryrnd \land \lnot \receiverrnd) }_i = 0 | \kv{\receiverrnd}_i = \kv{\receiver}_i) \nonumber \\
    &\stackrel{(d)}{=}& P( [\kv{\queryrnd \land \receiverrnd}_i=0] \cap [\kv{ \queryrnd \land \lnot \receiverrnd }_i = 0 ]| \kv{\receiverrnd}_i = \kv{\receiver}_i) \nonumber \\
       &\stackrel{(e)}{=}& \left\{ \begin{array}{cc}  P( [\kv{\queryrnd \land \receiverrnd}_i=0]| \kv{\receiverrnd}_i=1)  & \mbox{ if } \kv{\receiver}_i = 1\\ 
     P( [\kv{\queryrnd \land \lnot \receiverrnd}_i=0]| \kv{\receiverrnd}_i=0)  & \mbox{ if } \kv{\receiver}_i = 0 \end{array}\right. \\
    &\stackrel{(f)}{=}& \left\{ \begin{array}{cc} 1 - p_{q^*}/p_r & \mbox{ if } \kv{\receiverrnd}_i = 1\\ 
     1 - p_{q^{**}}/(1-p_r)& \mbox{ if } \kv{\receiverrnd}_i = 0 \end{array}\right.
\end{eqnarray*}
where (a) follows from the Definition $X_i = \psi( \kv{\senderrnd}_i, \kv{\queryrnd}_i )$ and the independence assumption and (b) follows from Lemma \ref{lem:complexabc}. 
To go from (b) to (c) we establish that $\algebraicset{\queryrnd} = \algebraicset{(\queryrnd \land \receiverrnd) \lor (\queryrnd \land \lnot \receiverrnd)}$ using the conditions for being a proper Logic System (Definition \ref{def:proper-logic-system}). By condition \ref{def:proper-logic-system}.\ref{kr3}, $\algebraicset{\lnot \receiverrnd} = \algebraicset{\receiverrnd}^c$. Thus, 
\begin{eqnarray}
    \algebraicset{\queryrnd} &=& \algebraicset{\queryrnd} \cap (\algebraicset{\receiverrnd} \cup \algebraicset{\lnot \receiverrnd}) \notag \\
    &=&(\algebraicset{\queryrnd} \cap \algebraicset{\receiverrnd}) \cup \algebraicset{(\queryrnd} \cap \algebraicset{\lnot \receiverrnd}) \label{dm} \\
    &=& \algebraicset{\queryrnd \land \receiverrnd} \cup \algebraicset{\queryrnd \land \lnot \receiverrnd} \label{pls-step1}\\
    &=& \algebraicset{(\queryrnd \land \receiverrnd) \lor (\queryrnd \land \lnot \receiverrnd)}, \label{pls-step2}
\end{eqnarray}
where (\ref{dm}) follows by the DeMorgan Laws for sets, (\ref{pls-step1}) follows by condition \ref{def:proper-logic-system}.\ref{kr2}, and (\ref{pls-step2}) follows by condition \ref{def:proper-logic-system}.\ref{kr1}. Then, to go from (c) to (d),
note that
\begin{eqnarray}
\algebraicset{\lnot((\queryrnd \land \receiverrnd) \lor (\queryrnd \land \lnot \receiverrnd))} &=& \algebraicset{(\queryrnd \land \receiverrnd) \lor (\queryrnd \land \lnot \receiverrnd)}^c \label{cd-step1}\\
&=& (\algebraicset{\queryrnd \land \receiverrnd} \cup \algebraicset{\queryrnd \land \lnot \receiverrnd})^c \label{cd-step2}\\
&=& \algebraicset{\queryrnd \land \receiverrnd}^c \cap  \algebraicset{\queryrnd \land \lnot \receiverrnd}^c \label{cd-step3}\\
&=& \algebraicset{\lnot(\queryrnd \land \receiverrnd)} \cap  \algebraicset{\lnot(\queryrnd \land \lnot \receiverrnd)}, \label{cd-step4}
\end{eqnarray}
where equality (\ref{cd-step1}) follows from condition \ref{def:proper-logic-system}.\ref{kr3}, (\ref{cd-step2}) follows by condition \ref{def:proper-logic-system}.\ref{kr1}, (\ref{cd-step3}) follows by DeMorgan, and (\ref{cd-step4}) follows by condition \ref{def:proper-logic-system}.\ref{kr3}. Putting these together, it follows that $\kv{(\queryrnd \land \receiverrnd) \lor (\queryrnd \land \lnot \receiverrnd) }_i = 0$ iff $[\kv{\queryrnd \land \receiverrnd}_i=0] \cap [\kv{ \queryrnd \land \lnot \receiverrnd }_i = 0 ]$. Here we use the fact that if $\mathtt{s} \in \surfacespace$ is a sentence, then $\kv{\mathtt{s}}_i = 0$ iff $\kv{\lnot \mathtt{s}}_i = 1$, which follows from condition \ref{def:proper-logic-system}.\ref{kr1}, $\algebraicset{\lnot \mathtt{s}} = \algebraicset{\mathtt{s}}^c$, since taking the complement of the kernel corresponds to flipping 1s to 0s and 0s to 1s in the vector notation.

To obtain (e), note that 
\begin{eqnarray*}
    [\kv{ \queryrnd \land \lnot \receiverrnd }_i = 0 ] \cap [ \kv{\receiverrnd}_i=1] &=& [ \kv{\receiverrnd}_i=1] \\
    \left[\kv{ \queryrnd \land \receiverrnd }_i = 0 \right] \cap [ \kv{\receiverrnd}_i=0] &=& [ \kv{\receiverrnd}_i=0]
\end{eqnarray*}
From this observation one then may trivially write
\begin{eqnarray*}
    [\kv{\queryrnd \land \receiverrnd}_i=0] \cap [\kv{ \queryrnd \land \lnot \receiverrnd }_i = 0 ] \cap [ \kv{\receiverrnd}_i=1] &=& [\kv{\queryrnd \land \receiverrnd}_i=0] \cap [ \kv{\receiverrnd}_i=1] \\
    \left[\kv{\queryrnd \land \receiverrnd}_i=0 \right] \cap [ \kv{ \queryrnd \land \lnot \receiverrnd }_i = 0 ] \cap [ \kv{\receiverrnd}_i=0] &=& [\kv{\queryrnd \land \lnot \receiverrnd}_i=0] \cap [ \kv{\receiverrnd}_i=0 ]
\end{eqnarray*}
from which the assertion follows. Finally (f) follows from the expected average kernel sizes as given in the theorem, in conjunction with the ``identically distributed'' assumption. Specifically,
\begin{eqnarray*}
    P( [ \kv{\queryrnd \land \receiverrnd }_i = 0 | \kv{\receiverrnd}_i = 1 ] ) &=& 1 - P( [ \kv{\queryrnd \land \receiverrnd }_i = 1 | \kv{\receiverrnd}_i = 1 ] ) \\ 
   &=& 1 - \frac{   P( [ \kv{\queryrnd \land \receiverrnd }_i = 1 ] ) P( \kv{\receiverrnd}_i = 1 |  \kv{\queryrnd \land \receiverrnd }_i = 1  ) }{P(\kv{\receiverrnd}_i = 1 )} \\
   &=& 1 - \frac{   P( [ \kv{\queryrnd \land \receiverrnd }_i = 1 ] ) }{P(\kv{\receiverrnd}_i = 1 )} \\
   &=& 1 - p_{q^*}/p_r
\end{eqnarray*}
with a parallel derivation applicable for the other equation in (f).

Similarly, one can write
\begin{eqnarray*}
    P(X_i = 1 | \kv{\receiverrnd}=\kv{\receiver} ) &\stackrel{(a)}{=}& P(X_i = 1 | \kv{\receiverrnd}_i=\kv{\receiver}_i) \nonumber \\
    &\stackrel{(b)}{=}& P( \kv{\senderrnd}_i = 1 | \kv{\receiverrnd}_i = \kv{\receiver}_i) \nonumber \\
   &\stackrel{(c)}{=}& P( \kv{(\senderrnd \land \receiverrnd ) \lor (\senderrnd \land \lnot \receiverrnd)}_i = 1 | \kv{\receiverrnd}_i = \kv{\receiver}_i) \nonumber \\
   &\stackrel{(d)}{=}& P( [ \kv{\senderrnd \land \receiverrnd }_i = 1 ] \cup [ \kv{\senderrnd \land \lnot \receiverrnd}_i  = 1 ]| \kv{\receiverrnd}_i = \kv{\receiver}_i) \nonumber \\
   &\stackrel{(e)}{=}& P(  \kv{\senderrnd \land \receiverrnd }_i = 1 | \kv{\receiverrnd}_i = \kv{\receiver}_i) + P( \kv{\senderrnd \land \lnot \receiverrnd}_i  = 1 | \kv{\receiverrnd}_i = \kv{\receiver}_i) \nonumber \\
          &\stackrel{(f)}{=}& \left\{ \begin{array}{cc}  P( [\kv{\senderrnd \land \receiverrnd}_i=1]| \kv{\receiverrnd}_i=1)  & \mbox{ if } \kv{\receiver}_i = 1\\ 
     P( [\kv{\senderrnd \land \lnot \receiverrnd}_i=1]| \kv{\receiverrnd}_i=0)  & \mbox{ if } \kv{\receiver}_i = 0 \end{array}\right. \\
    & \stackrel{(g)}{=} &  \left\{ \begin{array}{cc} p_{s^{*}}/p_r & \mbox{ if } \kv{\receiver}_i = 1\\ 
     p_{s^{**}}/(1-p_r) & \mbox{ if } \kv{\receiver}_i = 0
     \end{array}\right.
\end{eqnarray*}
Steps (a-d) can be justified using arguments very similar to those of the previous derivation. Step (e) follows from the observation that 
\begin{eqnarray*}
    [ \kv{\senderrnd \land \receiverrnd }_i = 1 ] \cap [ \kv{\senderrnd \land \lnot \receiverrnd}_i  = 1 ] = \emptyset
\end{eqnarray*}
Step (f) can be seen by noting that if $\kv{\receiver}_i = 1$, then the second term of (e) is zero, and if $\kv{\receiver}_i=0$, then the first term of (e) is zero. Step (g) follows from the expected average kernel sizes and the identically distributed assumption, using arguments similar to those of in step (f) of the earlier derivation.

By construction, the $X_i$s are mutually independent conditional on $\kv{\receiverrnd}=\kv{\receiver}$. Using this fact, we can now write
\begin{eqnarray*}
\lefteqn{I(\kv{\senderrnd}, \kv{\queryrnd};  \kv{g(f(\senderrnd,\queryrnd,\receiver),\receiver)} |\receiverrnd=\receiver)} \\
& \geq & I(X_{1}^{|\kernelspace|}; Z_1^{|\kernelspace|} | \kv{\receiverrnd}=\kv{\receiver}  ) \\
&=& \sum_i H(X_i | \kv{\receiverrnd}=\kv{\receiver} ) - \sum_i H(X_i | Z_1^{|\kernelspace|} X_1^{i-1}, \kv{\receiverrnd}=\kv{\receiver} ) \\ 
&\geq   & \sum_{i} I(X_i ; Z_i | \kv{\receiverrnd}=\kv{\receiver}) \\
    & = & \sum_{ i : \kv{\receiver}_i = 1} I(X_i; Z_i | \kv{\receiverrnd}=\kv{\receiver}) + \sum_{i : \kv{\receiver}_i = 0} I(X_i; Z_i | \kv{\receiverrnd}=\kv{\receiver}) \\
      & \geq & |\kv{\receiver}| \min_{X \sim (1-p_{q^{*}}/p_r, p_{s^{*}}/p_r, p_{q^{*}}/p_r - p_{s^{*}}/p_r  ), Q_{Z|X}:  E[\rho(X,Z)] = 0} I(X;Z) \\
    & & + |\kv{\receiver}^c| \min_{X \sim ( 1-p_{q^{**}}/(1-p_r), p_{s^{**}}/(1-p_r), p_{q^{**}}/(1-p_r) - p_{s^{**}}/(1-p_r)   ), Q_{Z|X}:  E[\rho(X,Z)] = 0} I(X;Z) \\ 
    & \geq & |\kv{\receiver}| \Lambda( p_{s^{*}}/p_r, 1-p_{q^{*}}/p_r) + |\kv{\receiver}^c| \Lambda( p_{s^{**}}/(1-p_r), 1-p_{q^{**}}/(1-p_r)   ).
\end{eqnarray*}
Finally note that $E_{\receiverrnd} \left[ |\kv{\receiverrnd}| \right] = |\kernelspace| p_r$
and thus putting everything together
\begin{eqnarray*}
\lefteqn{E_{\senderrnd,\queryrnd,\receiverrnd} \left[ \len{f(\senderrnd,\queryrnd,\receiverrnd)}  \right]} \\
&\geq& |\kernelspace| p_r \Lambda( p_{s^{*}}/p_r, 1-p_{q^{*}}/p_r) + |\kernelspace| (1-p_r) \Lambda( p_{s^{**}}/(1-p_r), 1-p_{q^{**}}/(1-p_r)   ) \\
&=& |\kernelspace| \Lambda( p_{s^{*}}, p_r-p_{q^{*}}) + |\kernelspace| \Lambda( p_{s^{**}}, 1-p_r-p_{q^{**}} ).
\end{eqnarray*}

We now prove the upper bound, which we emphasize, will be developed under the weaker assumption that $\senderrnd, \queryrnd, \receiverrnd$ follow a $(p_r,p_{s^*}, p_{s^{**}},p_{q^*},p_{q^{**}})$-law. Let $f$ and $g$ be the encoder and decoder functions that we will be constructing. Let $\sender, \query, \receiver \in \surfacespace$ and define
\begin{eqnarray*}
\sender^{*} &=& \surfaceand{\sender}{\receiver} \\
\sender^{**} &=& \surfaceand{\sender}{\surfacenot{\receiver}}  \\
\query^{*} &=& \surfaceand{\query}{\receiver} \\
\query^{**} &=& \surfaceand{\query}{\surfacenot{\receiver}} .
\end{eqnarray*}
We similarly define random versions of the above by replacing the lower case letters with upper case letters. We can compute, for the random versions, the corresponding normalized expected kernel sizes:
\begin{eqnarray*}
p_r &=&|\kernelspace|^{-1}E|\algebraicset{ \receiverrnd }| \\
p_{s^*} &=& |\kernelspace|^{-1}E|\algebraicset{\senderrnd^*}| \\
p_{s^{**}} &=& |\kernelspace|^{-1}E|\algebraicset{\senderrnd^{**} }| \\
p_{q^*} &=& |\kernelspace|^{-1}E|\algebraicset{ \queryrnd^*}| \\
p_{q^{**}} &=&|\kernelspace|^{-1}E|\algebraicset{ \queryrnd^{**} }|.
\end{eqnarray*}
Also define
\begin{eqnarray*}
  j_0 &=&  \{ i : \kv{\receiver}_i = 0 \} \\  
  j_1 &=&  \{ i : \kv{\receiver}_i = 1 \}  \\
    x^*_1 \cdots x^*_{|\kv{\receiver}|} &=& \left[ \psi( \kv{\sender^*} , \kv{\query^*}) \right]_{j_1}\\
    x^{**}_1 
    \cdots x^{**}_{|\kernelspace| - |\kv{\receiver}|}&=& [\psi( \kv{\sender^{**}} , \kv{\query^{**}}) ]_{j_0}
\end{eqnarray*}
and as before, define random versions by replacing lower case letters with upper case letters.
Using Theorem 
\ref{thm:shannonpartition} twice we can guarantee the existence of $f^{*}$, $g^{*}$, $f^{**}$, $g^{**}$ such that
\begin{eqnarray*}
E\left[ \len{f^{*}(X^*_1 \cdots X^*_{|\kv{\receiverrnd}|})} | \kv{\receiverrnd}\right] &\leq& |\kv{\receiverrnd}| \Lambda(p_{s^*}/p_r,1-p_{q^*}/p_r) +\\
& &+ 2 \log_2 \left( |\kv{\receiverrnd}| \Lambda(p_{s^*}/p_r,1-p_{q^*}/p_r) \right)  + 3\\
\sum_{i=1}^{|\kv{\receiverrnd}|} \rho(X_i^*,g^{*}(f^{*}(X_1^* \cdots X_{|\kv{\receiverrnd}|}^*))_i) &=& 0 \\
E\left[ \len{f^{**}(X^{**}_1 \cdots X^{**}_{|\kernelspace|-|\kv{\receiverrnd}|})} | \kv{\receiverrnd}\right] &\leq& (|\kernelspace|-|\kv{\receiverrnd}|) \Lambda(p_{s^{**}}/(1-p_r),1-p_{q^{**}}/(1-p_r)) \\
& & + 2 \log_2 \left( (|\kernelspace|-|\kv{\receiverrnd}| )\Lambda(p_{s^{**}}/(1-p_r),1-p_{q^{**}}/(1-p_r)) \right)  + 3\\
\sum_{i=1}^{|\kernelspace| - |\kv{\receiverrnd}|} \rho(X_i^{**},g^{**}(f^{**}(X_1^{**} \cdots X_{|\kernelspace| - |\kv{\receiverrnd}|}^{**}))_i) &=& 0 .
\end{eqnarray*}
 Next we  assemble an encoder and decoder for our setup. Given two binary strings $c_1, c_2 \in \{0,1\}^*$, let $c_1 c_2$ denote the string resulting from concatenating the two individual strings. We define our encoder as
\begin{eqnarray*}
f(\sender,\query,\receiver) = f^*(x^*_1 \cdots x^{**}_{|\kv{\receiver}|}) f^{**}(x^{**}_1 \cdots x^{**}_{|\kernelspace|-|\kv{\receiver}|}).
\end{eqnarray*}
From our earlier estimates,
\begin{eqnarray*}
E_{\senderrnd,\queryrnd,\receiverrnd} \left[ \len{f(\senderrnd,\queryrnd,\receiverrnd)} | \kv{\receiverrnd}\right] &\leq& |\kv{\receiverrnd}| \Lambda(p_{s^*}/p_r,1-p_{q^*}/p_{r}) \\
&& +(|\kernelspace|-|\kv{\receiverrnd}| )\Lambda(p_{s^{**}}/(1-p_r),1-p_{q^{**}}/(1-p_r)) + 6\\
&& + 
2 \log_2 |\kv{\receiverrnd}| \Lambda(p_{s^{*}}/p_r,1-p_{q^{*}}/p_r) \\
&& +
2 \log_2 \left( ( |\kernelspace| - |\kv{\receiverrnd}|) \Lambda(p_{s^{**}}/(1-p_r),1-p_{q^{**}}/(1-p_r))  \right)
\end{eqnarray*}
and using the law of total expectations, and the concavity $\cap$ of the logarithm, we obtain
\begin{eqnarray*}
|\kernelspace|^{-1} E_{\senderrnd,\queryrnd,\receiverrnd} \left[ \len{f(\senderrnd,\queryrnd,\receiverrnd)} \right] &\leq& p_r \Lambda(p_{s^*}/p_r,1-p_{q^*}/p_{r}) \\
&& + (1-p_r)  \Lambda(p_{s^{**}}/(1-p_r),1-p_{q^{**}}/(1-p_r)) + 6  |\kernelspace|^{-1}\\
&& + 
2 |\kernelspace|^{-1}\log_2 \left( p_r|\kernelspace| \Lambda(p_{s^{*}}/p_r,1-p_{q^{*}}/p_r) \right) \\
&& +
2 |\kernelspace|^{-1} \log_2 \left( (1-p_r)|\kernelspace| \Lambda(p_{s^{**}}/(1-p_r),1-p_{q^{**}}/(1-p_r))  \right).
\end{eqnarray*}
Now we construct the decoder. We define a function $\textit{lift}$, on two arguments: a \textit{filter} and a \textit{pattern}. The function lifts a binary \textit{pattern} to into a larger binary vector using the positions indicated by \textit{filter}, maintaining the order of the bits in the \textit{pattern}. For a vector $\textit{filter} \in \{0,1\}^{|\kernelspace|}$ and $\textit{pattern} \in \{0,1\}^{ |\textit{filter}|}$ (where $|\textit{filter}|$ indicates the number of ones in the vector $\textit{filter}$), let $a_i$, for $i=1,\ldots,|\textit{filter}|$, be the positions such that $\textit{filter}_{a_i} = 1$. We then define
\begin{eqnarray*}
    \textit{lift}(\textit{filter},\textit{pattern})_{a_i} = \textit{pattern}_i
\end{eqnarray*}
and define $\textit{lift}$ to be $0$ in all other indices.

The code that Theorem \ref{thm:shannonpartition} guarantees existence of is prefix free, and thus the output of $f(\senderrnd,\queryrnd,\receiverrnd)$ can be decoded sequentially. Define
\begin{eqnarray*}
\mathbf{a} &=&  
\textit{lift}\left( \kv{\receiver} , 
g^{*} \left( f^{*}\left( x^*_1 \cdots x^{*}_{|\kv{\receiverrnd}|} \right)  \right) \right) \\
\mathbf{b} &=& \textit{lift} \left( 1 - \kv{\receiver},g^{**} \left( f^{**}\left( x^{**}_1 \cdots x^{**}_{|\kernelspace|-|\kv{\receiver}|} \right)  \right)  \right).
\end{eqnarray*}
Now, assume that $\query$ is such that $\sender \entails \query$. Then by we know that
\begin{eqnarray*}
\kv{\sender} \kv{\receiver} \leq \mathbf{a} \leq \kv{\query} \kv{\receiver},  \\
\kv{\sender}(1- \kv{\receiver})  \leq \mathbf{b}  \leq \kv{\query}(1 - \kv{\receiver}) 
\end{eqnarray*}
where the products and inequalities are interpreted element-wise. And as a consequence 
\begin{eqnarray*}
\kv{\sender} \leq \mathbf{a} + \mathbf{b} \leq \kv{\query}.
\end{eqnarray*} 
The output of the decoder $g$ is defined to be
\begin{eqnarray*}
\ellv{ \mathbf{a} + \mathbf{b}}.
\end{eqnarray*}
This ensures that the requirement $g(f(\sender,\query,\receiver)) \entails \query$ is satisfied. \qed

\subsection{General setup when Alice does not know what Bob knows}
\label{ss:alice-does-not-know}

To conclude our sequence of theorems, we study a situation where Alice does not know the specific sentence $\receiverrnd$ knows, and is trying to ensure that Bob knows how to prove a targeted query $\queryrnd$. This setup can be seen as a combination of the setups in Theorem \ref{thm:less_is_more_simple} and Theorem \ref{thm:slepian_wolf_sender_unaware}, and it is the most complex situation that we analyze from the standpoint of the mathematical machinery used to prove it.

In a manner similar to the discussion leading to Theorem~\ref{thm:slepian_wolf_sender_unaware}, we rely on Table \ref{tab:mt_general} in order to define the communication pattern that we analyze. A total of 6 turns are allowed, starting with Bob making the first communication and concluding with Bob being able to prove a targeted query  $\queryrnd$. The outputs for all the encoders are denoted by $B^0,A^1,B^2,A^3,B^4, A^5$. The cost of communication is given by
\begin{eqnarray*}
    \frac{1}{|\kernelspace|} E_{\senderrnd, \queryrnd, \receiverrnd} \left[ \len{B^0} + \len{A^1} + \len{B^2} + \len{A^3} + \len{B^4} + \len{A^5} \right] .
\end{eqnarray*}
The image of all of the encoders is assumed to be prefix-free, assuming that whatever is the common context is kept fixed when considering the image. For a more detailed discussion of this, note the comments associated with Table \ref{tab:mt}. A 6-turn code $\{f^i,g^i\}_{i=0}^5$ as defined by Table \ref{tab:mt_general} and this discussion is called a 6-turn code for targeted queries.  The reason for the word ``targeted'' is because the goal for the communication can be restricted to be a subset of Alice's knowledge, rather than its entirety.

\begin{table}
\begin{center}
\begin{tabular}{ccccccc}
Alice's & common       &Alice's & direction & Bob's & common & Bob's  \\
private & context  & function &  & function & context & private  \\ 
context &     & &  & &  & context  \\ \hline 
$\sender, \query$ & & $g^0$ & $\stackrel{b^0}{\longleftarrow}$ & $f^0$ & & $\receiver$ \\
$\sender, \query$ & $b^0$ & $f^1$ & $\stackrel{a^1}{\longrightarrow}$ & $g^1$ & $b^0$ & $\receiver$ \\
$\sender, \query$ & $b^0, a^1$ & $g^2$ & $\stackrel{b^2}{\longleftarrow}$  & $f^2$ & $b^0, a^1$ & $\receiver$\\
$\sender, \query$ & $b^0, a^1, b^2$ & $f^3$ & $\stackrel{a^3}{\longrightarrow}$  & $g^3$ & $b^0,a^1,b^2$ & $\receiver$ \\
$\sender, \query$ & $b^0, a^1, b^2, a^3$ & $g^4$ & $\stackrel{b^4}{\longleftarrow}$  & $f^2$ & $b^0, a^1,b^2,a^3$ & $\receiver$\\
$\sender, \query$ & $b^0, a^1, b^2, a^3,b^4$ & $f^5$ & $\stackrel{a^5}{\longrightarrow}$  & $g^3$ & $b^0,a^1,b^2,a^3,b^4$ & $\receiver$ \\
& & & & $\downarrow$ & \\
& & & & $\senderhat$ &  \\
& & & &  $\senderhat \entails \query$& 
\end{tabular}
\end{center}
\caption{A 6-turn code for targeted queries. It is assumed that $\sender \entails \query$ and $\query \entails \receiver$. The actual step of Bob proving queries is not shown. }
\label{tab:mt_general}
\end{table}

\begin{restatable}{thm}{master}
Let $(L,  \kernelspace, \kappa, \ell)$ be a Logic System. Let $\senderrnd, \queryrnd, \receiverrnd  \in \surfacespace$ represent the sender, query and receiver logic sentences, with $\senderrnd, \queryrnd$ only known to Alice and $\receiverrnd$ only known to Bob. Assume that  $\senderrnd, \queryrnd, \receiverrnd$ have kernels that follow a ($p_s,p_q,p_r$)-law. Then
\begin{eqnarray*}
\lefteqn{\min_{\{f^i,g^i\}_{i=0}^5} \frac{1}{|\kernelspace|} E_{\senderrnd, \queryrnd, \receiverrnd} \left[ \len{B^0} + \len{A^1} + \len{B^2} + \len{A^3} + \len{B^4} + \len{A^5} \right]} \\
&\leq & \Lambda(p_s,p_r-p_q) + O\left( \frac{\log_2|\kernelspace|}{|\kernelspace|} \right) ,
\end{eqnarray*}
where the minimization is for all 6-turn codes for targeted queries (c.f.\ Table \ref{tab:mt_general}). Furthermore, if $\{(\kv{\senderrnd},\kv{\queryrnd}, \kv{\receiverrnd})_i\}_{i=1}^{|\kernelspace|}
$ are i.i.d.\ and $\receiverrnd \rightarrow \algebraicset{\receiverrnd} \rightarrow (\algebraicset{\senderrnd},\algebraicset{\queryrnd})$ then
\begin{eqnarray*}
\Lambda(p_s,p_r-p_q) &\leq& \min_{\{f^i,g^i\}_{i=0}^5} \frac{1}{|\kernelspace|} E_{\senderrnd, \queryrnd, \receiverrnd} \left[ \len{B^0} + \len{A^1} + \len{B^2} + \len{A^3} + \len{B^4} + \len{A^5} \right]
\end{eqnarray*}
\label{thm:master}
\end{restatable}


 \textbf{Proof.}  We start the proof using the beginning arguments from the proof of the lower bound of Theorem~\ref{thm:slepian_wolf_sender_unaware}. We assume a 6-turn code for targeted queries $\{f^i,g^i\}_{i=0}^5$,  and let $B^0, A^1, B^2, A^3, B^4, A^5 \in \{0,1\}^*$ be the binary strings output by the encoders $f^0,f^1,f^2,f^3,f^4,f^5$ that are employed by Bob and Alice. Using the same type of argument that led to (\ref{eq:first_step_lower_bound}), we have
 \begin{eqnarray*}
E_{\senderrnd,\queryrnd,\receiverrnd} \left[ \len{B^0} + \len{A^1} + \len{B^2} + \len{A^3} + \len{B^4} + \len{A^5} \right] \geq \Hreg{ \senderrndhat | \receiverrnd } .
\end{eqnarray*}
We now continue by using a more complex variant of the arguments found in the proof of Theorem~\ref{thm:less_is_more_simple}. We write
\begin{eqnarray*}
H( \senderrndhat | \receiverrnd ) & \stackrel{(a)}{\geq}& \Hreg{\kv{\senderrndhat } | \receiverrnd}  \\ 
& \stackrel{(b)}{=} & \Hreg{\kv{\senderrndhat} | \receiverrnd}  - \Hreg{\kv{\senderrndhat} | \senderrnd, \queryrnd, \receiverrnd}  \\ 
& \stackrel{(c)}{=} & I( \senderrnd, \queryrnd ; \kv{\senderrndhat} | \receiverrnd ) \\
& \stackrel{(d)}{\geq} & I( \kv{\senderrnd}, \kv{\queryrnd} ; \kv{ \senderrndhat} | \receiverrnd ) \\
& \stackrel{(e)}{=} & \Hreg{ \kv{\senderrnd}, \kv{\queryrnd} | \receiverrnd } - \Hreg{\kv{\senderrnd}, \kv{\queryrnd} | \receiverrnd,\kv{\senderrndhat} } \\
& \stackrel{(f)}{=} & \Hreg{ \kv{\senderrnd}, \kv{\queryrnd} | \kv{\receiverrnd} } - \Hreg{ \kv{\senderrnd}, \kv{\queryrnd}| \receiverrnd,  \kv{\senderrndhat}  } \\
& \stackrel{(g)}{\geq} & \Hreg{ \kv{\senderrnd}, \kv{\queryrnd} | \kv{\receiverrnd} } - \Hreg{\kv{\senderrnd}, \kv{\queryrnd} | \kv{\receiverrnd}, \kv{\senderrndhat}}  \\
&\stackrel{(h)}{=}& I(\kv{\senderrnd},\kv{\queryrnd};  \kv{\senderrndhat} | \kv{\receiverrnd})  
\end{eqnarray*}
where (a) follows from the fact that deterministic functions of random quantities cannot increase entropy, (b) holds because in the second term, the conditioning is over all randomness, rendering the corresponding entropy equal to zero, (c) is the definition of mutual information, (d) follows from the data processing inequality, (e) expands mutual information back into a difference of entropies, (f) follows from the assumption that
\begin{eqnarray*}
\receiverrnd  \rightarrow \kv{\receiverrnd} \rightarrow  (\kv{\senderrnd}, \kv{\queryrnd})
\end{eqnarray*}
as well as the data processing inequality, (g) follows from the fact that a weaker conditioning cannot reduce conditional entropy, and (h) is from the definition of mutual information. 

In reference to the proof of Theorem~\ref{thm:less_is_more_simple},  recall that for a given $\mathtt{a} \in \surfacespace$, $\kv{\mathtt{a}}$ can be regarded as an indicator vector for the kernel $\algebraicset{\mathtt{a}}$. Additionally, we use the same definition for $\psi(\cdot,\cdot)$:
\begin{eqnarray*}
\psi(\mathtt{a},\mathtt{b}) = 0 \times (\underline{1} - \mathtt{b}) + 1 \times \mathtt{a} + 2 \times ( \underline{1} - \mathtt{a} ) \times \mathtt{b} .
\end{eqnarray*}
where $( \underline{1} - \mathtt{a} ) \times \mathtt{b}$ is interpreted to be product element-wise. We now define
\begin{eqnarray*}
    X_1^{|\kernelspace|} &=& \psi(\kv{\senderrnd},\kv{\queryrnd}) 
    \\
    Z_1^{|\kernelspace|} &=& \kv{ \senderrndhat } \nonumber \\
    Y_1^{|\kernelspace|} &=& \kv{\receiverrnd} . \nonumber
\end{eqnarray*}

\begin{figure}
\begin{center}
\begin{tikzpicture}  
\clip (0,0) rectangle (10,7);
\draw (0,0)[fill=pink] rectangle (10,7);
\draw(0.25,0.25) rectangle (7,5);
\draw(5,-1.5)[rotate=45,fill=white] ellipse(2.5 and 1.5);
\draw(5.6,4.25) node {$\algebraicset{\queryrnd}$};
\draw(4,2)[fill=green!35!white] circle(1.0);  
\draw(4.0,2.6) node {$\algebraicset{\senderrnd}$};
\draw(1.0,4.5) node {$\algebraicset{\receiverrnd}$};
\end{tikzpicture}
\end{center}
\caption{Reference diagram for the proof of the lower bound in Theorem~\ref{thm:master}}
\label{fig:fourregions_master}
\end{figure}
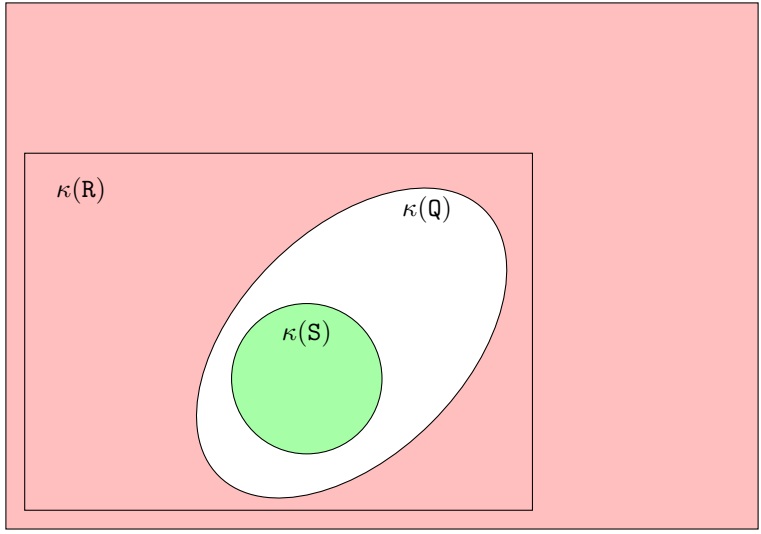

To interpret the first definition we refer the reader to Figure~\ref{fig:fourregions_master}, which is a counterpart to Figure~\ref{fig:less_is_more_simple}-b where we have added $\algebraicset{\receiverrnd}$. As in that figure,  the red region is assigned ``0'', the green  region ``1'' and the white region ``2''. 

The next sequence of arguments is a simplified version of those in Theorem \ref{thm:less_is_more_background}. For any kernel $k \in \{0,1\}^{|\kernelspace|}$:
\begin{eqnarray}
    P(X_i = 0 | Y_1^{|\kernelspace|}= k  ) &\stackrel{(a)}{=}& P(X_i = 0 | Y_i=k_i) \nonumber \\
    &\stackrel{(b)}{=}& P( \kv{\queryrnd}_i = 0 | Y_i = k_i) \nonumber \\
    &=& 1 - P( \kv{\queryrnd}_i = 1 | Y_i = k_i) \nonumber \\
    &=& 1 - \frac{P(\kv{\queryrnd}_i = 1) P( Y_i = k_i | \kv{\queryrnd}_i = 1 )}{P(Y_i = k_i)} \nonumber \\
    & \stackrel{(c)}{=} & \left\{ \begin{array}{cc}
        1 &  \mbox{ if } k_i = 0    \\
       1 - p_q/p_r &  \mbox{ if } k_i = 1 
\end{array} \right.
\label{eq:x_given_y_0}
\end{eqnarray}
where (a) follows from the Definition $X_i = \psi( \kv{\senderrnd}_i, \kv{\queryrnd}_i )$ and the independence assumption, (b) follows from Lemma \ref{lem:complexabc}, and (c) follows from the identically distributed assumption and the $(p_s,p_q,p_r)$-law. Similarly, one can deduce
\begin{eqnarray}
   P(X_i = 1 | Y_1^{|\kernelspace|} = k ) = P(X_i = 1 | Y_i = 1 ) = p_s/p_r
   \label{eq:x_given_y_1}
\end{eqnarray}
and therefore
\begin{eqnarray}
 P(X_i = 2 | Y_1^{|\kernelspace|} = k ) =     P(X_i = 2 | Y_i = 1 ) = p_q/p_r - p_s/p_r
    \label{eq:x_given_y_2}
\end{eqnarray}
One consequence of this analysis is that, conditional on $Y_1^{|\kernelspace|}=k$, the $\{X_i\}$ are independent although not generally identically distributed. Recall we are assuming that whenever using the 6-turn code for targeted queries, $\senderrnd \entails \senderrndhat$ and that  $\senderrndhat \entails \queryrnd$  and recall the definition of the distortion metric $\rho$ from Equations (\ref{eq:distortion_metric}) and (\ref{eq:distortion_metric_matrix}).  Then for all $1 \leq i \leq |\kernelspace|$,
\begin{eqnarray*}
    P(Z_i=0 | X_i=0, Y_1^{|\kernelspace|} = k ) &=& 1 \\
    P(Z_i=1 | X_i=1, Y_1^{|\kernelspace|} = k ) &=& 1
\end{eqnarray*}
and therefore as a consequence, for all $1 \leq i \leq |\kernelspace|$, under the conditioning $Y_1^{|\kernelspace|}=k$, with probability 1,
\begin{eqnarray}
\rho( X_i, Z_i) = 0 .
\label{eq:master_fact}
\end{eqnarray}

Following in the footsteps of the proof of the lower bound for Theorem~\ref{thm:less_is_more_simple},  we write:
\begin{eqnarray*}
I(\kv{\senderrnd},\kv{\queryrnd};  \kv{\senderrndhat} | \kv{\receiverrnd}) &=& I(\kv{\senderrnd},\kv{\queryrnd};  Z_1^{|\kernelspace|}| Y_1^{|\kernelspace|}) \\
 &\stackrel{(i)}{\geq}& I(X_{1}, \ldots, X_{|\kernelspace|}; Z_1, \ldots, Z_{|\kernelspace|} | Y_1^{|\kernelspace|} ) \\
&\stackrel{(j)}{\geq}& \sum_{k} P(Y_1^{|\kernelspace|}=k) I(X_{1}, \ldots, X_{|\kernelspace|}; Z_1, \ldots, Z_{|\kernelspace|} | Y_1^{|\kernelspace|} = k ) .
\end{eqnarray*}
The step (i) follows from the fact that $X_1^{|\kernelspace|}$ is a deterministic function of $\kv{\senderrnd}$ and $\kv{\queryrnd}$, and thus conditional on  $Y_1^{|\kernelspace|}$, the following Markov chain holds:
\begin{eqnarray*}
    X_1^{|\kernelspace|} 
    \rightarrow (\kv{\senderrnd}, \kv{\queryrnd}) \rightarrow Z_1^{|\kernelspace|}.
\end{eqnarray*}
Step (j) follows from the definition of conditional mutual information. We then focus on each individual term:
\begin{eqnarray*}    
  \lefteqn{I(X_{1}, \ldots, X_{|\kernelspace|}; Z_1, \ldots, Z_{|\kernelspace|} | Y_1^{|\kernelspace|}=k )} \\
  &=& \Hreg{ X_{1}, \ldots, X_{|\kernelspace|} |  Y_1^{|\kernelspace|}=k } - \Hreg{ X_{1}, \ldots, X_{|\kernelspace|} | Z_1, \ldots, Z_{|\kernelspace|},  Y_1^{|\kernelspace|}=k  } \\
 &\stackrel{(k)}{=}&  \sum_{i=1}^{|\kernelspace|} \Hreg{X_i| X_1^{i-1}, Y_1^{|\kernelspace|}=k} - \Hreg{X_i | Z_1, \ldots, Z_{|\kernelspace|}, 
  X_{1}^{i-1}, Y_1^{|\kernelspace|}=k}\\
 &\stackrel{(l)}{=}& \sum_{i=1}^{|\kernelspace|} \Hreg{X_i|Y_i = k_i} - \Hreg{X_i | Z_1, \ldots, Z_{|\kernelspace|}, 
  X_{1}^{i-1} , Y_1^{|\kernelspace|}=k}\\
  & \stackrel{(m)}{\geq} & \sum_{i=1}^{|\kernelspace|} \Hreg{X_i|Y_i = k_i} - \Hreg{X_i | Z_i, Y_i=k_i}   \\
    & = & \sum_{i=1}^{|\kernelspace|} I(X_i ; Z_i|Y_i=k_i)
\end{eqnarray*}
where (k) follows from the chain rule for entropy, (l) follows from the fact that the $|\kernelspace|$ tuples $\{(X,Y)_i\}$ are statistically independent as per the Theorem's assumption. Finally (m) follows from the fact that eliminating conditioning random variables cannot decrease entropy. We continue by splitting the last summation in two:
\begin{eqnarray}
    \sum_{i=1}^{|\kernelspace|} I(X_i ; Z_i|Y_i=k_i)  &=& \sum_{ i : k_i = 0 }I(X_i ; Z_i|Y_i=0) + \sum_{ i : k_i = 1 }I(X_i ; Z_i|Y_i=1)  \nonumber \\
     &=& \sum_{ i : k_i = 1 }I(X_i ; Z_i|Y_i=1) \label{eq:rd_trip_master}
\end{eqnarray}
where we have made the observation that conditioning on $Y_i=0$, $X_i$ is constant (and equal to zero) and therefore the corresponding mutual information is zero.

As before, to complete the lower bound, we note that (\ref{eq:rd_trip_master}) is an averaging of mutual informations where the marginal for $X_i$ is identical for all $\{i: k_i = 1\}$, as given by Equations (\ref{eq:x_given_y_0},\ref{eq:x_given_y_1},\ref{eq:x_given_y_2}), but the conditionals $Q_{Z_i|X_i,Y_i=1}$ are in general different. Assume temporarily that $|k| = |\{i:k_i=1\}| \geq 1$. Define a conditional distribution by averaging all those conditionals:
\begin{eqnarray}
Q_{Z^{\prime}|X^{\prime}}(z|x) \stackrel{\Delta}{=} \frac{1}{|k|} \sum_{ \{i: k_i=1\} } Q_{Z_i|X_i, Y_i}(z|x,y=1).
\label{eq:master_updated_conditional}
\end{eqnarray}
In reference to Equations (\ref{eq:x_given_y_0},\ref{eq:x_given_y_1},\ref{eq:x_given_y_2}), let $X^{\prime}$ be distributed according to $\{1-p_q/p_r, p_s/p_r, p_q/p_r-p_s/p_r\}$ and let the joint $X^{\prime},Z^{\prime}$ be defined by having $Z^{\prime}$ be the effect of passing $X^{\prime}$ through the channel defined by (\ref{eq:master_updated_conditional}). Because of (\ref{eq:master_fact}), it is the case that 
\begin{eqnarray*}
E_{X^{\prime}, Z^{\prime}}\left[ \rho(X^{\prime},Z^{\prime}) \right] = 0.
\end{eqnarray*}

Temporarily imagine $I(X_i; Z_i|Y_i=1)$ as a function of two distributions: a distribution on $X_i$ given $Y_i=1$ and a conditional distribution $Q_{Z_i | X_i, Y_i } ( z | x,y=1)$. Given our earlier arguments, in (\ref{eq:rd_trip_master}), the distribution of $X_i$ given $Y_i=1$ is identical for all those $\{i :  k_i = 1\}$, and precisely equal to that of $X^{\prime}$. Furthermore, it is known that mutual information is convex  $\cup$ if one keeps the marginal distribution of $X$ fixed as one varies the conditional distribution of $Z$ given $X$ and therefore the following bound holds:
\begin{eqnarray}
\sum_{ \{i: k_i=1\} } I(X_i ; Z_i|Y_i=1)  \geq |k| I(X^{\prime};Z^{\prime}) \geq |k| \min_{P(X,Z) \in \mathcal{D}} I(X;Z) \label{eq:master_rd}
\end{eqnarray}
where the domain $\mathcal{D}$ for the minimization is defined by joint distributions for $X,Z$ with $X \sim (1-p_q/p_r,p_s/p_r,p_q/p_r-p_s/p_r )$ and $E[\rho(X,Z)] = 0$. Also note that even though we assumed $|k| \geq 1$, the bound (\ref{eq:master_rd}) is trivially true for $|k|=0$ due to the nonnegativity of mutual information. Such minimization, which can be obtained using standard variational methods, results in the following expression
\begin{eqnarray*}
    |k| \Lambda(p_s/p_r, 1 - p_q/p_r).
\end{eqnarray*}
We now bring back the distribution for $k$, and write
\begin{eqnarray*}
    \sum_{k} P( Y_1^{|\kernelspace|} = k ) |k| \Lambda(p_s/p_r, 1 - p_q/p_r) = p_r \Lambda(p_s/p_r, 1 - p_q/p_r) = \Lambda(p_s,p_r-p_q)
\end{eqnarray*}
where the last equality follows from basic properties of $\Lambda$ (Lemma \ref{lem:elementarylambda}). This concludes the proof of the lower bound. 

The proof of the upper bound is adapted from the proof of Theorem~\ref{thm:slepian_wolf_sender_unaware}. We stress that this result will hold under the very general assumption that $\senderrnd, \queryrnd, \receiverrnd$ follow a $(p_s, p_q, p_r)$-law with $p_s \leq p_q \leq p_r$. As before, we establish the precise domain/image of the encoder and decoders for our 6-turn mode:
\begin{eqnarray*}
f^{ 0 } : \surfacespace & \rightarrow & \{0,1\}^*, \hspace{0.1in} g^{0} : \{0,1\}^{*} \rightarrow \mathbb{Z}, \\
f^{ 1 } : \surfacespace^{2} & \rightarrow & \{0,1\}^*, \hspace{0.1in} g^{1} : \{0,1\}^{*} \rightarrow \mathbb{Z}^{2}, \\
f^{ 2 } : \mathbb{Z} & \rightarrow & \{0,1\}^*, \hspace{0.1in} g^{2} : \{0,1\}^{*} \rightarrow \mathbb{Z}, \\
f^{ 3 } : \surfacespace^{2}  \times \mathbb{Z}^{3} & \rightarrow & \{0,1\}^*, \hspace{0.1in} g^{3} : \{0,1\}^*\times \surfacespace   \rightarrow  \surfacespace \times \{ \text{success}, \text{failure} \}, \\
f^4 : \{ \text{success}, \text{failure} \} &\rightarrow& \{0,1\}, \hspace{0.1in} g^4 : \{0,1\} \rightarrow \{ \text{success}, \text{failure} \}, \\
f^{5} : \surfacespace & \rightarrow & \{0,1\}^{*}, \hspace{0.1in}  g^{5} : \{0,1\}^{*} \times \{ \text{success}, \text{failure} \} \times \surfacespace  \rightarrow  \surfacespace .
\end{eqnarray*}

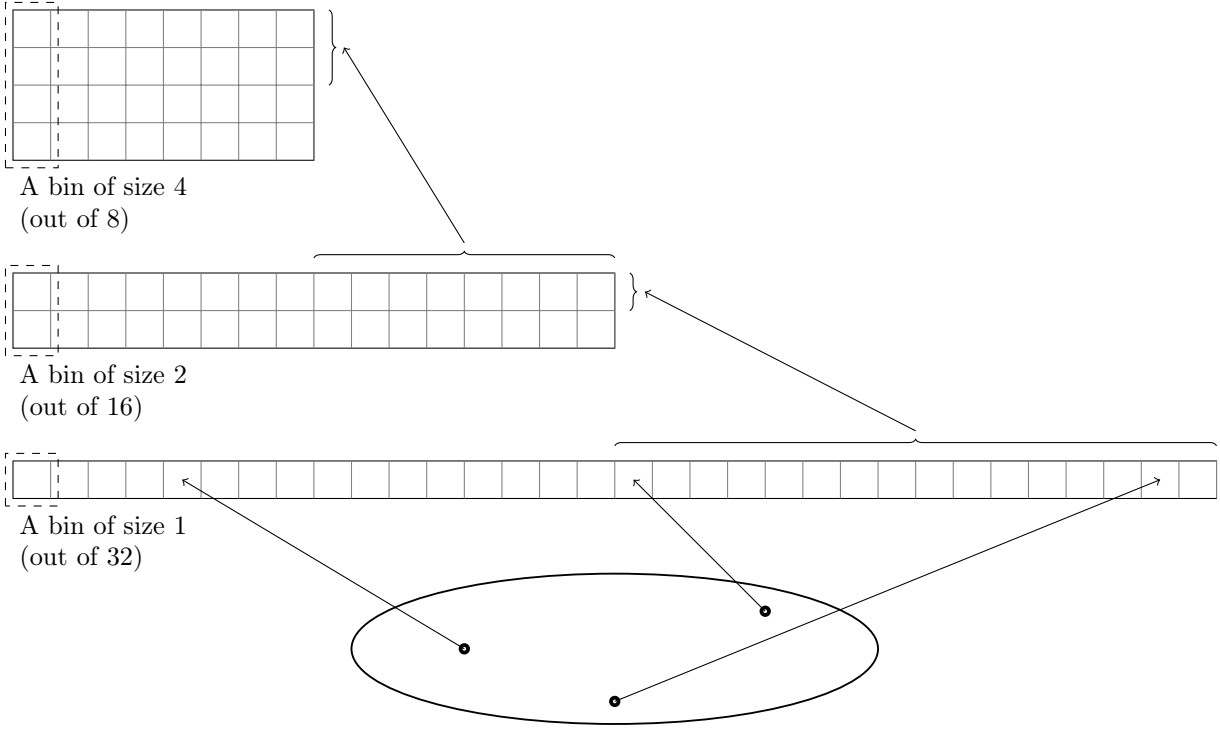
\begin{figure}[h]
\begin{center}
\begin{tikzpicture}
\draw[rotate=0, line width = 0.25mm] (9, 3) ellipse (3.5cm and 1cm);
\draw[rotate=0, line width = 0.5mm] (7, 3) circle (0.05cm);
\draw[rotate=0, line width = 0.5mm] (11, 3.5) circle (0.05cm);
\draw[rotate=0, line width = 0.5mm] (9, 2.3) circle (0.05cm);
\draw (1,5)[color=black,fill=white] rectangle (17,5.5);
\draw[step=0.5cm,color=gray] (1,5) grid (17,5.5);
\draw (1,7)[color=black,fill=white] rectangle (9,8);
\draw[step=0.5cm,color=gray] (1,7) grid (9,8);
\draw (1,9.5)[color=black,fill=white] rectangle (5,11.5);
\draw[step=0.5cm,color=gray] (1,9.5) grid (5,11.5);
\draw [decorate, decoration = {brace}] (9,5.7) --  (17,5.7);
\draw [decorate, decoration = {brace}] (9.2,8) --  (9.2,7.5);
\draw [decorate, decoration = {brace}] (5,8.2) --  (9,8.2);
\draw [decorate, decoration = {brace}] (5.2,11.5) --  (5.2,10.5);
\draw[->]        (13,5.9)   -- (9.4,7.75);
\draw[->]        (7,8.4)   -- (5.4,11.0);
\draw[->]        (7,3)   -- (3.25,5.25);
\draw[->]        (11,3.5)   -- (9.25,5.25);
\draw[->]        (9,2.3)   -- (16.25,5.25);
\draw(0.9,9.4)[dashed] rectangle(1.6,11.6);
\draw(0.9,6.9)[dashed] rectangle(1.6,8.1);
\draw(0.9,4.9)[dashed] rectangle(1.6,5.6);
\node at (2.2,4.4) [align=left] {A bin of size 1\\ (out of 32)};
\node at (2.2,6.4) [align=left] {A bin of size 2\\ (out of 16)};
\node at (2.2,8.9) [align=left] {A bin of size 4\\ (out of 8)};
\end{tikzpicture}
\end{center}
\caption{An adaptive rate hash function. An initial hash function, depicted at the bottom, can be used to create a family of hash functions by stacking halves iteratively. A total of three hash functions are shown; an exemplary bin for each is illustrated using a dashed box.  }
\label{fig:adaptive_rate}
\end{figure}

In reference to the proof of Theorem~\ref{thm:slepian_wolf_sender_unaware}, a key step is the definition of a hash function. For the present result we will require an \emph{adaptive rate} hash function which allows the number of bins to be flexibly changed in response to specific requirements. We refer the reader to Figure~\ref{fig:adaptive_rate} for this construction. A hash function with $2^i$ bins for some integer $i$, requiring $i$ bits to specify a specific bin, can also behave as a hash function with $2^{i-1}$ bins by stacking the second half of the bins for the former on top of the left half, creating half the bins with twice the number of elements and thus requiring one bit less to fully specify a bin. This stacking procedure can be continued, creating a family of hash functions with fewer bit requirements from a single hash function. Given any desired bit rate, provided that it is equal or lower than the maximum bit rate that the hash function supports, the adaptive hash function will be able match that desired bit rate within at most 1 bit of inefficiency. Define
\begin{eqnarray*}
T(w)=\{c \in \{0,1\}^{|\mathcal{M}|}: ~ |c|=w\}.
\end{eqnarray*}
Alice and Bob first interact to agree on the following functions for integers $0 < s \leq q \leq r$
\begin{eqnarray}
    \text{weight}(s,q,r) &=& \left\lceil \frac{r s}{s + r - q } \right\rceil \nonumber \\
 \text{code\_size}(s,q,r)  &=& \left\lceil \log_2|\kernelspace | \left( \begin{array}{c} |\kernelspace| \\ \text{weight}(s,q,r) \end{array} \right) / \left( \begin{array}{c} q - s\\ \text{weight}(s,q,r) - s   \end{array} \right)  \right\rceil  \label{eq:code_size_def}
\end{eqnarray}
as well as
\begin{itemize}
\item a set of codewords $\text{code}(s,q,r) \subseteq T(\text{weight}(s,q,r))$, indexed using integers $1,\ldots,\text{code\_size}(s,q,r)$
\item an adaptive rate hash function  $\text{bin}_{s,q,r,\text{rate}} : \text{code}(s,q,r) \rightarrow \mathbb{Z}$.
\end{itemize}
From this definition it should be clear that a codeword is an element of $\{0,1\}^{|M|}$. In our paper,  a codeword is equivalent to a kernel, since it can be seen as the indicator vector of the kernel. Thus a code can be seen as a set of kernels. How these are constructed will be described shortly. For two vectors $a,b$ of the same length, we will write $a \leq b$ whenever $a_i \leq b_i$ for every $i$. The protocol is then as follows:
\begin{enumerate}
\item The receiver sends $|\kv{ \receiverrnd}|$ to the sender, which she decodes. This step defines $f^{0}$ and $g^{0}$. \label{step:master_rec}
\item The sender sends $|\kv{\senderrnd}|$ and $|\kv{\queryrnd}|$, which are decoded at the receiver side. This defines $f^1$ and $g^1$ \label{step:master_send}.
\item Both sender and receiver compute a target weight and a code book size using the functions  (\ref{eq:code_size_def}) evaluated on $|\kv{\senderrnd}|, |\kv{\queryrnd}|$,  and $|\kv{\receiverrnd}|$.
\item The receiver calculates the average bit rate it wants from the receiver using the expression
\begin{eqnarray}
\text{RATE} = \frac{1}{|\kernelspace|}\log_2 \left\lceil | \left\{ a \in \text{code}(|\kv{\senderrnd}|,|\kv{\queryrnd}|,|\kv{\receiverrnd}|): 
a \leq \kv{\receiverrnd} \right\} | \right\rceil + \frac{\log_2 |\kernelspace|}{|\kernelspace|}\label{eq:def_rate}.
\end{eqnarray}
Note the ``round up'' operator inside of the logarithm. This implies, in particular, that $2^{|\kernelspace| \text{RATE}}$ will be a proper power of 2, and hence the resulting rate can be used in the context of the adaptive rate hash function. The integer $|\kernelspace| \text{RATE}$ is sent back to the sender, defining $f^2$ and $g^2$. At this point they both agree on target weight, code book size and target rate to be used in the adaptive rate hash function. \label{step:adaptive_rate}
\item The sender attempts to find $A \in \text{code}(|\kv{\senderrnd}|,|\kv{\queryrnd}|,|\kv{\receiverrnd}|)$ with the property that
 \begin{eqnarray}
    \kv{\senderrnd} \leq A \leq \kv{\queryrnd}.
   \label{eq:master_sandwich_proprty}
 \end{eqnarray}
If this fails, then the sender signals the failure to the receiver and executes step \ref{step:master_fallback}. If it succeeds, the sender signals success and sends to the receiver the bin to which $A$ was mapped, for the hash function with the agreed upon rate.  \label{step:master_bin}
    \item The receiver attempts to retrieve $A$ using the bin index, its knowledge of $\kv{\receiverrnd}$ and the algorithm described below. The outcome of this attempt results in an element of $\surfacespace$ and a ``success'' or a default, dummy element from $\surfacespace$ and a ``failure''. Together with the previous step, this completes $f^3, g^3$.      
    \label{step:master_dec}
    \item The success/failure of the attempt is signaled back to the sender using a single bit. This defines both $f^4$ and $g^4$.
    \item If successful, the sender has nothing to do anymore, as the receiver expects no further communication. The receiver simply outputs the element from $\surfacespace$ computed by $g^3$, designating it as the output $\senderrndhat$, partly defining $g^5$.
    \item If unsuccessful, the sender sends $\kv{\senderrnd}$ as a binary vector of length $|\kernelspace|$, which is decoded by the receiver and becomes $\senderrndhat$, the designated output of the protocol, completing the definition of $f^5$ and $g^5$. \label{step:master_fallback}
\end{enumerate} 
The algorithm guarantees that the receiver at the end will have in possession enough information to reproduce a message $\senderrndhat$ that can be used to prove $\queryrnd$. We now demonstrate that the normalized total bidirectional cost in bits is asymptotically $\Lambda(p_s,p_r-p_q)$.

Instead of constructing a specific $\text{code}(s,q,r)$ and hash function  $\text{bin}_{s,q,r,\text{rate}}$ we will define a probability distribution over each of these, evaluate the corresponding expected performance, and demonstrate that in average the performance as desired, implying the existence of deterministic versions of these with at least the same performance. We shall refer to the random constructions as $\text{CODE}(s,q,r)$ and $\text{BIN}_{s,q,r,\text{rate}}$, respectively.

We construct $\text{CODE}(s,q,r)$ by drawing uniformly and independently at random from $T(\text{weight}(s,q,r))$ a total of $\text{code\_size}(s,q,r)$ codewords. We construct $\text{BIN}_{s,q,r,\text{rate}}$ by mapping every element of $\text{CODE}(s,q,r)$ uniformly and independently at random to an integer in the range $\{1, \ldots, 2^{\lceil \log_2 |\kernelspace| \rceil} \}$ (each representing a bin), and then iteratively stacking halves of these bins to produce an adaptive rate hash function, as illustrated by Figure~\ref{fig:adaptive_rate}. 

For brevity, define 
\begin{eqnarray*}
W=\text{weight}(|\kv{\senderrnd}|,|\kv{\queryrnd}|,|\kv{\receiverrnd}|) \\
\mathcal{C} = \text{CODE}(|\kv{\senderrnd}|,|\kv{\queryrnd}|,|\kv{\receiverrnd}|) .
\end{eqnarray*}

In the algorithm above, there are two key failure events for which we want to estimate probabilities. First we focus on the probability of a failure in step \ref{step:master_bin}.
The probability that any one element of $\mathcal{C}$  satisfies (\ref{eq:master_sandwich_proprty}) is
\begin{eqnarray*}
P_{succ} = \frac{\left( \begin{array}{c} |\kv{\queryrnd}| - |\kv{\senderrnd}| \\ W - |\kv{\senderrnd}|   \end{array} \right) }{\left( \begin{array}{c} |\kernelspace| \\ W \end{array} \right)} .
\end{eqnarray*}

The probability that none of the elements of $\mathcal{C}$ meet the condition (\ref{eq:master_sandwich_proprty}) is then
\begin{eqnarray*}
(1-P_{succ})^{|\mathcal{C}|} &\leq& \exp\left( - P_{succ} |\mathcal{C}|\right) \\
& \leq & \frac{1}{|\kernelspace|}
\label{eq:master_failure_prob}
\end{eqnarray*}
where we used the definition (\ref{eq:code_size_def}) to obtain the latter. Now we focus on the failure probability in step \ref{step:master_dec}. Define 
\begin{eqnarray}
    \text{hypotheses}(\senderrnd,\queryrnd,\receiverrnd) \stackrel{\Delta}{=} \left\{ \alpha \in \mathcal{C} :  \alpha \leq \kv{\receiverrnd}, \text{BIN}_{|\kv{\senderrnd}|,|\kv{\queryrnd}|,|\kv{\receiverrnd}|}(\alpha) = \text{BIN}_{|\kv{\senderrnd}|,|\kv{\queryrnd}|,|\kv{\receiverrnd}|}(A) \right\} .
    \label{eq:hypelast}
\end{eqnarray} 
To estimate the failure probability, we upper bound the probability of this event:
\begin{eqnarray*}
P\left(\left[ | \text{hypotheses}(\senderrnd,\queryrnd,\receiverrnd)| \geq 2 \right]\right).
\end{eqnarray*}
Let $I_{\alpha}$ be equal to 1 if $\alpha \in \text{hypotheses}(\senderrnd,\queryrnd, \receiverrnd)$ and 0 otherwise. Then we can upper bound the error probability as
\begin{eqnarray*}
P\left(\left[ | \text{hypotheses}(\senderrnd,\queryrnd,\receiverrnd)| \geq 2 \right]\right) &=& P\left(  \sum_{\alpha} I_{\alpha} \geq 2 \right) .
\end{eqnarray*}
We note that when conditioning on $\senderrnd,\queryrnd,\receiverrnd,A, \mathcal{C}$, the only randomness that remains in (\ref{eq:hypelast}) is that of the randomness of bin assignments. We then write
\begin{eqnarray}
    P\left(  \sum_{\alpha} I_{\alpha} \geq 2 \lvert \senderrnd, \queryrnd, \receiverrnd, A, \mathcal{C}  \right) &=& 1-P\left(  \sum_{\alpha \in \mathcal{C},  \alpha \neq A, \alpha \leq \kv{\receiverrnd} } I_{\alpha} = 0  \lvert \senderrnd, \queryrnd, \receiverrnd, A, \mathcal{C}  \right) \label{eq:master_first} \\
    &=& 1 - \Pi_{\alpha \in \mathcal{C}, \alpha \neq A,   \alpha \leq \kv{\receiverrnd} } P(I_{\alpha} = 0 | \senderrnd, \queryrnd, \receiverrnd, A, \mathcal{C}) \label{eq:master_second} \\
       &=& 1 - \Pi_{\alpha \in \mathcal{C}, \alpha \neq A,   \alpha \leq \kv{\receiverrnd} } \left( 1 - 2^{-|\kernelspace| \text{RATE}}  \right) 
       \nonumber \\
        &\leq& 1 - \Pi_{\alpha \in \mathcal{C},   \alpha \leq \kv{\receiverrnd} } \left( 1 - 2^{-|\kernelspace| \text{RATE}}  \right) 
        \nonumber \\
       & = & 1 - \left( 1 - 2^{-|\kernelspace| \text{RATE}}  \right)^{ |\{ \alpha \in \mathcal{C},   \alpha \leq \kv{\receiverrnd}   \}|  } 
       \nonumber \\
           & \leq & 2^{-|\kernelspace| \text{RATE}} |\{ \alpha \in \mathcal{C},   \alpha \leq \kv{\receiverrnd}   \}|  \nonumber \\
           & = & 2^{-|\kernelspace| \text{RATE} + \log_2 |\{ \alpha \in \mathcal{C},   \alpha \leq \kv{\receiverrnd}   \}|} \nonumber  \\
       & = & \frac{1}{|\kernelspace|}. \label{eq:final_conclusion}
\end{eqnarray}
As in the proof of Theorem~\ref{thm:slepian_wolf_sender_unaware}, the derivation above is the essence of the theorem. The most delicate step is the one that leads from (\ref{eq:master_first}) to (\ref{eq:master_second}), where independence of the $\{ I_{\alpha} \} \setminus \{ A \}$ events under the given conditioning is invoked to rewrite the probability of the event as a product of probabilities. The core reason why this independence holds is because under the given conditioning, the bins to which any one $\alpha$ is mapped are chosen independently from each other over the set of possible bins. Since these steps are very similar to those in Theorem~\ref{thm:background_log_info}, we skip most explanations and simply point out that the final expression (\ref{eq:final_conclusion}) comes from the definition (\ref{eq:def_rate}). In summary, we now have that 
\begin{eqnarray*}
P\left(\left[ | \text{hypotheses}(\senderrnd,\queryrnd,\receiverrnd)| \geq 2 \right]\right) &\leq& \frac{1}{|\kernelspace|}.
\end{eqnarray*}
We can now account for all of the bits sent in the protocol. The biggest contribution to the bit rate is that of step \ref{step:master_bin}:
\begin{eqnarray*}
    E\left[ \text{RATE} \right] &\leq& E\left[\frac{1}{|\kernelspace|}\log_2 | \left\{ A \in \text{CODE}(|\kv{\senderrnd}|,|\kv{\queryrnd}|,|\kv{\receiverrnd}|): 
A \leq \kv{\receiverrnd} \right\} |\right] +  \frac{1}{|\kernelspace|} + \frac{\log_2|\kernelspace|}{|\kernelspace|}\\
 &=& E\left[ \left. E \left[\frac{1}{|\kernelspace|}\log_2 | \left\{ A \in \text{CODE}(|\kv{\senderrnd}|,|\kv{\queryrnd}|,|\kv{\receiverrnd}|): 
A \leq \kv{\receiverrnd} \right\} | \right\vert \senderrnd, \queryrnd, \receiverrnd \right] \right] + O\left( \frac{\log_2|\kernelspace|}{|\kernelspace|} \right) \\
 &\leq & E\left[ \frac{1}{|\kernelspace|} \left. \log_2 E \left[ | \left\{ A \in \text{CODE}(|\kv{\senderrnd}|,|\kv{\queryrnd}|,|\kv{\receiverrnd}|): 
A \leq \kv{\receiverrnd} \right\} | \right\vert \senderrnd, \queryrnd, \receiverrnd \right] \right]  +  O\left(\frac{\log_2|\kernelspace|}{|\kernelspace|} \right).
\end{eqnarray*}
We recall that the way $\text{CODE}(|\kv{\senderrnd}|,|\kv{\queryrnd}|,|\kv{\receiverrnd}|)$ is constructed is by choosing uniformly at random 
\begin{eqnarray}   
 \text{code\_size}(|\kv{\senderrnd}|,|\kv{\queryrnd}|,|\kv{\receiverrnd}|)
\label{eq:total_codewords}
\end{eqnarray}
elements from $T(W)$. Next note that
\begin{eqnarray}
  |  \{A : A \leq \kv{\receiverrnd} \} \cap T(W) | =  \left( \begin{array}{c} |\kv{\receiverrnd}| \\ W \end{array} \right).
\label{eq:subset_of_interest}
\end{eqnarray}
Any one element of $T(W)$ has a probability of
\begin{eqnarray}
    \left( \begin{array}{c} |\kernelspace| \\ W \end{array} \right)^{-1}
\label{eq:individual_probability}
\end{eqnarray}
The average number of elements of  $\{A: A \leq \kv{\receiverrnd}\}$ chosen to be part of $\text{CODE}(|\kv{\senderrnd}|,|\kv{\queryrnd}|,|\kv{\receiverrnd}|)$, conditional on $\kv{\senderrnd}, \kv{\queryrnd}, \kv{\receiverrnd}$ is then given by the product of (\ref{eq:total_codewords}), (\ref{eq:subset_of_interest}) and (\ref{eq:individual_probability}):
\begin{eqnarray*}
\lefteqn{  \log_2 E\left[ \left. | \left\{ A \in \text{CODE}(|\kv{\senderrnd}|,|\kv{\queryrnd}|,|\kv{\receiverrnd}|): 
A \leq \kv{\receiverrnd} \right\} |  \right\vert \senderrnd, \queryrnd, \receiverrnd \right]} \\ 
&=& \log_2 \left( \log_2 |\kernelspace |  \left( \begin{array}{c} |\kv{\receiverrnd}| \\ W \end{array} \right) / \left( \begin{array}{c} |\kv{\queryrnd}| - |\kv{\senderrnd}| \\ W - |\kv{\senderrnd}|   \end{array} \right) \right) \\
& \leq &  (|\kv{\senderrnd}| + |\kv{\receiverrnd}| - |\kv{\queryrnd}|) \Hbin{|\kv{\senderrnd}|/(|\kv{\senderrnd}| + |\kv{\receiverrnd}| - |\kv{\queryrnd}|)   } +  \log_2 \log_2 |\kernelspace |   \\
&=& \Lambda( |\kv{\senderrnd}|, |\kv{\receiverrnd}| - |\kv{\queryrnd}| ) +  \log_2 \log_2 |\kernelspace | .
\end{eqnarray*} 
Further taking the expectation over the remaining randomness, using the concavity $\cap$ of the $\Lambda$ function (see Lemma~\ref{lem:elementarylambda}), and normalizing, we obtain the estimate
\begin{eqnarray*}
    E\left[ \text{RATE} \right]   \leq \Lambda( p_s, p_r - p_q ) + O\left( \frac{\log_2 |\kernelspace|}{|\kernelspace|}\right) .
\end{eqnarray*}
The cost of steps \ref{step:master_rec}, \ref{step:master_send}, \ref{step:adaptive_rate} can be verified to be $O(|\kernelspace|^{-1}\log |\kernelspace|)$ by using 
$\delta$ Elias coding. Finally, in the case of failure when trying to find an $A$ meeting (\ref{eq:master_sandwich_proprty}) the cost is $O(|\kernelspace|^{-1})$ given (\ref{eq:master_failure_prob}). \qed

\section{Formal logical underpinnings} \label{sec:logical-underpinnings}

\smallskip

In this section we formally define what we mean by a logic, something that we only introduced informally in the description preceding Definition \ref{def:core} of a Logic System. To make the definition of a logic rigorous, we introduce several additional concepts, including: what a structure is (Definition \ref{def:structure}), what it means for a structure to model a set of logic sentences (Definition \ref{def:modeling}), what it means for a logic sentence to be true (Definition \ref{def:truth}), the notions of a logic being sound and complete (Definitions \ref{def:soundness} and \ref{def:completeness}), and several subtle variations on these latter two notions (Definitions \ref{def:strong-soundness}, \ref{def:strong-completeness}, and \ref{def:omega-strong-completeness}). Two key results of this section are Theorem \ref{thm:fundamental-equivalence}, which shows that condition (\ref{fundamental-rel}) of Definition \ref{def:core} is equivalent to strong soundness and a type of strong completeness of the underlying logic, and Lemma \ref{lem:kernel-rels}, which gives general conditions under which a Logic System can be guaranteed to satisfy conditions \ref{kr1}-\ref{kr3} of Definition \ref{def:proper-logic-system}, and hence be deemed a \emph{proper} Logic System. 

With the aim of developing as all-encompassing a notion of what constitutes a logic as possible, we will establish a set of definitions that set us up well for First-Order Logic and logics that ``extend'' First-Order Logic, such as second and higher-order logics \cite{sep-logic-higher-order}\footnote{Logics that we will, however, not consider formally, other than to say that they extend First-Order Logic in the sense of Definition \ref{def:logic_extension}.}. Towards the end of our development (Subsection~\ref{ss:pl}), we will see that one can fit Propositional Logic and other quantifier-free logics into this model as well. Among other things, at the end of this section we will be able to conclude that all the results of this paper apply to Propositional Logic on a fixed number of variables  as well as the First-Order Logic of structures of fixed finite sizes.

\subsection{The notion of a Logic} \label{ss:logic}
By a \emph{logic}, $L = (\tau, \mathbb{S}, \mathcal{T}, \sigma, \mathcal{P})$, we mean (i) a vocabulary, or set of symbols, $\tau$, (ii) a syntax, $\mathbb{S}$, or set of rules for combining the elements of $\tau$ together to form ``well-formed sentences'', (iii) a definition of truth, $\mathcal{T}$, or set of rules describing how truth propagates through logical symbols, such as $\land, \lor, \lnot, \forall$ and $\exists$, that are part of the vocabulary, (iv) a set of axioms, $\sigma$ (with, possibly, $\sigma = \varnothing$), or sentences that are assumed to be true without proof, and finally (v) a proof system, $\mathcal{P}$, comprising a set of rules of inference, for establishing when certain well-formed sentences follow from other well-formed sentences. 
As in the definition of a Logic System, we refer to the set of  well-formed sentences over $\tau$, given the syntax of $L$, by $\surfacespace$. We denote the entailment operator associated with the proof system $\mathcal{P}$ by $\entails$ (or by $\entails_\mathcal{P}$ if the associated proof system is not entirely clear). The meta-mathematical expression $\mathtt{s} \entails \mathtt{t}$ means that the sentence $\mathtt{t}$ can be proved from $\sigma \cup \{\mathtt{s}\}$ using $\mathcal{P}$. The symbol $\entails$ is not part of any logical vocabulary and hence the expression $\mathtt{s} \entails \mathtt{t}$ is not a sentence of any logical language -- hence our assertion that the expression $\mathtt{s} \entails \mathtt{t}$ is a ``meta-mathematical expression''. As noted in Subsection~\ref{subsec:logic_systems}, we can apply the entailment operator, $\entails$, to sets of sentences as well as to individual sentences.

We described the vocabulary and syntax of classical Propositional Logic in Subsection~\ref{subsec:logic_systems}.
A second important logic that we will consider is First-Order Logic. Although the vocabulary of First-Order Logic includes the logical connectives and grouping parentheses of Propositional Logic, the vocabularies of these two logics are otherwise quite distinct. The vocabulary of First-Order Logic does \emph{not} include propositional variables, but adds the universal and existential quantifiers, $\forall$ and $\exists$, as well as a countable number of variables, $x_i$, that can be associated with the quantifiers. The vocabulary of First-Order Logic also includes function and relation symbols of different arities, $f^{a_i}_i$ and $R^{\alpha_i}_i$, as well as constant symbols, $c_i$. As is customary, we will always assume that there is a distinguished binary equality relation symbol ${=}(\cdot,\cdot)$, and per convention, write $x = y$ for ${=}(x,y)$ and $ x \neq y$ for ${\lnot}{=}(x,y)$. A vocabulary that includes the logical connectives of Propositional Logic, parentheses, a countable number of variable symbols, the quantifiers $\forall$ and $\exists$, as well as a particular set of function, relation and constant symbols (the relations symbols necessarily including ${=}(\cdot,\cdot)$), is called a \textbf{First-Order vocabulary}.  When it is understood that a vocabulary, $\tau$, is First-Order, one typically writes $\tau = (\{f_i^{a_i}\},\{R_i^{\alpha_i}\}, \{c_i\})$, specifying just the non-logical symbols that distinguish $\tau$ from  other First-Order vocabularies. One of the simplest First-Order vocabularies is the vocabulary of directed graphs. In addition to the built-in equality symbol, this vocabulary consists of a single binary relation symbol, $E(\cdot,\cdot)$. In this case we would write $\tau = (\varnothing, \{E(\cdot,\cdot)\}, \varnothing)$, or, equivalently, $\tau = (\varnothing, \{E^2\}, \varnothing)$. In some cases when we work with graphs, we may be considering, say, shortest path, or connectivity questions between specified nodes. In this case it may be convenient to expand the vocabulary to include constants, which are typically denoted $s$ and $t$ (for ``source'' and ``target''). The vocabulary would then be denoted $\tau = (\varnothing, \{E(\cdot,\cdot)\}, \{s,t\})$.

The concept of a ``variable'' in First-Order Logic is completely different from the concept of a variable in Propositional Logic, as is illustrated by their vastly different syntax, or rules for sentence formation. To understand the syntax of First-Order Logic, we build up a set of definitions.

\begin{definition} \label{term}
A \textbf{term} is defined inductively as follows:
\begin{itemize}
    \item Each variable or constant symbol is a term;
    \item If $f^m$ is an $m$-ary function symbol, and $t_1,\ldots,t_m$ are terms, then $f^m(t_1,\ldots,t_m)$ is also a term.
\end{itemize}
\end{definition}

\begin{definition} \label{def:primitive-fol}
A \textbf{primitive formula}\footnote{Sometimes referred to as an atomic formula.} in First-Order Logic is a string of symbols of the form $R^m(t_1,\ldots,t_m)$ where $R^m$ is an $m$-ary relation symbol (possibly the binary equality symbol ${=}(t_1, t_2)$) and $t_1,\ldots,t_m$ are terms.  Analogously, we refer to a primitive formula that does not contain variables as a \textbf{primitive sentence}. 
\end{definition}

\begin{definition} \label{formula}
A string of symbols is a \textbf{formula} of First-Order Logic if it is either a primitive formula or can be constructed from primitive formulas by repeated application of the following rules:
\begin{enumerate}
    \item [R1.]\label{f-r1} If $\phi$ is a formula, then so is $\lnot \phi$; 
    \item[R2.]\label{f-r2} If $\phi$ and $\psi$ are formulas, then so are $\phi \lor \psi$ and $\phi \land \psi$; 
    \item[R3.]\label{f-r3} If $\phi$ is a formula, then so are $\exists x \phi$ and $\forall x \phi$.
\end{enumerate}
\end{definition}

The same symbol in a First-Order vocabulary can appear in multiple positions within a given formula. We refer to each appearance of the same symbol as an \textbf{occurrence} of the symbol.

\begin{definition} \label{bound-unbound-free}
We recursively define the notion of an occurrence of a variable being either \textbf{free} (equivalently, \textbf{unbound}) or \textbf{bound} to a quantifier within a formula. If an occurrence of $x$ in a formula $\phi$ is free, the variable $x$ is also said to \textbf{occur freely} in $\phi$. In a formula $\phi$ without quantifiers, all occurrences of all variables appearing in $\phi$ are considered to be free.  Then we have the following:
\begin{enumerate}
    \item [R1.]If a variable occurrs freely in $\phi$ then it also occurrs freely in $\lnot \phi$; 
    \item[R2.] If a variable occurrs freely in $\phi$ or $\psi$, then it occurrs freely in $\phi \lor \psi$ and $\phi \land \psi$; 
    \item[R3.] If a variable $x$ occurrs freely in $\phi$, then it is bound in the formula $\exists x\phi$ and in the formula $\forall x \phi$. If the variable $x$ does not occur in $\phi$, we also say that $x$ is bound in the formula $\exists x\phi$ and in the formula $\forall x \phi$.
\end{enumerate}
\end{definition}

Given a logic $L$, with language $\surfacespace$, if $\phi$ is a formula with free variables compatible with the syntax of $L$, with some abuse of notation we shall sometimes write $\phi \in \surfacespace$. A formula can have both free and bound occurrences of the same variable. Suppose the variable $x$ appears freely in both the formulas  $\phi$ and $\psi$. Then there are both free and bound occurrences of $x$ in the formula $\phi \vee \exists x \psi$.

\begin{definition} \label{sentence}
A \textbf{sentence} in First-Order Logic is a formula having no free variables.    
\end{definition}

\begin{remark}
Now that we have the full definitions of what it means for a string of logical symbols to be either a formula or a sentence, we can say that a primitive formula (respectively, primitive sentence) is a formula (respectively, sentence) that contains neither logical connectives nor quantifiers.
\end{remark}

Definitions \ref{term} through \ref{sentence}  provide a specification of the syntax of First-Order Logic. Since it is conceivable for other logics beside First-Order Logic to have this syntax, we call this syntax \textbf{First-Order syntax}. There are of course many equivalent ways to specify the same syntax -- meaning that starting with the same vocabularies one would apply the somewhat different rules and arrive at the same language. We will loosely refer to any such set of rules as First-Order syntax. A \textbf{First-Order language} is a language obtained from a First-Order vocabulary by applying First-Order syntax.

\medskip

Truth in every logic is just defined for the sentences of the logic.
To understand how truth is defined for a logic, we will first need to make an excursion into the branch of logic known as \emph{model theory}. We do so in the next subsection.

\subsection{Models of logic sentences and Truth in Models} \label{ss:models}
\begin{definition} \label{def:structure}
In mathematical logic, a \textbf{structure} $\mu = (\mathcal{U}, \{f^{a_i}_i\}, \{R^{\alpha_i}_i\}, \{c_i\})$ consists of a \emph{non-empty} set $\mathcal{U}$ (sometimes called a ``universe'' or ``universe of discourse''), 
together with collections (possibly empty) of functions $\{f^{a_i}_i\}$ and relations $\{R^{\alpha_i}_i\}$, each of finite arity defined on the elements of $\mathcal{U}$ (each function $f^{a_i}_i$ also having its image in $\mathcal{U}$ so that $f^{a_i}_i: \mathcal{U}^{a_i} \rightarrow \mathcal{U}$, and each relation $R^{\alpha_i}_i$ defining a subset of $\mathcal{U}^{\alpha_i}$, with $R^{\alpha_i}_i()$ returning True for a given $\alpha_i$-tuple of elements of $\mathcal{U}$ if the particular $\alpha_i$-tuple is in the given subset, and returning False otherwise), and again, optionally, a collection of constants $\{c_i\}$, which may be thought of as functions of $0$-arity picking out individual elements of $\mathcal{U}$. When we talk about the ``elements'' of the structure $\mu$, we mean the elements of the set $\mathcal{U}$. 
\end{definition} 

In First-Order Logic every structure is assumed to include an equality relation which is associated with the equality symbol, ${=}(\cdot,\cdot)$, and is required to behave in the accustomed manner so that two elements of the structure are equal iff they are the same element. 
An example of a structure in First-Order Logic is a directed graph viewed as a set of nodes together with (the equality relation and) a single binary relation, $E(\cdot, \cdot)$, defined on the nodes, which is true iff there is a directed edge going from the first node to the second node. A graph may also have constants defined, such as a specially designated ``source'' and ``terminal'' nodes, typically denoted $s$ and $t$, respectively.  We say that a structure ``interprets'' the function, relation and constant symbols of a given logical vocabulary if it contains functions and relations of the same arity as each of the function and relation symbols, and has elements, that we call constants, that it associates with each of the constant symbols in the vocabulary. 

\begin{definition} \label{def:tau-struc}
Given a logical vocabulary $\tau$ that includes some number of function, relation, and constant symbols, by a \textbf{$\tau$-structure} one means a structure that interprets each of the function, relation and constant symbols in $\tau$ via concrete functions and relations of the same arity and constants. 
\end{definition}

\medskip

We are now going to take up the subject of defining \emph{truth} in structures, which will become the basis of what we mean, more generally, by \emph{truth} in a logic. Recall that every logic must contain a definition of truth, $\mathcal{T}$, which comprises a set of rules that describe how truth propagates through the logical symbols, such as $\lor, \land, \lnot, \exists, \forall$, in its vocabulary. It will take us some time to develop this subject, but when a logic sentence, $\mathtt{s}$, is found to be true in a given structure, $\mu$, we shall designate this fact with the symbology $\mu \models \mathtt{s}$. If the sentence $\mathtt{s}$ is \emph{not} true in $\mu$, we will instead write $\mu \not \models \mathtt{s}$.

In First-Order Logic with some vocabulary $\tau$, consider the terms without free variables -- such terms are called \textbf{ground terms}. Note that if $\tau$ has no constant symbols then there are no ground terms.  If $\tau$ \emph{does} contain constant symbols, then in any structure $\mu$ that interprets the function, relation and constant symbols of $\tau$, the ground terms resolve to a specific element of $\mu$. If we then consider an $m$-ary relation symbol\footnote{Note that there is at least one such relation symbol, since we assume ${=}(\cdot,\cdot)$ is part of every First-Order vocabulary.} $R^m \in \tau$, and the expression $R^m(t_1,\ldots,t_m)$ where $t_1,\ldots,t_m$ are all ground terms, then either  $R^m(t_1,\ldots,t_m)$ or $\neg R^m(t_1,\ldots,t_m)$ holds in $\mu$, and we analogously write, $\mu \models R^m(t_1,\ldots,t_m)$ or $\mu \not \models R^m(t_1,\ldots,t_m)$ in these two cases.

\begin{definition} \label{def:logic_extension}
 We say that one logic, $L' = (\tau', \mathbb{S}', \mathcal{T}', \sigma', \mathcal{P}')$ with associated language $\surfacespace'$, \textbf{extends} another logic, $L = (\tau, \mathbb{S}, \mathcal{T}, \sigma, \mathcal{P})$ with associated language $\surfacespace$, if (i) $\tau \subseteq \tau', \surfacespace \subseteq \surfacespace', \sigma \subseteq \sigma'$, (ii) the rules for truth preservation through the common logical symbols of $\tau$ and $\tau'$ (in other words, through the logical symbols of $\tau$) are the same, and (iii) every sentence $\mathtt{s} \in \surfacespace$ that is provable in $L$ is provable in $L'$, in other words, for every $\mathtt{s} \in \surfacespace, \entails_\mathcal{P} \mathtt{s}$ implies $\entails_{\mathcal{P}'}\mathtt{s}$. The logic $L'$ is then said to be an \textbf{extension} of the logic $L$. 
\end{definition}

The First-Order Logic of directed graphs with vocabulary $\tau' = (\varnothing, \{E(\cdot,\cdot)\}, \{s,t\})$ thereby extends the First-Order Logic of directed graphs with vocabulary $\tau = (\varnothing, \{E(\cdot,\cdot)\}, \varnothing)$. Further, the First-Order Logic of Undirected Graphs can be viewed as an extension of the First-Order Logic of Directed Graphs, if in both logics we use the same edge relation symbol $E(\cdot,\cdot)$ and to the First-Order Logic of Undirected Graphs we add the single axiom
\begin{equation}
    \forall x\forall y(E(x,y) \implies E(y,x)). \label{eqn:undir-graph}
\end{equation}

\medskip

Let us return now to the question of defining truth in a given logic. The following definition, though somewhat refined since its original conception in 1935, is credited to the Polish logician Alfred Tarski and often referred to as Tarski's \emph{theory of truth} \cite{tarski-conception-of-truth1935, enwiki:1233551294}. It is a definition of truth that is based on the notion of truth in structures. 
The rules for truth propagation through the logical symbols $\lor, \land$ and $\lnot$ are very simple and require little explanation, but things get a bit more complicated when we get to the rules for truth propagation through the quantifiers, $\forall, \exists$ of First-Order Logic. To completely prescribe these rules, we need a couple of definitions.

\begin{definition} \label{def:var_assignment}
Given a logic $L = (\tau, \mathbb{S}, \mathcal{T}, \sigma, \mathcal{P})$, and $\tau$-structure $\mu$ with universe $\mathcal{U}$, denote the set of variables in $\tau$ by $\overline{X} = \{x_1,\ldots\}$. Then a \textbf{variable assignment} is a mapping $\alpha: \overline{X} \rightarrow \mathcal{U}$, taking each variable to a specific element of the universe, $\mathcal{U}$.  
\end{definition}

\begin{definition} \label{def:faf}
Suppose we have a formula $\phi$ with free variables $x_1,\ldots,x_k$, and a structure $\mu$ that interprets each of the function, relation and constant symbols in $\phi$, and let $a_1,\ldots,a_k$ denote elements of the universe of $\mu$. Then we write $\phi(x_1,\ldots,x_k)[x_1/a_1,\ldots,x_k/a_k]$, or, alternatively, $\phi(x_1,\ldots,x_k)[x_1/a_1],\ldots,[x_k/a_k]$, to denote the mapping of each of the freely occurring variables $x_i$ to corresponding elements $a_i$ under a particular variable assignment $\alpha$, such that $\alpha(x_i) = a_i$. We refer to $\phi(x_1,\ldots,x_k)[x_1/a_1,\ldots,x_k/a_k]$ in this case as a \textbf{fully assigned formula}. 
\end{definition}

It is important to note that in Definition \ref{def:faf} the elements of the structure $\mu$ that we have designated by $a_1,\ldots,a_k$ are \emph{not} part of any logical language (and, importantly, not constants) and hence a fully assigned formula is \emph{not} a sentence.  However, due to their resemblance to sentences, we shall typically use Roman lettering and the $\mathtt{typewriter~font}$ to denote fully assigned formulas, e.g., we will denote, say $\mathtt{r} = \phi(x_1,\ldots,x_k)[x_1/a_1,\ldots,x_k/a_k]$. For succinctness, we sometimes will write $\phi[x_1/a_1,\ldots,x_k/a_k]$ in lieu of $\phi(x_1,\ldots,x_k)[x_1/a_1,\ldots,x_k/a_k]$. Further, as long as $\phi(x_1,\ldots,x_k) \in \surfacespace$ we will say, with analogous abuse of notation, that $\mathtt{r} \in \surfacespace$. 
We will also have the need to consider formulas, where all of the variables but one are assigned, and the remaining variable is free. We denote such a formula using notation $\phi(x)[x_1/a_1,\ldots,x_k/a_k]$, with $x$ denoting the single unassigned free variable.
 
Note that given a term $t$ and a structure $\mu = (\mathcal{U}, \{f^{a_i}_i\}, \{R^{\alpha_i}_i\}, \{c_i\})$ that interprets each of the function and constant symbols appearing in $t$ (recall that terms do \emph{not} contain relation symbols), if we replace
each variable appearing in $t$ by an element of $\mathcal{U}$, the result evaluates to an element of $\mathcal{U}$. Then, given an $\mathtt{m}$-ary relation $R^\mathtt{m}$ and terms $t_1,\ldots,t_m$, whose collective set of free variables is $\{x_{i_1},\ldots,x_{i_k}\}$, we say that $\mu \models R^\mathtt{m}(t_1,\ldots,t_m)[x_{i_1}/a_1,\ldots,x_{i_k}/a_k]$, if when we replace each variable $x_{i_j}$ by the respective element $a_j$, the $\mathtt{m}$-tuple of elements of $\mathcal{U}$ given by $(t_1[x_{i_1}/a_1,\ldots,x_{i_k}/a_k],\ldots,t_m[x_{i_1}/a_1,\ldots,x_{i_k}/a_k])$ is an element of $R^\mathtt{m}$ (in other words, $R^\mathtt{m}$ evaluates to True on this tuple). Combining definitions \ref{def:primitive-fol} and \ref{def:faf} we say that $R^\mathtt{m}(t_1,\ldots,t_m)[x_{i_1}/a_1,\ldots,x_{i_k}/a_k]$ is a \textbf{primitive fully assigned formula}. 

\smallskip

Our goal will be to define truth in a given structure for all fully assigned formulas and, in so doing, define truth for all sentences, starting with the primitive fully assigned formulas.  Just like we did for sentences, we use the notation $\mu \models 
\mathtt{r}$ to denote the fact that the fully assigned formula $\mathtt{r}$ is true in $\mu$, or instead write $\mu \not \models \mathtt{r}$, if $\mathtt{r}$ is not true (equivalently, is false) in $\mu$. Although we have defined what we mean by the individual fully assigned formulas, $\mathtt{r} = \phi[x_{i_1}/a_{i_1},\ldots,x_{i_k}/a_{i_k}]$ and $\mathtt{s} = \psi[x_{j_1}/a_{j_1},\ldots,x_{j_\ell}/a_{j_\ell}]$, we will also need to define what we mean by expressions like $\lnot \mathtt{r}, \mathtt{r} \lor \mathtt{s}$ and $\mathtt{r} \land \mathtt{s}$. The extension to the logical not operator, $\lnot \mathtt{r} = \lnot \phi[x_{i_1}/a_{i_1},\ldots,x_{i_k}/a_{i_k}]$ is obvious. However, for $\mathtt{r} \lor \mathtt{s}$ if we were to write $(\phi \lor \psi)[x_{i_1}/a_{i_1},\ldots,x_{i_k}/a_{i_k}, x_{j_1}/a_{j_1},\ldots,x_{j_\ell}/a_{j_\ell}]$, this expression could be ambiguous if some of the variables used in $\phi$, namely the $x_{i_1},\ldots,x_{i_k}$ overlap with some of the variables used in $\psi$, namely the $x_{j_1},\ldots,x_{j_\ell}$, but the associated assigned values of $\mathcal{U}$ are different. Thus, to express $\mathtt{r} \lor \mathtt{s}$ as a fully assigned formula, we sequentially change each variable in $\psi$ that also appears in $\phi$ to a not yet used variable in \emph{either} $\phi$ or $\psi$ so that at the end we get a new expression with different variables. Let us designate by $\psi' = \psi'(x'_{j_1},\ldots,x'_{j_\ell})$ the formula that is otherwise identical to $\psi$, but  where the new set of variables, $\{x'_{j_1},\ldots,x'_{j_\ell}\}$, is disjoint from the set of variables in $\phi$. We then take $\mathtt{r} \lor \mathtt{s}$ to be the fully assigned (and unambiguous) formula $(\phi \lor \psi')[x_{i_1}/a_{i_1},\ldots,x_{i_k}/a_{i_k}, x'_{j_1}/a_{j_1},\ldots,x'_{j_\ell}/a_{j_\ell}]$. The fully assigned formula $\mathtt{r} \land \mathtt{s}$ is defined analogously.

\begin{definition} \label{def:truth-preservation} 
\textbf{Rules of Truth Preservation in Structures for First-Order Logic.}  
Suppose we are given a logic $L = (\tau, \mathbb{S}, \mathcal{T}, \sigma, \mathcal{P})$ that extends First-Order Logic, together with an associated logical language $\surfacespace$. Let $\mu$ be a $\tau$-structure and let $\mathtt{r}$ and $\mathtt{s}$ be two fully assigned formulas of $\surfacespace$. Then we have:

\begin{enumerate}[label=\roman*.]
    \item If $\mathtt{r}$ is a primitive fully assigned formula that holds in $\mu$, then $\mu \models \mathtt{r}$;  \label{tp1}
    \item $\mu \models \neg \mathtt{r}$ iff $\mu \not \models \mathtt{r}$; \label{tp2}
    \item $\mu \models \mathtt{r} \lor \mathtt{s}$ iff $\mu \models \mathtt{r}$ or $\mu \models \mathtt{s}$; \label{tp3}
    \item $\mu \models \mathtt{r} \land \mathtt{s}$ iff $\mu \models \mathtt{r}$ and $\mu \models \mathtt{s}$; \label{tp4}
    \item If $\mathtt{r} = \phi[x_{i_1}/a_1,\ldots,x_{i_k}/a_k], \mathtt{s} = \phi'[x'_{i_1}/a_1,\ldots,x'_{i_k}/a_k]$ and the formulas $\phi$ and $\phi'$ are identical up to a renaming of their variables, then $\mu \models \mathtt{r}$ iff $\mu \models \mathtt{s}$. \label{tp5}
\end{enumerate}
\forceindent For every formula $\phi(x) = \phi(x)[x_{i_1}/a_1,\ldots,x_{i_k}/a_k]$ with a single unassigned free variable $x$, we have:
\begin{enumerate}[label=\roman*.]
    \setcounter{enumi}{5}
    \item $\mu \models \forall x(\phi(x))$ iff for every element $a \in \mu$, $\mu \models \phi(x)[x/a]$; \label{tp6}
    \item $\mu \models \exists x(\phi(x))$ iff for some element $a \in \mu$, $\mu \models \phi(x)[x/a]$. \label{tp7}
\end{enumerate}
\end{definition}
  
By virtue of the sentence construction rules (a.k.a.\ syntax) of First-Order Logic, rules \ref{tp1} through \ref{tp7} are sufficient to define truth for all fully assigned formulas, and hence all sentences, with an arbitrary number of quantifiers -- one simply argues by induction on the quantifier rank\footnote{Also known as quantifier nesting depth.} of the associated fully assigned formula. Suppose, for example, that we have the sentence $\mathtt{s} = \exists x (\forall y E(x,y) \lor \forall y E(y,x))$, from the First-Order Logic of graphs, which is of quantifier rank $2$, and we are trying to determine the truth or falsity of this sentence for a given directed graph $\mu$. Rules \ref{tp1} through \ref{tp7} directly describe how to assign truth to all fully assigned formulas of quantifier rank 1. By rule \ref{tp7}, $\mu \models \mathtt{s}$ iff there is some $a \in \mathcal{U}$, such that $\mu \models (\forall y E(x,y) \lor \forall y E(y,x))[x/a]$. We are now down to the quantifier rank 1 case and rule \ref{tp3} applies, telling us that $\mu \models (\forall y E(x,y) \lor \forall y E(y,x))[x/a]$ iff $\mu \models \forall y E(x,y)[x/a]$ or $\mu \models \forall y E(y,x)[x/a]$. Rule \ref{tp6} now applies to each of these pieces, giving that $\mu \models \forall y E(x,y)[x/a]$ iff for every $b \in \mathcal{U}, \mu \models E(x,y)[x/a,y/b]$ and $\mu \models \forall y E(y,x)[x/a]$ iff for every $b \in \mathcal{U}, \mu \models E(y,x)[x/a,y/b]$.

\begin{definition} \label{def:modeling}
Given a logic $L$ with associated language $\surfacespace$, suppose we are given a set, $\mathcal{S} \subseteq \surfacespace$, of sentences. Then a structure $\md$ is said to be a \textbf{model} of, or for, $\mathcal{S}$ if the function, relation and constant symbols used in $\surfacespace$ are interpreted in $\md$, and each of the sentences of $\mathcal{S}$ are true in $\md$ (equivalently, for every $\mathtt{s} \in \mathcal{S}, \mu \models \mathtt{s}$). The set of sentences $\mathcal{S}$ is also said the be \textbf{modeled} by $\mu$.
\end{definition} 

Consider the First-Order Logic of directed graphs and the sentence:
\begin{equation}
    \exists x(\lnot E(x,x)).  \label{eq:some-non-self-loop}\\
\end{equation}
The logical sentence (\ref{eq:some-non-self-loop}) says that the relation designated by $E(\cdot,\cdot)$ is not reflexive.  A model of this sentence is any directed graph that has a single node without a self-loop (i.e., an edge starting and terminating at the same node). Quantification is over the vertices of the graph and $E(\cdot,\cdot)$ is interpreted as the directed edge relation between vertices.  Consider the subtly different sentence:
\begin{equation}
    \forall x(\lnot E(x,x)).  \label{eq:no-self-loops}\\
\end{equation}
This sentence says that the relation designated by $E(\cdot,\cdot)$ \emph{never} holds for one and the same element. A model of this sentence is any directed graph such that \emph{no} node has a self-loop.

\begin{definition} \label{def:truth}
Given a logic $L = (\tau, \mathbb{S}, \mathcal{T}, \sigma, \mathcal{P})$ with associated language $\surfacespace$, a sentence $\mathtt{s} \in \surfacespace$ is said to be \textbf{true} in $L$ if it is true in all $\tau$-structures that are models of $\sigma$. 
Analogously, the sentence $\mathtt{s} \in \surfacespace$ is said to be \textbf{false} in $L$ if the sentence $\neg \mathtt{s}$ is true in all models of $\sigma$ (equivalently, if $\mathtt{s}$ is false in all models of $\sigma$). 
\end{definition}

\begin{remark} \label{rmk:truth-and-falsity-in-models}
Suppose we are given a logic $L = (\tau, \mathbb{S}, \mathcal{T}, \sigma, \mathcal{P})$  that extends First-Order Logic and its associated language $\surfacespace$. Consider a logic sentence $\mathtt{s} \in \surfacespace$  and a $\tau$-structure $\mu$. By Condition \ref{def:truth-preservation}.\ref{tp2} for Truth Preservation, either $\mathtt{s}$ is true in $\mu$ or false in $\mu$. Though the sentence $\mathtt{s}$ is necessarily either true or false in $\mu$, it may be \emph{neither} true nor false in $L$, since to be true in $L$ it must be true in all $\tau$-structures that are models of $\sigma$, and similarly, to be false in $L$, it must be false in all such $\tau$-structures. Despite the fact that the sentence $\mathtt{s}$ may be neither true nor false in $L$, it is still the case that the sentence $\mathtt{s} \lor \lnot\mathtt{s}$ is true in $L$ since in every $\tau$-structure either $\mathtt{s}$ or $\lnot \mathtt{s}$ is true and hence in every $\tau$-structure, $\mathtt{s} \lor \lnot\mathtt{s}$ holds. 
\end{remark}

To make Remark \ref{rmk:truth-and-falsity-in-models} concrete, consider the First-Order Logic of graphs and let $\mathtt{s}$ be the sentence (\ref{eq:no-self-loops}) that says that there are no self-loops. Given any particular graph $\mu$, either the graph has a vertex with a self-loop or it does not, and thus either $\mathtt{s}$ or $\neg \mathtt{s}$ holds in $\mu$. While neither $\mathtt{s}$ nor $\neg \mathtt{s}$ holds for all graphs, certainly, $\mathtt{s} \lor \lnot\mathtt{s}$ does.

\medskip

\begin{definition}  \label{def:proof-system}\label{def:proof}
Given a logic $L = (\tau, \mathbb{S}, \mathcal{T}, \sigma, \mathcal{P})$, a \textbf{proof system},  $\mathcal{P}$, is a set of rules, known as \emph{rules of inference}, each rule specifying conditions under which, given the truth of certain known sentences, one can conclude the truth of one or more other sentences. Now, let $\surfacespace$ be the language associated with $L$, and let $\mathcal{S} \subseteq \surfacespace$, and $\mathtt{s} \in \surfacespace$. A \textbf{proof} of the sentence $\mathtt{s}$, under the assumption of the truth of the set of sentences $\mathcal{S}$ comprises a finite sequence of steps, including for each step, (i) a rule of inference from the proof system for $L$, (ii) a sentence or set of sentences to which the rule is applied, such sentence(s) either coming from the axioms, the set $\mathcal{S}$, or the conclusion of prior steps, and (iii) a concluded sentence, and where the final concluded sentence is the sentence $\mathtt{s}$. The sentence $\mathtt{s}$ is then said to be \textbf{proved} or \textbf{provable} from $\mathcal{S}$ in $L$, which we denote using the notation $\mathcal{S} \entails \mathtt{s}$. 
\end{definition}

\begin{definition}
Given a logic $L$ with associated language, $\surfacespace$, a \textbf{theorem} of $L$ is a sentence $\mathtt{s} \in \surfacespace$ that is provable without additional assumptions (i.e., where $\mathcal{S} = \varnothing$ in Definition \ref{def:proof}). By the \textbf{theorems of $L$}, one means the set of all such sentences.
\end{definition}

We have defined what we mean by a proof system  but, thus far, have not given an example of one. Our immediate objective is to give such an example for Propositional Logic. Before proceeding to that, let us give a preliminary definition.

\begin{definition}
In Propositional Logic on some number $\mathtt{m}$ of variables, a sentence $\mathtt{s} \in \surfacespace_\mathtt{m}$, is said to be a \textbf{tautology} or a \textbf{propositional tautology} if $\mathtt{s}$ is true for all truth-value assignments to the $\mathtt{m}$ propositional variables $\mathtt{X_1},\ldots,\mathtt{X_m}$.
\end{definition}

A key idea that motivates the upcoming definition, is that if we have a propositional tautology $\mathtt{s} \in \surfacespace_\mathtt{m}$ and now consider another logic $L$, with language $\surfacespace$, that contains the logical connectives of Propositional Logic and propagates truth through these connectives in the same way as Propositional Logic (in other words, in accordance with rules \ref{tp1}-\ref{tp4} of Definition \ref{def:truth-preservation}), then if we replace the propositional variables $\mathtt{X_1},\ldots,\mathtt{X_m}$ by arbitrary sentences $\mathtt{s_1},\ldots,\mathtt{s_m} \in \surfacespace$, we will get a new sentence $\mathtt{s'}$ that is necessarily true in $L$.

There are many so-called \emph{propositional proof systems}, aimed at codifying a set of rules of inference sufficient for proving all the propositional tautologies \cite{buss-intro-to-proof-theory1998}. Often these systems are studied from the vantage point of establishing  the polynomial-time provability of every propositional tautology from a finite set of rules of inference \cite{cook-reckhow1979} -- a fundamental open problem in theoretical computer science. In this work, we are not interested in questions involving polynomial time provability, or of finding a minimum set of rules of inference, and are just concerned with being able to derive all propositional tautologies from an arbitrary finite set of rules of inference. We therefore adopt the following definition.

\begin{definition} \label{def:classical-pps}
A logic $L$, with associated language $\surfacespace$ and a vocabulary that includes the logical symbols $\lor, \land$ and $\lnot$, is said to include a \textbf{classical propositional proof system} if the following conditions hold. For every triple of sentences $\mathtt{s, s_1, s_2} \in \surfacespace$ one has: 
\begin{enumerate}[label=\roman*.]
    \item If $\mathtt{s} \entails \mathtt{s_1}$ then $(\mathtt{s} \entails \mathtt{s_1} \lor \mathtt{s_2}$ and $\mathtt{s} \entails \mathtt{s_2} \lor \mathtt{s_1})$, \label{cpps-1}
    \item $\mathtt{s} \entails \mathtt{s_1} \land \mathtt{s_2}$ iff $(\mathtt{s} \entails \mathtt{s_1}$ and $\mathtt{s} \entails \mathtt{s_2})$, \label{cpps-2}
    \item $(\mathtt{s_1} \entails \mathtt{s}$ and $\mathtt{s_2} \entails \mathtt{s})$ iff $\mathtt{s_1} \lor \mathtt{s_2} \entails \mathtt{s}$, \label{cpps-3}
    \item If $(\mathtt{s_1} \entails \mathtt{s}$ or $\mathtt{s_2} \entails \mathtt{s})$ then $\mathtt{s_1} \land \mathtt{s_2} \entails \mathtt{s}$, \label{cpps-4}
    \item If $(\entails \neg \mathtt{s_1} \lor \mathtt{s_2})$ then $\mathtt{s_1} \entails \mathtt{s_2}$, \label{cpps-5}
    \item $\entails \mathtt{s} \lor \neg \mathtt{s}$. \label{cpps-6}
\end{enumerate}
\end{definition}

It is common to include the symbol $\implies$ in the logical vocabulary of $L$, where upon the condition $\neg \mathtt{s_1} \lor \mathtt{s_2}$ of \ref{def:classical-pps}.\ref{cpps-5} is typically  replaced by $\mathtt{s_1} \implies \mathtt{s_2}$, and known as modus ponens. Condition \ref{def:classical-pps}.\ref{cpps-6} is known as the Law of the Excluded Middle.

With the symbol $\implies$ in the logical vocabulary, a more common propositional proof system is to use the following set of named rules: \emph{And-introduction}: If $\mathtt{s} \entails \mathtt{s_1}$ and $\mathtt{s} \entails \mathtt{s_2}$ then  $\mathtt{s} \entails \mathtt{s_1} \land \mathtt{s_2}$, \emph{And-elimination}: If $\mathtt{s} \entails \mathtt{s_1} \land \mathtt{s_2}$ then $\mathtt{s} \entails \mathtt{s_1}$ and $\mathtt{s} \entails \mathtt{s_2}$, \emph{Or-introduction}: If $\mathtt{s} \entails \mathtt{s_1}$ or $\mathtt{s} \entails \mathtt{s_2}$ then  $\mathtt{s} \entails \mathtt{s_1} \lor \mathtt{s_2}$, \emph{Negation-elimination}: $\lnot\lnot\mathtt{s} \entails \mathtt{s}$, \emph{Modus ponens}: If $\entails \mathtt{s_1} \implies \mathtt{s_2}$ then $\mathtt{s_1} \entails \mathtt{s_2}$ and \emph{Resolution}: If $\entails \mathtt{s} \lor \mathtt{s_1}$ and $\entails \lnot \mathtt{s} \lor \mathtt{s_2}$ then $\entails \mathtt{s_1} \lor \mathtt{s_2}$. Note, however, that \ref{cpps-1}--\ref{cpps-4} give us the And/Or introduction and elimination rules, \ref{cpps-5} is Modus ponens, and \ref{cpps-6} (together with the other rules) allow us to prove Resolution and Negation-elimination.

Although there are logics with proof systems that don't include a classical propositional proof system (e.g., Intuitionistic Logic \cite{Heyting1956-HEYIAI-2}, which does not include \ref{cpps-6}), including a classical propositional proof system may be regarded as the bare minimum requirement of any proof system for doing classical mathematics.

With these definitions in hand, we have the following important notions:

\begin{definition} \label{def:soundness}
A logic $L$, with axioms $\sigma$ and language $\surfacespace$, is said to be \textbf{sound} if all of its theorems are true in all models of $\sigma$. 
\end{definition}

\begin{definition} \label{def:strong-soundness}
Suppose we are given a logic $L$, with axioms $\sigma$ and language $\surfacespace$. The logic $L$ is said to be \textbf{strongly sound} if for every sentence $\mathtt{s} \in \surfacespace$ and every set, $\mathcal{S} \subseteq \surfacespace$, of sentences, if the sentence $\mathtt{s}$ is provable from $\mathcal{S}$, then $\mathtt{s}$ is true in all models of $\sigma \cup \mathcal{S}$.
\end{definition}

\begin{definition} \label{def:completeness}
A logic $L$, with axioms $\sigma$ and language $\surfacespace$, is said to be \textbf{complete} if every sentence that is true in all models of $\sigma$ is a theorem of $L$. 
\end{definition}

\begin{definition} \label{def:strong-completeness}
Suppose we are given a logic $L$, with axioms $\sigma$ and language $\surfacespace$. The logic $L$ is said to be \textbf{strongly complete} if for every sentence $\mathtt{s} \in \surfacespace$ and every set, $\mathcal{S} \subseteq \surfacespace$, of sentences, if $\mathtt{s}$ is true in all models of $\sigma \cup \mathcal{S}$, then $\mathtt{s}$ can be proven from $\mathcal{S}$ in $L$.
\end{definition}

\noindent In the case of completeness, we will actually need a notion that is in between completeness and strong completeness as follows.

\begin{definition} \label{def:omega-strong-completeness}
Suppose we are given a logic $L$, with axioms $\sigma$ and language $\surfacespace$. The logic $L$ is said to be \textbf{$\omega$-strongly complete} if for every sentence $\mathtt{s} \in \surfacespace$ and every \emph{finite} set, $\mathcal{S} \subseteq \surfacespace$, of sentences, if $\mathtt{s}$ is true in all models of $\sigma \cup \mathcal{S}$, then $\mathtt{s}$ can be proven from $\mathcal{S}$ in $L$.
\end{definition}

\begin{definition} \label{def:consistentcy}
Given a logic $L$ with associated language $\surfacespace$, a set $\mathcal{S} \subseteq \surfacespace$ of sentences is said to be \textbf{consistent} if there is a structure $\mu$ in which all of the sentences of $\mathcal{S}$ are true. The set of sentences $\mathcal{S}$ is said to be \textbf{inconsistent} otherwise.
\end{definition} 

\medskip

For the statement of the fundamental equivalence theorem (Theorem \ref{thm:fundamental-equivalence}) that is one of the center-pieces of this subsection, we will need one additional definition.

\begin{definition} \label{def:isomorphic}
Suppose we are given a logical vocabulary $\tau$ and two $\tau$-structures ${\mu_1, \mu_2}$, such that $\mu = \{\mathcal{U}, f^{a_1}_1,\ldots,f^{a_k}_k, R^{\alpha_1}_1,\ldots,R^{\alpha_{\ell}}_{\ell}, c_1,\ldots,c_m\}$ and $\mu' = \{\mathcal{U}', f'^{a_1}_1,\ldots,f'^{a_k}_k, R'^{\alpha_1}_1,\ldots,R'^{\alpha_{\ell}}_{\ell}, c'_1,\ldots,c'_m\}$, where $\mathcal{U}$ and $\mathcal{U}'$ are the underlying universes, $f^{a_i}, R^{\alpha_i}_i(\cdot), c_i$ and $f'^{a_i}_i, R'^{\alpha_i}_i(\cdot), c'_i$ are the function, relation and constants associated with the same function, relation and constant symbols, and $a_i, \alpha_i$ the associated arities of those functions/function symbols and relations/relation symbols. Then $\mu$ and $\mu'$ are said to be \textbf{isomorphic} (equiv. \textbf{$\tau$-isomorphic}) iff there is a function, relation and constant preserving bijection $\pi:\mathcal{U} \rightarrow \mathcal{U}'$. In other words, there is a bijective map $\pi$ such that: 
\begin{enumerate}[label=\roman*.]
    \item For every function symbol $f_i^{a_i} \in \tau$ and every $a_i-$tuple of elements $(e_1,\ldots,e_{a_i})$ from $\mathcal{U}_1$, one has $\pi(f_i(e_1,\ldots,e_{a_i})) = f'_i(\pi(e_1),\ldots,\pi(e_{a_i}))$, 
    \item For every relation symbol $R_i^{\alpha_i} \in \tau$ and every $\alpha_i-$tuple of elements $(e_1,\ldots,e_{\alpha_i})$ from $\mathcal{U}_1$, one has $R_i(e_1,\ldots,e_{\alpha_i})$ iff $R'_i(\pi(e_1),\ldots,\pi(e_{\alpha_i}))$.
    \item For $1 \leq i \leq m, \pi(c_i) = c'_i$.
\end{enumerate}
\end{definition}

\begin{restatable}[fundamental equivalence]{thm}{fund-equiv]} 
Let $L = (\tau, \mathbb{S}, \mathcal{T}, \sigma, \mathcal{P})$ be a logic with associated language $\surfacespace$, and, moreover, such that the vocabulary $\tau$ contains the logical symbol $\land$, the following hold:
\begin{enumerate}
    \item Given sentences $\mathtt{r}, \mathtt{s} \in \surfacespace$, the syntax $\mathbb{S}$ supports the formation of the sentence $\mathtt{r} \land \mathtt{s}$. \label{fe1}
    \item Truth propagates through the symbol $\land$ in accordance with how truth propagates through this symbol in Propositional Logic (Definition \ref{def:truth-preservation}.\ref{tp4}), in other words, for every $\tau$-structure $\mu$, and every pair of sentences $\mathtt{r}, \mathtt{s} \in \surfacespace$, $\mu \models \mathtt{r} \land \mathtt{s}$ iff both $\mu \models \mathtt{r}$ and $\mu \models \mathtt{s}$, \label{fe2}
    \item The proof system of $L$ is sufficient to utilize rules \ref{cpps-2} and \ref{cpps-4} of being a classical Propositional Proof System. In other words, for sentences, $\mathtt{s_1}, \mathtt{s_2}, \mathtt{s} \in \surfacespace$, (i) $\mathtt{s} \entails \mathtt{s_1} \land \mathtt{s_2}$ iff $(\mathtt{s} \entails \mathtt{s_1}$ and $\mathtt{s} \entails \mathtt{s_2})$, and (ii) if $\mathtt{s_1} \entails \mathtt{s}$ or $\mathtt{s_2} \entails \mathtt{s}$, then $\mathtt{s_1} \land \mathtt{s_2} \entails \mathtt{s}$. \label{fe3}
\end{enumerate}
Suppose further that $\kernelspace$ consists of all $\tau$-structures up to the isomorphism.
Let $\kappa: \surfacespace \rightarrow \mathcal{P}(\kernelspace)$ map each sentence to the collection of structures in $\kernelspace$ that satisfy (i.e., model) it as well as satisfy each of the sentences of $\sigma$\footnote{It is possible that $\kernelspace$ is not actually a set but rather a \emph{proper class}, as would be the case if $\kernelspace$ were the set of all directed graphs up to isomorphism. Then $\mathcal{P}(\kernelspace)$ would not be a set either, and is properly referred to as the \emph{power class} of $\kernelspace$, rather than the power set of $\kernelspace$.}. Then, for all $\mathtt{s_1, s_2} \in \surfacespace$, 
\begin{equation} \label{eq:fe1}
    \mathtt{s_1} \entails \mathtt{s_2} \mbox{ implies } \algebraicset{\mathtt{s_1}} \subseteq \algebraicset{\mathtt{s_2}}
\end{equation}
iff $L$ is strongly sound. Further, for all $\mathtt{s_1, s_2} \in \surfacespace$,
\begin{equation} \label{eq:fe2}
     \algebraicset{\mathtt{s_1}} \subseteq \algebraicset{\mathtt{s_2}} \mbox{ implies } \mathtt{s_1} \entails \mathtt{s_2} 
\end{equation}
iff $L$ is $\omega$-strongly complete.
\label{thm:fundamental-equivalence}
\end{restatable}

\textbf{Proof.} Suppose first that $L$ is strongly sound. Then, under the assumption that $\mathtt{s_1} \entails \mathtt{s_2}$, in any structure in which $\{\mathtt{s_1}\} \cup \sigma$ is true, $\mathtt{s_2}$ is true. It follows that $\algebraicset{\mathtt{s_1}} \subseteq \algebraicset{\mathtt{s_2}}$. On the other hand, suppose $L$ is \emph{not} strongly sound. 
Then there is some set of sentences $\mathcal{S}$ from which $L$ can prove a sentence $\mathtt{s}$ even though $\mathtt{s}$ is not true in all models of $\sigma \cup \mathcal{S}$. However, by virtue of the rules for what constitutes a proof in a logic (Definition \ref{def:proof}), in the purported proof of $\mathtt{s}$, only finitely many sentences $\mathtt{s_1},\ldots,\mathtt{s_n} \in \mathcal{S}$ are used. Hence, starting from just the axioms of $L$, we can write $\{\mathtt{s_1} ,\ldots,\mathtt{s_n}\} \entails \mathtt{s}$. Then, since $\mathtt{s_1} \entails \mathtt{s_1}$, by repeated application of condition \ref{fe3}.(ii) in the statement of the lemma, we get that $\mathtt{s_1} \land \cdot\cdot\cdot \land \mathtt{s_n} \entails \mathtt{s_1}$. Since the same argument can be made for each the sentences $\mathtt{s_i}$, it follows that  $\mathtt{s_1} \land \cdot\cdot\cdot \land \mathtt{s_n} \entails \{\mathtt{s_1} ,\ldots,\mathtt{s_n}\}$. Hence, be transitivity of the entailment relation, $\entails$\footnote{The fact that the meta-mathematical relation $\entails$ is transitive follows from how proofs are defined in Definition \ref{def:proof}.}, it follows that $\mathtt{s_1} \land \cdot\cdot\cdot \land \mathtt{s_n} \entails \mathtt{s}$. But $\mathtt{s_1},\ldots, \mathtt{s_n} \in \mathcal{S}$, so the sentences $\mathtt{s_1},\ldots, \mathtt{s_n}$ are true in every model of $\sigma \cup \mathcal{S}$, and so too, by condition \ref{fe2} in the statement of the lemma, $\mathtt{s_1} \land \cdot\cdot\cdot \land \mathtt{s_n}$ is true in every model of $\sigma \cup \mathcal{S}$. On the other hand, $\mathtt{s}$ is \emph{not} true in every model of $\sigma \cup \mathcal{S}$ and so we have $\mathtt{s_1} \land \cdot\cdot\cdot \land \mathtt{s_n} \entails \mathtt{s}$ but $\kappa(\mathtt{s_1} \land \cdot\cdot\cdot \land \mathtt{s_n}) \not \subseteq \kappa(\mathtt{s})$. The first part of the lemma, involving implication (\ref{eq:fe1}) and strong soundness is therefore established.

Next, assume $L$ is $\omega$-strongly complete and suppose $\algebraicset{\mathtt{s_1}} \subseteq \algebraicset{\mathtt{s_2}}$. Then the sentence $\mathtt{s_2}$ is true in every structure in which $\sigma$ and $\mathtt{s_1}$ are true. It follows, by $\omega$-strong completeness, that $\mathtt{s_1} \entails \mathtt{s_2}$. Suppose, on the other hand, that $L$ is \emph{not} $\omega$-strongly complete. There is then some \emph{finite} set of sentences $\mathcal{S} = \{\mathtt{s_1},\ldots,\mathtt{s_n}\} \subseteq \surfacespace$ and a sentence $\mathtt{s}$ such that $\mathtt{s}$ is true in all models of $\sigma \cup \mathcal{S}$ but for which we don't have $\mathcal{S} \entails \mathtt{s}$. Since $\mathcal{S} \entails \mathtt{s_i}$ for $1 \leq i \leq n$, by condition \ref{fe3}.(i) it follows that $\mathcal{S} \entails \mathtt{s_1} \land \cdot\cdot\cdot \land \mathtt{s_n}$. Hence, by the transitivity of $\entails$, it must be that $\mathtt{s_1} \land \cdot\cdot\cdot \land \mathtt{s_n} \not \entails \mathtt{s}$. By condition \ref{fe2} of the lemma, $\kappa(\mathcal{S}) = \kappa(\{\mathtt{s_1},\ldots,\mathtt{s_n}\}) = \kappa(\mathtt{s_1} \land \cdot\cdot\cdot \land\mathtt{s_n}) \subseteq \kappa(\mathtt{s})$. Hence, if $L$ is not $\omega$-strongly complete, condition (\ref{eq:fe2}) does not hold with respect to the pair of sentences $\mathtt{s_1} \land \cdot\cdot\cdot \land\mathtt{s_n}$ and $\mathtt{s}$.
This establishes the second half of the lemma and the overall proof is complete.
\qed

\smallskip

In addition to First-Order Logic, a logic that satisfies the assumptions of Theorem \ref{thm:fundamental-equivalence} is All Positive First-Order Logic, a fragment of First-Order Logic that has no negation symbol and no universal quantifier \cite{kozen_1981}.

\medskip

Returning to our previously established terminology around the word \emph{kernel} (Definition \ref{def:core} from Section \ref{sec:preliminaries}), we see that under the mapping defined in Theorem \ref{thm:fundamental-equivalence}, the kernel of a sentence $\mathtt{s}$, i.e., $\kappa(\mathtt{s})$, is the collection of structures in $\kernelspace$ which are models of $\mathtt{s}$, equivalently, the collection of structures in $\kernelspace$ for which $\mathtt{s}$ is true.

\begin{restatable}[kernel relations]{lem}{kernel-rels]}
Let $\lambda = (L,  \kernelspace, \kappa, \ell)$, be a Logic System with associated vocabulary $\tau$ and language $\surfacespace$, and suppose $\lambda$ satisfies the following conditions:
\begin{enumerate}[label=\roman*.]
    \item The vocabulary $\tau$ contains the logical symbols $\lor, \land$ and $\lnot$, and truth propagates through these symbols in accordance with how truth propagates in classical Propositional Logic (in other words, in accordance with rules \ref{tp1}-\ref{tp4} of Definition \ref{def:truth-preservation}).
    \item The collection $\kernelspace$ consists of an arbitrary collection of $\tau$-structures.
    \item The function $\kappa: \surfacespace \rightarrow \mathcal{P}(\kernelspace)$ maps each sentence, $s \in \surfacespace$, to the collection of all structures in $\kernelspace$ that model it.
\end{enumerate} 
Then for every $\mathtt{s,t} \in \surfacespace$, the following hold:
\begin{enumerate}
\item $\kappa(\mathtt{s} \lor \mathtt{t}) = \kappa(\mathtt{s}) \cup \kappa(\mathtt{t})$, \label{kr1prime}
\item $\kappa(\mathtt{s} \land \mathtt{t}) = \kappa(\mathtt{s}) \cap \kappa(\mathtt{t})$, \label{kr2prime}
\item $\kappa(\lnot \mathtt{s}) = \kappa(\mathtt{s})^C$. \label{kr3prime} 
\end{enumerate}
Thus $\lambda$ is a proper Logic System.
\label{lem:kernel-rels}
\end{restatable} 
\textbf{Proof.} 
By truth preservation rule \ref{def:truth-preservation}.\ref{tp3}, for every $\tau$-structure $\mu$, $\mu \models \mathtt{s} \lor \mathtt{t}$ iff $\mu \models \mathtt{s}$ or $\mu \models \mathtt{t}$. Given the definition of $\kappa$, equality 1.\ immediately follows. Analogously, by truth preservation rule \ref{def:truth-preservation}.\ref{tp4}, for every $tau$-structure $\mu$, $\mu \models \mathtt{s} \land \mathtt{t}$ iff $\mu \models \mathtt{s}$ \emph{and} $\mu \models \mathtt{t}$, so equality 2.\ follows. Finally, by rule \ref{def:truth-preservation}.\ref{tp2}, $\mu \models \lnot \mathtt{s}$ iff $\mu \not \models \mathtt{s}$, which gives equality 3.
\qed

\subsection{Propositional Logic} 
\label{ss:pl}

In this section we show that Propositional Logic on a fixed number, $\mathtt{m}$, of variables, can fit within the same basic framework for being a logic as those logics that extend First-Order Logic, with the exception that Propositional Logic does not admit quantifiers. Fulfilling our promise from Subsection~\ref{subsec:logic_systems}, we will also show that with appropriate choices for the set $\kernelspace$ and the functions $\kappa, \ell$, the Propositional Logic of $\mathtt{m}$ variables can be turned into a proper Logic System.

The fundamental difficulty in trying to fit Propositional Logic into the framework we have elucidated for extensions of First-Order Logic is that there is no analog of the propositional variables $\mathtt{X_1},\ldots,\mathtt{X_m}$ in the vocabularies of logics that extend First-Order Logic, and the accepted terminology of calling the $\mathtt{X_i}$ ``variables'' adds a certain amount of confusion. It is therefore useful to think of the following non-standard vocabulary, $\tau_{\textrm{PL}}$, for Propositional Logic. In addition to the usual logical connectives $\lor, \land$ and $\lnot$, $\tau_{\textrm{PL}}$ includes a unary relation symbol $One(\cdot)$, a binary relation symbol ${<} (\cdot,\cdot)$, and a set of $\mathtt{m}$ constant symbols $c_1,\ldots,c_\mathtt{m}$. There are no variable symbols and no quantifiers. Per convention, we write $c_i < c_j$ in lieu of ${<}(c_i, c_j)$. In addition, the formal proscription of this logic includes the following set of axioms, $\sigma_\textrm{PL}$. For every $i,j$ with $1 \leq i,j \leq \mathtt{m}$,
\begin{eqnarray}
    c_i < c_j~&&\textrm{if i < j}, \label{co1}\\
    \lnot(c_i < c_j)~&&\textrm{otherwise}. \label{co2}
\end{eqnarray}
We may then regard the propositional variable $\mathtt{X_i}$ as shorthand for the expression  $One(c_i)$. The rules of sentence formation are now just the same as in First-Order Logic (Definitions \ref{term} -- \ref{sentence}) with the exception that there is no rule R3 in Definition \ref{formula}. The only terms are the constant symbols, and every formula is a sentence. There are no variables, so Definition \ref{bound-unbound-free} does not apply.  The primitive sentences are the sentences $One(c_i)$ for $\mathtt{1} \leq i \leq \mathtt{m}$ and sentences are just Boolean combinations of these, in other words Boolean combinations of the $\mathtt{X_i}$ -- as we'd expect.

Recall that a logic, $L = (\tau, \mathbb{S}, \mathcal{T}, \sigma, \mathcal{P})$, consists of a vocabulary $\tau$, a syntax $\mathbb{S}$, or rules for sentence formation, rules, $\mathcal{T}$, for truth propagation through the logical symbols of $\tau$, a set of axioms, $\sigma$, and a proof system $\mathcal{P}$. For Propositional Logic we have described $\tau, \mathbb{S}$ and $\sigma$. The rules of $\mathcal{T}$ are just rules \ref{tp1}-\ref{tp4} of Definition \ref{def:truth-preservation}, and the proof system, $\mathcal{P}$, of Propositional Logic is just the classical propositional proof system given in Definition \ref{def:classical-pps}.

The axioms $\sigma_\textrm{PL}$ (the family of sentences (\ref{co1}) and (\ref{co2}), completely determine how the relation ${<}(\cdot,\cdot)$ is defined on the constants, so, with one caveat that we shall get to in a moment, $\tau_\textrm{PL}$-structures -- the structures of Propositional Logic on $\mathtt{m}$ variables -- are completely determined by how the relation $One(\cdot)$ is defined on the $\mathtt{m}$ constants $c_\mathtt{1},\ldots,c_\mathtt{m}$. Any particular definition of $One(\cdot)$ corresponds to a choice of which values on $One(c_i)$ are true, and hence to truth value assignment to the $\mathtt{m}$ propositional variables $\mathtt{X_1},\ldots,\mathtt{X_m}$, in accordance with how we defined the set $\kernelspace$ back in Example \ref{ex:prop_logic} of Subsection~\ref{subsec:logic_systems}.
Although, as we have pointed out, the accepted terminology of calling the expressions $\mathtt{X_i}$ (a.k.a.\ $One(c_i)$) ``propositional variables'' is misleading,  we shall by and large keep to the accepted terminology. Note that a structure $\mu = (\mathcal{U}, \varnothing, \{One(\cdot), {<}(\cdot,\cdot)\}, \{c_1,\ldots,c_\mathtt{m}\})$ can have a universe, $\mathcal{U}$, with more than the $\mathtt{m}$ elements associated with the constants $c_1,..,c_\mathtt{m}$ -- this is the caveat that we alluded to earlier. These elements, however, can have no bearing on the truth or falsity of any sentence, since they cannot be addressed, and we call them \textbf{atoms}. There are no atoms in First-Order Logic. The possibility of atoms requires just one change to the definitions we have assembled in the prior section, namely to the definition of what it means for two structures to be isomorphic ($\tau$-isomorphic) -- Definition \ref{def:isomorphic}. Instead of demanding that the function, relation and constant-preserving map $\pi: \mathcal{U} \rightarrow \mathcal{U}'$ be a bijection, we must require instead that the map $\pi$ be a bijection with respect to the \emph{non-atoms} in $\mathcal{U}$ and $\mathcal{U}'$. 

\medskip

Let us now revert to the more customary way of thinking about Propositional Logic with $\mathtt{m}$ ordered propositional variables $\mathtt{X_1,\ldots,X_m}$ and $\kernelspace = \{\mathtt{False},\mathtt{True}\}^{\mathtt{m}}$ defined to be the set of all truth value assignments to the $\mathtt{m}$ ordered propositional variables. Given a logic sentence $\sender \in \sls{\mathtt{m}}$, we define the kernel function $\kappa:\surfacespace_\mathtt{m} \rightarrow \mathcal{P}(\kernelspace)$ such that  $\algebraicset{\sender}$ is the subset of $\kernelspace$ that makes $\sender$ true. 
For brevity, in what follows, we will generally write $1$ for $\mathtt{True}$ and $0$ for $\mathtt{False}$.  As an illustrative example, suppose the propositional variables are properties of objects in the popular game of twenty questions. For simplicity, let us take $\mathtt{m} = 3$ and let $\mathtt{X_1}$ denote the property `is a country', let $\mathtt{X_2}$ denote the property `is a place', and let $\mathtt{X_3}$ denote the property `has a population of over one million people'. Sample sentences of the associated Propositional language $\surfacespace_3$ are $\mathtt{X_1} \lor \lnot \mathtt{X_2}$ and $\neg \mathtt{X_3}$. One then has $\algebraicset{\mathtt{X_1} \lor \lnot \mathtt{X_2}}= \{000,001,100,101,110,111\}$ and $\algebraicset{\neg\mathtt{X_3}}= \{000,010,100,110\}$.

The following Lemma formalizes the fact that Propositional Logic on a fixed number of variables, with $\kernelspace$ and $\kappa$ defined as above, satisfies the condition (\ref{fundamental-rel}) for being a Logic System (Definition \ref{def:core}).

\begin{restatable}[entailment and kernels]{lem}{imp_ker}
 Suppose $L$ is Propositional Logic on $\mathtt{m}$ variables. Then, given two logic sentences, $\mathtt{a}, \mathtt{b} \in \sls{\mathtt{m}}$, $\mathtt{a} \entails \mathtt{b}$ if and only if $\algebraicset{\mathtt{a}} \subseteq \algebraicset{\mathtt{b}}$.
 \label{lem:pl_implication_kernel}
\end{restatable}

\textbf{Proof.} The lemma follows immediately from Theorem \ref{thm:fundamental-equivalence} by virtue of the fact that $\kernelspace$ contains a model of every consistent sentence in $\surfacespace_\mathtt{m}$ and Propositional Logic is both strongly sound and strongly complete (and therefore $\omega$-strongly complete). See \cite{enwiki:1200422835} for details concerning the soundness and completeness of Propositional Logic.
\qed

\medskip

As an illustration, one can easily check that the kernel subset relation predicted by the Lemma works for the example
\begin{eqnarray*}
(\mathtt{X_1} \lor \lnot \mathtt{X_2})\land \lnot \mathtt{X_3} \entails \mathtt{X_1} \lor \lnot \mathtt{X_2}.
\end{eqnarray*}

Fundamental to the results in this manuscript is the idea that one can recover from a kernel $\kappa(\sender)$ a logic sentence $\senderhat$ that is functionally equivalent to $\kappa(\sender)$, in the sense that $\sender \entails \senderhat$ and $\senderhat \entails \sender$. Since $\kappa(\cdot)$ is a many-to-one function, it doesn't have an inverse, but we will show that a sentence in Disjunctive Normal Form (DNF) can be canonically constructed from the kernel.  Denote by $\mathcal{P}(\{0,1\}^\mathtt{m})$ the collection of all possible sets of $\mathtt{m}$-bit strings -- equivalently, the set of all possible kernels of logic sentences on $\mathtt{m}$ propositional variables. Note that since there are $2^\mathtt{m}$ possible $\mathtt{m}$-bit strings there are $2^{2^\mathtt{m}}$ sets of $\mathtt{m}$-bit strings. 

\begin{definition} [from kernels to logic sentences and back]
We define $\ell: \mathcal{P}(\{0,1\}^\mathtt{m}) \rightarrow \sls{\mathtt{m}}$ to be the function that maps an element $\Mds \in \mathcal{P}(\{0,1\}^\mathtt{m})$ to a canonical DNF sentence that has $\Mds$ as its kernel. Let $\Mds = \{e_\mathtt{1}, e_\mathtt{2}, \ldots, e_\mathtt{k}\}$, with each $e_\mathtt{i} = \langle\epsilon_{\mathtt{i1}},\ldots,\epsilon_{\mathtt{im}}\rangle \in \{0,1\}^\mathtt{m}$. Then to obtain $\ell(\Mds)$, for each $e_\mathtt{i}$ we create a conjunctive clause including $\mathtt{X_j}$ whenever $\epsilon_{ij}=1$ and including $\lnot\mathtt{X_j}$ whenever $\epsilon_{ij}=0$. We then take $\ell(\Mds)$ to be the disjunction of all these conjunctive clauses.
\label{def:ell}
\end{definition}

We generally write an $m$-bit string more compactly as $\epsilon_{\mathtt{i1}}\cdot\cdot\cdot\epsilon_{\mathtt{im}}$ rather than $\langle \epsilon_{\mathtt{i1}},\ldots,\epsilon_{\mathtt{im}}\rangle$, so $101$ rather than $\langle 1,0,1\rangle$. With this more compact notation, an example of the above definition is: $\ell( \{000,100,110\} ) = (\lnot \mathtt{X_0} \land \lnot \mathtt{X_1} \land \lnot \mathtt{X_2}) \lor (\mathtt{X_0} \land \lnot \mathtt{X_1} \land \lnot \mathtt{X_2}) \lor (\mathtt{X_0} \land \mathtt{X_1} \land \lnot \mathtt{X_2})$. 

\begin{restatable}{lem}{fund_rel} 
For every $\Mds \in \mathcal{P}(\{0,1\}^\mathtt{m})$, the functions $\kappa$ and $\ell$ satisfy the relation $\kappa(\ell(\Mds)) = \Mds$. 
\label{lem:pl_fund_rel}
\end{restatable}

\textbf{Proof.} Let $\Mds \in \mathcal{P}(\{0,1\}^\mathtt{m})$ with $\Mds = \{e_\mathtt{1}, e_\mathtt{2}, \ldots, e_\mathtt{k}\}$ and each $e_\mathtt{i} = \langle\epsilon_{\mathtt{i1}},\ldots,\epsilon_{\mathtt{im}}\rangle \in \{0,1\}^\mathtt{m}$. It is clear by construction that $\Mds \subseteq \kappa(\ell(\Mds))$. So let $e \in \{0,1\}^\mathtt{m}$ be such that $e \notin \Mds$. We must show that $e \notin \kappa(\ell(\Mds))$, in other words that $e$ is not satisfied by the DNF sentence given by $\ell(\Mds)$. Since $e \notin \Mds$, $e$ differs from each $e_\mathtt{i} = \langle\epsilon_{\mathtt{i1}},\ldots,\epsilon_{\mathtt{im}}\rangle$ in at least one position. But it then follows that $e$ does not satisfy any of the clauses in $\ell(\Mds)$ and therefore $e$ is not satisfied by $\ell(\Mds)$ -- the disjunction of these clauses. The lemma follows.
\qed

\medskip

Lemmas \ref{lem:pl_implication_kernel} and \ref{lem:pl_fund_rel} enable us to conclude that:

\begin{cor}
Propositional Logic on a fixed set of $m$ variables, where $\kernelspace$ is the set of all truth-value assignments to the $m$ variables, $\kappa$ is the function taking each sentence to the set of truth-value assignments for which the sentence is true, and $\ell$ is the function defined in Definition \ref{def:ell}, is a well defined Logic System.  
\label{cor:pl-is-a-ls}
\end{cor} 

Henceforth, we shall call the Logic System for Propositional Logic on $m$ variables that includes the definitions of $\kernelspace, \kappa$, and $\ell$, as given in the above corollary, the \emph{standard Logic System} for Propositional Logic.  Since $\kernelspace$ and $\kappa$ as defined in Corollary \ref{cor:pl-is-a-ls} satisfy the conditions of Lemma \ref{lem:kernel-rels}, by the conclusion of that lemma, we immediately have:

\begin{cor}
The standard Logic System for Propositional Logic on a fixed set of $m$ variables is a proper Logic System.
\label{cor:pl-is-a-pls}
\end{cor} 

\subsubsection{Propositional Logic: Synopsis of Notation and Terminological Conventions}
\begin{figure} [ht]
\centerline{\scalebox{0.55}{\includegraphics{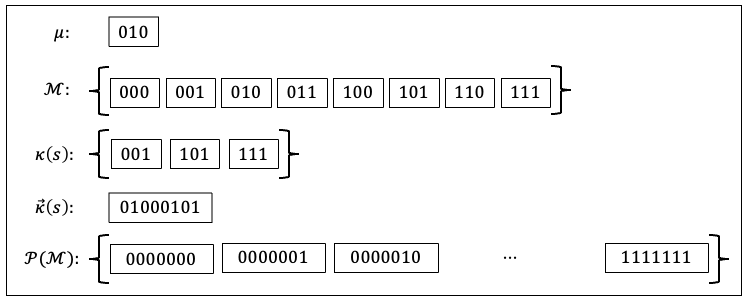}}}
\caption{Pictorial representation of the objects $\mu, \mathcal{M}, \algebraicset{\mathtt{s}}, \vec{\kappa}(\mathtt{s})$ and $\mathcal{P}(\mathcal{M})$ for the case of Propositional Logic on 3 variables. The displayed kernels, $\algebraicset{\mathtt{s}}$ and $\vec{\kappa}(\mathtt{s})$ are for the sentence $\mathtt{s} = (\mathtt{X_1} \lor \lnot \mathtt{X_2}) \land \mathtt{X_3}$.}
\label{fig:PropLogicExamples}
\end{figure}

Suppose we are considering Propositional Logic on $\mathtt{m}$ variables, with vocabulary $\tau$, language $\surfacespace_\mathtt{m}$ and $\mathtt{s} \in \surfacespace_\mathtt{m}$. Let $\lambda = (L,  \kernelspace, \kappa, \ell)$ denote the \textbf{standard Logic System} for this logic. Then we have the following:
\begin{itemize}
    \item $\mu$ = structure = truth-value assignment to the $\mathtt{m}$ propositional variables $\mathtt{X_1},\ldots,\mathtt{X_m}$. May be thought of as a bit-string of length $\mathtt{m}$. More formally, $\mu = (\mathcal{U}, {<}(\cdot, \cdot),One(\cdot), c_\mathtt{1},\ldots,c_\mathtt{m})$ for a particular choice of the unary relation $One(\cdot)$. $\mathtt{X_i}$ is then shorthand for $One(c_i)$.
    \item $\mathcal{M}$ = set of all structures up to isomorphism = set of all truth-value assignments to the $\mathtt{m}$ propositional variables.
    \item $\kappa(\mathtt{s})$ = ``kernel'' of the sentence $\mathtt{s}$ = set of truth value assignments to the $\mathtt{m}$ variables that make the sentence $\mathtt{s}$ true = structures in which $\mathtt{s}$ is true.
    \item $\vec{\kappa}(\mathtt{s})$ = indicator vector representation for the kernel of $\mathtt{s}$. We enumerate the $2^\mathtt{m}$ possible truth-value assignments to the $\mathtt{m}$ variables as binary strings and indicate which assignments make the sentence $\mathtt{s}$ true using a length-$2^\mathtt{m}$ indicator vector.
    \item $\mathcal{P}(\mathcal{M})$ = all possible sets of truth-value assignments = all possible kernels. There are $2^{2^\mathtt{m}}$ such sets. This is the image of the function $\kappa$ in the definition of a Logic System (Definition \ref{def:core}).
\end{itemize}
Figure \ref{fig:PropLogicExamples} depicts the objects described above for the case of Propositional Logic on 3 variables and the sentence $\mathtt{s} = (\mathtt{X_1} \lor \lnot \mathtt{X_2}) \land \mathtt{X_3}$     .

\subsection{First-Order Logic and Models of Bounded Size}
\label{ss:fol}

In this subsection we consider the  First-Order Logic (FOL) of directed graphs on a fixed set of $m$ vertices and show that, by picking $\mathcal{M}, \kappa$ and $\ell$ appropriately, this logic can be turned into a well defined proper Logic System. 
In order to define the inference rules for FOL above and beyond those for Propositional Logic, we need some additional definitions.

\begin{definition}
Suppose we are given a FO language $\surfacespace$ with associated FO vocabulary $\tau$. Let $x$ be a variable in $\tau$, $t$ a term comprised of symbols from $\tau$, and let $\phi \in \surfacespace$ be a formula with a single free variable $x$. Then by $\mathrm{SUBST}(\{x/t\},\phi)$, we mean the result of replacing every \emph{free} occurrence of the variable $x$ in $\phi$ by $t$.    
\end{definition}

Note the slight subtlety in the above definition, that the formula $\phi$ may have bound occurrences of the variable $x$ in addition to the free occurrences. The substitution called out in $\mathrm{SUBST}(\{x/t\},\phi)$ just replaces the free occurrences of $x$ with $t$.

We then have the following: 

\begin{definition} \label{def:fol-proof-system}
Suppose we are given a logic $L = (\tau, \mathbb{S}, \mathcal{T}, \sigma, \mathcal{P})$ such that $\tau$ is a FO vocabulary, $\mathbb{S}$ is FO Syntax, so that the associated language, $\surfacespace$, is a FO language. When we say that $L$ contains a \textbf{proof system for FOL}, we mean that the proof system, $\mathcal{P}$, includes a classical propositional proof system and, in addition, includes the following rules of inference. For every formula $\phi \in \surfacespace$ with one free variable,
\begin{enumerate}[label=\roman*.]
    \item \textbf{Universal Instantiation.} For every ground term $g$, if $\entails \forall x\phi(x)$ then $\entails \mathrm{SUBST}(\{x/g\},\phi(x))$, \label{fop1}
    \item \textbf{Existential Instantiation.} Suppose there is a constant symbol $k \in \tau$ that is not mentioned in any sentence of $\sigma$ or, if this is not the case, augment $\tau$ with an additional constant symbol, $k$, and extend the associated FO language, $\surfacespace$, accordingly. If  $\entails \exists x \phi(x)$, then $\entails \mathrm{SUBST}(\{x/k\},\phi(x))$, \label{fop2} 
    \item \textbf{Universal Generalization.} If, for an arbitrary constant symbol $c \in \tau$ that is not mentioned in any sentence of $\sigma$, $\entails p(c)$ then $\entails \forall x \phi(x)$, \label{fop3}
    \item \textbf{Existential Generalization.} If for some constant symbol $c \in \tau$, and for a sentence $\sigma$ in which $c$ appears, suppose $\entails \sigma$. Then, replacing some (possibly all) occurrences of $c$ by a variable $x$ not appearing in $\sigma$, we have that $\entails \exists x \sigma(x)$. \label{fop4}
\end{enumerate}
\end{definition}

\bigskip

Henceforth, in this section, by a graph we mean a directed graph.  Our logic, $L$, is the FOL of such graphs, $\surfacespace$ is the associated language, having, in addition to the equality relation symbol $=(\cdot,\cdot)$, the single additional relation symbol $E(\cdot,\cdot)$, and there is a single axiom,  $\sigma_\mathtt{m}$, saying that a graph must contain exactly $m$ vertices:

\begin{equation} \label{m-elt-sentence}
    \sigma_\mathtt{m} = \exists \mathtt{x_1} \cdot\cdot\cdot \exists \mathtt{x_\mathtt{m}} \forall \mathtt{y} \bigg( \bigwedge_{1 \leq i\neq  j \leq m}\mathtt{x_i} \neq \mathtt{x_j}~\wedge~\bigvee_{1 \leq i \leq \mathtt{m}}\mathtt{y} = \mathtt{x_i}\bigg). 
\end{equation}
This sentences says, firstly, that there are some $\mathtt{m}$ distinct nodes, and, secondly, that any additional node must be equal to one of the $\mathtt{m}$ distinct nodes. 
The collection, $\kernelspace$, of structures, is the set of all directed graphs on $m$ vertices, and, following Definition \ref{def:fol-proof-system}, $\entails$ is the usual logical entailment in FOL, but for which we can additionally assume the truth of the sentence $\sigma_\mathtt{m}$. We will show that we can define the functions $\kappa$ and $\ell$ in such a way that $\lambda = (L, \kernelspace, \kappa, \ell)$ becomes a proper Logic System.

For the function $\kappa$, and $\mathtt{s} \in \surfacespace$, we define $\kappa(\mathtt{s})$ to be the set $\Mds$ of all $\mathtt{m}$ vertex graphs satisfying $\mathtt{s}$.  For a given graph $G = (V,E) \in \kernelspace$ we next exhibit a sentence $\mathtt{s}_G$ that is true for $G$ but false for every other graph in $\kernelspace$.  For this purpose, let $V = \{v_i\}_{i=1}^\mathtt{m}$. Then write:
\begin{equation} \label{eqn:s_sub_G}
    \mathtt{s}_G = \exists \mathtt{x_1} \cdot\cdot\cdot \exists \mathtt{x_m}\bigg(\bigwedge_{i \neq j} \mathtt{x_i} \neq \mathtt{x_j} ~\land~ \bigwedge_{(v_i, v_j) \in E} E(\mathtt{x_i}, \mathtt{x_j}) ~\land~ \bigwedge_{(v_i, v_j) \notin E} \lnot E(\mathtt{x_i}, \mathtt{x_j})\bigg).
\end{equation}
Now, given $\Mds \subseteq \kernelspace$, let
\begin{equation} \label{eqn:ell_of_mu_folg}
    \ell(\Mds) = \bigvee_{G \in \Mds}\mathtt{s}_G. 
\end{equation}

\begin{restatable}{lem}{fund_rel_folg} 
With $\kappa$ and $\ell$ as defined above, and $\Mds \subseteq \kernelspace$ an arbitrary set of $m$ vertex graphs, the functions $\kappa$ and $\ell$ satisfy the relation $\kappa(\ell(\Mds)) = \Mds$. 
\label{lem:folg_fund_rel}
\end{restatable}

\textbf{Proof.} To prove the lemma we show that the sentence $\ell(\Mds)$, as given by (\ref{eqn:ell_of_mu_folg}), is satisfied by every graph $G \in \Mds$ but by no other $m$-vertex graph. Each sentence $\mathtt{s}_G$, as in (\ref{eqn:s_sub_G}), satisfies some $G \in \Mds$, so their disjunction, $\bigvee_{G \in \Mds}\mathtt{s}_G$, is satisfied by \emph{every} $G \in \Mds$. 
On the other hand, if $G \notin \Mds$ then $G$ satisfies none of the $\mathtt{S}_G$ for $G \in \Mds$, and so $G$ does not satisfy the disjunction of all these sentences, which is $\ell(\Mds)$, and so the lemma is established.
\qed

\begin{restatable}{lem}{imp_ker_folg}
 Let $L$ be the FOL of $\mathtt{m}$-node directed graphs and let $\surfacespace$ be the associated FO language. Given $\mathtt{a}, \mathtt{b} \in \surfacespace$, we have that $\mathtt{a} \entails \mathtt{b}$ if and only if $\algebraicset{\mathtt{a}} \subseteq \algebraicset{\mathtt{b}}$.
 \label{lem:folg_implication_kernel}
\end{restatable}

\textbf{Proof.} A sentence in the First-Order theory of graphs given $\sigma_\mathtt{m}$ is true iff it is true for all $m$-vertex graphs. The rules of inference of FOL above and beyond those of Propositional Logic, are easily seen to preserve truth, so FOL is strongly sound. Strong completeness (and hence $\omega$-strong completeness) follows as a consequence of G\"{o}del's Completeness Theorem \cite{shoenfield_2001}\footnote{Though strong completeness can be established considerably more simply in this case via the method of quantifier elimination. See, for example, \cite{hodges_1997}.}. The result therefore follows by Theorem \ref{thm:fundamental-equivalence}. 
\qed

\begin{restatable}{thm}{fol-on-graphs-isa-pls} 
   Consider $\lambda = (L, \kernelspace, \kappa, \ell)$, where $L$ is the First-Order Logic of Directed Graphs on a fixed number, $m$, of vertices, $\kernelspace$ is the set of all directed graphs on $m$ vertices, $\kappa$ is the function mapping each sentence in $\kernelspace$ to the $m$-vertex directed graphs that satisfy it, and $\ell$ is defined as in (\ref{eqn:s_sub_G}), (\ref{eqn:ell_of_mu_folg}). Then $\lambda$ is a proper Logic System.
\label{thm:fol-on-graphs-isa-pls}
\end{restatable}

\textbf{Proof.}
By lemmas \ref{lem:folg_fund_rel} and \ref{lem:folg_implication_kernel}, $\lambda$ gives rise to a well defined Logic System. By Lemma \ref{lem:kernel-rels}, $\lambda$ is also a proper Logic System.
\qed

\medskip

The same result (Theorem \ref{thm:fol-on-graphs-isa-pls}) holds for undirected graphs on $m$ vertices.  We simply add the sentence (\ref{eqn:undir-graph}) as an additional axiom. Analogous to Propositional Logic, we refer to the Logic System for the First-Order Logic of Directed Graphs described in Theorem \ref{thm:fol-on-graphs-isa-pls} as the \textbf{standard Logic System for Directed Graphs}, and if we add sentence \ref{eqn:undir-graph}, it is then the \textbf{standard Logic System for Undirected Graphs}.

\medskip

We make three further observations: (1) We could just as well have considered graphs of size less than or equal to a fixed $m$, rather than just those of size exactly $m$. (2) There is no set of FO sentences such that if we were to take these sentences as axioms, the set of models would be precisely the set of all finite graphs (a simple consequence of the Compactness Theorem of FOL \cite{shoenfield_2001}). (3) We could consider the generic (non-size-limited) theory of graphs and take $\kernelspace$ to be the set of all finite graphs, but then there would be some subsets of $\kernelspace$ (of countably infinite cardinality) that one could not capture with a sentence from the First-Order language of graphs, and hence we wouldn't be able to establish condition (\ref{pseduo-inverse-rel}) for being a Logic System for such a system. 

\subsubsection{First-Order Logic of Directed Graphs: Synopsis of Notation and Terminological Conventions}

\begin{figure} [ht]
\centerline{\scalebox{0.55}{\includegraphics{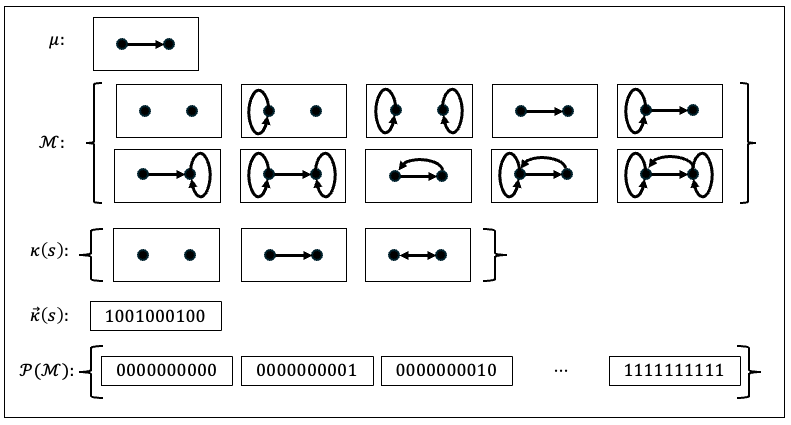}}}
\caption{Pictorial representation of the objects $\mu, \mathcal{M}, \algebraicset{\mathtt{s}}, \vec{\kappa}(\mathtt{s})$ and $\mathcal{P}(\mathcal{M})$ for the case of the First-Order Logic of Directed Graphs on 2 nodes. The displayed kernels, $\algebraicset{\mathtt{s}}$ and $\vec{\kappa}(\mathtt{s})$ are for the sentence $\mathtt{s} = \forall x (\neg E(x,x))$.}
\label{fig:DirGraphExamples}
\end{figure}

Suppose we are considering the FOL of Directed Graphs on $\mathtt{m}$ vertices, with vocabulary $\tau$, language $\surfacespace$ and $\mathtt{s} \in \surfacespace$. Let $\lambda = (L,  \kernelspace, \kappa, \ell)$ denote the \textbf{standard Logic System} for this logic. Then we have the following:
\begin{itemize}
    \item $\mu$ = structure = an $\mathtt{m}$-vertex directed graph.
    \item $\mathcal{M}$ = set of all $\mathtt{m}$-vertex directed graphs.
    \item $\kappa(\mathtt{s})$ = ``kernel'' of the sentence $\mathtt{s}$ = set of all $\mathtt{m}$-vertex directed graphs for which the sentence $\mathtt{s}$ is true.
    \item $\vec{\kappa}(\mathtt{s})$ = indicator vector representation for the kernel of $\mathtt{s}$. We enumerate the  possible directed graphs on the $\mathtt{m}$ vertices and indicate which of the graphs make the sentence $\mathtt{s}$ true using an indicator vector.  
    \item $\mathcal{P}(\mathcal{M})$ = all possible sets of directed $\mathtt{m}$-vertex graphs = all possible kernels. This is the image of the function $\kappa$ in the definition of a Logic System (Definition \ref{def:core}).
\end{itemize}
Figure \ref{fig:DirGraphExamples} depicts the objects described above for the case of the FOL of Directed Graphs on 2 vertices and the sentence $\mathtt{s} = \forall x (\lnot E(x,x))$, which says there are no self-loops.

\section{A linkage with polynomial algebra}
\label{sec:linkage_algebra}

Thus far we have explored the connections between logic and information theory. In this section we explore a connection to a third area of mathematics, which is the algebra of multivariate polynomials with variables and coefficients belonging to a finite field.  The fundamental insight in doing so is that
Propositional 
Logic sentences can be represented using polynomials, which in turn provides us access to powerful mathematical tools such as Gr\"obner bases. We will then use these mathematical tools to establish a fundamental result on the ``optimality of increments''; to paraphrase, synchronizing the knowledge between a receiver and a sender in a communication optimal way can be accomplished by having the receiver add new non-trivial sentences in its knowledge base of logic sentences (without changing its old sentences). 
Hence, our general approach consists of first converting the original logic sentences into polynomials, then exploiting the foregoing mathematical tools to perform reduction and decomposition transformations in the polynomial domain, and finally converting the resulting polynomials back to logic expressions.
The technique of our general approach presented in this portion of the article is largely independent of the rest, and can be read in isolation. It is also a very general technique that has applicability beyond this article. 

Although we speak of a ``logic sentence'' in possession of the sender Alice or the receiver Bob throughout other parts of the article, we will slightly change this perspective throughout the present section. Instead of talking about a single logic sentence, we will talk about collections of logic sentences that are implicitly combined using  conjunction; having explicit smaller logic sentences will significantly help in our exposition herein.
In addition,
we overload our previous notation and 
allow $\senderhat, \receiver$ to represent finite sets of logical expressions.
Thus, we introduce the notation $\mathcal{P}_F(\cdot)$ to be the collection of all finite subsets of $\mathcal{P}(\cdot)$, and thus
$\mathcal{P}_F(\sls{m})$
represents the set of finite subsets of Propositional Logic sentences and we assume that $\senderhat, \receiver \in \mathcal{P}_F(\sls{m})$. The main setup is that we have a collection of logic sentences $\senderhat$ that are being presented to a receiver as a new ``truth'' that should replace whatever the receiver previously had in his mind, which is the collection of logic sentences $\receiver$. It is assumed that $\senderhat \entails \receiver$. The receiver Bob is not keen to completely replace the contents of his mind with the new $\senderhat$, and would rather extract from $\senderhat$ a $\Delta \in \mathcal{P}_F(\sls{m})$ that can be more surgically added to his existing knowledge base $\receiver$ and that will have the effect that $\Delta \cup \receiver$ is logically equivalent to $\senderhat$ in the sense that the same logic sentences can be proved from either. The receiver would want $\Delta$ to have some kind of ``minimality'' or ``orthogonality'' with respect to $\receiver$, in particular he does not want any sentence in $\Delta$ to be provable using what he already knew ($\receiver$).
Moreover, we would like to use the receiver's knowledge to reduce the size of the logic expressions in $\Delta$ when possible. 
For measuring size we can assume that logic sentences are expressed as an exclusive-or ($\oplus$) of a collection of conjunctions of logical variables.
This representation allows us to preserve size when we convert between logic sentences and polynomials, since the logic operation $\oplus$ corresponds to the polynomial operation $+$ and logical conjunction $(\land)$ corresponds to polynomial multiplication $(\cdot)$.

The ultimate theoretical result for this section in the context of the above setup is our post-processing result with respect to incremental communications at the level of logic expressions, which we present next where the size of a logic expression is its length in characters using the ($\oplus$, $\land$) representation.

\begin{restatable}[Post-processing at logic level to obtain incremental communications]{thm}{postproc}
    There exists a function
\begin{equation*}
    \Delta : \mathcal{P}_F(\sls{m})^2 \rightarrow \mathcal{P}_F(\sls{m})
\end{equation*}
such that, for any $\senderhat,\receiver,\mathtt{w} \in \mathcal{P}_F(\sls{m})$, the function satisfies the property that, if $\receiver \entails \mathtt{w}$, then $\mathtt{w} \notin \Delta(\senderhat,\receiver)$ and $\Delta(\senderhat,\receiver) \cup \receiver \entails \senderhat$ and $\senderhat \entails \Delta(\senderhat,\receiver) \cup \receiver$.  Furthermore, the size of every logic sentence in
$\Delta(\senderhat,\receiver)$ is bounded by $O(m |\kappa(\receiver)|)$.
\label{thm:postproc}
\end{restatable}

\begin{example}[Illustration of Theorem~\ref{thm:postproc}]\label{postproc-ex} 
Consider
a situation where we have three logical variables $\mathtt p$, $\mathtt q$ and $\mathtt r$. 
Assume the sender wants to send $\{ \mathtt p , \mathtt q , \mathtt r \}$, but the receiver already knows that 
$\mathtt p \implies \mathtt q$. So, taking advantage of the receiver's information, the sender's information can be reduced to $\{ \mathtt p , \mathtt r\}$, i.e., the sender's original sentence is a logical consequence of $\{ \mathtt p , \mathtt r\}$ using the receiver's knowledge.
\end{example}

For the simple illustration of Theorem~\ref{thm:postproc} in the above example, the reduction and decomposition steps are obvious.
In this section, however, we present our more general approach that can always be used to make these transformations through a sequence of converting logic sentences to polynomials, performing reduction and decomposition transformations in the polynomial domain, and converting the resulting polynomials back to logic expressions.
We first present in this context multivariate polynomials with variables and coefficients belonging to a general finite field, which includes defining key mathematical properties and tools, and establishing mathematical results toward the goal of this section with respect to such multivariate polynomials and general finite fields.
We then make connections between these multivariate polynomials over general finite fields to corresponding logic sentences over a specific finite field comprising binary truth values in order to formally prove our main post-processing result in Theorem~\ref{thm:postproc}.
Lastly, we discuss some applications of the incremental communications provided by this theorem, which can be used in settings throughout this article and beyond.

\subsection{Multivariate polynomials over finite fields}
Throughout our article, we have assumed for simplicity that propositions have binary truth value assignments (truth or false), and we will continue to adopt this perspective here.
However, the results and proofs of this subsection will be provided for general finite fields, or Galois fields, denoted by $K$.
The entire section is devoted to setting up our proof of Theorem \ref{thm:postproc} based on related results within the context of multivariate polynomials with coefficients belonging to a general finite field $K$.
We focus in this subsection on these related results, which require us to establish the machinery of algebras of multivariate polynomials over finite fields, and then return to the proof of Theorem \ref{thm:postproc} in the next subsection.
The set of all polynomials over variables $x_1,\ldots,x_m$ with coefficients in a finite field $K$ is denoted by $\Kpolyring$. 

Before proceeding to our proofs of the corresponding mathematical results, we provide definitions of key properties employed in some of the proofs.
For more information on these mathematical properties, we refer the reader to~\cite{CLO:Ideals_Varieties}.

\begin{definition}[Ideal]
 Given polynomials $f_i \in \Kpolyring$ for $i=1,\ldots,n$, the set of all polynomials of the form $\sum_{i=1}^n a_i f_i$ with $a_i \in \Kpolyring$ is called the \emph{ideal} generated by $f_1, \ldots, f_n$ and denoted by $(f_1, \ldots, f_n)$.
\end{definition}

Since we want to restrict our solutions to lie within the general finite field $K$ of cardinality $|K|$, it is assumed throughout this section that these variables satisfy the conditions
\begin{eqnarray}
    x_i^{|K|} - x_i = 0 .
\label{eq:fieldpolynomials}
\end{eqnarray}
These conditions are called the \emph{field polynomials}, and they imply that $x_i \in K$.

We had earlier defined the kernel $\algebraicset{\mathtt{a}}$ in propositional logic to be the function that maps each sentence $\mathtt{a} \in \mathcal{L}$ to the set of
truth-value assignments that make $\mathtt{a}$ true.

With abuse of notation, we define a parallel notion for a polynomial as follows.

\begin{definition}
    For a given $a \in \Kpolyring$, the kernel $\algebraicset{a}$ is defined as
\begin{eqnarray*}
\algebraicset{a} \stackrel{\Delta}{=} \left\{ x_1 x_2 \ldots x_m \in K^m : a(x_1,\ldots,x_m)=0 \right\} .
\end{eqnarray*}
\end{definition}

Given an arbitrary subset of $K^m$, or equivalently, an element of $\mathcal{P}(K^m)$, it is possible to construct a polynomial whose kernel coincides exactly with that subset as is shown in the following result, after presenting a preliminary lemma used in its proof.
For consistency with the rest of the paper, when operating on finite sets such as $K^m$, we will use the notation $\mathcal{P}$ instead of $\mathcal{P}_F$.
With abuse of notation, we use $\ell$ defined earlier for logical expressions to apply herein for polynomials.

\begin{restatable}[Product of finite field elements]{lem}{fin_field_prod}
~The product of all nonzero elements in any finite field is -1. 
\label{lem:find_field_prod}
\end{restatable}

\textbf{Proof.}
First note that the only elements such that $x x = 1$ are 1 and -1.
Thus each element different from 1 and -1 can be paired with its inverse in the product and cancels. In finite fields with an odd number of elements 1 and -1 are different, and so the reduced product is $1 (-1) = -1$. In finite fields with an even number of elements $1 = -1$ and the result follows. 
\qed

\begin{restatable}[Reconstruction of a polynomial from a proposed kernel]{lem}{recon}
~There exists a function $\ell : \mathcal{P}(K^m) \rightarrow \Kpolyring$ such that, for every set $\psi \subseteq K^m$, $\algebraicset{\ell(\psi)}=\psi$. 

\label{lem:recon}
\end{restatable}

\textbf{Proof.}
Since $K$ is a finite field,  $\psi \subseteq K^m$ is a finite set.
We first observe that, given a point $p\in K$, there exists a polynomial $I_p \in  K[x]$ such that $I_p(p) = 1$ and $I_p(q) = 0$ for all $q \in K$ where $q \neq p$. 
Recall from~\eqref{eq:fieldpolynomials} that $x^{|K|}-x$ is a field polynomial whose roots are all the elements of $K$.
Then $(x^{|K|}-x)/(x-p)$ is a polynomial that vanishes at all points other than $p$, but takes the value $-1$ at $p$, since its value at $p$ is just the product of all nonzero elements of the finite field and Lemma~\ref{lem:find_field_prod} applies.
We define $I_p \stackrel{\Delta}{=} - (x^{|K|}-x)/(x-p)$. Note that when $|K|=2$,
in which case $-1 = 1$, this construction simplifies to $I_p = x + p + 1$.
Now let $c = (c_1, \ldots, c_m) \in \psi$; then the polynomial $P_c = \prod_{i=1}^m I_{c_i}(x_i)$ takes the value 1 at the point  $c$, and takes the value 0 at all other points of $K^m$. 
Thus the polynomial $\ell(\psi) = -1 + \sum_{c\in \psi} P_c $ takes the value zero at each point of $\psi$ and is nonzero at all other points of $K^m$. 
\qed

Next, we want to be able to reduce a polynomial modulo an ideal representing known information, for which we need a notion of division with remainder for multivariate polynomials. To define the notion of a leading term of a polynomial, we require a total ordering on the monomials in $\Kpolyring$ that respects multiplication; i.e., for all monomials $m_1$, $m_2$ and $m_3$ with $1 \le m_3$, if $m_1 < m_2$, then $m_3 m_1 < m_3 m_2$. 
Since any monomial can be expressed as $\prod_{i=1}^m {x_i}^{e_i}$, one typical ordering is a lexicographic ordering on the exponents $e_i$.
Any such ordering allows us to define the notion of reduction modulo a set of polynomials, together with the related notion of Gr\"obner bases.

\begin{definition}[Polynomial Reduction]
Given polynomials $p$ and $q$, if some monomial of $q$ is divisible by the leading term of $p$, we can remove that monomial by subtracting a multiple of $p$ from $q$. We can continue this process until no monomials of $q$ are divisible by the leading term of $p$, at which point we say that $q$ is \emph{reduced with respect to} $p$.  Similarly, given a finite collection of polynomials $\hat{p} \stackrel{\Delta}{=} \{ p_1 , \ldots, p_k \}$, we call a polynomial $q$ \emph{reduced with respect to} $\hat{p}$ if no monomial of $q$ is divisible by a leading monomial of some $p_i \in \hat{p}$. Note that this reduced form with respect to a collection of polynomials can depend on the order reductions are done and thus is not necessarily unique.
\end{definition}

\begin{definition}[Gr\"obner Bases]
Among all sets of polynomials $G_I$ that generate an ideal $I$, those which have the property that $p \in I$ if and only if $p$ can be reduced to zero by $G_I$ are called a \emph{Gr\"obner basis} for the ideal $I$. The notion of Gr\"obner Basis depends on the notion of the leading term of a polynomial which depends on the choice of monomial ordering. Note that the reduced form modulo a Gr\"obner basis is uniquely determined.
\end{definition}

Next, we present some key results of ideals which contain all the field polynomials~\eqref{eq:fieldpolynomials}.

\begin{restatable}[Decomposition of field polynomial ideal]{lem}{field-poly-ideal}
Let $I$ be the ideal generated by the field polynomials 
$(x_1^{|K|} - x_1, \ldots, x_m^{|K|}-x_m)$, and let $M_i$ be the collection of ideals of the form $M_i = (x_1 - c_{i,1},\ldots, x_m - c_{i,m})$ as the coordinates $c_{i,j}$ range over all m-tuples of elements of $K$. Then, we have $I = \cap M_i$.
\label{lem:field-poly-ideal}
\end{restatable}
\textbf{Proof.} Recall that $x^{|K|}-x = \prod_{c_i\in K}(x - c_i)$. Since the product of ideals generated by relatively prime polynomials in one variable is the same as their intersection, we have the desired result when $m=1$. The general case can then be shown by induction on the number of variables.
\qed

\begin{restatable}[Kernel property of field polynomial ideal]{lem}{kernel-ideal}
Given $v, w \in \Kpolyring$, let $I$ be the ideal generated by $v$ and the field polynomials~\eqref{eq:fieldpolynomials}. Then $w \in I$ if and only if $\kappa(v) \subseteq \kappa(w)$. 
In particular, if the zeros of $v$ in $K$ have coordinates $c_{i,1},\ldots, c_{i,m}$,
then $I = \cap (x_1 - c_{i,1}, \ldots, x_m - c_{i,m})$.
\label{lem:kernel-ideal}
\end{restatable}
\textbf{Proof.} The second claim is shown by adding $v$ to the ideal decomposition in Lemma~\ref{lem:field-poly-ideal}. The first claim is an application of the second, since $w \in I$ if and only if $w$ is contained in each component ideal of $I$ if and only if $\kappa(v) \subseteq \kappa(w)$.
\qed\\
Exploiting these key properties, we now prove our post-processing result with respect to incremental communications at the level of multivariate polynomials over general finite fields.

\begin{restatable}[Post-processing at polynomial level to obtain incremental communications]{lem}{postproc2} There exists a function
\begin{equation*}
    \Delta : \mathcal{P}_F(\Kpolyring)^2 \rightarrow \mathcal{P}_F(\Kpolyring)
\end{equation*}
such that, for any $u,v,w \in \mathcal{P}_F(\Kpolyring)$, 
the function satisfies the property that, if $\kappa(v) \subseteq \kappa(w)$, then $w \notin \Delta(u,v)$ and  $\algebraicset{ \Delta(u,v) \cup v } = \algebraicset{u \cup v}$. Furthermore, the polynomials in $\Delta(u,v)$ are reduced with respect to a Gr\"obner basis for the ideal generated by $v$ and the field polynomials~\eqref{eq:fieldpolynomials}.
\label{lem:postproc2}
\end{restatable}

\textbf{Proof.}
We seek to remove, from $u$, any polynomials $w$ that satisfy the supposition $\kappa(v) \subseteq \kappa(w)$. First, we observe that $\kappa(v) \subseteq \kappa(w)$ if and only if  $w$ is contained in the ideal $(v,  x_1^{|K|}-x_1, \ldots, x_m^{|K|}-x_m)$ by Lemma~\ref{lem:kernel-ideal}.

Let $G_v$ be a Gr\"obner basis for the ideal $(v,  x_1^{|K|}-x_1, \ldots, x_m^{|K|}-x_m)$,
and define $\Delta(u,v)$ to be the reduction of the elements of $u$ with respect to $G_v$.
The reduction process guarantees that all polynomials $w$ such that $\kappa(v) \subseteq \kappa(w)$ will reduce to zero, and thus they are no longer contained in $\Delta(u,v)$, i.e., $w \notin \Delta(u,v)$. Since all solutions of $v$ lie in $K^m$, a vector space of dimension $m$ over $K$, we obtain $\algebraicset{v} = \algebraicset{v,  x_1^{|K|}-x_1, \ldots, x_m^{|K|}-x_m}$, and therefore $\algebraicset{ u \cup (v,  x_1^{|K|}-x_1, \ldots, x_m^{|K|}-x_m)} = \algebraicset{u \cup v}$.
But $\Delta(u,v)$ and $u$ are the same modulo $(v,  x_1^{|K|}-x_1, \ldots, x_m^{|K|}-x_m)$,
and hence we have
\begin{eqnarray*}
\algebraicset{ \Delta(u,v) \cup v } &=&
\algebraicset{ \Delta(u,v) \cup (v,  x_1^{|K|}-x_1, \ldots, x_m^{|K|}-x_m)  } \\
&=&\algebraicset{ u \cup (v,  x_1^{|K|}-x_1, \ldots, x_m^{|K|}-x_m)  } \\
&=& \algebraicset{ u \cup v }.
\end{eqnarray*} 
By construction, the polynomials in $\Delta(u,v)$ are reduced with respect to the ideal generated by $v$ and the field polynomials.
\qed 

Now, we provide an explicit bound on the number of monomials in a reduced polynomial.

\begin{restatable}[Bound on size of reduced polynomials]{lem}{monom-count} Given $v, w \in \Kpolyring$,
the number of monomials in the reduced representation of $w$ with respect to the ideal generated by $v$ and the field polynomials~\eqref{eq:fieldpolynomials} is bounded by $|\kappa(v)|$.
\end{restatable}

\textbf{Proof.}
Let $I$ be the ideal generated by $v$ and the field polynomials.
The reduction mapping sends $w$ to $\Kpolyring/I$, a finite dimensional vector space over $K$, which we can assume is generated by monomials. Let $z_1, \ldots, z_n$ represent the set of zeros of $v$ in $K$. For each zero $z_i = c_{i,1}, \ldots, c_{i,m}$, let $M_i$ be the ideal generated by $(x_1 - c_{i,1}, \ldots, x_m - c_{i,m})$, and thus by Lemma~\ref{lem:kernel-ideal} we obtain $I = \cap_{i=1}^n M_i$.
Since $\dim (\Kpolyring/ M_i ) =1$, we have that $\dim (\Kpolyring/I ) = n$, which is the number of zeros of $I$, i.e., $|\kappa(v)|$.
\qed\\

Finally, we show that polynomials over an arbitrary finite field can be considered to be a proper Logic System, i.e., they satisfy Definition~\ref{def:core} and Definition~\ref{def:proper-logic-system}.

\begin{restatable}[Proper logic of polynomials]{lem}{proper-logic-polynomials} 
The set of all polynomials over variables $x_1,\ldots,x_m$ with coefficients in the general finite field $K$, i.e., $\Kpolyring$, forms a Proper Logic system as specified by Definitions \ref{def:core} and \ref{def:proper-logic-system}.
\end{restatable}

\textbf{Proof.}
Given polynomials $s$ and $t$, we say that
$s \entails t$ if and only if $t$ is contained in the ideal generated by $s$ and the field polynomials~\eqref{eq:fieldpolynomials}. 
Lemma~\ref{lem:recon} shows the existence of $\ell$ in Definition~\ref{def:core}
(c.f.~Definition~\ref{def:ell}),
and Lemma~\ref{lem:kernel-ideal} shows that
condition (\ref{fundamental-rel})
of Definition~\ref{def:core} holds.

We define the meaning of the operators $\lor, \land, \lnot$ for polynomials using the following formulas with respect to $\ell$ and $\kappa$:
\begin{enumerate}
\item ${s} \lor {t} = \ell (\kappa({s}) \cup \kappa({t}))$;
\item $ s \land t = \ell(\kappa({s}) \cap \kappa({t}))$;
\item $\lnot {s} = \ell(\kappa({s})^c)$. 
\end{enumerate}
Then, applying $\kappa$ to both sides of these three equations and using condition (\ref{pseduo-inverse-rel}) of Definition~\ref{def:core}, we see that Definition~\ref{def:proper-logic-system} holds and the desired result follows. Note that a simpler definition of the operator $\lor$ is ${s} \lor {t} = s t$, since $\kappa (s t) = \kappa({s}) \cup \kappa({t})$.
\qed

\subsection{Propositional Logic sentences}
We now return to Propositional Logic sentences and the special case $K=GF(2)$, the finite field of size $2$, which comprises the binary alphabet $\{0,1\}$ together with multiplication ($\cdot$) and addition ($+$) corresponding to the logic operators $\land$ and $\oplus$, respectively.
In this case, 
we associate $0$ with the logic value false and $1$ with the logic value true, and thus the variables $x_1,\ldots,x_m$ denote whether the corresponding properties are false or true. Throughout this subsection, all references to logic are intended to mean Propositional Logic.

Any logic sentence in $\sls{m}$ can be written as an equation involving a polynomial in $\GFpolyring$ and vice versa.
To establish this, we will rely on Table \ref{tab:transform}. We assume that a logic sentence in $\sls{m}$ makes use of parentheses to ensure that at most two operands are clearly associated with any operation; then the logic sentence may be parsed so as to obtain a tree representation, where every node denotes an operation from 
the list $\{\lnot, \lor, \land, \oplus, \implies\}$ and where the branches flowing downwards from the node (one or two, depending on the operator) represent the operands being passed to the operator. The tree is unique as per our earlier assumption that parentheses have been used to eliminate any possible ambiguity. Obtaining a polynomial equation representation for this logic sentence can be done through the following four steps:
\begin{enumerate}
    \item Replacing every symbol $\mathtt{X}_i$ with its corresponding variable $x_i$.
    \item Replacing every operator with operands fully described in terms of variables from $\{x_1,\ldots,x_m\}$ with the corresponding mathematical expression as described in Table~\ref{tab:transform}, repeating until all operators have been replaced.
    \item Adding 1 to the resulting polynomial expression.  The reason for this is that we want the kernel of the polynomial form to correspond to its set of zeros, not its set of ones.  
    \item Equating to zero the resulting expression.
\end{enumerate}

\begin{table}[htbp]
\begin{eqnarray*}
\begin{array}{cc}
\mbox{logic form} & \mbox{polynomial form} \\ \hline
\lnot \mathtt{a} & a+1 \\
\mathtt{a} \lor  \mathtt{b} &  a + b + a \cdot b  \\
\mathtt{a} \land  \mathtt{b} & a \cdot b\\
\mathtt{a} \oplus \mathtt{b} & a + b \\
\mathtt{a} \implies \mathtt{b} & a\cdot (1+b) + 1
\end{array}
\end{eqnarray*}
\caption{Conversion between logic sentences and polynomial expressions. Here  $\mathtt{a}, \mathtt{b}$ denote logic sentences and $a,b$ their corresponding polynomial expressions.}
\label{tab:transform}
\end{table}

As a simple example to illustrate this, the translation of the truth of the logic sentence that $``(\mathtt{ X_2} \implies \mathtt{X_1})\land \lnot \mathtt{X_3}''$ is given by $(x_2 \cdot (1+x_1) + 1)\cdot(1+x_3) + 1= 0$. Obtaining an expression in $\sls{m}$ from a polynomial in $\GFpolyring$
(which is assumed to have been equated to zero) can be done similarly. In this case, we assume only two arithmetic operators are present $\{\cdot, +\}$, which are simply swapped with the logic operators $\{\land, \oplus\}$.

For notational purposes, we adopt the following convention.

\begin{convention}
For a given logic sentence $\mathtt{a} \in \sls{m}$,
the corresponding polynomial is
$a \in \GFpolyring$, and vice versa. We will use the notation
\L$( a ) = \mathtt{a}$.
In contrast to the use of the bold Courier font for logic sentences throughout the article, we use the Times New Roman font in this section for the corresponding polynomials.
\label{con:ouronlyconvention}
\end{convention}

The following basic result helps us transition between logic expressions and polynomials, thus providing a parallel with Lemma~\ref{lem:pl_implication_kernel}.

\begin{restatable}[Duality of kernels]{lem}{duality} For a given $\mathtt{a} \in \sls{m}$ and its corresponding polynomial $a \in \GFpolyring$, we have $\algebraicset{\mathtt{a}} = \algebraicset{a}$. In particular, $\mathtt{a} \entails \mathtt{b}$ if and only if $\algebraicset{a} \subseteq \algebraicset{b}$. 
\label{lem:duality}
\end{restatable}

\textbf{Proof.} 
The first statement follows from the property that the mapping between logic sentences and polynomials (given by the four steps above) sends points where the logic sentence is true to points where the associated polynomial is zero (false).
The second result then follows from Lemma~\ref{lem:pl_implication_kernel}.
\qed

It is important to note that we can use ideal membership to decide whether $\mathtt{a} \entails \mathtt{b}$.
From Lemma \ref{lem:kernel-ideal}, given $a, b \in \GFpolyring$, 
we have that
\L$(a) \entails$ \L$(b)$ 
if and only if $b \in (a, x_1^2-x_1, \ldots, x_m^2-x_m$). 
More generally, when $f_1, \ldots, f_n$ are the polynomial representations of the logic expressions $\mathtt{f_1}, \ldots, \mathtt{f_1}$, then $g \in (f_1, \ldots, f_n)$ shows
\L$(g)$
will be a logical consequence of $\{\mathtt{f_1}, \ldots, \mathtt{f_n}\}$.

Using Lemma~\ref{lem:recon} and converting the constructed polynomial back to a logic formula, we immediately obtain the
following result.

\begin{restatable}[Reconstruction of a logic sentence from a proposed kernel]{lem}{recon_logic}
~There exists a function $\ell : \mathcal{P}(K^m) \rightarrow \sls{m}$ such that, for every set $\psi \subseteq K^m$, 
$\kappa(\ell(\psi))=\psi$. 
\label{lem:recon_logic}
\end{restatable}

Finally, building on the above results at the polynomial level, we now prove our main post-processing result with respect to incremental communications at the logic level, which we restate for convenience.

\postproc*

 \textbf{Proof.}
 Using the transformations provided in Table~\ref{tab:transform}, we can convert all our logic expressions to polynomials over $GF(2)$.
 We can then use the construction in Lemma~\ref{lem:postproc2} to compute $\Delta(\sender,\receiver)$ in polynomial form.
 Since the
 multiplication ($\cdot$) and addition ($+$) operations
 on polynomials over $GF(2)$ correspond to the logic operations $\oplus$ and $\land$, respectively, each monomial corresponds to a conjunction of variables and the number of monomials in our polynomials is the same as the number of conjuncts in our logic expressions. 
Hence, the reduction of $\Delta(u,v)$ with respect to the ideal generated by $v$ and the field polynomials established in Lemma~\ref{lem:postproc2} implies a corresponding reduction of logic formulas in $\Delta(\sender,\receiver)$ with respect to the logic information contained in $\receiver$. For each monomial $a$,
\L$( a )$ 
is a logic conjunction involving at most $m$ variables. Hence, the bound $|\kappa(v)|$ on the number of monomials in a reduced polynomial yields the bound $O(m|\kappa(v)|)$ on the size of a reduced logic expression.

If $\receiver \entails \mathtt{w}$, then the polynomial corresponding to $\mathtt{w}$ is contained in the ideal generated by the polynomials corresponding to $\receiver$ and the field polynomials~\eqref{eq:fieldpolynomials}, and thus it can be reduced to zero;
moreover, as in Lemma~\ref{lem:postproc2}, we have that $\mathtt{w} \notin \Delta(\sender,\receiver)$.
Finally, the same lemma shows that 
$\algebraicset{ \Delta(\sender,\receiver) \cup \receiver } = \algebraicset{\sender \cup \receiver}$.
\qed

\begin{example}[Illustration of Lemma~\ref{lem:postproc2} and Theorem~\ref{thm:postproc}]\label{postproc-ex2} 
Within the context of Example~\ref{postproc-ex} above, the polynomial version of the sender's sentence, $\mathtt p \land \mathtt q \land \mathtt r$,  is $p q r + 1$ using Table~\ref{tab:transform} and adding 1 as specified in step 3 above. 
The polynomial version of the receiver's knowledge, $\mathtt p \implies \mathtt q$, is $p (1 + q) + 1 + 1$ or simply $p (1 + q)$ since $1 + 1 = 0$ in $Z/2$, again using Table~\ref{tab:transform} and complementing.
Since we are operating over $Z/2$, the field equations associated with the polynomial variables $p,q,r$ are $\{p^2+p,q^2+q,r^2+r\}$.
The Gr\"obner basis for the ideal of the receiver's knowledge and field equations is $G_v = (p q + p, p^2+p, q^2+q, r^2+r)$. 
If we reduce $p q r +1$ by $G_v$ we obtain $p r + 1$, which is simply the remainder after dividing by $p q + p$. Hence, in this case, the function $\Delta$ returns $(pr + 1)$.
One can also take an additional step to replace $\Delta$ with a collection of smaller polynomials.
If we next compute a Gr\"obner basis for $p r + 1$ along with the field equations, we obtain $(p+1, r+1, q^2+q)$.
The polynomial $q^2+q$ vanishes on both elements of $Z/2$, and thus corresponds to a logical tautology and can be omitted.
We therefore have a reduced $\Delta$ of the form
$(p+1, r+1)$, which converts to the set of logic expressions $\{\mathtt p, \mathtt r\}$ by complementing and reducing $1+1$ to $0$.
\end{example}

\subsection{Applications of Incremental Communications}

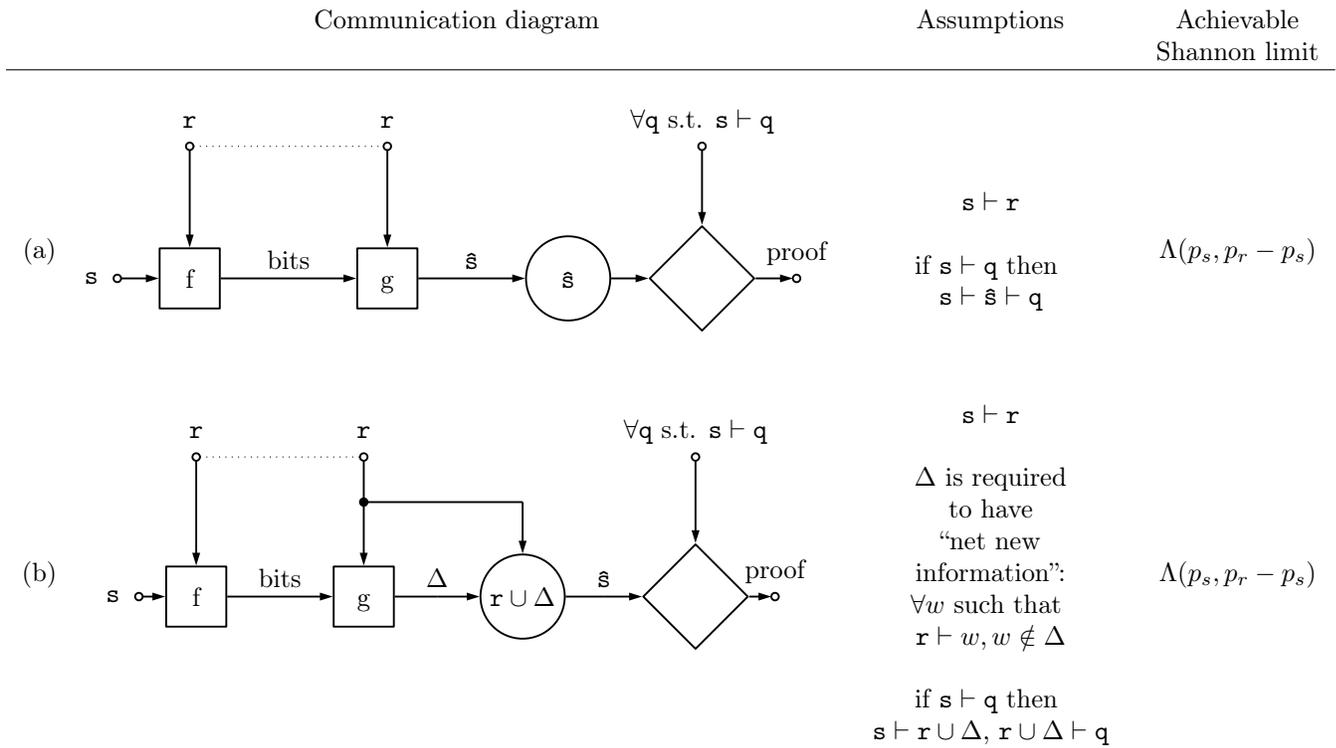
\begin{figure}[htbp]
{ \begin{tabular}{cccc} 
& Communication diagram & Assumptions & Achievable\\ 
& & & Shannon limit \\ \hline \\ \\ 
(a) &
\begin{tabular}{l}
\begin{tikzpicture}
	\matrix[row sep=2.5mm, column sep=5mm]
	{tm
            \node[coordinate]          (hm00) {}; & 
		\node[dspnodeopen,dsp/label=above]         (hm01) {$\receiver$};     &
		\node[coordinate] (hm02) {};  &
		\node[dspnodeopen,dsp/label=above]          (hm03) {$\receiver$};     &
		\node[coordinate]          (hm03p5) {};     &
            \node[coordinate]          (hm04) {}; &
		\node[dspnodeopen,dsp/label=above]         (hm05) {$\forall \query \mbox{ s.t. } \sender \entails \query$};     &
        \node[coordinate]          (hm06) {};  \\ \\
  
		\node[coordinate]         (rm00) {};  &
        \node[coordinate]         (rm01) {};  &
		\node[coordinate]         (rm02) {};  &
        \node[coordinate]         (rm03) {};  &
		\node[coordinate]         (rm04) {};     &
		\node[coordinate]         (rm05) {};  \\ \\
  
		\node[dspnodeopen,dsp/label=left] (mm00) {$\sender$};  &
		\node[dspsquare]         (mm01) {f};     &
            \node[coordinate,label=bits] (mm02) {asdf}; &
		\node[dspsquare]          (mm03) {g};     &
		\node[coordinate,label=$\senderhat$]          (mm03p5) {};     &
            \node[dspcircle,dsp/label=above] (mm04) {$\senderhat$}; &
            \node[dspdiamond] (mm05) {}; &
            \node[dspnodeopen,dsp/label=above] (mm06) {proof}; \\ 
		\node[coordinate]         (tm00) {};     &
		\node[coordinate]         (tm01) {};     &
		\node[coordinate] (tm02) {}; &
		\node[coordinate]          (tm03) {};     &
		\node[coordinate]          (tm04) {}; \\
	};
 
        \begin{scope}[start chain]
            \chainin (hm05);
            \chainin (mm05) [join=by dspconn];
        \end{scope}
        
        \begin{scope}[start chain]
            \chainin (hm03);
            \chainin (mm03) [join=by dspconn];
        \end{scope}
        
	\begin{scope}[start chain]
		\chainin (mm00);
		\chainin (mm01) [join=by dspconn];
		\chainin (mm03) [join=by dspconn];
		\chainin (mm04) [join=by dspconn];
		\chainin (mm05) [join=by dspconn];
		\chainin (mm06) [join=by dspconn];
    \end{scope}

    \begin{scope}[start chain]
       \chainin (hm01);
       \chainin (mm01) [join=by dspconn];
    \end{scope}

\draw[dotted] (hm01)--(hm03);

\end{tikzpicture}
\end{tabular} & 
\begin{tabular}{c}  $\sender \entails \receiver$  \\ \\ $\mbox{if } \sender \entails \query \mbox{ then }$ \\ $\sender \entails \senderhat \entails \query$
\end{tabular} & $\Lambda(p_s,p_r-p_s)$\\ \\
(b) &
\begin{tabular}{l}
\begin{tikzpicture}
	\matrix[row sep=2.5mm, column sep=3mm]
	{tm
 		\node[coordinate]          (hm00) {};  &
		\node[dspnodeopen,dsp/label=above]         (hm01) {$\receiver$};     &
		\node[coordinate] (rm02) {};  &
		\node[dspnodeopen,dsp/label=above]          (hm03) {$\receiver$};     &
		\node[coordinate]          (hm04) {};     &
            \node[coordinate]          (hm05) {};  &
            \node[coordinate]         (hm06) {}; & 
             \node[dspnodeopen,dsp/label=above]         (hm07) {$\forall \query \mbox{ s.t. } \sender \entails \query$};     &
            \node[coordinate] (hn08) {};  \\ \\

		\node[coordinate]         (rm00) {};     &
		\node[coordinate]         (rm01) {};     &
          \node[coordinate]         (rm02) {};  &
		\node[dspnodefull]         (rm03) {};     &
		\node[coordinate]         (rm04) {};     &
            \node[coordinate]          (rm05) {};  &
		\node[coordinate]         (rm06) {}; \\ \\
  
		\node[dspnodeopen,dsp/label=left] (mm00) {$\sender$};  &
		\node[dspsquare]         (mm01) {f};     &
            \node[coordinate,label=bits] (mm02) {}; &
		\node[dspsquare]          (mm03) {g};     &
            \node[coordinate,label=$\Delta$] (mm04) {}; &
		\node[dspcircle] (mm06) {$\receiver \cup \Delta$};     &
            \node[coordinate,label=$\senderhat$] (mm07) {$asd$};  &
            \node[dspdiamond] (mm08) {};  &
            \node[dspnodeopen,dsp/label=above] (mm09) {proof}; \\

		\node[coordinate]         (tm00) {};     &
		\node[coordinate]         (tm01) {};     &
		\node[coordinate,label=] (tm02) {};  &
		\node[coordinate]          (tm03) {};     &
		\node[coordinate]          (tm04) {};  &
	    \node[coordinate]          (tm05) {};  &
		\node[coordinate]          (tm06) {}; \\
 };

\draw[dotted] (hm01)--(hm03);

	\begin{scope}[start chain]
		\chainin (mm00);
		\chainin (mm01) [join=by dspconn];
		\chainin (mm03) [join=by dspconn];
		\chainin (mm04) [join=by dspline];
          \chainin (mm06) [join=by dspconn];
          \chainin (mm07) [join=by dspline];
          \chainin (mm08) [join=by dspconn];
          \chainin (mm09) [join=by dspconn];
        \end{scope}

        \begin{scope}[start chain]
            \chainin (hm01);
            \chainin (mm01) [join=by dspconn];
        \end{scope}

        \begin{scope}[start chain]
            \chainin (hm07);
            \chainin (mm08) [join=by dspconn];
        \end{scope}

        \begin{scope}[start chain]
            \chainin (rm03);
            \chainin (rm05) [join=by dspline];
            \chainin (mm06) [join=by dspconn];
        \end{scope}

        \begin{scope}[start chain]
            \chainin (hm03);
            \chainin (mm03) [join=by dspconn];
        \end{scope}
        
\end{tikzpicture} \end{tabular} & \begin{tabular}{c}  $\sender \entails \receiver$ \\ \\ $\Delta$ is required \\ to have \\ ``net new \\ information'': \\ $\forall w \mbox{ such that }$\\$\receiver \entails w, w \notin \Delta$ \\ \\ $\mbox{if } \sender \entails \query \mbox{ then }$\\$\sender \entails \receiver \cup \Delta$, $\receiver \cup \Delta \entails \query$ 
\end{tabular}  & $\Lambda(p_s,p_r-p_s)$  \\ \\ \\

\end{tabular} 
}
\caption{ Motivation for incremental communications.}
\label{fig:summary_incremental}
\end{figure}

\begin{figure}[htbp]
\begin{center}
     \begin{tikzpicture}
	\matrix[row sep=2.5mm, column sep=5mm]
	{tm		

             \node[coordinate] (f00) {}; &
            \node[coordinate] (f01) {}; &
            \node[coordinate] (f02) {}; &
            \node[coordinate] (f03) {} ; &
            \node[coordinate] (f04) {}; &
            \node[coordinate] (f05) {}; & \\
            
            \node[dspnodeopen,dsp/label=left] (y00) {$\receiver$}; &
            \node[dspsquare] (y01) {$\kappa$}; &
            \node[coordinate,label=$\algebraicset{\receiver}$] (y02) {}; &
            \node[coordinate] (y03) {} ; &
            \node[coordinate] (y04) {}; &
            \node[coordinate] (y05) {}; & 
            \node[coordinate,label=$\algebraicset{\receiver}$] (y06) {}; & 
            \node[dspsquare] (y07) {$\kappa$}; & 
            \node[dspnodeopen,dsp/label=above] (y08) {$\receiver$}; &
            \node[coordinate] (y09) {}; 
            \\

		\node[dspnodeopen,dsp/label=left] (mm00) {$\sender$};  &
		\node[dspsquare]         (mm01) {$\kappa$};     &
            \node[coordinate,label=$\algebraicset{\sender}$] (mm02) {}; &
		\node[dspelongated]          (mm03) { ennumerative \\ code };     &
		\node[coordinate,label=]          (mm03p5) {};     &
            \node[dspbigsquare] (mm04) {decoder};    & 
            \node[coordinate,label=$\algebraicset{\sender}$] (mm05) {}; &
            \node[dspsquare]         (mm06) {$\ell$};  &
		\node[dspnodeopen,dsp/label=above] (mm07) {$\senderhat$};   & 
            \node[dspbigsquare] (mm08) {algebraic\\reduction}; & 
            \node[dspnodeopen,dsp/label=above] (mm09) {$\Delta$}; \\
            
            \node[coordinate] (k00) {}; &
            \node[coordinate] (k01) {}; &
            \node[coordinate,label=$\len{\algebraicset{\sender}}$] (k02) {}; &
            \node[dspelongated]  (k03) { $\delta$ Elias \\ code } ; &
            \node[coordinate] (k04) {}; &
            \node[coordinate] (k05) {}; & \\

            \node[coordinate] (b00) {}; &
            \node[coordinate] (b01) {}; &
            \node[coordinate] (b02) {}; &
            \node[coordinate] (b03) {} ; &
            \node[coordinate] (b04) {}; &
            \node[coordinate] (b05) {}; & \\
  };

        \begin{scope}[start chain]
            \chainin (y00);
            \chainin (y01) [join=by dspconn];
            \chainin (y03) [join=by dspline];
            \chainin (mm03) [join=by dspconn];
        \end{scope}
        \begin{scope}[start chain]
            \chainin (y08);
            \chainin (y07) [join=by dspconn];
            \chainin (y05) [join=by dspline];
            \chainin (mm04) [join=by dspconn];
        \end{scope}
        \begin{scope}[start chain]
        \chainin (y08);
        \chainin (y09) [join=by dspline];
        \chainin (mm08)[join=by dspconn];
        \end{scope}
	\begin{scope}[start chain]
		\chainin (mm00);
		\chainin (mm01) [join=by dspconn];
		\chainin (mm03) [join=by dspconn];
		\chainin (mm04) [join=by dspconn];
		\chainin (mm05) [join=by dspline];
		\chainin (mm06) [join=by dspconn];
            \chainin (mm07) [join=by dspconn];
            \chainin (mm08) [join=by dspconn];
            \chainin (mm09) [join=by dspconn];
        \end{scope}

        \begin{scope}[start chain]
            \chainin (mm01);
            \chainin (k01) [join=by dspline];
            \chainin (k03) [join=by dspconn];
            \chainin (k05) [join=by dspline];
            \chainin (mm04) [join=by dspconn];
        \end{scope}

        \begin{scope}[start chain,dotted]
            \chainin(f04);
            \chainin(b04) [join=by dspline];
        \end{scope}
\end{tikzpicture}
\end{center}
\caption{Proof strategy for adding incremental communications to the setup of Theorem \ref{thm:background_log_info}, with $\ell$ from Lemma~\ref{lem:recon_logic} and $\Delta$ from Theorem~\ref{thm:postproc}.}
\label{fig:incremental_strategy}
\end{figure}
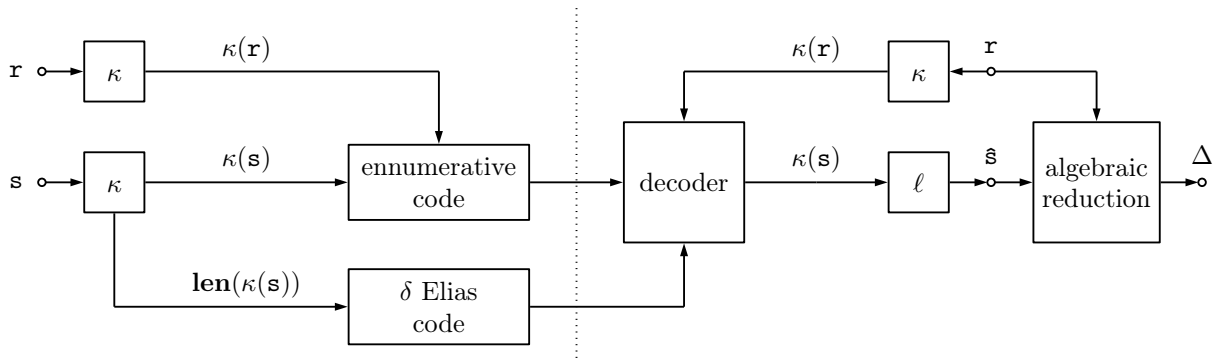

The construction presented in Theorem~\ref{thm:postproc} can then be used in any number of settings; in this article, it is relevant to Theorems \ref{thm:background_log_info}, \ref{thm:less_is_more_background} and \ref{thm:master}. As a representative example, we illustrate in Figures~\ref{fig:summary_incremental} and~\ref{fig:incremental_strategy} how the device implied by Theorem~\ref{thm:postproc} is used in the case of Theorem \ref{thm:background_log_info}. In Figure~\ref{fig:summary_incremental} we repeat the communication diagram associated with Theorem \ref{thm:background_log_info} and below it we demonstrate an alternative, in principle more restrictive, setup where the goal of the decoder $g$ is to produce an increment that needs to be incorporated into the existing knowledge base of logic sentences $\receiver$. As suggested by the achievable Shannon limit column on the right, the fundamental limits in both setups are exactly the same; that is, the restriction introduced by the incremental communication requirement does NOT make the Shannon limit worse. The fact that this is the case is a simple corollary of the upper bound in Theorem \ref{thm:background_log_info} coupled with Theorem \ref{thm:postproc}, and thus no further formal statement or proof is given; instead we refer the reader to Figure~\ref{fig:incremental_strategy} which is an updated version of Figure~\ref{fig:background_log_info} where the algebraic reduction step is implemented by Theorem \ref{thm:background_log_info}.

\section{Speculative Future Directions}
\label{sec:speculations}

The fact that logic is so foundational and deeply intertwined in computer science and mathematics leads one to be able to imagine many possibilities.  In this section, we will allow ourselves to speculate on some ways that this work might be elaborated upon to address various important topics, in the hopes of inspiring future authors.  We discuss these in four subsections: practical methods and applications; 
extensions of the core setup, including the logic foundations;
additional social scenarios; and, lastly, other perspectives on semantic information not treated in our article.

\subsection{Practical methods and applications}
{\bf Improvements on practical coding techniques.} In our article we introduced practical methods, based on both linear and nonlinear codes,  for both the ``less is more'' and ``no need to know'' scenarios. These codes nonetheless are optimal in the sense of approaching the Shannon bounds only for a very limited set of parameters for these scenarios. It is an open problem to design optimal codes for even the scenarios introduced in our article, let alone other variants as discussed in this section.

{\bf Artificial intelligence.}
Some practical usage possibilities lie in AI, where logic has long provided one of its most fundamental representations, for expressing human knowledge and allowing reasoning upon it \cite{russell2016artificial}.  A direct practical realization of the situation described above is possible due to the steady rise over the years in the efficacy of the technologies of {\em semantic parsing} \cite{kamath2019surveysemanticparsing}, or translating natural language sentences into logic sentences which represent their underlying meaning, {\em e.g.}, \cite{Abdelaziz2021}.  The canonical situation above is motivated by the long-standing idea of a {\em theory of mind} \cite{thrymind}, a model of the mind of an agent in terms of his/her knowledge (and possibly goals) in terms of logic sentences.  In this context, the ability to determine the most informative logic sentence to send would provide a solution to the question {\em what is the next best thing to say?} to a formal and quantifiable degree for which we know no equivalent in AI to date.  This would lie in sharp contrast to the lack of intentionality in current approaches to text generation based on large language models, which represent a kind of stream of consciousness-like random walk that generates statistically likely combinations of words, though progress in prompting technologies can make the output appear more intentional \cite{intention}.


Machine learning has a rich landscape of logic-based models, including decision trees \cite{cart} and the emerging area of neuro-symbolic AI \cite{nsss24}.  \cite{CALVANESESTRINATI2021107930} calls for a new brand of ``semantic machine learning'' that has a world model, which is one of the emphases within neuro-symbolic AI.  Information theory provides key tools for learning theory \cite{Hellstrm2023GeneralizationBP}, suggesting that these results could be used to deeply characterize the effect of logically-expressed knowledge on sample complexity.  \cite{belfiore2021logicalinformationcellsi} shows a use of semantic information to characterize logical expressivity in a neural network.  

{\bf Data transmission and compression.}
As mentioned earlier, a recent batch of vision papers such as \cite{Popovski2019SemanticEffectivenessFA} have argued for the great potential of ``much lower data transmission requirements'' \cite{9864327} held by semantic approaches to communication, typically sketching nominal communication architectures with additional semantic components \cite{CALVANESESTRINATI2021107930}.  This has been fueled by some theory showing that focusing on preserving meaning has the potential for savings over focusing on preserving bits, notably \cite{basu:semantic}, which also showed some empirical gains in a limited setting.

As part of this recent surge of interest in semantic communication ideas, a number of approaches have rushed to show empirical gains, as described in many surveys such as \cite{engsemsurvey}.  A prominent thread showing some gains using deep learning is exemplified by \cite{weng2021semantic}.  Some approaches such as \cite{Liu2022EnhancingCR}, leveraging the intuition that one does not need to get the bits right to get the meaning right, have shown some gains particularly in the low-SNR regime.

With our approach, one can imagine potential direct impact for at least compression scenarios involving logic or programs, both in software and hardware.  The extent to which the presumed gains can transfer to other forms of data depends on their ease of translation to equivalent forms in terms of some form of logic or program  (see additional discussion below).  

{\bf Automated theorem proving and mathematics.}
The process of reasoning, or automated theorem proving \cite{atp-book}, itself contains a difficult problem, premise selection (or which logic sentence should be operated upon next, by an inference rule, in order to prove the desired sentence) or proof guidance in general, which could perhaps be aided by a time-efficient or approximate version of our model scenario.

\subsection{Core setup}

{\bf More expressive logics.}
As one can see, our conditions for treatable logic systems are fairly general.
This includes First-Order Logic of structures of fixed finite sizes, or of structures of sizes up to a given maximum size.  It also includes the All Positive First-Order Logic for structures of fixed size / up to a given fixed size.

Future researchers may consider extensions to First-Order Logic with Counting, Multi-Valued and Probabilistic Logics such as \cite{nilsson86, cozman00, fagin24}, upon which many of the techniques of the fast-emerging area of neuro-symbolic AI are founded, and Second and Higher-order logics.
Extension to Higher-order logics has a particular significance with respect to programs, discussed next.

{\bf Programs.}  Algorithmic information theory (AIT), whose roots were laid in the 1960s \cite{li2008kolmogorov}, generalizes information theory and relates it to computation.  
The {\em algorithmic complexity} or {\em Kolmogorov complexity} of a string $x$ is defined as the length of the shortest program that computes or outputs $x$, where the program is run on some fixed reference universal computer. 
 AIT's powerful ideas have long sparked the imagination around many possibilities for practical implementations of the theory.  However, AIT's core concept, the length of the shortest program, is incomputable, making AIT largely impractical.  Though various schemes exist for approximating the quantity, such as Levin search \cite{levin1973universal} and its later improvements, they are generally still not efficient enough for practical usage, and may yield programs far from the theory's bound.  

We suggest that generalizing semantic information theory to a semantic AIT could open new doors.
Noting the Curry-Howard isomorphism \cite{curry-howard}, which states an equivalence between programs and (higher-order) logic, this might bear some resemblance to the current notion of sending the minimum amount of logic (in the form of bits) to the receiver.  One can similarly imagine sending program statements to allow the receiver's existing program to be extended to compute results of interest.  A notion that could correspond to that of the information value of a logical sentence that can already be deduced given existing logic statements being zero is one of the information value of a computation that can already be performed given existing program elements being zero.  In this case, we would invoke formal notions of {\em program semantics}, which are fully grounded in logic, going back to the work of Hoare \cite{hoare69}.  Some of the ideas of \cite{juba:universal_semantic} may be relevant here.

In the same setting as standard information theory, for a given distribution, AIT achieves exactly the same bounds. We have shown that a semantic information theory can provide increased compression over such bounds.  This is a consequence of AIT being ultimately semantics-free, as is standard information theory.  We have also shown practically computable codes which achieve our stated bounds.  It is thus possible that semantic information theory applied to programs could, for some purposes, serve as a practically computable version of Kolmogorov complexity.  Further, it can be thought of as extending the idea of program shortness to account for the amount of program capability (e.g., libraries of functions) that the receiver already has.

{\bf Reasoning power and computation.}
The assumption that the dialog counterpart has a reasoning engine at hand accounts for intelligence on the part of the recipient -- preventing the need for the speaker or teacher to communicate things that will be obvious since they can be deduced from the recipient’s existing knowledge.  However, the computational cost needed to make those deductions is not accounted for in the current model, while in more complex instances, it may be more realistic not to assume the receiver can always perform any reasoning needed.  For example, one could account for limited reasoning time/resources, as in bounded rationality \cite{boundedrat13} or teaching a child.

{\bf Sharper results for more general distributions.} Early in our paper (see the expression (\ref{eq:alternate}) and Theorem \ref{thm:logicinfo}) we remarked that the optimal code length to fully convey what a sentence represents is given by $-\log_2 P( \kappa(\sender))$. Our results are mostly focused nonetheless on kernel sizes, which give us a simpler, but coarser theory. What if there are some structures in $\kernelspace$ that are more likely than others, or what if there are correlations in their occurrences? What if the number of structures that model a logic sentence is so small that effectively we are in the setting where the parameters $p_s, p_r, p_q$ are zero? In all these scenarios, a much more refined theory is necessary.

{\bf Uncertainty in logic sentences.} Throughout the article, we have assumed that Alice's logic sentence is believed to be true in all cases. Similarly Bob's sentence is believed to be true in most cases except the ones where we explicitly state that there could be conflict or misinformation. But what if this belief was only partial? How would this change the results and algorithms?

\subsection{Additional scenarios}

Throughout our paper, we mostly centered around the concept that Alice holds some version of truth that Bob wants to take advantage of. In what follows we postulate other scenarios that are interesting in their own right.

{\bf  Unawareness of possible conflict.} In the situation modeled by Theorem \ref{thm:background_log_info}, Alice's sentence does not necessarily entail Bob's (there's a possible conflict), however Alice is fully aware of what sentence Bob has. The proposed extension lifts this last assumption -- what is the total minimum expected communication cost?

{\bf  Collaboration.} Here Alice and Bob each have a sentence that is presumed true but unknown to each other and they wish to pool their logic sentences with a minimal amount of total communication. This type of extension to our problem appears to be directly in line with problems studied under communication complexity \cite{10.1145/800135.804414} and  interactive communication \cite{orlitsky:interactive}.
 
{\bf Adversarialness and disinformation.} Bob, who holds a sentence he believes to be true, does not necessarily trust Alice, who may or may not hold a true sentence and is unaware of what Bob knows. Assuming Bob's goal is to sharpen his sentence without accepting something that is not true, and Alice's goal is to have Bob accept something that is not true, what are good strategies for either?

{\bf Consultancy.} In our current model, at communication time Alice is provided the query. What if Bob is the one given the query, instead of Alice? How does that change the problem?

{\bf Teaching.} What would be the solution if there are multiple Bobs and one Alice, and a single message will be transmitted to all Bobs? Building on logical theories of teaching/learning such as \cite{hintikka-teaching}, this work may enable the practical implementation of a sense of optimal (or at least principled) teaching in the long-standing area of intelligent tutoring systems \cite{its}.  One could imagine its use at the stage of curricular design or even at the granular level during live interaction, particularly in combination with a logical model of natural language (described next).  

{\bf Scientific inquiry.} Scientific inquiry can be thought of asking questions of nature \cite{hintikka-scimethod}, suggesting analogous usage as a basis for optimal/principled automated scientific experimentation \cite{autosci-pnas}. Extensions of our model to account for query design could help with some of these problems. 

\subsection{Other perspectives on semantic information}

{\bf Foundations of probability and information theory.} 
Ellerman \cite{ellerman:logic_entropy_2021} observes a  duality that can be stated using the deep relationships between logic and probability.  He examines the mathematical foundations of information and constructs a theory around the concept of ``information as distinctions''; Ellerman goes on to define ``logic entropy'' and shows how Shannon's entropy can be seen as a special case of logic entropy. 
At a high level, our conclusions are natural implications of the ability of logic to compactly describe sets.  It can be shown that standard information theory can be derived from this logical standpoint.  In fact, our framework can be related to Shannon's own work on ``lattice information theory'' \cite{shannon:lattice} which similarly describes a more general abstract formalism for information theory from the starting point of discrete sets, though it did not connect to logic.

\section{Concluding remarks}
\label{sec:concluding}
This article introduces what we believe are the first collection of sharp results on an information theory for the communication of logic sentences under with an assumption of a deductive mechanism at the receiver side, including a rigorous development of the type logics for which our results apply. A diverse set of communication situations are treated in this work, including settings where the goal of the communication is to efficiently allow a receiver to deduce all or a subset of the logic statements that a sender can deduce, as well as settings where a receiver may already be in possession of a related logic statement which the sender may or may not be aware of. Practical codes based on linear coding techniques are developed and experimental results are offered demonstrating potential significant gains compared to classical communication techniques.

\section*{Acknowledgements}
The authors acknowledge helpful conversations with the following individuals: Ron Fagin, Phokion Kolaitis, Jason Rute, Kush Varshney and Mark Wegman. Work of W. Szpankowski was partially supported by the NSF Center for Science of Information (CSoI) Grant CCF-0939370, and also by NSF Grants CCF-2006440, and CCF-2211423.

\bibliographystyle{abbrv}
\bibliography{main}

\begin{thebibliography}{10}

\bibitem{Abdelaziz2021}
I.~Abdelaziz, S.~Ravishankar, P.~Kapanipathi, S.~Roukos, and A.~Gray.
\newblock {A Semantic Parsing and Reasoning-Based Approach to Knowledge Base
  Question Answering}.
\newblock In {\em {Proceedings of the AAAI Conference on Artificial
  Intelligence}}, volume~35 of {\em AAAI 2021}, pages 15985--15987, May 2021.

\bibitem{autosci-pnas}
D.~Adam.
\newblock The automated lab of tomorrow.
\newblock {\em Proceedings of the National Academy of Sciences},
  121(17):e2406320121, 2024.

\bibitem{its}
J.~K. Ali~Alkhatlan.
\newblock Intelligent tutoring systems: A comprehensive historical survey with
  recent developments.
\newblock {\em International Journal of Computer Applications}, 181(43):1--20,
  Mar 2019.

\bibitem{bao:semantic}
J.~Bao, P.~Basu, M.~Dean, C.~Partridge, A.~Swami, W.~Leland, and J.~A. Hendler.
\newblock Towards a theory of semantic communication.
\newblock In {\em IEEE Network Science Workshop}, pages 110--117, Los Alamitos,
  CA, USA, jun 2011. IEEE Computer Society.

\bibitem{bao:semantic_extended}
J.~Bao, P.~Basu, M.~Dean, C.~Partridge, A.~Swami, W.~Leland, and J.~A. Hendler.
\newblock Towards a theory of semantic communication.
\newblock In {\em Extended Technical Report}, 2011.

\bibitem{basu:semantic}
J.~Bao, P.~Basu, M.~Dean, C.~Partridge, A.~Swami, W.~Leland, and J.~A. Hendler.
\newblock Towards a theory of semantic communication.
\newblock In {\em IEEE Network Science Workshop}, pages 110--117, Los Alamitos,
  CA, USA, jun 2011. IEEE Computer Society.

\bibitem{carnap53}
Y.~Bar-Hillel and R.~Carnap.
\newblock Semantic information.
\newblock {\em The British J. Philosophy of Science}, 4:147--157, 1953.

\bibitem{barwise-model-theoretic-logics-1985}
J.~Barwise.
\newblock Model theoretic logics: Concepts and aims.
\newblock In J.~Barwise and S.~Feferman, editors, {\em Model Theoretic Logics},
  pages 1--24. Cambridge University Press, 1985.

\bibitem{Barwise1997}
J.~Barwise and J.~Seligman.
\newblock {\em Information Flow: The Logic of Distributed Systems}.
\newblock Cambridge University Press, 1997.

\bibitem{BASU2014188}
P.~Basu, J.~Bao, M.~Dean, and J.~Hendler.
\newblock Preserving quality of information by using semantic relationships.
\newblock {\em Pervasive and Mobile Computing}, 11:188--202, 2014.

\bibitem{belfiore2021logicalinformationcellsi}
J.-C. Belfiore, D.~Bennequin, and X.~Giraud.
\newblock Logical information cells i, 2021.

\bibitem{berger1971rate}
T.~Berger.
\newblock {\em Rate Distortion Theory: A Mathematical Basis for Data
  Compression}.
\newblock Prentice-Hall electrical engineering series. Prentice-Hall, 1971.

\bibitem{cart}
L.~Breiman, J.~H. Friedman, R.~A. Olshen, and C.~J. Stone.
\newblock {\em Classification and Regression Trees}.
\newblock {Chapman and Hall/CRC}, 1984.

\bibitem{breitbart:size_of_binary_decision_diagrams}
Y.~Breitbart, H.~Hunt, and D.~Rosenkrantz.
\newblock On the size of binary decision diagrams representing boolean
  functions.
\newblock {\em Theoretical Computer Science}, 145(1):45--69, 1995.

\bibitem{buss-intro-to-proof-theory1998}
S.~Buss.
\newblock {\em Handbook of Proof Theory}, chapter An introduction to proof
  theory, pages 1--78.
\newblock Elsevier, New York, 1998.

\bibitem{CALVANESESTRINATI2021107930}
E.~{Calvanese Strinati} and S.~Barbarossa.
\newblock 6{G} networks: Beyond shannon towards semantic and goal-oriented
  communications.
\newblock {\em Computer Networks}, 190:107930, 2021.

\bibitem{Carnap1942}
R.~Carnap.
\newblock {\em Introduction to Semantics}.
\newblock Harvard University Press, Cambridge, 1942.

\bibitem{nsss24}
{Centaur AI Institute}.
\newblock {3rd Neuro-Symbolic AI Summer School}.
\newblock \url{https://neurosymbolic.github.io/nsss2024}, 2024.

\bibitem{cook-reckhow1979}
S.~Cook and R.~A. Reckhow.
\newblock The relative efficiency of propositional proof systems.
\newblock {\em Journal of Symbolic Logic}, 44(1):36--50, 1979.

\bibitem{cover:enumerative}
T.~Cover.
\newblock Enumerative source encoding.
\newblock {\em IEEE Transactions on Information Theory}, 19(1):73--77, 1973.

\bibitem{CLO:Ideals_Varieties}
D.~Cox, J.~Little, and D.~O’Shea.
\newblock {\em Ideals, Varieties, and Algorithms: An Introduction to
  Computational Algebraic Geometry and Commutative Algebra}.
\newblock Springer, 2015.

\bibitem{cozman00}
F.~Cozman.
\newblock Credal networks.
\newblock {\em Artificial Intelligence}, 120(2):199–233, 2000.

\bibitem{thrymind}
F.~Cuzzolin, A.~Morelli, B.-I. Cîrstea, and B.~Sahakian.
\newblock Knowing me, knowing you: Theory of mind in ai.
\newblock {\em Psychological Medicine}, 50:1057--1061, 04 2020.

\bibitem{Devlin1991-DEVLAI}
K.~Devlin.
\newblock {\em Logic and Information}.
\newblock Cambridge University Press, 1991.

\bibitem{dalfonso}
S.~D’Alfonso.
\newblock On quantifying semantic information.
\newblock {\em Information}, 2(1):61--101, 2011.

\bibitem{elias:integers}
P.~Elias.
\newblock Universal codeword sets and representations of the integers.
\newblock {\em IEEE Transactions on Information Theory}, 21(2):194--203, 1975.

\bibitem{ellerman:logic_entropy_2021}
D.~Ellerman.
\newblock {\em New Foundations for Information Theory: Logical Entropy and
  Shannon Entropy}.
\newblock Springer, 2021.

\bibitem{fagin24}
R.~Fagin, R.~Riegel, and A.~Gray.
\newblock Foundations of reasoning with uncertainty via real-valued logics.
\newblock {\em Proceedings of the National Academy of Sciences}, 121(21), 2024.

\bibitem{feynman_1965_flp}
R.~Feynman, R.~Leighton, M.~Sands, and E.~Hafner.
\newblock {\em {The Feynman Lectures on Physics; Vol. I}}.
\newblock Addison-Wesley, 1965.

\bibitem{floridi:semantic}
L.~Floridi.
\newblock Outline of a theory of strongly semantic information.
\newblock {\em Minds and Machines}, 14, 05 2004.

\bibitem{intention}
J.~Grindrod.
\newblock Large language models and linguistic intentionality.
\newblock {\em Synthese}, 204(2):71, 2024.

\bibitem{boundedrat13}
B.~Grosof and T.~Swift.
\newblock Radial restraint: a semantically clean approach to bounded
  rationality for logic programs.
\newblock In {\em Proceedings of the Twenty-Seventh AAAI Conference on
  Artificial Intelligence}, AAAI'13, page 379–386. AAAI Press, 2013.

\bibitem{guo2022semantic}
T.~Guo, Y.~Wang, J.~Han, H.~Wu, B.~Bai, and W.~Han.
\newblock Semantic compression with side information: A rate-distortion
  perspective, 2022.

\bibitem{gunduz:beyond_bits}
D.~Gündüz, Z.~Qin, I.~E. Aguerri, H.~S. Dhillon, Z.~Yang, A.~Yener, K.~K.
  Wong, and C.-B. Chae.
\newblock Beyond transmitting bits: Context, semantics, and task-oriented
  communications.
\newblock {\em IEEE Journal on Selected Areas in Communications}, 41(1):5--41,
  2023.

\bibitem{Hellstrm2023GeneralizationBP}
F.~Hellstr{\"o}m, G.~Durisi, B.~Guedj, and M.~Raginsky.
\newblock Generalization bounds: Perspectives from information theory and
  pac-bayes.
\newblock {\em ArXiv}, abs/2309.04381, 2023.

\bibitem{Heyting1956-HEYIAI-2}
A.~Heyting.
\newblock {\em Intuitionism: An Introduction}.
\newblock North-Holland Pub. Co., Amsterdam,, 1956.

\bibitem{hintikka-scimethod}
J.~Hintikka.
\newblock On the logic of an interrogative model of scientific inquiry.
\newblock {\em Synthese}, 47(1):69--83, 1981.

\bibitem{hintikka-teaching}
J.~Hintikka.
\newblock A dialogical model of teaching.
\newblock {\em Synthese}, 51(1):39--59, 1982.

\bibitem{hoare69}
C.~A.~R. Hoare.
\newblock An axiomatic basis for computer programming.
\newblock {\em Commun. ACM}, 12(10):576–580, Oct. 1969.

\bibitem{hodges_1997}
W.~Hodges.
\newblock {\em A Shorter Model Theory}.
\newblock Cambridge University Press, 1997.

\bibitem{curry-howard}
W.~A. Howard.
\newblock The formulae-as-types notion of construction.
\newblock In J.~P. Seldin and J.~R. Hindley, editors, {\em To H.B. Curry:
  Essays on Combinatory Logic, Lambda Calculus and Formalism}, pages 479--490.
  Academic Press, 1980.
\newblock original paper manuscript dated 1969.

\bibitem{juba:universal_semantic}
B.~Juba and M.~Sudan.
\newblock Universal semantic communication {I}.
\newblock In {\em Proceedings of the Fortieth Annual ACM Symposium on Theory of
  Computing}, STOC '08, page 123–132, New York, NY, USA, 2008. Association
  for Computing Machinery.

\bibitem{kamath2019surveysemanticparsing}
A.~Kamath and R.~Das.
\newblock A survey on semantic parsing, 2019.

\bibitem{ks08}
J.~Konorski and W.~Szpankowski.
\newblock What is information?
\newblock {\em 2008 IEEE Information Theory Workshop}, pages 269--270, 2008.

\bibitem{kozen_1981}
D.~Kozen.
\newblock Positive first-order logic is {NP}-complete.
\newblock {\em IBM Journal of Research and Development}, 25(4):327--332, 1981.

\bibitem{Kripke1963-KRISCO}
S.~Kripke.
\newblock Semantical considerations on modal logic.
\newblock {\em Acta Philosophica Fennica}, 16:83--94, 1963.

\bibitem{levin1973universal}
L.~A. Levin.
\newblock Universal sequential search problems.
\newblock {\em Problems of Information Transmission}, 9(3):115--116, 1973.

\bibitem{li2008kolmogorov}
M.~Li, P.~Vit{\'a}nyi, et~al.
\newblock {\em An introduction to Kolmogorov complexity and its applications},
  volume~3.
\newblock Springer, 2008.

\bibitem{liu_shao_zhang_poor:indirect_rd_semantic}
J.~Liu, S.~Shao, W.~Zhang, and H.~V. Poor.
\newblock An indirect rate-distortion characterization for semantic sources:
  General model and the case of gaussian observation.
\newblock {\em IEEE Transactions on Communications}, 70(9):5946--5959, 2022.

\bibitem{liu:rd_semantic}
J.~Liu, W.~Zhang, and H.~V. Poor.
\newblock A rate-distortion framework for characterizing semantic information.
\newblock In {\em IEEE International Symposium on Information Theory}, 05 2021.

\bibitem{Liu2022EnhancingCR}
Y.~Liu, Y.~Zhang, P.~Luo, S.~Jiang, K.~Cao, H.~Zhao, and J.~Wei.
\newblock Enhancing communication reliability from the semantic level under low
  snr.
\newblock {\em Electronics}, 2022.

\bibitem{mehta:decsion_tree_representation_of_boolean_functions}
D.~Mehta and V.~Raghavan.
\newblock Decision tree approximations of boolean functions.
\newblock {\em Theoretical Computer Science}, 270(1):609--623, 2002.

\bibitem{nilsson86}
N.~J. Nilsson.
\newblock Probabilistic logic.
\newblock {\em Artificial Intelligence}, 28(1):71--87, 1986.

\bibitem{9864327}
K.~Niu, J.~Dai, S.~Yao, S.~Wang, Z.~Si, X.~Qin, and P.~Zhang.
\newblock A paradigm shift toward semantic communications.
\newblock {\em IEEE Communications Magazine}, 60(11):113--119, 2022.

\bibitem{niu2024mathematicaltheorysemanticcommunication}
K.~Niu and P.~Zhang.
\newblock A mathematical theory of semantic communication, 2024.

\bibitem{odonnell:decision_trees_influential_variable}
R.~O'Donnell, M.~Saks, O.~Schramm, and R.~Servedio.
\newblock Every decision tree has an influential variable.
\newblock In {\em 46th Annual IEEE Symposium on Foundations of Computer Science
  (FOCS'05)}, pages 31--39, 2005.

\bibitem{orlitsky:interactive}
A.~Orlitsky.
\newblock Interactive communication: balanced distributions, correlated files,
  and average-case complexity.
\newblock In {\em Proceedings 32nd Annual Symposium of Foundations of Computer
  Science}, pages 228--238, 1991.

\bibitem{atp-book}
M.~Pantsar.
\newblock Theorem proving in artificial neural networks: new frontiers in
  mathematical ai.
\newblock {\em European Journal for Philosophy of Science}, 14(1):4, 2024.

\bibitem{10.1145/800070.802192}
C.~H. Papadimitriou and M.~Sipser.
\newblock Communication complexity.
\newblock In {\em Proceedings of the Fourteenth Annual ACM Symposium on Theory
  of Computing}, STOC '82, page 196–200, New York, NY, USA, 1982. Association
  for Computing Machinery.

\bibitem{Popovski2019SemanticEffectivenessFA}
P.~Popovski, O.~Simeone, F.~Boccardi, D.~G{\"u}nd{\"u}z, and O.~Sahin.
\newblock Semantic-effectiveness filtering and control for post-5g wireless
  connectivity.
\newblock {\em Journal of the Indian Institute of Science}, 100:435--443, 2019.

\bibitem{riordanlower}
J.~Riordan and C.~E. Shannon.
\newblock The number of two-terminal series-parallel networks.
\newblock {\em Journal of Mathematics and Physics}, 21(1-4):83--93, 1942.

\bibitem{russell2016artificial}
S.~J. Russell and P.~Norvig.
\newblock {\em Artificial intelligence: a modern approach}.
\newblock Pearson, 2016.

\bibitem{shannon:lattice}
C.~Shannon.
\newblock The lattice theory of information.
\newblock {\em Transactions of the IRE Professional Group on Information
  Theory}, 1(1):105--107, 1953.

\bibitem{shannon:BST48}
C.~E. Shannon.
\newblock A mathematical theory of communication.
\newblock {\em Bell System Technical Journal}, 27:379–423, 623–656, 1948.

\bibitem{shannon:rd}
C.~E. Shannon.
\newblock Coding theorems for a discrete source with a fidelity criterion.
\newblock In {\em Claude E. Shannon: Collected Papers}, pages 325--350.
  Wiley-IEEE Press, 1993.

\bibitem{Shao2022ATO}
Y.~Shao, Q.~Cao, and D.~G{\"u}nd{\"u}z.
\newblock A theory of semantic communication.
\newblock {\em IEEE Transactions on Mobile Computing}, 23:12211--12228, 2022.

\bibitem{shoenfield_2001}
J.~R. Shoenfield.
\newblock {\em Mathematical Logic}.
\newblock Addison-Wesley, 1st edition, 2001.

\bibitem{slepian_wolf:coding}
D.~Slepian and J.~Wolf.
\newblock Noiseless coding of correlated information sources.
\newblock {\em IEEE Transactions on Information Theory}, 19(4):471--480, 1973.

\bibitem{stavrou:goal_semantic_rd}
P.~A. Stavrou and M.~Kountouris.
\newblock The role of fidelity in goal-oriented semantic communication: A rate
  distortion approach.
\newblock {\em IEEE Transactions on Communications}, 71(7):3918--3931, 2023.

\bibitem{ayg-spa}
W.~Szpankowski and A.~Grama.
\newblock Frontiers of science information: Shannon meets turing.
\newblock {\em IEEE Computer}, (51):32--42, 2018.

\bibitem{0ea4add0-94e6-3877-bd4e-0c67171ac249}
A.~Tarski.
\newblock The semantic conception of truth and the foundations of semantics.
\newblock {\em Philosophy and Phenomenological Research}, 4(3):341--376, 1944.

\bibitem{tarski-conception-of-truth1935}
A.~Tarski.
\newblock The concept of truth in formalized languages.
\newblock In {\em A. Tarski: Logic, Semantics, Metamathematics: papers from
  1923-1938}, pages 152--278. Oxford: Clarendon Press, 1956.

\bibitem{sep-logic-higher-order}
J.~Väänänen.
\newblock {Second-order and Higher-order Logic}.
\newblock In E.~N. Zalta, editor, {\em The {Stanford} Encyclopedia of
  Philosophy}. Metaphysics Research Lab, Stanford University, {F}all 2021
  edition, 2021.
\newblock
  \url{https://plato.stanford.edu/archives/fall2021/entries/logic-higher-order/}.

\bibitem{weaver:commmentary_math_comm}
{W. Weaver}.
\newblock Recent contributions to the mathematical theory of communication.
\newblock In {\em ETC A Review of General Semantics}, volume~10, pages
  261--281, 1953.

\bibitem{weng2021semantic}
Z.~Weng, Z.~Qin, and G.~Y. Li.
\newblock Semantic communications for speech signals.
\newblock In {\em ICC 2021-IEEE International Conference on Communications},
  pages 1--6. IEEE, 2021.

\bibitem{engsemsurvey}
D.~Wheeler and B.~Natarajan.
\newblock Engineering semantic communication: A survey.
\newblock {\em IEEE Access}, PP:1--1, 01 2023.

\bibitem{wiki:logicincs}
{Wikipedia contributors}.
\newblock Logic in computer science --- {W}ikipedia{,} the free encyclopedia.
\newblock \url{https://en.wikipedia.org/wiki/Logic_in_computer_science}, 2024.

\bibitem{wiki:mathlogic}
{Wikipedia contributors}.
\newblock Mathematical logic --- {W}ikipedia{,} the free encyclopedia.
\newblock \url{https://en.wikipedia.org/wiki/Mathematical_logic}, 2024.

\bibitem{enwiki:1200422835}
{Wikipedia contributors}.
\newblock Propositional calculus --- {Wikipedia}{,} the free encyclopedia.
\newblock \url{https://en.wikipedia.org/wiki/Propositional_calculus}, 2024.

\bibitem{enwiki:1233551294}
{Wikipedia contributors}.
\newblock Semantic theory of truth --- {Wikipedia}{,} the free encyclopedia.
\newblock \url{https://en.wikipedia.org/wiki/Semantic_theory_of_truth}, 2024.

\bibitem{wyner_ziv:coding}
A.~Wyner and J.~Ziv.
\newblock The rate-distortion function for source coding with side information
  at the decoder.
\newblock {\em IEEE Transactions on Information Theory}, 22(1):1--10, 1976.

\bibitem{10.1145/800135.804414}
A.~C.-C. Yao.
\newblock Some complexity questions related to distributive
  computing(preliminary report).
\newblock STOC '79, page 209–213, New York, NY, USA, 1979. Association for
  Computing Machinery.

\bibitem{yu:information_lattice_learning}
H.~Yu, J.~A. Evans, and L.~R. Varshney.
\newblock Information lattice learning.
\newblock {\em J. Artif. Intell. Res.}, 77:971--1019, July 2023.

\bibitem{Yu2024SemanticCW}
H.~Yu and L.~R. Varshney.
\newblock Semantic compression with information lattice learning.
\newblock {\em 2024 IEEE International Symposium on Information Theory
  Workshops (ISIT-W)}, pages 1--6, 2024.

\end{thebibliography}

\appendix

\section{Proofs of miscellaneous results}

\elementarylambda*

\textbf{Proof.} 
By definition, it is clear that $\Lambda(a,b) \geq 0$ for $a,b \geq 0$. The gradient and Hessian of $\Lambda(a,b)$ are respectively given by
\begin{eqnarray*}
\nabla \Lambda(a,b) = \left[ \begin{array}{c}
\log_2 \frac{a+b}{a} \\
\log_2 \frac{a+b}{b}
\end{array} \right], \qquad\mbox{Hess}(\Lambda(a,b) ) = \frac{1}{\log 2} \left[
\begin{array}{cc}
-\frac{b}{a(1+b)} & \frac{1}{a+b} \\
\frac{1}{a+b} & -\frac{a}{b(a+b)}\\
\end{array}
\right].
\end{eqnarray*}
Over $a,b \in (0,\infty)$, the gradient is strictly positive. As a consequence, the function is monotonically increasing when one of the arguments is fixed, and thus $\Lambda(a,b) \leq \Lambda(a,b+\Delta_b) \leq \Lambda(a+\Delta_a,b+\Delta_b)$ with at least one of those inequalities being strict due to the assumption. Moreover, the Hessian is positive semi-definite since its eigenvalues are $0$ and $-\frac{a^2+b^2}{ab(a+b)}$, the latter always being strictly negative in the same domain. From this we conclude that the function is concave $\cap$. Because $\Lambda(a,0)=\Lambda(0,b)=0$, the Lemma assertions can be extended to the full domain $[0,+\infty) \times [0,+\infty)$. To prove the last statement, first note that $\Lambda(a,b) = \Lambda(b,a)$. For a given $\lambda \in [0,1]$, let $a_{\lambda} = \lambda a + (1-\lambda)b$ and $b_{\lambda} = \lambda b + (1-\lambda)a$. Due to the concavity $\cap$ of the $\Lambda$ function, we deduce that
\begin{eqnarray*}
\Lambda(a,b) = \lambda \Lambda(a,b) + (1-\lambda) \Lambda(b,a) \leq \Lambda( a_{\lambda} , b_{\lambda} ) .
\end{eqnarray*}
Next note that, since $a+b<1$, then $a_{\lambda} + b_{\lambda} < 1$, and from the monotonicity property proved earlier we have
\begin{eqnarray*}
\Lambda( a_{\lambda} , b_{\lambda} )  < \Lambda( a_{\lambda}, 1- a_{\lambda} ) = \Hbin{a_{\lambda}}.
\end{eqnarray*}
The final statement in the Lemma is an elementary consequence of the definition of $\Lambda$.  \qed

\section{Formalization of the nonlinear code of subsection \ref{ss:nonlinear}  within Propositional Logic} \label{app:nasa_formalization}

To make the nonlinear code of subsection \ref{ss:nonlinear} more memorable, we crafted a problem statement involving a space mission where the spacecraft has six possible actions, many of which can be applied potentially simultaneously (but not sequentially). One action saves the mission, another results in a catastrophe, and the other four have no effect. Houston knows which action saves the mission and which action results in a catastrophe. What is the minimum number of bits that Houston needs to send to the spacecraft? The answer is 2 bits and the associated algorithm is implied by small code. 

It turns out that we can encode the 6-action scenario entirely with three Boolean variables, call the variables $\mathtt{b_1, b_2}$ and $\mathtt{b_3}$.  There are several ways to do this, but rather than encoding which action is the remedial action, let us instead encode which action is the \textit{catastrophic} action. We choose the following encoding:
\begin{eqnarray*}
    \textrm{$A_1$ is the catastrophic action}~(000)&:& \chi_1 = \mathtt{\neg b_1 \wedge \neg b_2 \wedge \neg b_3} \\
    \textrm{$A_2$ is the catastrophic action}~(001)&:& \chi_2 =\mathtt{\neg b_1 \wedge \neg b_2 \wedge b_3} \\
    \textrm{$A_3$ is the catastrophic action}~(010)&:& \chi_3 = \mathtt{\neg b_1 \wedge b_2 \wedge \neg b_3} \\
    \textrm{$A_4$ is the catastrophic action}~(011)&:& \chi_4 = \mathtt{\neg b_1 \wedge b_2 \wedge b_3} \\
    \textrm{$A_5$ is the catastrophic action}~(100)&:& \chi_5 = \mathtt{b_1 \wedge \neg b_2 \wedge \neg b_3} \\
    \textrm{$A_6$ is the catastrophic action}~(101)&:& \chi_6 = \mathtt{b_1 \wedge \neg b_2 \wedge b_3} \\
\end{eqnarray*}
The symbols $\chi_1,...,\chi_6$ are shorthand names for the respective propositional sentences, and not additional variables.  The three digit binary numbers in parentheses are mnemonics for how we are doing the encoding with the three Boolean variables.  

With this encoding in hand, the objective of the Sender can be thought of as giving the Receiver just enough information so that for the actual remedial action, $A_k$, the Receiver can prove that $A_k$ is \textit{not} the catastrophic action. As long as the Sender can convey this information then they can be assured that the Receiver will eventually take the needed remedial action per the agreed upon protocol. 

Consider the $4 \times 6$ partition matrix from subsection \ref{ss:nonlinear}, which we have copied in below,
\begin{center}
 \texttt{\\
0 0 0 1 1 1 \\
0 1 1 0 1 0 \\
1 0 1 1 0 0 \\
1 1 0 0 0 1 \\
}
\end{center}
where we think of the $i$th action as being associated with the $i$th column. In any given scenario faced by the astronauts, exactly one of the rows will apply. If, say, the 2nd row applies and there is a $1$ in cell $(2,i)$ the meaning is that action $A_i$ either is the remedial action or harmless, and so \textit{not} the catastrophic action. On the other hand, if cell $(2,i)$ contains a $0$, the meaning is that action $A_i$ is either the catastrophic action or harmless, and so not the remedial action. The important feature of this partition matrix is that for any distinct pair of indices $i,j$, where we can think of $A_i$ as being the remedial action and $A_j$ the catastrophic action, there is a row of this matrix such that the number in the $i$th entry is $1$ and the number in the $j$th entry is $0$. Hence the name ``partition matrix''.

Row 1 (the row consisting of $0~0~0~1~1~1$) in the partition matrix is then given by the Boolean expression, which in English reads ``One of the actions \{1, 2, 3\} is the catastrophic action and each the actions \{4, 5, 6\} is not the catastrophic action'':
\begin{eqnarray*}
\sigma_1 &=& \mathtt{((\neg b_1 \wedge \neg b_2 \wedge \neg b_3) \vee (\neg b_1 \wedge \neg b_2 \wedge  b_3) \vee (b_1 \wedge b_2 \wedge \neg b_3))~~~\wedge}\\
&&\mathtt{\neg(\neg b_1 \wedge b_2 \wedge b_3) \wedge \neg(b_1 \wedge \neg b_2 \wedge \neg b_3) \wedge \neg(b_1 \wedge \neg b_2 \wedge b_3)}.
\end{eqnarray*}
The other rows of the matrix are then analogously given by the Boolean expressions:
\begin{eqnarray*}
\sigma_2 &=& \mathtt{((\neg b_1 \wedge \neg b_2 \wedge \neg b_3) \vee (\neg b_1 \wedge b_2 \wedge b_3) \vee (\wedge \neg b_2 \wedge b_3))~~~\wedge}\\
&&\mathtt{\neg(\neg b_1 \wedge \neg b_2 \wedge  b_3) \wedge \neg(\neg b_1 \wedge b_2 \wedge \neg b_3)  \wedge \neg(b_1 \wedge \neg b_2 \wedge \neg b_3)}, \\
\sigma_3 &=& \mathtt{((\neg b_1 \wedge \neg b_2 \wedge  b_3) \vee (b_1 \wedge \neg b_2 \wedge \neg b_3) \vee (b_1 \wedge \neg b_2 \wedge b_3))~~~\wedge}\\
&&\mathtt{\neg(\neg b_1 \wedge \neg b_2 \wedge \neg b_3) \wedge \neg(\neg b_1 \wedge b_2 \wedge \neg b_3)  \wedge \neg(\neg b_1 \wedge b_2 \wedge b_3)}, \\
\sigma_4 &=& \mathtt{((\neg b_1 \wedge b_2 \wedge \neg b_3) \vee (\neg b_1 \wedge b_2 \wedge b_3) \vee (b_1 \wedge \neg b_2 \wedge \neg b))~~~\wedge}\\
&&\mathtt{\neg(\neg b_1 \wedge \neg b_2 \wedge \neg b_3) \wedge \neg(\neg b_1 \wedge \neg b_2 \wedge  b_3)  \wedge \neg(b_1 \wedge \neg b_2 \wedge b_3)}.
\end{eqnarray*}

The sender and receiver can then agree to transmit 2 bits   according to which expression $\sigma_i$ applies, for example $00 \mapsto \sigma_1, 01 \mapsto \sigma_2, 10 \mapsto \sigma_3, 11 \mapsto \sigma_4$.

\section{Algorithms Used for the Experiments} \label{app:experimental_algs}

\subsection{Generation of Kernels} \label{app:zero-sets}

Pseudocode for the heuristic to generate kernels consistent with the receiver's knowledge, the query, and the sender's knowledge is given in Algorithm \ref{alg:gen_zeroes}.
\begin{small}
\begin{algorithm} [ht]
\caption{\texttt{GENERATE\_KERNELS(NUM\_VARS}, $p_r, p_q, p_s$)} \label{alg:gen_zeroes}
\begin{algorithmic} 
\State \texttt{NUM\_VARS} \Comment{Given as input. Set to $10$ for all of our test cases.}
\State $p_r, p_q, p_s$ \Comment{Given as input.}\\
\State \texttt{sender\_kernel =} [$\mt{2}^\mt{NUM\_VARS}$]
\State \texttt{query\_kernel =} [$\mt{2}^\mt{NUM\_VARS}$]
\State \texttt{receiver\_kernel =} [$\mt{2}^\mt{NUM\_VARS}$]
\For{$\mt{i = 1}, \dots, \mt{2}^{\mt{NUM\_VARS}}$}
    \State $r$ = \texttt{random()}
    \If{$r \leq p_r$} 
        \State \texttt{receiver\_kernel[i] = True}
        \If{$r \leq p_q$}
            \State \texttt{query\_kernel[i] = True}
            \If{$r \leq p_s$}
                \State \texttt{sender\_kernel[i] = True}
            \EndIf
        \EndIf
    \EndIf 
\EndFor
\\
\State \Return \texttt{sender\_kernel, query\_kernel, receiver\_kernel}
\end{algorithmic}
\end{algorithm} 
\end{small}
The values $p_r, p_q, p_s$ refer respectively to the probability that a randomly generated truth value assignment is consistent with  the receiver's knowledge, the query, and the sender's knowledge.

\subsection{Generation of Generalized Decision Tree Sentences} \label{app:decision-trees}

Pseudocode for implementing the generalized decision tree algorithm, given a kernel, is given in Algorithm \ref{alg:gen_tnf}. We say that the resulting sentence is a Generalized Decision Tree, or GDT for short. The algorithm runs by recursively calling the method \texttt{GENERATE\_GDT\_FOR\_KERNEL()} with the kernel as an argument. The output is a Boolean expression in  Generalized Decision Tree form. We adopt the convention that the empty Boolean expression, \texttt{""}, is true for all variable assignments. The first step in the recursive routine is to check if there any true values (equivalently, any $\mt{1}$s) in the kernel. Since we maintain the kernel as an indicator array this is not quite the same thing as checking that the kernel is empty, though that is what is being done conceptually. If there are none then we have a contradiction so the simplest contradictory sentence is output. This condition only happens, as we shall see, at the highest level of the recursive stack.  The next step is to extract constants, in other words to extract any variables that appear only positively or only negatively in every satisfying truth value assignment in the kernel. Say two such variables are found and they are $\mt{X_i}$ and $\mt{X_j}$ with $\mt{X_i}$ appearing only positively and $\mt{X_j}$ appearing only negatively.  Then the kernel is reduced to remove the variables $\mt{X_i}$ and $\mt{X_j}$ and the output string ($\mt{gdt}$ in the pseudocode) is initialized to $\mt{X_i \wedge -X_j}$. Next, a check it made to see if either the number of remaining variables is zero (which would also mean that there are no true values in the kernel) or the number of true values is equal to $2$ to the power of the number of remaining variables. If this is the case, the function exits, just outputting the string of conjuncted constants, if any. (Note that this immediate return prevents the routine from ever being invoked with no zeroes except on initial invocation.) Otherwise, the function finds a variable, let us call it $\mt{X_b}$, that is most balanced in terms of its positive and negative occurrences among the true values associated with the kernel. With this variable identified, two further reduced kernels are computed, one consisting of the true values associated with all variables minus $\mt{X_b}$, when $\mt{X_b}$ is true, and one one consisting of the true values associated with all variables minus $\mt{X_b}$, when $\mt{X_b}$ is false. In the pseudocode, these reduced kernels are identified as $\mt{pos\_kernel}$ and $\mt{neg\_kernel}$ respectively. Finally, the psdeudocode distinguishes two case depending on whether $\mt{tnf}$ is non-empty (the truth set associated with the original kernel contained constants) or otherwise. If $\mt{tnf}$ is not empty then it is a conjunction of (one or more) variables or their negations. We take the conjunction of this with \begin{small}$\mt{((X_b \wedge GENERATE\_GDT\_FOR\_KERNEL(pos\_kernel)) \wedge (-X_b \wedge GENERATE\_GDT\_FOR\_KERNEL(neg\_kernel)))}$\end{small}. If $\mt{tnf}$ is empty than we do not need to conjunct in anything and we also don't need the outer parentheses.

We illustrate the recursive calling of $\mt{GENERATE\_GDT\_FOR\_KERNEL()}$ through a small example. Suppose we have $5$ Boolean variables that we  call $\mt{X_1,...,X_5}$, and the truth set associated with the kernel consists of the values $\{00100, 00101, 10101, 10111, 11001\}$. Thus, in the first of the true values (equivalently, satisfying truth value assignments), $00100$, $\mt{X_1 = X_2 = X_4 = X_5 = False}$ and $\mt{X_3 = True}$. In the outer call to $\mt{GENERATE\_GDT\_FOR\_KERNEL()}$, there are no constants, the number of true values is not $2^5$ so we proceed to finding a most balanced variable, in this case there is exactly one most balanced variable and it is $\mt{X_1}$. We then prepare the reduced positive kernel, which is $\mt{pos\_kernel =} \{0101, 0111, 1001\}$, and the reduced negative kernel which is $\mt{neg\_kernel =} \{0100, 0101\}$.

We thus set 
\begin{eqnarray}
    \mt{gdt} &=& \mt{(X_1 \wedge GENERATE\_GDT\_FOR\_KERNEL(pos\_kernel)) ~~\wedge} \notag \\ 
    && \mt{(-X_1 \wedge GENERATE\_GDT\_FOR\_KERNEL(neg\_kernel))}. \label{subexp}
\end{eqnarray} 
Now, in the call to $\mt{GENERATE\_GDT\_FOR\_KERNEL(pos\_kernel)}$ we see that for the reduced kernel $\mt{pos\_kernel}$ $= \{0101, 0111, 1001\}$, the last variable, which is named $\mt{X_5}$, appears only positively and is therefore constant.  We can therefore pull out this variable and reduce the kernel further to $\{010, 011, 100\}$. The number of true values that remain is not $2^3$ so we proceed to finding a most balanced variable, one such is the first of the remaining variables, which is named $\mt{X_2}$. We omit the remaining details, but the result is the following expression:
\begin{equation*}
    \mt{gdt = (X_1 \wedge X_5 \wedge ((X_2 \wedge -X_3 \wedge -X_4) \vee (-X_2 \wedge X_3))) \vee (-X_1 \wedge -X_2 \wedge X_3 \wedge -X_4)}.
\end{equation*}
It is worth noting that the subexpression for $\mt{GENERATE\_GDT\_FOR\_KERNEL(neg\_kernel)}$ in (\ref{subexp}) is especially simple since for the truth set, $\mt{neg\_kernel =} \{0100, 0101\}$, the first three variables, which have original names $\mt{X_2, X_3}$ and $\mt{X_4}$, are all constant and can be pulled out. We are then left with the (much) reduced kernel $\{0,1\}$, and here, since $\mt{NUM\_VARS = 1}$ we do have that $\mt{num\_zeroes\_to\_match == 2^{NUM\_VARS}}$ and so can omit the last variable ($\mt{X_5}$).

\begin{small}
\begin{algorithm} [ht]
\caption{\texttt{GENERATE\_GDT\_FOR\_KERNEL(kernel)}} \label{alg:gen_tnf}
\begin{algorithmic} 
\State $\mt{kernel[2^{NUM\_VAR}}]$ \Comment{Given as input.}
\State $\mt{NUM\_VARS}$ \Comment{Inferred from the size of the input array. Set to $10$ for all test cases.}
\State $\mt{num\_trues = COUNT\_TRUES(zeroes\_to\_match)}$ \Comment{The number of 1s in the input array.}

\\ 

\If{$\mt{num\_trues == 0}$} \Comment{Can only happen at the highest level;} 
    \State \Return $\textrm{"}\mt{X1} \wedge \mt{-X1}\textrm{"}$\Comment{Return simple contradiction}
\EndIf
\\
\State $\mt{constants = EXTRACT\_CONSTANTS(kernel)}$
\State $\mt{num\_trues = REDUCE\_KERNEL\_MOD\_CONSTANTS(kernel, constants)}$
\State $\mt{NUM\_VARS = NUM\_VARS - len(constants)}$
\State $\mt{gdt = GET\_CONJUNCTION\_FOR\_CONSTANTS(constants)}$
\If{$\mt{NUM\_VARS == 0~\OR~num\_trues == 2^{NUM\_VARS}}$}
\State \Return \texttt{gdt}
\EndIf
\\
\State $\mt{var = FIND\_MOST\_BALANCED\_VAR(kernel)}$
\State $\mt{pos\_kernel = REDUCE\_KERNEL\_MOD\_POS\_VARIABLE(kernel, var)}$
\State $\mt{neg\_kernel = REDUCE\_KERNEL\_MOD\_NEG\_VARIABLE(kernel, var)}$
\\
\If{$\mt{gdt \neq \textrm{""}}$}
\State $\mt{gdt = gdt + \textrm{"}\wedge(( \textrm{"} + var + \textrm{"}\wedge\textrm{"} + GENERATE\_GDT\_FOR\_KERNEL(pos\_kernel) + \textrm{"})~\vee\textrm{"}}$
\State $~~~~~~~~~~~~~~~~~+~\textrm{"}(- \textrm{"} + \mt{var} + \textrm{"}\wedge\textrm{"} + \mt{GENERATE\_GDT\_FOR\_KERNEL(neg\_kernel)} + \textrm{"}))\textrm{"}$
\Else
\State $\mt{gdt =  \textrm{"}( \textrm{"} + var + \textrm{"}\wedge\textrm{"} + GENERATE\_GDT\_FOR\_KERNEL(pos\_kernel) + \textrm{"})~\vee\textrm{"}}$
\State $\mt{~~~~~~~~~~~~~~~~~+~\textrm{"}(- \textrm{"} + var + \textrm{"}\wedge\textrm{"} + GENERATE\_GDT\_FOR\_KERNEL(neg\_kernel) + \textrm{"})\textrm{"}}$
\EndIf
\\
\State \Return \texttt{gdt}
\end{algorithmic}
\end{algorithm} 
\end{small}

\end{document}